\documentclass[12pt]{article}

\usepackage{amsmath,amssymb}
\usepackage{graphicx}
\usepackage{amscd}
\usepackage{theorem}

\newcommand{\R}{{\mathbb{R}}}
\newcommand{\Z}{{\mathbb{Z}}}
\newcommand{\C}{{\mathbb{C}}}

\newcommand{\bea}{\begin{eqnarray}}
\newcommand{\eea}{\end{eqnarray}}

\newcommand{\nn}{\nonumber}

\newcommand{\bp}{\begin{pmatrix}}
\newcommand{\ep}{\end{pmatrix}}

\newcommand{\bps}{\begin{smallmatrix}}
\newcommand{\eps}{\end{smallmatrix}}

\newcommand{\ti}{\tilde}
\newcommand{\la}{\langle}
\newcommand{\ra}{\rangle}

\def \cH{{\cal H}}
\def \V{{\cal V}}

\def \qed{\hfill $\blacksquare$}

\def \raw{\rightarrow}

\def \lgraw{\longrightarrow}

\def \Lglraw{\Longleftrightarrow}

\def \Id{\mathrm{Id}}

\def \gf{\mathit{gf}}
\def \rank{\mathrm{rank}}

\def \dim{\mathrm{dim}}

\def \mes{{\cal D}}

\def \fb{{\bf f}}

\def \half{\frac{1}{2}}
\def \ov#1{\frac{1}{#1}}

\def \fpartial#1{\frac{\partial}{\partial {#1}}}
\def \fpart#1#2{\frac{\partial #1}{\partial #2}}
\def \flpartial#1{\frac{\overleftarrow{\partial}}{\partial #1}}
\def \frpartial#1{\frac{\overrightarrow{\partial}}{\partial #1}}
\def \flpart#1#2{\frac{#1 \overleftarrow{\partial}}{\partial #2}}
\def \frpart#1#2{\frac{\overrightarrow{\partial} #1}{\partial #2}}
\def \fd#1{\frac{d}{d{#1}}}

\def \g{{\frak g}}
\def \l{{\frak l}}
\def \m{{\frak m}}

\def \cF{{\cal F}}

\def \cO{{\cal O}}
\def \cM{{\cal M}}

\def \l({\left(}
\def \r){\right)}

\def \0{{\bf 0}}
\def \1{{\bf 1}}

\def \eb{{\bf e}}
\def \ebb{{\bar {\bf e}}}
\def \m{{\frak m}}
\def \tri{\triangle}
\def \phib{{\bar \phi}}

\def \-{-\hspace*{-0.2cm}-}
\def \ie{{\it i.e.}\ }
\def \ib{{\bar i}}
\def \jb{{\bar j}}
\def \Qt{{\tilde Q}}

\def \mm{\supset\hspace*{-0.2cm}-}
\def \D{-\hspace*{-0.05cm}\framebox{$\delta$}\hspace*{-0.05cm}-}
\def \f{-\hspace*{-0.05cm}\framebox{$\fb$}\hspace*{-0.05cm}-}
\def \Q{-\hspace*{-0.05cm}\framebox{$\m$}\hspace*{-0.05cm}-}
\def \F{-\hspace*{-0.05cm}\framebox{$\cF$}\hspace*{-0.05cm}-}

\def \mmp{\supset\hspace*{-0.2cm}-\hspace*{-0.08cm}\circ}
\def \mms{\circ\hspace*{-0.08cm}-\hspace*{-0.2cm}\subset}

{\theorembodyfont{\rmfamily}
 \newtheorem{defn}{Definition}[section]
 \newtheorem{thm}{Theorem}[section]
 \newtheorem{lem}{Lemma}[section]
 
 \newtheorem{prop}{Proposition}[section]
 
 \newtheorem{rem}{Remark}[section]
}

\numberwithin{equation}{section}

\begin{document}

\begin{titlepage}
\thispagestyle{empty}
\begin{flushleft}
\hfill hep-th/0112228\\
UTMS 2001-34\hfill December, 2001 \\
\end{flushleft}

\vskip 1.5 cm

\begin{center}
\noindent{\Large \textbf{Homotopy Algebra Morphism and 
}}\\

\vspace*{0.5cm}

\noindent{\Large \textbf{Geometry of Classical String Field Theories
 }}\\
\renewcommand{\thefootnote}{\fnsymbol{footnote}}

\vskip 2cm
{\large 
Hiroshige Kajiura 
\footnote{e-mail address: kuzzy@ms.u-tokyo.ac.jp}\\

\noindent{ \bigskip }\\

\it
Graduate School of Mathematical Sciences, University of Tokyo \\
Komaba 3-8-1, Meguro-ku, Tokyo 153-8914, Japan\\
\noindent{\smallskip  }\\
}

\bigskip
\end{center}
\begin{abstract}

We discuss general properties of classical string field theories 
with symmetric vertices in the context of deformation theory. 
For a given conformal background there are many string field theories 
corresponding to different decomposition of moduli space of Riemann 
surfaces. 
It is shown that any classical open string field 
theories on a fixed conformal background are $A_\infty$-quasi-isomorphic 
to each other. 
This indicates that they have isomorphic moduli space of classical 
solutions. 
The minimal model theorem in $A_\infty$-algebras plays a key role 
in these results. Its natural and geometric realization on 
formal supermanifolds is also given. 
The same results hold for classical closed string field theories, 
whose algebraic structure is governed by $L_\infty$-algebras.

\end{abstract}
\vfill

\end{titlepage}
\vfill
\setcounter{footnote}{0}
\renewcommand{\thefootnote}{\arabic{footnote}}
\newpage

\tableofcontents

\section{Introduction and Summary}

The present paper is motivated to make clear the complicated structures 
of string field theories (SFTs) 
in terms of homotopy algebras. 
We assume the existence of well-defined SFTs on a conformal background, 
and discuss the general properties which they should possess. 
We show that any classical open SFTs, which 
are constructed so as to reproduce the open string correlation functions 
on-shell, are quasi-isomorphic to each other. 
Moreover, when such SFT actions are given, 
$A_\infty$-quasi-isomorphisms between them are constructed. 
These results guarantee that there is one-to-one correspondence of 
the equations of motions corresponding to marginal deformation between 
such family of SFTs on the same conformal background. 
This gives an answer about the issue for the relation between SFTs with 
different decomposition of moduli spaces. 
These arguments also give the minimal model theorem in deformation theory 
a geometric insight. 
These arguments are applicable for classical closed SFT similarly, 
and it can be shown that 
all such consistent classical closed SFTs are $L_\infty$-quasi-isomorphic 
to each other.

SFT has been investigated as a candidate for 
string theory which describes nonperturbative effects. 
SFT gives one of the way which extends 
on-shell two dimensional string theory to off-shell theories. 
Many Lorentz covariant SFTs have been constructed. 
The covariant open or closed SFT with light cone type vertices
(HIKKO's SFT)\cite{HIKKO}, 
a very simple open SFT which consists of only a three-point vertex 
(Witten's open SFT or cubic SFT)\cite{W1}, and so on. 
Witten's SFT can be treated in the context of 
Batalin-Vilkovisky (BV)-formalism\cite{Th}. 
HIKKO's closed SFT is also 
extended to quantum SFT by 
employing the quantum BV-master equation\cite{H}. 
The quantum master equation is moreover applied to construct quantum closed 
SFT with symmetric vertices\cite{Z1}. 
Though this theory has infinite sort of 
vertices of higher punctures and higher genus, it has 
a very beautiful algebraic structure. For instance for the classical part, 
the set of the tree vertices has the structure of 
a $L_\infty$-algebra. 
Open-closed SFT is also considered in this direction \cite{Z2}. 
The open-closed symmetric vertices have relations 
from quantum BV-master equation, where `symmetric' means cyclic for 
open string punctures and commutative for closed string punctures. 
Several subalgebras of subsets of the vertices can be 
considered : disk (tree) vertices with punctures only on the boundary, 
which has the structure of an $A_\infty$-algebra, sphere vertices with 
punctures (an $L_\infty$-algebra), 
disk vertices with both open and closed vertex insertions 
(though the algebraic structure of which does not have a particular name), 
all vertices with no boundaries (the theory of which is the above 
quantum closed SFT\cite{Z1}), and so on. 
Recently a classical open SFT, 
which possesses the $A_\infty$-structure, is constructed explicitly\cite{N} 
by deforming the cubic SFT\cite{W1}.

All the above SFTs satisfy the (classical or quantum) BV-master equation. 
In constructing a SFT action, any types of vertices, 
which are written by the powers of string fields and their coefficients, 
are considered and the master equation are used in order to decide the 
coefficients
\footnote{The use of BV-formalism is different from the original 
use of BV-formalism for gauge theories (see subsection \ref{ssec:42}), but 
the similar treatment is done for instance for 
topological theories in superfield formalisms \cite{AKSZ,I,Park,CF2,Ba}. }. 
Moreover, in the two-dimensional world sheet picture, 
the fact that the SFT action satisfies the BV-master equation corresponds 
to that the moduli spaces of Riemann surfaces are 
single-covered\cite{SZ}. 
Hereafter in this paper, 
we treat only (bosonic) SFTs with the symmetric 
vertices and those algebraic structures 
are discussed.

There have been mainly two issues for the SFTs constructed as above : 
i) the realization about the relation between SFTs constructed 
by different decomposition of moduli space of Riemann surfaces 
on a fixed conformal background ;\ and 
ii) the background independence \cite{Sbg1,Sbg2,Sbg3}. 
In order to assert that the SFT gives an nonperturbative 
definition of string theory the second issue is necessary. 
For the first issue, it might be believed that the SFTs derived 
with different decomposition of the moduli space are physically equivalent 
in some sense. Indeed in \cite{HZ} for quantum closed SFT it is shown that 
any infinitesimal variation of the decomposition leads 
infinitesimal field redefinition preserving the SFT actions and 
the BV-symplectic structures. 
Analogous consequence is expected for open SFTs. 
In fact in \cite{N} an one parameter family of the classical open SFT 
is discussed and the infinitesimal field redefinition preserving the action 
is found. 
However the relation between SFTs which differ finitely in the 
decomposition of the moduli space has not ever been discussed.

Roughly speaking, 
these issues are equivalent to that of looking for the map between SFTs 
preserving the BV-symplectic structures. 
Restricting the arguments to the classical theory, as were mentioned above, 
the classical closed SFT satisfying the classical BV-master equation 
has the structure of a $L_\infty$-algebra, and similarly 
the classical open SFT has the structure of an $A_\infty$-algebra. 
As will be explained in section \ref{sec:4}, 
the algebraic structure of classical open (resp. closed) SFT 
is an $A_\infty$-algebra (resp. $L_\infty$-algebra) which possesses 
the graded cyclic (resp. commutative) symmetry through 
the BV-symplectic structure. 
$A_\infty$-algebra has appeared for the first time in \cite{Sta1}. 
It is a deformation of an associative graded algebra with differential 
(DGA), and consists of a differential $Q$, a product $\bullet$, and higher 
products. $L_\infty$-algebra is its graded commutative-symmetrized 
version. It is given as a deformation of a differential graded Lie algebra 
(DGLA), and consists of a differential $Q$, a Lie bracket $[\ ,\ ]$, and 
higher brackets\cite{LS}. 
In SFT, $Q$ is the BRST-operator \cite{KO}, the product 
$\bullet$ (resp. the Lie bracket $[\ ,\ ]$ )corresponds to the 
trivalent vertex of classical open SFT (resp. classical closed SFT), and 
higher products (resp. higher brackets) correspond to higher 
vertices of classical open (resp. closed) SFT.

In SFT or others, one of the important advantage to finding out the 
$A_\infty$ or $L_\infty$-structures is presumably that they have $A_\infty$ 
or $L_\infty$-morphisms which transform 
an $A_\infty$ or $L_\infty$-algebra to another one. 
For example, the existence of the deformation quantization\cite{BFFLS} 
on general Poisson manifolds is proved as a consequence of constructing a 
$L_\infty$-morphism between certain two $L_\infty$-algebras
\footnote{DGLAs are $L_\infty$-algebras whose 
higher brackets are set to be zero. In fact, 
the two $L_\infty$-algebras considered in \cite{K} are both DGLAs. 
The reason that these DGLAs have to be treated in the context of 
$L_\infty$-algebras is that a $L_\infty$-morphism, which is a 
{\it nonlinear map} between these two DGLAs 
which preserves the solutions of the Maurer-Cartan equations, is needed. 
}
\cite{K}. 
Here when a Poisson structure on a manifold $M$ is given, 
the deformation quantization on $M$ 
means that the associative product is constructed 
as the power expansion of a 
deformation parameter $\hbar$, the leading term of which is the usual 
commutative product of functions on $M$, and next term of which is the 
Poisson bracket. 
The constructed $L_\infty$-morphism induces the isomorphism between the 
cohomologies of these two algebras with respect to the differentials $Q$. 
Such morphism is called a {\it quasi-isomorphism}. 
The fact that the above two algebras are quasi-isomorphic to each other 
is called {\it formality}\cite{K}, which has been conjectured originally 
in \cite{K2}. 
Moreover when an $A_\infty$ or $L_\infty$-algebra is given, 
one can define its Maurer-Cartan equation, the solution space of which 
gives moduli space in the context of deformation theory. 
In addition, an $A_\infty$ or $L_\infty$-morphism preserves 
the solution space of the Maurer-Cartan equations. 
In the above case of the deformation quantization problem\cite{K}, 
the solution space of the Maurer-Cartan equation for one side 
of the two $L_\infty$-algebras gives the space of the Poisson structures, 
and that for another side gives the space of the structures of 
the associative products. 
Therefore constructing the $L_\infty$-morphism has led 
the existence of the deformation quantization
\footnote{In \cite{CF}, the $L_\infty$-morphism defined in \cite{K} 
is explicitly 
derived as a BV-quantization of Poisson-sigma model on a disk. 
The condition that the morphism is actually a $L_\infty$-morphism is 
identified with the Ward identity in BV-formalism. However the relation 
between the $L_\infty$-algebra and BV-formalism is different from that 
in SFT. In the former case 
the BV-bracket corresponds to a Lie bracket in the $L_\infty$-algebra on one 
side. }. 

For classical SFT, the Maurer-Cartan equation is nothing but 
the equation of motion, and the $A_\infty$ or $L_\infty$-morphisms, which, 
by definition, preserve the Maurer-Cartan equation, correspond to the 
field transformation. 
Thus the problem of 
considering the family of SFTs satisfying the classical BV-formalism 
is translated to that of considering the family of $A_\infty$ or 
$L_\infty$-algebras (with the BV-symplectic structure), and 
one can employ the $A_\infty$ or $L_\infty$-morphisms in order to 
realize the relation between two different SFTs in the family. 
In such reason, we want to find $A_\infty$ or $L_\infty$-morphisms 
in various situations in SFT. 
However unfortunately, 
in general when any two $A_\infty$ or $L_\infty$-algebras are 
given, the canonical way of constructing the morphism between them 
does not exist, 
though in the deformation quantization problem, the $L_\infty$-morphism 
is exactly found (from the insight of the two-dimensional topological 
field theory). 

Recently there is a development at this point. 
In \cite{K2,K} the following fact is mentioned : 
when an $A_\infty$ (resp. $L_\infty$)-algebra $\cH$ with nonvanishing $Q$ is 
given, by restricting the algebra to the kernel (or the cohomology class) 
of $Q$, the vector space $\cH^p$ (with $Q=0$) is obtained, 
and then there exists an $A_\infty$ (resp. $L_\infty$)-structure on $\cH^p$ 
(with $Q=0$) 
which is quasi-isomorphic to the original $A_\infty$ ($L_\infty$)-algebra 
$\cH$. This is called {\it minimal model theorem} in \cite{K2,K}. 
Moreover in \cite{KS}
the canonical $A_\infty$-structure 
on $\cH^p$ and a canonical $A_\infty$-quasi-isomorphism 
from $\cH^p$ to $\cH$ are constructed explicitly using Feynman graphs. 
The same argument holds for $L_\infty$-algebra. 
In \cite{KS} this is applied to the homological mirror conjecture, 
which states that 
the derived category of $A_\infty$-category \cite{Fukaya} in $A$-model 
and the derived category of coherent sheaves in $B$-model 
are equivalent\cite{mirror}. 
Recently, for example, the minimal theorem is applied in \cite{Laz,Tom,LazR} 
to transform the topological Chern-Simons SFT action \cite{W2} 
to so-called D-brane superpotential. 

However the minimal model theorem is very important not only for 
topological theories but for usual SFT, where 
$Q$ is the BRST-operator and $\cH^p$ is the physical state space. 
The fact that an $A_\infty$-structure is constructed on $\cH^p$ 
means that the two-dimensional string theory has the structure of 
an $A_\infty$-algebra. 
This statement is essentially already known. 
In \cite{WZ} it is described that the tree closed string theory 
has the structure 
of the $L_\infty$-algebra (and extended this result to quantum case 
\cite{V}). 
The fact holds similar for open string theory and 
which implies that the $A_\infty$-structure of a classical open SFT 
on a conformal background is $A_\infty$-quasi-isomorphic to the 
$A_\infty$-structure of the two-dimensional theory. 

\vspace*{0.2cm}

Using the above argument, we get the results stated at the beginning. 
We will explain in the classical {\it open} SFT case 
in the following two reason. 
First, open SFT has the cyclic symmetry of its vertices and closed 
SFT has the graded commutative symmetry, which includes the 
cyclic symmetry but is much larger symmetry than it. Therefore essentially 
we can get the algebraic structure of closed SFT by graded commutative-
symmetrizing the cyclic open string vertices. Second, 
since $A_\infty$-morphisms transform the classical solutions of classical 
open SFT to those of another classical open SFT, it may be applicable to 
the problem of tachyon condensation\cite{tac} in SFT\cite{SZn,Oh}, 
though directly can not as will be commented in Discussions. 

\vspace*{0.2cm}

This paper is organized as follows. 

In section \ref{sec:2}, the definitions and some known facts about 
$A_\infty$-algebras are summarized. 

In section \ref{sec:3}, the way of constructing consistent SFT which 
satisfies classical BV-master equation is reviewed, because the 
main idea of them is essential for the later arguments. 
We give only the formal construction of SFT, 
and its concrete realization is omitted. 
For more details, see for example \cite{Z1,N}.

In section \ref{sec:4}, the algebraic structures of SFT constructed in 
section \ref{sec:3} are discussed. 
In subsection \ref{ssec:41} it is clarified that the algebraic 
structure of the SFTs, which has 
cyclic vertices and satisfies classical BV-master equation, 
is an $A_\infty$-algebra with cyclic symmetry through the 
BV-symplectic structure. 
Moreover, we describe that 
the gauge transformation for the $A_\infty$-structure is the gauge 
transformation of BV-formalism\cite{BV1,BV2,HT} 
in subsection \ref{ssec:42}. 
Finally in subsection \ref{ssec:43} the notations and definitions 
are summarized for later arguments 
including those already used in section \ref{sec:3} and \ref{sec:4}. 
The result in this section might also be known essentially, but 
there might not be literatures which gives the similar explanations.  

In section \ref{sec:5}, 
the minimal model theorem\cite{KS} is introduced and 
applied to classical open SFTs. 
In order to understand its meaning, 
we demonstrate that it arises from the issue of finding 
the solutions of the Maurer-Cartan equation for an $A_\infty$-algebra. 
For the original $A_\infty$-algebra, another canonical $A_\infty$-algebra 
and an $A_\infty$-quasi-isomorphism between them are derived naturally, and 
they are expressed using some Feynman diagrams. 
Moreover we give a proof of the minimal model theorem in this direction. 
Its geometrical realization on formal noncommutative supermanifolds is 
also given. 
In subsection \ref{ssec:5gf}, it is clarified that the Feynman graph 
defined in the previous subsection is actually the Feynman graph in SFT. 
In order to see the propagator explicitly, 
we discuss mainly the case of Siegel gauge and 
clarify that the usual propagator in SFT can be applied to the procedure 
in the previous subsection. 
In subsection \ref{ssec:53}, it is shown that the canonical 
$A_\infty$-algebra generated graphically 
gives the correlation functions of open strings (Lem.\ref{lem:main}). 
Moreover in subsection \ref{ssec:main} we show that 
all SFTs on a fixed conformal 
field theory are quasi-isomorphic to each other(Thm.\ref{thm:main}). 
This immediately follows from the fact that 
any SFT constructed consistently as will be explained 
in section \ref{sec:3} 
coincides with the open string correlation functions on-shell. 
The quasi-isomorphism is described in terms of Feynman diagram in SFT, 
and it gives a finite field transformation on certain subspace. 
Furthermore, a boundary SFT like action which is isomorphic to the original 
SFT action is proposed.

Since all the above arguments are very formal, 
in section \ref{sec:6} we apply them to the classical open SFT 
explicitly constructed in \cite{N}. 
In \cite{N} an one parameter family of the classical open SFTs 
is discussed from the viewpoint of the renormalization group\cite{P}, 
which states that the variation of the fields and the coefficients of the 
vertices with respect to the renormalization scale cancel each other and 
the action is invariant. 
The flow of the fields is then derived, which 
gives an infinitesimal field redefinition between 
the SFTs with different renormalization scales. 
After reviewing a part of the arguments in \cite{N}, 
we show that the infinitesimal field redefinition gives an 
$A_\infty$-isomorphism on the Siegel gauge. 
Moreover it is observed that 
the infinitesimal version of the finite $A_\infty$-quasi-isomorphism 
discussed in the above section coincides with that given in \cite{N} 
on the subspace. 
Finally various viewpoints in this paper is summarized 
in this explicit model.

In Conclusions and Discussions, the issue of 
the background independence\cite{SZ1,SZ2} is rearranged. 
The tachyonic solution in cubic SFT\cite{SZn} is also argued from the 
viewpoint of this paper. 
The relation to the boundary SFT\cite{W3} and 
the application to other SFTs are also commented. 

In Appendix \ref{sec:A}, the precise meaning of taking the dual of the 
$A_\infty$-algebras is presented. 
The correspondence between an $A_\infty$-algebra and its dual means the 
correspondence between a SFT in the operator language and its field theory 
representation, and 
it is used implicitly in the body of this paper in order to simplify 
some explanations.  
In Appendix \ref{ssec:A1}, the dual is defined with an inner product, 
and its graphical explanation is also described. 
The dual picture is used in many literatures, but there are 
less literatures where the explicit relation between them is presented. 
In Appendix \ref{ssec:A2}, 
$A_\infty$-algebras are realized geometrically in the dual picture. 
It will be seen that those are described in terms of noncommutative 
formal supermanifolds. 

In Appendix \ref{sec:B} some detail calculations for string 
vertices are presented.

%%%%%%%%%%%%%%%%%%%%%%%%%%%%%%%%%%%%%%%%%%%%%%%%%%%%%%%%
%
%%%%%%%%%%%%%%%%%%%%%%%%%%%%%%%%%%%%%%%%%%%%%%%%%%%%%%%%

 \section{$A_\infty$-algebra}
\label{sec:2}

It is known that classical open (resp. closed) SFTs have the structure 
of $A_\infty$-algebras\cite{Z2,GZ,N} (resp. $L_\infty$-algebras\cite{Z1}). 
Here summarizes some basic facts about 
$A_\infty$-algebras ((strong) homotopy associative algebras). 
The facts are applicable for 
$L_\infty$-algebras ((strong) homotopy Lie algebras).

$A_\infty$-algebras are defined in terms of coalgebras. 
As will be defined below, for $\cH$ a graded vector space, 
we consider $C(\cH):=\oplus_{k\geq 1}{\cH^{\otimes k}}$ 
as a coalgebra. 
In the terminology of SFT, $\cH$ is the Hilbert space of string states and 
the degree (grading) of $\cH$ is related to the ghost number. 
Though coalgebras may be unfamiliar, it seems natural to the many 
body system (of strings), and it is useful to control 
$A_\infty$- (or $L_\infty$-)algebras formally very simple. 
Intuitively, or geometrically, the dual picture of coalgebras is suitable 
to realize them. 
For $\Phi\in\cH$ a string field of degree zero, splitting $|\Phi\rangle$ 
as $|\Phi\rangle=|\eb_i\rangle\phi^i$ where $\{|\eb_i\rangle\}$ are 
the basis of the string state $\cH$ and 
$\{\phi^i\}$ are the corresponding supercoordinate
\footnote{Here `super' means `graded', and `graded' means having degree. }
, the `dual picture' means the picture on the supercoordinates. 
The supercoordinates are the string fields in SFT. 
Later we define a `coproduct' on a coalgebra. The `coproduct' is natural 
structure for field theory because, 
in the dual picture, it is equivalent to that 
the fields $\{\phi^i\}$ possesses an associative product as an 
algebra.

 \subsection{Coalgebra, coderivation, and cohomomorphism}
\label{ssec:21}

Since 
$A_\infty$- (and $L_\infty$-) algebras are coalgebras with some additional 
structures, 
here introduce it. 
\begin{defn}[coalgebra, coassociativity]
Let $C$ be (generally infinite dimensional) graded vector space.  
When a {\it coproduct} 
$\triangle: C\lgraw C\otimes C$ is defined on $C$ and 
it is {\it coassociative}, {\it i.e.}
\[
 (\triangle\otimes{\bf 1})\triangle=({\bf 1}\otimes\triangle)\triangle
\]
then $C$ is called a {\it coalgebra}. 
\end{defn}
\begin{defn}[coderivation]
A linear operator $\m: C\raw C$ raising the degree of $C$ by one is 
called {\it coderivation} when   
\[
 \triangle \m= (\m\otimes{\bf 1})\triangle+({\bf 1}\otimes \m)\triangle
\]
is satisfied. Here, for $x, y\in C$, 
the sign is defined 
$({\bf 1}\otimes \m)(x\otimes y)=(-1)^{x}(x\otimes\m(y))$ 
through this operation where the $x$ on $(-1)$ denotes the degree of 
$x$. 
\label{defn:coder}
\end{defn}
\begin{defn}[cohomomorphism]
Given two coalgebras $C$ and $C'$, an {\it cohomomorphism 
(coalgebra homomorphism)} $\cF$ from $C$ to $C'$ is a 
map of degree zero which satisfies the condition
\begin{equation}
 \tri\cF=(\cF\otimes\cF)\tri\ .
\end{equation}
\end{defn}
\begin{rem}
Coassociativity of $\tri$, the condition of coderivations and 
cohomomorphisms are equal 
to that the following diagrams commute
\begin{equation*}
\begin{CD}
 C @>{\triangle}>> C\otimes C\\
 @V{\triangle}VV   @V{\triangle\otimes{\bf 1}}VV\\
 C\otimes C @>{{\bf 1}\otimes\triangle}>> C\otimes C\otimes C
\end{CD}\ ,\quad 
\begin{CD}
 C @>\m>> C\\
 @V{\triangle}VV   @V{\triangle}VV\\
 C\otimes C @>{{\bf 1}\otimes \m+\m\otimes{\bf 1}}>> C\otimes C
\end{CD}\ ,\quad
\begin{CD}
 C @>{\cF}>> C'\\
 @V{\tri}VV   @V{\tri}VV\\
 C\otimes C @>{\cF\otimes\cF}>> C'\otimes C'
\end{CD} \ .
\end{equation*}
If the orientation of these map are reversed and the coproduct is 
replaced by a product, then the coassociativity, the coderivation, and the 
cohomomorphism take place to associativity, a derivation, and 
a homomorphism of the corresponding algebra, respectively. 

Reversing the orientation of the maps corresponds to taking the dual of 
the coalgebra. The precise meaning of the dual in the present 
paper is given in Appendix \ref{ssec:A1}. 
 \label{rem:dual}
\end{rem}
Here, for any graded vector space $\cH$, on can consider its tensor coalgebra
\[
 C(\cH)=\oplus_{k\geq 1}{\cH^{\otimes k}}
\] 
as a coalgebra. 
Note that $C(\cH)$ 
does not contain the summand $\cH^{\otimes 0}=\C$. In particular, 
it does not have a counit. 

For this coalgebra, 
the coassociative coproduct on $C(\cH)$ is uniquely determined as 
\begin{equation}
 \tri(\eb_1\cdots\eb_n)=\sum_{k=1}^{n-1}
  (\eb_1\cdots\eb_k)\otimes(\eb_{k+1}\cdots\eb_n)\ .
\end{equation}
The form of the coderivation corresponding to this coproduct is 
also given as follows : let $\{m_k\}_{k\ge 1}$ be the set of multilinear 
maps of degree one 
\begin{equation}
 \begin{array}{cccc}
  m_k\ :&\cH^{\otimes k}&\lgraw& \cH\\
               &\eb_1\otimes\cdots\otimes\eb_k&\mapsto & 
                              m_k(\eb_1\cdots\eb_k)
 \end{array}\quad ,\qquad 
\end{equation}
and define 
 \begin{equation*}
  \m_k(\eb_1\cdots\eb_n)=\sum_{p=1}^{n-k}
 (-1)^{\eb_1+\cdots+\eb_{p-1}}\eb_1\cdots\eb_{p-1}
 m_k(\eb_p\cdots\eb_{p+k-1})
 \eb_{p+k}\cdots\eb_n \ ,\quad \eb_i\in\cH\ .
 \end{equation*}
Here $\eb_1+\cdots+\eb_{p-1}$ on $(-1)$ denotes the degree of 
$\eb_1\cdots\eb_{p-1}$. The sign factor appears when $m_k$, which 
has degree one, passes through the $\eb_1\cdots\eb_{p-1}$. 
Then summing up this $\m_k$ for $k\ge 1$, 
\begin{equation}
 \m=\m_1+\m_2+\m_3+\cdots\ ,
\end{equation}
and this $\m$ is the coderivative. 
The coderivative on the coalgebra $C(\cH)$ is always written in this 
form. 

Moreover, the form of a cohomomorphism $\cF :C(\cH)\raw C(\cH')$ 
is determined by a collection of degree zero multilinear maps 
$f_n: \cH^{\otimes n}\rightarrow \cH' (n\ge 1)$ 
which are homogeneous of degree zero as the following form
\begin{equation}
 \cF(\eb_1\cdots\eb_n)=\sum_{1\leq k_1<k_2\cdots <k_i=n}
f_{k_1}(\eb_1\cdots \eb_{k_1})\otimes 
f_{k_2-k_1}(\eb_{k_1+1}\cdots \eb_{k_2})\otimes\cdots\otimes f_{n-k_{i-1}}
(\eb_{k_{i-1}+1}\cdots \eb_n)\ ,
 \label{cohom}
\end{equation}
where each $f(\cdots)$ belongs to $\cH'$.

 \subsection{$A_\infty$-algebra and $A_\infty$-morphism}
\label{ssec:22}

\begin{defn}[$A_\infty$-algebra] 
Let $\cH$ be a graded vector space and 
$C(\cH)=\oplus_{k\geq 1}{\cH^{\otimes k}}$ be its tensor coalgebra. 
An $A_\infty$-algebra is a coalgebra $C(\cH)$ with a coderivation $\m$ which 
satisfies 
\[
 (\m)^2=0\ .
\]
\end{defn}
If we act $(\m)^2=(\m_1+\m_2+\cdots)^2$ on $\eb_1\cdots\eb_n\in C(\cH)$, 
its image belongs to $\cH^{\otimes 1}\oplus\cdots\oplus\cH^{\otimes n}$, and 
the condition that the $\cH$ part of the image equal zero is 
sufficient to the condition $(\m)^2=0$
\footnote{in the same reason that the differential $d$ on differential 
forms or BRST-operator $\delta$ on polynomials of fields 
and ghosts (and antifields) is nilpotent. }. 
The equation becomes 
\begin{equation}
\sum_{\substack{k+l=n+1\\j=0,\cdots,k-1}}
{(-1)^{\eb_1+\cdots+\eb_j}
 m_k(\eb_1,\cdots,\eb_j,m_l(\eb_{j+1},\cdots,\eb_{j+l}),
 \eb_{j+l+1},\cdots,\eb_n)}=0
 \label{ainf}
\end{equation}
for $n\geq 1$,  and $\eb_i$ on $(-1)$  
denotes the degree of $\eb_i$.

The first three constraints in eq.(\ref{ainf}) read:
\bea
\label{a3}
& &m_1^2=0~~\nn\ ,\\
& &m_1(m_2(\eb_1,\eb_2))+m_2(m_1(\eb_1),\eb_2)
+(-1)^{\eb_1}m_2(\eb_1,m_1(\eb_2))=0~~\ ,\\
& & m_2(\eb_1, m_2(\eb_2, \eb_3))+m_2(m_2(\eb_1, \eb_2), \eb_3)\nn\\
& &~~~~~~~
+m_1(m_3(\eb_1,\eb_2,\eb_3))+
m_3(m_1(\eb_1), \eb_2, \eb_3)+(-1)^{\eb_1}m_3(\eb_1, m_1(\eb_2), \eb_3)
\nn\\
& &~~~~~~~+(-1)^{\eb_1+\eb_2}m_3(\eb_1, \eb_2, m_1(\eb_3))=0~~.\nn
\eea
The first equation indicates $m_1$ is nilpotent 
and $(\cH, m_1)$ makes a complex.
The second equation implies differential $m_1$ satisfies Leibniz rule for 
the product $m_2$. 
The third equation means product $m_2$ is associative up to the 
term including $m_3$. 
\begin{rem}
In the case $m_n=0$ for $n\geq 3$, an $A_\infty$-algebra 
reduces to a differential graded (associative) algebra (DGA). 
The differential $d$ and the product $\bullet$ 
of DGA correspond to $m_1$ and $m_2$, respectively. However, the product 
$\bullet$ of DGA preserves the degree and $m_2$ in $A_\infty$-algebras 
raises the degree by one. In this reason, when a DGA $\g$ is included in 
an $A_\infty$-algebra $(\cH,\m)$, 
the degree in the $A_\infty$-algebra is defined 
as the degree of the DGA minus one. Let $s : \g^k\raw\cH^{k-1}[1]$ 
be the inclusion map. The $[1]$ `eats' one degree of $\cH$, and the degree 
of $\cH^{k-1}[1]$ is defined as $k-1$ through the operation. 
Then the following diagram commutes 
\begin{equation*}
 \begin{CD}
  \g^k\otimes\g^l @>{\ \bullet\ }>> \g^{k+l}\\
   @V{s}VV               @V{s}VV\\
  \cH^{k-1}[1]\otimes\cH^{l-1}[1] @>{m_2(\ ,\ )}>> \cH^{k+l-1}[1]\ .
 \end{CD}
\end{equation*}
The degree of Witten's open SFT\cite{W1} is usually defined as the degree 
of DGA explained above. There are many literatures where the degree of 
$A_\infty$-algebras are defined with the DGA degree. 
However, when higher products $m_3, m_4,\cdots$ are introduced, 
the degree given in (Def.\ref{defn:coder}) 
is simpler for $A_\infty$-algebras. 
In this reason, we use this convention in the present paper. 
The precise relation between these two conventions can be found in \cite{GJ}.
 \label{rem:degree}
\end{rem}
\begin{rem}
We have mentioned in (Rem.\ref{rem:dual}) about 
the dual picture of coalgebras. 
In this dual picture, the nilpotent coderivation $\m$ is replaced to a 
differential on a formal (noncommutative) supermanifold. 
Let $\Phi=\eb_i\phi^i\in\cH$ be an elements of $\cH$ 
with supercoordinates $\{\phi^i\}$. The degree of $\phi^i$ is minus the 
degree of $\eb_i$ so that the degree of $\Phi$ is zero. Define 
\begin{equation}
 m_k(\eb_1,\cdots,\eb_k)=\eb_j c^j_{1\cdots k}\ ,
 \qquad c^j_{1\cdots k}\in\C\ .
 \label{c}
\end{equation}
In this representation, 
$m_k(\Phi,\cdots,\Phi)=\Phi\flpartial{\phi^j} c^j_{i_1\cdots i_k}
\phi^{i_k}\cdots\phi^{i_1}$, and the differential $\delta$ 
in the dual picture is written as 
\begin{equation}
 \delta=\sum_{k=1}^\infty\flpartial{\phi^j}c^j_{i_1\cdots i_k}
 \phi^{i_k}\cdots\phi^{i_1}\ ,
 \label{dualD}
\end{equation}
where the operation of $\flpartial{\phi^j}\cdots$ is defined as 
$(\phi^3\phi^2\phi^1)\flpartial{\phi^j}c^j_{i_1\cdots i_k}
 \phi^{i_k}\cdots\phi^{i_1}=\phi^3\phi^2c^1_{i_1\cdots i_k}
 \phi^{i_k}\cdots\phi^{i_1}+(-1)^{\eb_1}\phi^3c^2_{i_1\cdots i_k}
 \phi^{i_k}\cdots\phi^{i_1}\phi^1+(-1)^{\eb_1+\eb_2}c^3_{i_1\cdots i_k}
 \phi^{i_k}\cdots\phi^{i_1}\phi^2\phi^1$. 
The sign arises when the $\delta$ with degree one 
passes through some elements 
which have their degree. 
The consistency of this operation is explained in Appendix \ref{ssec:A1}. 
The condition that $\delta$ is differential \ie $(\delta)^2=0$ is 
equal to the $A_\infty$-condition(\ref{ainf}) rewritten using (\ref{c}).
This actually corresponds to the BV-BRST transformation as will be 
seen in subsection \ref{ssec:41}. 
Note that in this paper we denote $\{\phi^i\}$ 
as both fields and antifields in the terminology of the BV-formalism. 
The geometry on this dual picture is explained in Appendix \ref{ssec:A2}.
\end{rem}

\begin{defn}[$A_\infty$-morphism]
Given two $A_\infty$ algebras 
$(\cH, \m)$ and $(\cH',\m')$, 
{\it an $A_\infty$-morphism} $\cF: (\cH, \m)\raw (\cH',\m')$ as a 
cohomomorphism from $C(\cH)$ to $C(\cH')$ satisfying 
\begin{equation*}
 \cF\m=\m'\cF\ .
\end{equation*}
\label{defn:amorp}
\end{defn}
If we act this equation on $\eb_1\cdots\eb_n\in C(\cH)$ for $n\ge 1$, 
its image belongs to $\sum_{n'=1}^n{\cH'}^{\otimes n'}$, and 
picking up the ${\cH'}^{\otimes 1}$ part of the equation yields 
\begin{equation}
 \begin{split}
&\sum_{1\leq k_1<k_2\dots <k_i=n}{m'_i(f_{k_1}(\eb_1,\cdots,\eb_{k_1}),
f_{k_2-k_1}(\eb_{k_1+1},\cdots,\eb_{k_2})\cdots f_{n-k_{i-1}}
(\eb_{k_{i-1}+1},\cdots,\eb_n))}\\
&\qquad=\sum_{k+l=n+1}\sum_{j=0}^{k-1}{(-1)^{\eb_1+\dots +\eb_j}
f_k(\eb_1,\cdots,\eb_j,m_l(\eb_{j+1},\cdots,\eb_{j+l}), 
\eb_{j+l+1},\cdots,\eb_n)}\ .
\end{split}
\label{amorphism}
\end{equation}
The first two constraints in (\ref{amorphism}) read:
\begin{equation}
 \begin{split}
 m'_1(f_1(\eb_1))&=f_1(m_1(\eb_1))\ ,\\
m'_2(f_1(\eb_1),f_1(\eb_2))&=
f_1(m_2(\eb_1,\eb_2))\\
&+m'_1(f_2(\eb_1,\eb_2))+f_2(m_1(\eb_1),\eb_2)
+(-1)^{\eb_1}f_2(\eb_1, m_1(\eb_2))\ .\nn
 \end{split}
\end{equation}
In particular, the first equation implies 
$f_1$ induces a degree zero linear map $f_{1*}$ between the 
cohomologies $H_{m_1}(\cH)$ and $H_{m'_1}(\cH')$. 
The dual representation of $\cF$ will be mentioned in the next subsection. 
In SFT this $\cF$ corresponds to a field transformation between two 
different SFTs and the condition of the $A_\infty$-morphism 
(Def.\ref{defn:amorp}) indicates that 
the field transformation $\cF$ is compatible 
with the BV-BRST transformations.
 
\begin{defn}[quasi-isomorphism]
An $A_\infty$-morphism $\cF$ is called a {\it quasi-isomorphism} if
$f_{1}$ is a degree zero 
isomorphism between the cohomology spaces $H_{m_1}(\cH)$ and 
$H_{m_1'}(\cH')$. 
\end{defn}
It is known that if $\cF$ is quasi-isomorphism, there is a  
inverse quasi-isomorphism $\cF^{-1}: (\cH', \m')\raw (\cH,\m)$
\cite{K2,K}, which will be shown in (Rem.\ref{rem:inverse}).

 \subsection{Maurer-Cartan equation}
\label{ssec:23}

Here we define Maurer-Cartan equations for $A_\infty$-algebras. 
It corresponds to the equation of motions in SFT (eq.(\ref{eomsft})). 
Consider formally the following exponential map
\footnote{Note that this $e^\Phi$ does not belong to $C(\cH)$ 
because $\1\in\cH^{\otimes 0}$. $e^\Phi-\1$ is then an element of $C(\cH)$. 
However as will be seen it is 
convenient to include $\1$ for some formulation. } of $\Phi\in \cH$ 
\begin{equation}
 e^\Phi:=\1+\Phi+\Phi\otimes\Phi+\Phi\otimes\Phi\otimes\Phi+\cdots\ .
\end{equation}
$e^\Phi-\1\in C(\cH)$ satisfies 
$\tri (e^\Phi-\1)=(e^\Phi-\1)\otimes (e^\Phi-\1)$ and such element 
is called an {\it grouplike element}. 
If we define 
\begin{equation*}
 \m_*(e^\Phi):=m_1(\Phi)+m_2(\Phi\otimes\Phi)
 +m_3(\Phi\otimes\Phi\otimes\Phi)+\cdots\ ,
\end{equation*}
then $\m(e^{\Phi})=e^\Phi\m_*(e^\Phi)\cdot e^\Phi$, and 
$\m(e^\Phi)=0$ is equivalent to $\m_*(e^\Phi)=0$, 
where $\1$ is defined as 
$\cH^{\otimes m}\otimes\1\otimes\cH^{\otimes n}=
\cH^{\otimes (m+n)}$ for $m,n\ge 0$ and $m+n\ge 1$. 
$\m_*(e^\Phi)=0$ is called Maurer-Cartan equation for $A_\infty$-algebras. 
When an $A_\infty$-algebra is DGA, {\it i.e.} $m_3=m_4=\cdots=0$, 
its Maurer-Cartan equation takes the form 
$m_1(\Phi)+m_2(\Phi\otimes\Phi)=0$. It is nothing but the condition of 
a flat connection. 

Now we explain that any $A_\infty$-morphisms preserve the solution of two 
Maurer-Cartan equations. The fact means that the $A_\infty$-morphisms 
preserve the equations of motions for SFTs. 
Let $(\cH,\m)$ and $(\cH',\m')$ be two 
$A_\infty$-algebras and $\cF : (\cH,\m)\raw (\cH',\m')$ be a 
$A_\infty$-morphism. 
$\Phi'$ is constructed from $\Phi$ as the pushforward of $\cF$ 
\begin{equation}
 \Phi'=\cF_*(\Phi)=\sum_{n=1}^{\infty}f_n(\Phi\cdots\Phi)\ .
 \label{fredef}
\end{equation}
Direct calculation using eq.(\ref{cohom}) then yields that $\cF$ satisfies 
\begin{equation}
 \cF(e^{\Phi}-\1)=e^{\Phi'}-\1 \ . \label{Uexpalpha}
\end{equation}
To show that $\cF$ preserves the solutions of Maurer-Cartan equations, 
it is sufficient to say that $\m'(e^{\Phi'})=0$ if $\m(e^{\Phi})=0$, 
which can be immediately shown because
\begin{equation*}
 \m'(e^{\Phi'})=\m'\cF(e^{\Phi})=\cF\m(e^{\Phi})=0\ . 
\end{equation*}
\begin{rem}
In the dual picture of these coalgebras, 
for $\Phi=\eb_i\phi^i\in\cH$ and $\Phi'=\eb_{i'}\phi^{i'}\in\cH'$, 
define 
\begin{equation*}
 f_k(\eb_1\cdots\eb_k)=\eb_{j'} f^{j'}_{1\cdots k}\ ,
 \qquad f^{j'}_{1\cdots k}\in\C\ ,\quad \eb_{j'}\in\cH'
\end{equation*}
and eq.(\ref{fredef}) can be expressed as 
\begin{equation*}
 \phi^{i'}=f^{i'}_{j}\phi^j+f^{i'}_{j_1 j_2}\phi^{j_2}\phi^{j_1}
 +f^{i'}_{j_1 j_2 j_3}\phi^{j_3}\phi^{j_2}\phi^{j_1}+\cdots\ .
\end{equation*}
This can be regarded as a nonlinear coordinate transformation between 
the two formal noncommutative supermanifolds. 
This statement is also explained in Appendix \ref{ssec:A2}.
Moreover in this expression one can easily seen that 
when the $f^i_j$ has its inverse, this transformation is locally 
diffeomorphism, and the map 
$\Phi'=\cF_*(\Phi)$ has its inverse. 
\end{rem}
\begin{rem}
There are gauge transformations which preserve the Maurer-Cartan equations. 
Its infinitesimal representation is of the form 
\begin{equation*}
 \delta_\alpha\Phi=\m_*(e^\Phi\alpha e^\Phi)
\end{equation*}
where $\alpha=\eb_i\alpha^i$ is a gauge parameter of degree minus one, 
therefore 
the degree of $\alpha^i$ is minus the degree of $\eb_i$ minus one. 
The fact that this transformation preserves the space of the solution 
of the Maurer-Cartan equation can be directly checked as 
\begin{equation*}
 \delta_\alpha\m_*(e^\Phi)=\m_*(e^\Phi(\delta_\alpha\Phi) e^\Phi)
 =\m_*(e^\Phi\m_*(e^\Phi\alpha e^\Phi)e^\Phi)
 =\m_*(\m(e^\Phi\alpha e^\Phi))=0\ .
\end{equation*}
In the third equality, the Maurer-Cartan equation $\m_*(e^\Phi)=0$ is used. 
Moreover, if any two $A_\infty$-algebras $(\cH,\m)$, $(\cH',\m')$ 
and an $A_\infty$-morphism $\cF$ between them are given, 
then $A_\infty$-morphism restricted to the spaces of the solutions of 
the Maurer-Cartan equations is equivariant under 
the gauge transformations on both sides. 
In other words 
$\delta_{\alpha'}\Phi'=\cF(\delta_\alpha\Phi)$ holds 
on the Maurer-Cartan equations 
where $\alpha'$ is defined by $\cF$ as 
$\alpha'=\cF_*(e^\Phi\alpha e^\Phi):=f_1(\alpha)+f_2(\alpha,\Phi)+
f_2(\Phi,\alpha)+\cdots$. 
This obeys from the condition of 
the $A_\infty$-morphism (Def.\ref{defn:amorp})
\begin{equation*}
 \m'\cF(e^\Phi\alpha e^\Phi)|_{{\cH'}^{\otimes 1}}
 =\cF\m(e^\Phi\alpha e^\Phi)|_{{\cH'}^{\otimes 1}}\ .
\end{equation*}
The space of Maurer-Cartan equation over this gauge action is considered 
as the moduli space in the terminology of deformation theory. 
The above fact then means that 
the moduli space is transformed by $A_\infty$-morphisms. In particular, 
if the $A_\infty$-morphism is quasi-isomorphism, the moduli space is 
isomorphically transformed by the quasi-isomorphism. 
This gauge transformation exactly corresponds to the gauge transformation 
in SFT as will be explained in subsection \ref{ssec:42}. 
\label{rem:gauge}
\end{rem}

 \section{Moduli space of Riemann surfaces and BV-formalism}
\label{sec:3}

In this section, the relevance of BV-formalism in SFT will be reviewed. 
When SFTs are controlled by BV-formalism, 
classical open SFTs\cite{Z2,GZ,N} have $A_\infty$-structures and 
classical closed SFTs\cite{SZ,Z1} have $L_\infty$-structures. 
These algebraic structure of SFT is summarized in subsection \ref{ssec:41} 
and \ref{ssec:42}. 
We attempt to explain those as simple as possible, and in order to do so 
we will transfer various representations for SFT or $A_\infty$-algebras 
to each other.
The precise definition and convention used there are summarized 
in subsection \ref{ssec:43}. 

When constructing SFT, the sum of the Feynman graphs 
of the same topology in the sense of Riemann surfaces must reproduce 
the correlation function of corresponding Riemann surface on-shell. 
The correlation functions in string theory are calculated by integrating 
out over the moduli space of Riemann surfaces. 
Each vertex in SFT is constructed by integrating 
out over the subspace of the moduli space of Riemann surfaces. 
In order that the Feynman rule reproduces the correlation function 
of string theory on-shell, the sum of each Feynman graphs with the 
same topology fill all the moduli space of Riemann surface 
without crevices and without double covered. 
As will be seen below, 
this condition restricts the way of creating the vertices of SFT, 
and produces recursion relations for vertices, which is often called 
the string factorization equations\cite{SZ} (equation (\ref{sfeq})).

Here we review them briefly in the case of classical open 
SFT\cite{Z2,N}. The argument is similar for the 
classical closed SFT\cite{SZ,Z1}. 

Let $\Phi\in\cH$ be a string field. More precisely, 
$\Phi=\eb_i\phi^i$ where $\{\eb_i\}$ are the basis 
of the string Hilbert space $\cH$ and $\{\phi^i\}$ are the string fields. 
The aim is constructing an open SFT action 
of the following form, 
\begin{equation}
 S=S_0+\V\ ,\qquad S_0:=\half\omega(\Phi,Q\Phi),\quad 
 \V=\sum_{k\ge 3}\V_k\ ,
\end{equation}
where $S_0$ is the kinetic term, and $\V_k$ is the $k$-point vertex 
(the term of $k$ powers of string fields $\Phi$). 
$\omega(\ ,\ )$ denotes BPZ-inner product (in CFT). 
In order for the action $S$ to be consistent for string theory, 
the $n$ point amplitude which is calculated by using the Feynman rule 
with the action $S$ must reproduce the corresponding 
$n$-point correlation function of string theory 
when the $n$ external states $\eb_i, (n=1,\cdots,n)$ 
are physical, {\it i.e.} $Q|\eb_i\rangle=0$
\footnote{Here we assume for simplicity 
that the basis $\eb_i$ are taken so that the subbasis of $\{\eb_i\}$ 
span the physical Hilbert space.
}. 
Here we concentrate to review for classical open SFT, so the 
$n$ point amplitude which should be considered is the $n$ point 
tree graph amplitude for open string, and the corresponding 
correlation function in string theory is the disk amplitude with $n$ 
external states on the boundary of the disk. 
(The arguments are same for other case like 
classical and quantum closed SFT\cite{Z1}, 
quantum open-closed SFT\cite{Z2}). 

Let $\cM_n$ be the compactified moduli space of disk with $n$ punctures. 
The dimension of $\cM_n$ is $\dim \cM_n=n-3$. 
Suppose that the vertices are now constructed, and consider the $n$ point 
tree amplitude. It consists of the sum of every Feynman graph of 
$n$ point tree graphs. Here $n$ point tree graphs are the graphs 
which are produced by connecting the vertices with the propagators 
and the topology of which are $n$ point tree graph of open strings. 
As has been explained briefly, in order for the vertices to be made 
consistently, the sum of this Feynman graphs must reproduce 
the single-covered moduli space $\cM_n$
\begin{equation}
 \cM_n=\cM_n^0\cup\cM_n^1\cup\cM_n^2\cup\cdots\ ,
 \label{decomp}
\end{equation}
where $\cM_n^I$ denotes the subspace of the moduli space $\cM_n$ 
which corresponds to the Feynman graphs with $I$ propagators. 
Now the tree graphs are considered, therefore the number of the vertices 
of the $n$ point tree Feynman graphs with $I$ propagators is $I+1$. 
Because 
\begin{equation}
 n=\sum_{m=1}^{I+1}v_m-2I\label{dim}
\end{equation} 
where $v_m\ge 3$ are the numbers of the external 
legs of the vertices, $\cdots$ in eq.(\ref{decomp}) does not continue 
infinitely. 
\begin{figure}[h]
 \hspace*{1.3cm}\includegraphics{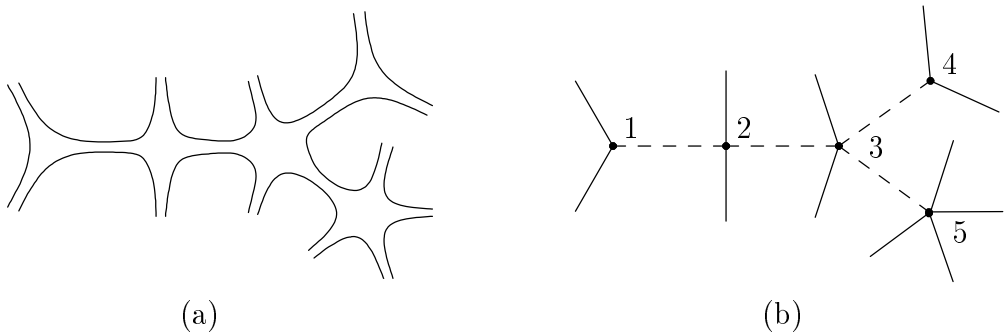}
 \caption{Consider for example the diagram of the string interaction as 
Fig.(a). We represent such diagrams as Fig.(b). The dashed lines denote the 
propagators. Here the vertices are labeled by $1\cdots 5$. 
The numbers of legs for the vertices are 
$v_1=3$, $v_2=4$, $v_3=5$, $v_4=3$, $v_5=5$. 
The number of the propagators equal $I=4.$ The graph has 
twelve external legs, and eq.(\ref{dim}) holds because 
$12=3+4+5+3+5-2\cdot 4$.}
 \label{fig:3}
\end{figure}

Let us consider to construct the vertices inductively and 
suppose that the vertices $\V_k$ with $3\le k\le n-1$ are constructed 
in the following form
\begin{equation}
 \V_k:=\ov{k}\int_{\cM_k^0}\la\Omega|_{1\cdots k}
 |\Phi\ra_1\cdots |\Phi\ra_k=:\ov{k}\la V_k||\Phi\ra\cdots |\Phi\ra\ .
 \label{vertex}
\end{equation}
Here $\Omega$ denotes the volume form on the Moduli space $\cM_k$. 
A point on $\cM_k$ characterizes a 
conformal structure of the disk with $k$ punctures at its boundary. 
The $k$ boundary insertions are symmetrized cyclic, \ie 
$\la\Omega|_{1\cdots k}=\la\Omega|_{2\cdots k 1}$. 
Thus, when $\eb_{i_1},\cdots,\eb_{i_k}$ are physical states, 
$\la\Omega|_{1\cdots k}|\eb_{i_1}\ra_1\cdots |\eb_{i_k}\ra_k$ 
on a point on $\cM_k$ is the correlation function of the $k$-point disk 
with corresponding conformal structure, and 
$\int_{\cM_k}\la\Omega|_{1\cdots k}|\eb_{i_1}\ra_1\cdots |\eb_{i_k}\ra_k$ 
gives the on-shell S-matrix element. 
By construction, the dimension which the vertex 
$\V_k$ has is $\dim\cM_k^0=k-3$ for $k\le n-1$. Besides, 
$\cM_k^I$ is the moduli space reproduced by vertices 
$\V_{v_1},\cdots, \V_{v_{I+1}}$ and $I$ propagators. 
The vertex $\V_{v_m}$ has its dimension of moduli 
$v_m-3$ and the propagator has the dimension one, which is the parameter of 
the length of the evolution of the open string $\R_+$ 
(see Fig.\ref{fig:pv}.(a)). 
Therefore, the dimension of $\cM_k^I$ is 
$(v_1-3)+\cdots+(v_{I+1}-3)+I=(k+2I-3(I+1))+I=k-3$ because 
$k+2I=\sum_{m=1}^{I+1}v_m$. Note that the dimension of $\cM_k^I$ is 
independent of the number of the propagators $I$. Thus, the moduli 
space is decomposed consistently as eq.(\ref{decomp}) for $k\le n-1$. 
Here we want to determine the decomposition of the moduli space $\cM_n$ 
as in eq.(\ref{decomp}) in order to construct the vertex $\V_n$. 
$\cM_n$ is of course given, 
because it is determined only from the Riemann surface. 
Alternatively, for $I\ge 1$ 
$\cM_n^I$ are determined by the induction hypothesis, that is, 
they are determined by the vertices $\V_k$ with $k\le n-1$ and the 
propagators. 
Thus one gets $\cM_n^0$ and 
consequently $\V_n$ of the form in eq.(\ref{vertex}). 

Next, it will be explained that 
the SFT action $S=S_0+\V$ of which vertices are constructed as 
above in eq.(\ref{vertex}) satisfies 
the classical master equation (\ref{meq}). 
Let us consider the infinitesimal variation of the decomposition 
of Riemann surfaces. More precisely consider to take the boundary of 
each $\cM_n^I$, and  denote the operation as $\partial$, 
and write the integral of 
$\ov{n}\la\Omega||\Phi\ra\cdots|\Phi\ra$ over $\cM_n$ as 
\begin{equation}
 \ov{n}\int_{\cM_n}\la\Omega||\Phi\ra\cdots|\Phi\ra
 =\ov{n}\int_{\cM_k^0}\la\Omega||\Phi\ra\cdots|\Phi\ra
 +\ov{n}\int_{\cM_k^1}\la\Omega||\Phi\ra\cdots|\Phi\ra
 +\cdots\ .
 \label{dcorr}
\end{equation}
Taking its boundary yields 
\begin{equation}
 0=\partial(\V_n)+\sum_{\substack{k_1+k_2=n+2\\k_1,k_2\ge 3}}
 \half\l(\begin{split}
    &\partial(\V_{k_1})\-(\V_{k_2})\\
   +&(\V_{k_1})\-\partial(\V_{k_2})\\
   +&(\V_{k_1})\partial(\-)(\V_{k_2})
 \end{split}\r)
 +\sum_{\substack{k_1+k_2+k_3=n+4\\k_1,k_2,k_3\ge 3}}\l(\cdots\r)+\cdots\ .
 \label{recur1} 
\end{equation}
The first equality in the above equation follows from the fact that 
the left hand side of the equation (\ref{dcorr}) does not depend 
on the way of the decomposition of $\cM_n$, \ie $\partial\cM_n=0$. 
This equation exists for $n\ge 3$, and the constraint for $n=3, 4$ read 
\begin{equation}
 n=3\ :\ 0=\partial\V_3\ ,
 \qquad n=4\ :\  0=\partial\V_4+\V_3\partial(\-)\V_3\ .
\end{equation}
The first equation ($n=3$) means $\cM_3$ has no moduli (a point). 
As will be clear later, 
the vertex $\V_{k+1}$ corresponds to the $A_\infty$-structure 
$m_k$ (eq.(\ref{defc}) or eq.(\ref{opvm})), 
and the first equation and the second equation 
corresponds to the second equation and the third equation in eq.(\ref{a3}), 
respectively. 
The equation (\ref{recur1}) is, in fact, equivalent to 
\begin{equation}
 0=\partial(\V_n)+\sum_{\substack{k_1+k_2=n+2\\k_1,k_2\ge 3}}
 \half(\V_{k_1})\partial(\-)(\V_{k_2})\ , 
 \label{sfeq}
\end{equation}
which is the first term and one of the second term on the right hand side 
of the identity(\ref{recur1}). The reason why these are equivalent is that 
the other parts of eq.(\ref{recur1}) cancel by induction. 
For example, $\partial(\V_{k_1})\-(\V_{k_2})$ in the second term 
cancels with one of the third term $(\cdots)$ of the form 
$\sum_{\substack{k+l=k_1+2\\k,l\ge 3}}\l((\V_k)\partial(\-)(\V_l)\r)
\-(\V_{k_2})$. 
The recursion equation (\ref{sfeq}) is called {\it the 
string factorization equation}\cite{SZ}. 

Finally we will rewrite the string factorization equation (\ref{sfeq}) 
in BV-formalism. The $(\V_{k_1})\partial(\-)(\V_{k_2})$ means 
sewing up these vertices with the shortest propagator length, where 
the corresponding moduli is a subspace of the boundary of $\cM_n^1$ 
which is common with the boundary of $\cM_n^0$. 
This sewing $\partial(\-)$ is given by the inverse reflection operator 
$|\omega\ra$, and as will be clarified later in subsection \ref{ssec:43}, 
$|\omega\ra$ is equivalent to the BV-bracket 
$\flpartial{\phi^i}\omega^{ij}\frpartial{\phi^j}=(\ ,\ )$. 
This leads 
\begin{equation}
 (\V_k)\partial(\-)(\V_l)
 =\V_k\flpartial{\phi^i}\omega^{ij}\frpartial{\phi^j}\V_l
 =(\V_k, \V_l)\ . 
 \label{rwv}
\end{equation}
On the other hands, after some calculation in conformal field theory
one obtains\cite{Z1,N} 
\begin{equation}
 2\partial\V_n=2\int_{\partial\cM_n^0}\la\Omega||\Phi\ra\cdots|\Phi\ra
                    =2(S_0,\V_n)\ .
 \label{rws} 
\end{equation}
Rewriting eq.(\ref{sfeq}) with (\ref{rwv}) and (\ref{rws}) 
for $n\ge 3$ and summing up these, 
we obtain the classical BV-master equation
\begin{equation}
 (S, S)=0\ .\label{meq}
\end{equation}
The precise definitions for the notation used here are summarized in 
subsection \ref{ssec:43}. 
After preparing those and other identities the recursion relation 
(\ref{sfeq}) is derived again explicitly in Appendix \ref{ssec:B1}. 
One explicit example of constructing SFT in this procedure will be given 
in section \ref{sec:6}.

 \section{$A_\infty$-structure and BV-formalism}
\label{sec:4}

Continuing the argument in the previous section, 
the relation between an $A_\infty$-structure and BV-formalism will be 
discussed for classical open SFT. 
In subsection \ref{ssec:41}, it is explained that the SFT constructed above, 
which satisfies the BV-master equation (\ref{meq}), has an 
$A_\infty$-structure. In subsection \ref{ssec:42}, 
the BV-gauge transformations for the SFTs 
are identified with 
the gauge transformation for the 
Maurer-Cartan equations for $A_\infty$-algebras. 
Finally in subsection \ref{ssec:43}, 
the notation and definitions, including those 
which are used implicitly in the previous section and this section, 
are summarized. The relation between the operator language of SFT, 
$A_\infty$-language, and its dual representation is clarified and 
their graphical representation is also presented.

 \subsection{$A_\infty$-structure in SFT}
\label{ssec:41}

Represent the string state as 
$|\Phi\rangle=|\eb_i\rangle\phi^i$ where $\{\eb_i\}$ is the basis of the 
string Hilbert space $\cH$ 
and $\phi^i$ are its coordinate whose degree is minus 
the degree of $\eb_i$, and take a component representation of the 
above constructed SFT action as 
\begin{equation}
 \begin{split}
  &\V_k=\ov{k}
 \la V_k||\eb_{i_1}\phi^{i_1}\ra\cdots|\eb_{i_k}\phi^{i_k}\ra
 =\ov{k}\V_{i_1\cdots i_k}\phi^{i_k}\cdots\phi^{i_1}\\
  &\half\omega(\Phi, Q\Phi)
 =\half\V_{i_1 i_2}\phi^{i_2}\phi^{i_1}\ .  
 \end{split}\label{opdvert}
\end{equation} 
Moreover we define for $k\ge 2$, 
\begin{equation}
 c^j_{i_1\cdots i_k}:=(-1)^{\eb_l}\omega^{jl}\V_{li_1\cdots i_k}\ .
 \label{defc}
\end{equation}

On the other hand, the BV-BRST transformation is defined as 
\begin{equation}
 \delta =(\ ,S)\ .
 \label{BVt}
\end{equation}
With this $\delta$, the classical master equation (\ref{meq}) is written 
as $\delta S=0$ and by using the Jacobi identity of the BV-bracket, 
$\delta^2=(\ ,\half(S, S))$ is satisfied. These facts read that the following 
three statements are equivalent : the action $S$ satisfies the BV-master 
equation (\ref{meq}), the action $S$ is invariant under the BV-BRST 
transformation (\ref{BVt}), and the BV-BRST transformation $\delta$ is 
nilpotent. 

When the action is written with $\{c^j_{i_1\cdots i_k}\}$ defined in 
eq.(\ref{defc}), 
the BV-BRST transformation becomes 
\begin{equation}
 \delta=(\ ,S)=\sum_{k=1}^\infty\flpartial{\phi^j}c^j_{i_1\cdots i_k}
 \phi^{i_k}\cdots\phi^{i_1}\ .
\end{equation}
Because $S$ satisfies the BV-master equation(\ref{meq}), this $\delta$ is 
nilpotent, and the fact means that $\{c^j_{i_1\cdots i_k}\}$ define an 
$A_\infty$-algebra in the dual picture as explained in (Rem.\ref{rem:dual}). 
Note however that the $\{c^j_{i_1\cdots i_k}\}$ does not only define an 
$A_\infty$-structure but has the cyclic structure by lowering the upper 
index by the symplectic structure $\omega_{ij}$. 
The algebraic structure of classical open SFT 
is the $A_\infty$-structure with cyclic symmetry through 
an appropriate inner product, 
where the symplectic structure of BV-formalism plays the role 
of the inner product.

 \subsection{BV-gauge transformation}
\label{ssec:42}

The BV-BRST transformation for $\Phi$ is 
$\delta\Phi=\sum_{k\ge 1}m_k(\Phi)=\m_*(e^\Phi)$. 
The corresponding gauge transformation is then written as 
\begin{equation}
 \begin{split}
 \delta_\alpha\Phi&=\m_*(e^\Phi\alpha e^\Phi)\\
 &=Q\alpha+m_2(\alpha,\Phi)+m_2(\Phi,\alpha)
 +m_3(\alpha,\Phi,\Phi)+m_3(\Phi,\alpha,\Phi)+m_3(\Phi,\Phi,\alpha)+\cdots\ ,
 \end{split}
 \label{gauge}
\end{equation}
where $\alpha=\eb_i\alpha^i$ is a gauge parameter of degree minus one, 
therefore 
the degree of $\alpha^i$ is minus the degree of $\eb_i$ minus one. 
This is exactly the gauge transformation for Maurer-Cartan equations given 
by (Rem.\ref{rem:gauge}). Therefore if any two SFT action with 
$A_\infty$-structures and an $A_\infty$-morphism between them are given, 
the gauge transformations eq.(\ref{gauge}) on both sides are 
compatible with the $A_\infty$-morphism in the solution spaces of the 
Maurer-Cartan equations. 
The gauge transformation is written as 
$\delta_{\alpha}\Phi=m_*(e^\Phi)\flpartial{\phi^i}\alpha^i$, and in the 
language of the component fields, it is 
\begin{equation*}
 \delta_\alpha=\sum_{k=1}^\infty\flpartial{\phi^j}c^j_{i_1\cdots i_k}
 \l(\phi^{i_k}\cdots\phi^{i_1}\flpartial{\phi^i}\alpha^i\r)\ .
\end{equation*}
The action is invariant under this $\delta_\alpha$ because 
$0=(S,S)\flpartial{\phi^i}\alpha^i$ leads $\delta_\alpha S=0$
\footnote{It holds even if the Poisson structure $\omega^{ij}$ is 
non-constant. It follows from the Jacobi identity of $\omega^{ij}$ 
and $0=((S,S),\phi^i\omega_{ij}\alpha^j)$ (See comments in 
{\it cyclic algebra with BV-Poisson structure} in the next 
subsection).}. 
Thus the gauge transformation for $A_\infty$-algebras fit 
the standard argument of BV-formalism\cite{BV2,HT}. Moreover, 
\begin{equation}
 0=\omega_{kj}\frpartial{\phi^j}(S,S)\flpartial{\phi^i}\alpha^i
 |_{\fpart{S}{\phi}=0}
 \label{hessian}
\end{equation}
indicates that the generator of the gauge transformation is degenerate 
and the rank of the Hessian for the quadratic part 
of the action $S_0$ is less than half of the number of the basis 
$\{\eb_i\}$ on the space $\{\phi |\fpart{S}{\phi}=0\}$, 
though the number of the basis is infinity. 
The origin $\phi=0$ is also the solution for $\{\phi |\fpart{S}{\phi}=0\}$, 
and in SFT case eq.(\ref{hessian}) at the origin 
is nothing but the condition $(Q)^2=0$. 
When the ratio of the rank of the Hessian 
over the number of the basis is just half, 
the action is called {\it proper}. 
SFT is just the case. The Hessian at the origin is $\V_{i_1 i_2}$ 
in eq.(\ref{opdvert}), which is determined by $Q$. (The reducibility of 
the gauge group of SFT action then comes from the Virasoro symmetry of $Q$.)
The above arguments lead that the rank of the Hessian is equal to 
the rank of unphysical states $\cH^u$ which generate the gauge 
transformations. It is much larger than the rank of physical states $\cH^p$, 
which is the cohomology class with respect to $Q$. 
Let $\cH^t$ be $Q$-trivial states then $\cH=\cH^t\cup\cH^u\cup\cH^p$ and 
$\rank\ \cH^u=\rank\ \cH^t$. {}From these it can be seen that 
actually $\rank\ \{\V_{i_1 i_2}\}/\rank\ \cH=\half$. 
and SFT actions are proper at the origin of $\cH$. 

Though SFT is treated in the context of the BV-formalism, 
the use is different from that in the original context of 
BV-formalism\cite{BV1,BV2,HT}, 
where beginning with the gauge invariant action 
which does not include antifields, 
the terms including antifields are added to the original action 
so that the action satisfies the master equation and is proper. 
Restricting the antifields to zero recovers the original action, 
where the rank of the Hessian is less than the rank of the fields. 
We call it the trivial gauge. 
The gauge fixing is then performed by shifting the trivial gauge and 
restricting the antifields so that the rank of the Hessian is 
equal to the rank of the fields, \ie half of the rank of the total space 
including antifields ($\cH$ in SFT). 
In SFT, however, the antifields are originally included 
in the quadratic term $S_0$ 
and BV-master equation is used in order to determine the form of higher 
vertices. Therefore the trivial gauge fixing 
can be consistent gauge fixing in SFT. 
Actually the quadratic term $S_0$ reads that 
the rank of the Hessian of the gauge fixed quadratic part 
$\sim \half\rank\ \cH$ 
at least as perturbation theory around the origin $\phi=0$. 
This trivial gauge is called {\it Siegel gauge} in SFT and used in more 
explicit arguments in subsection \ref{ssec:5gf} and section \ref{sec:6}. 

Coming back to the property of the gauge transformation, 
we mention two remarks about it. 
The gauge transformation makes Lie algebra on-shell. 
In the case of classical closed SFT ($L_\infty$-algebra), 
the fact can be found in \cite{Z1}.  
Moreover, if two classical open SFT and the $A_\infty$-morphism between them 
are given, the gauge transformations on both SFTs are compatible with 
the $A_\infty$-morphism on-shell, where on-shell means the solution of the 
Maurer-Cartan equations. This fact follows from (Rem.\ref{rem:gauge}).

 \subsection{Operator language, $A_\infty$-algebra, its dual, 
and their graphical representation}
\label{ssec:43}

In this subsection some notation used before and later is summarized. 
We identify the operator language and the coalgebraic representation as 
\begin{equation*}
 |\eb_{i_1}\ra_1\cdots|\eb_{i_n}\ra_n
 =\eb_{i_1}\otimes\cdots\otimes\eb_{i_n}\ .
\end{equation*}
This leads, for instance, $|\Phi\ra_1\cdots|\Phi\ra_n=
|\eb_{i_1}\ra_1\cdots|\eb_{i_n}\ra_n\phi^{i_n}\cdots\phi^{i_1}$.

$\bullet$\ \ {\it Symplectic form and Poisson structure}

First, a (constant) symplectic structure $\omega : \cH\otimes\cH\raw \C$ 
induced from the BPZ-inner product is defined. 
In the operator language, the symplectic structure is defined as 
\begin{equation*}
 \omega(\eb_i,\eb_j):=\la\omega|_{12}|\eb_i\ra_1|\eb_j\ra_2\ .
\end{equation*}
Here the $\la\omega|_{12}$ is the reflection operator, which denotes 
the sewing of the state $|\ \ra_2$ at the origin of the upper half plane 
with the state $|\ \ra_1$ at infinity on it. 
Define the property of the exchanging the labels for kets as 
\begin{equation*}
 \la\omega|_{21}=-\la\omega|_{12}\ ,
\end{equation*}
and then $\la\omega|_{21}|\eb_j\ra_2|\eb_i\ra_1
=(-1)^{\eb_i\eb_j}\la\omega|_{21}|\eb_i\ra_1|\eb_j\ra_2
=-(-1)^{\eb_i\eb_j}\la\omega|_{12}|\eb_i\ra_1|\eb_j\ra_2$. 
The symplectic structure is translated into its component expression as 
\begin{equation*}
 \la\omega|_{12}|\eb_i\ra_1|\eb_j\ra_2=\omega_{ij}\ ,
\end{equation*}
and the above calculation reads 
\begin{equation*}
 \omega_{ji}=-(-1)^{\eb_i\eb_j}\omega_{ij}\ .
\end{equation*}
Thus $\omega$ determines a graded symplectic structure. 
$\omega_{ij}$ is by definition constant on $\cH$, \ie 
it is independent of $\{\phi\}$. Now 
$\omega_{ij}\ne 0$ iff the degree of $\eb_i$ plus the degree of $\eb_j$ is 
equal to one. Therefore one always gets $(-1)^{\eb_i\eb_j}=1$ and then 
$\omega_{ji}=-\omega_{ij}$. 
Moreover in this reason the degree of $\la\omega|_{12}$ is minus one 
so that the degree of $\omega_{ij}\in\C$ has degree zero.

The inverse reflector is defined as the inverse of $|\omega\ra_{12}$ 
\begin{equation}
 \la\omega|_{12}|\omega\ra_{23}= _3\hspace*{-0.1cm}\1_1\ ,
 \label{poisson}
\end{equation}
where $_3\1_1$ denotes the identity operator which maps 
from $|\ \ra_1$ to $|\ \ra_3$. The degree of $|\omega\ra_{12}$ is plus one. 
Expand it as 
\begin{equation}
 |\omega\ra_{23}=|\eb_j\ra_2|\eb_k\ra_3\omega^{jk}(-1)^{\eb_k}
 \label{reflexp}
\end{equation}
and eq.(\ref{poisson}) is rewritten as 
$_3\1_1=\la\omega|_{12}|\eb_j\ra_2|\eb_k\ra_3\omega^{jk}(-1)^{\eb_k}
=|\eb_k\ra_3\la\omega|_{12}|\eb_j\ra_2\omega^{jk}(-1)^{\eb_j\eb_k}$. 
We then define the dual state as 
\begin{equation}
 \la\eb^k|_1:=\la\omega|_{12}|\eb_j\ra_2\omega^{jk}(-1)^{\eb_j\eb_k}\ .
 \label{dstate}
\end{equation}
By definition, the degree of $\la\eb^k|_1$ is minus the degree of $\eb_k$. 
The equation (\ref{poisson}) is then expressed as 
$_3\1_1=|\eb_k\ra_3\la\eb^k|_1$. The dual states must have 
the following inner product
\begin{equation}
 \la\eb^k|_1|\eb_i\ra_1=\delta^k_i\ . 
 \label{opip}
\end{equation}
This gives a condition for $\omega^{jk}$ in eq.(\ref{dstate}), which is 
\begin{equation}
 \omega_{ij}\omega^{jk}=\delta_i^k=\omega^{kj}\omega_{ji}\ .
 \label{sp}
\end{equation}
Moreover this identity leads 
\begin{equation}
 \omega^{kj}=-(-1)^{\eb_j\eb_k}\omega^{jk}
 =-\omega^{jk}\ \ \ (\ =-(-1)^{(\eb_j+1)(\eb_k+1)}\omega^{jk}\ )\ .
 \label{pantisym}
\end{equation} 
Using this, one can see from (\ref{reflexp}) that the inverse reflector 
is symmetric with respect to the labels for bras
\begin{equation*}
 |\omega\ra_{23}=|\omega\ra_{32}\ ,
\end{equation*}
and the dual state (\ref{dstate}) is rewritten as 
\begin{equation}
 \la\eb^k|_1:=\omega^{kj}\la\omega|_{01}|\eb_j\ra_0\ .
 \label{dstate2} 
\end{equation}
The complete set of $\1$ is then defined by this dual state as 
\begin{equation}
 \1=|\eb_i\ra\la\eb^i|\ . \label{one}
\end{equation}
The equations (\ref{sp}) and (\ref{pantisym}) indicate that $\omega^{ij}$, 
the inverse of the symplectic structure $\omega_{ij}$, 
gives a Poisson structure. Explicitly  
\begin{equation*}
 \flpartial{\phi^i}\omega^{ij}\frpartial{\phi^j}=:(\ ,\ )
\end{equation*}
is the {\it BV-Poisson bracket}. 
This gives the identification of the inverse reflection operator 
$|\omega\ra_{23}$ and the BV-bracket assumed in eq.(\ref{rwv}). 
Its compatibility with the string vertices will be checked after 
introducing the vertices in 
{\it cyclic algebra with BV-Poisson structure}. 

The symplectic structure is also expanded with the dual basis as 
\begin{equation*}
 \la\omega|_{12}=\omega_{ij}\la\eb^j|_2\la\eb^i|_1\ .
\end{equation*}
Note that the dual state (\ref{dstate2}) and the inner product (\ref{opip}) 
can be regarded as those used in Appendix \ref{ssec:A1}.

$\bullet$\ \ {\it String vertex and $A_\infty$-structure}

The vertex with the operator representation and its component representation 
were related in eq.(\ref{opdvert}) as 
\begin{equation*}
 \la V_{k+1}| |\Phi\ra\cdots|\Phi\ra=
\la V_{k+1}| |\eb_{i_1}\ra\cdots|\eb_{i_{k+1}}\ra
\phi^{i_{k+1}}\cdots\phi^{i_1}
=\V_{i_1\cdots i_{k+1}}\phi^{i_{k+1}}\cdots\phi^{i_1}\ ,
\end{equation*}
where the indices which label the bras and kets are omitted. 
On the other hand, the $A_\infty$-structures in both representations 
were related by eq.(\ref{c}) as 
\begin{equation*}
  m_k(\eb_{i_1},\cdots,\eb_{i_k})=\eb_j c^j_{i_1\cdots i_k}\ ,
\end{equation*}
and this $\{c^j_{i_1\cdots i_k}\}$ and $\{\V_{i_1\cdots i_{k+1}}\}$ were 
connected to the component representation by eq.(\ref{defc}) as 
\begin{equation}
 c^j_{i_1\cdots i_k}:=(-1)^{\eb_l}\omega^{jl}\V_{li_1\cdots i_k}\ .
 \label{defc2}
\end{equation}
The corresponding one in the operator representation is 
\begin{equation}
 \la V_{k+1}| |\Phi\ra\cdots|\Phi\ra=\la\omega|_{12}|\Phi\ra_1
 |m_k(\Phi,\cdots,\Phi)\ra_2\ 
 \label{opvm}
\end{equation}
for $k\ge 1$
\footnote{The coefficient for $\phi^{i_{k+1}}\cdots\phi^{i_1}$ reads 
$\la V_{k+1}| |\eb_{i_1}\ra\cdots|\eb_{i_{k+1}}\ra
=(-1)^{\eb_{i_1}}\la\omega|_{12}|\eb_{i_1}\ra_1 
|m_k(\eb_{i_2},\cdots,\eb_{i_{k+1}})\ra_2$. }. 
Actually, by using eq.(\ref{defc2}), this identity can be checked 
\begin{equation*}
 \begin{split}
 \omega(\Phi,m_k(\Phi))&=\omega(\eb_{i_1}\phi^{i_1},m_k(\eb_{i_2}\phi^{i_2},
 \cdots,\eb_{i_{k+1}}\phi^{i_{k+1}}))\\
 &=\phi^{i_1}\omega(\eb_{i_1},\eb_{j}c^j_{i_2\cdots i_{k+1}})
 \phi^{i_{k+1}}\cdots\phi^{i_2}\\
 &=(-1)^{\eb_{i_1}}\omega_{i_1 j}c^j_{i_2\cdots i_{k+1}}
 \phi^{i_{k+1}}\cdots\phi^{i_2}\cdot\phi^{i_1}\\
 &=\V_{i_1\cdots i_{k+1}}\phi^{i_{k+1}}\cdots\phi^{i_1}
  =\la V_{k+1}| |\Phi\ra\cdots|\Phi\ra\ 
 \end{split}
\end{equation*}
where $m_k(\Phi):=m_k(\Phi,\cdots,\Phi)$. 
Thus the action can be rewritten in this form, 
\begin{equation}
 S=\half\omega(\Phi,Q\Phi)+\sum_{k\ge 2}\ov{k+1}\omega(\Phi,m_k(\Phi))\ .
 \label{actionf}
\end{equation}

On the other hand, one can define $m_k$ by $\la V_{k+1}|$ and the inverse 
reflector $|\omega\ra$ as follows 
\begin{equation}
|m_k(\Phi,\cdots,\Phi)\ra_0=
\la V_{k+1}|_{1\cdots k+1}|\omega\ra_{01}|\Phi\ra_2\cdots |\Phi\ra_{k+1}\ .
 \label{opmv} 
\end{equation} 
In fact, the coefficient for $\eb_j\cdot\phi^{i_k}\cdots\phi^{i_1}$ 
reproduces eq.(\ref{defc2}), and acting $\la\omega|_{a0}|\Phi\ra_a$ on 
both sides of the eq.(\ref{opmv}) from left reproduces eq.(\ref{opvm}) 
using the identity (\ref{poisson}).

$\bullet$ {\it cyclic algebra with BV-Poisson structure}

The vertices $\ov{k}\V_{i_1\cdots i_k}\phi^{i_k}\cdots\phi^{i_1}$ is 
cyclic, \ie 
\begin{equation*}
 \begin{split}
 \ov{k}\V_{i_1\cdots i_k}\phi^{i_k}\cdots\phi^{i_1}
 &=(-1)^{(\eb_{i_k}+\cdots+\eb_{i_2})\eb_{i_1}}
 \ov{k}\V_{i_1\cdots i_k}\phi^{i_1}\phi^{i_k}\cdots\phi^{i_2}\\
 &=\ov{k}\V_{i_2\cdots i_k i_1}\phi^{i_1}\phi^{i_k}\cdots\phi^{i_2}
 \end{split}
\end{equation*}
This cyclic symmetry is the consequence of the property of trace in 
the terminology of \cite{W1}. 
The BV-bracket of two cyclic vertices are then rewritten as 
the sewing of the vertices 
\begin{equation*}
 \begin{split}
 &\ov{k}\V_{i_1\cdots i_k}\phi^{i_k}\cdots\phi^{i_1}
 \flpartial{\phi^i}\omega^{ij}\frpartial{\phi^j}
  \ov{l}\V_{j_1\cdots j_l}\phi^{j_l}\cdots\phi^{j_1}\\
 &=\ov{k}\la V_k|_{1\cdots k}|\Phi\ra_1\cdots |\Phi\ra_k
 \flpartial{\phi^i}\omega^{ij}\frpartial{\phi^j}
 \ov{l}\la V_l|_{1'\cdots l'}|\Phi\ra_{1'}\cdots |\Phi\ra_{l'}\\
 &=\la V_k|_{1\cdots k}|\Phi\ra_1\cdots |\Phi\ra_{k-1}|\eb_i\ra_k
 \omega^{ij}(-1)^{\eb_j}
 \la V_l|_{1'\cdots l'}|\eb_j\ra_{1'}|\Phi\ra_{2'}\cdots |\Phi\ra_{l'}\\
 &=\la V_k|_{1\cdots k}\la V_l|_{1'\cdots l'}
  |\omega\ra_{k1'}(|\Phi\ra)^n\ ,
 \end{split}
\end{equation*}
where $k+l=n+2$. This gives the explicit calculation in eq.(\ref{rwv}). 
Here $(|\Phi\ra)^n$ are inserted on the boundary of the disk $S^1$, 
so the last line of the above equation is also cyclic and the cyclic vertices 
close as a Lie algebra 
with respect to Lie bracket $(\ ,\ )$. In component language 
it is of the form 
\begin{equation*}
 \la V_k|_{1\cdots k}\la V_l|_{1'\cdots l'}
  |\omega\ra_{k1'}(|\Phi\ra)^n
 =\ov{n}\l(\la V_k|_{1\cdots k}\la V_l|_{1'\cdots l'}
  |\omega\ra_{k1'}|\eb_{i_1}\ra\cdots |\eb_{i_n}\ra + cyclic \r)
 \phi^{i_n}\cdots\phi^{i_1}
\end{equation*}
where $cyclic$ means $|\eb_{i_1}\ra\cdots |\eb_{i_n}\ra$ is moved 
cyclic with appropriate sign and each term is summed up. 
Using this we can also rewrite the recursion relation directly and 
arrive at the condition for the $A_\infty$-structure(\ref{ainf}). 
Furthermore if we define the free tensor algebra of this cyclic algebra, 
the BV-bracket $(\ ,\ )$ then defines a Poisson structure on the 
free tensor algebra. We can also extend the BV-Poisson structure 
$\omega^{ij}$ non-constant as the Poisson structure on a 
formal noncommutative supermanifold. However in the present paper 
we discuss mainly in the case that $\omega^{ij}$ is constant, so 
we want to report the issue separately elsewhere. Some related concepts are 
found in \cite{K3}.

Note that $\la V_k|$ is written as 
\begin{equation*}
 \la V_k|=\la V_k||\eb_{i_1}\ra\la\eb^{i_1}|\cdots |\eb_{i_k}\ra\la\eb^{i_k}|
 =\V_{i_1\cdots i_k}\la\eb^{i_k}|\cdots\la\eb^{i_1}|\ , 
\end{equation*}
which is the same form as $\V_{i_1\cdots i_k}\phi^{i_k}\cdots\phi^{i_1}$. 
Thus the coordinates of the states and the dual states are 
identified as a vector space (Appendix \ref{ssec:A1}). 
Similarly $\fpartial{\phi^i}$ can be identified with $\eb_i$.

$\bullet$\ \ {\it other algebraic relations used later}

In addition to the BRST charge $Q$, in section \ref{sec:5} 
the propagator $Q^+$ which has degree minus one is introduced. 
Here the algebraic properties of these 
operators together with $\omega$ are summarized. 
We impose the propagator satisfies the following relation 
\begin{equation}
 (Q^+)^2=0\ ,\qquad Q^+QQ^+=Q^+\ .
 \label{qdag}
\end{equation}
{}From the properties of the BPZ-inner product, 
$Q$ and $Q^+$ operate on $\la\omega|$ and $|\omega\ra$ as 
\begin{equation}
 \la\omega|_{12}Q^{(2)}=-\la\omega|_{12}Q^{(1)}\ ,\quad\ 
 \la\omega|_{12}(Q^+)^{(2)}=\la\omega|_{12}(Q^+)^{(1)}\ ,
\quad\ (Q^+)^{(2)}|\omega\ra_{12}=(Q^+)^{(1)}|\omega\ra_{12}
 \label{qqdag}
\end{equation}
where the indices $^{(1)}$ or $^{(2)}$ denote the bras or kets 
where the operators act on. The justification for these relations are 
clarified in subsection \ref{ssec:5gf}. 

By employing the property of $Q$ acting on $\la\omega|$ (\ref{qqdag}), 
here the orthogonal decomposition of the inner product $\omega$ is 
examined. Let us define the Hodge-Kodaira decomposition of string Hilbert 
space $\cH$ as 
\begin{equation}
 QQ^u+Q^uQ+P^p=\1\ ,
\end{equation}
where degree minus one operator $Q^u$ is defined so that 
$QQ^u$, $Q^uQ$ and $P^p$ are the projections onto $Q$-trivial states, 
unphysical states and physical states, respectively. 
Here define the space of physical states as 
\begin{equation}
 \cH^p:=P^p\cH\ .
\end{equation}
Similarly define the projections onto $Q$-trivial states and 
unphysical states as $QQ^u=P^t$ and $Q^uQ=P^u$ and express the decomposition 
of $\cH$ as 
\begin{equation*}
 \cH^t:=P^t\cH\ ,\quad \cH^u:=P^u\cH\ ,\qquad \cH=\cH^t\cup\cH^u\cup\cH^p\ .
\end{equation*}
There are ambiguities of the choice of $Q^u$. 
$Q$-trivial states $\cH^t$ is unique but $\cH^p$ is unique up to 
the $Q$-trivial states $\cH^t$, and unphysical states $\cH^u$ is 
unique up to $Q$-trivial and physical states $\cH^t\cup\cH^p$. 
Consider the inner product 
\begin{equation*}
 \omega_{ij}=\omega(\eb_i,\eb_j)\ .
\end{equation*}
{}From the first equation in (\ref{qqdag}), we can see the following 
properties without fixing the ambiguities : 
when $\eb_j\in\cH^t$ 
then $\omega_{ij}=0$ for $\eb_i\in\cH^t\cup\cH^p$, 
when $\eb_j\in\cH^t\cup\cH^p$ then $\omega_{ij}=0$ for 
$\eb_i\in\cH^t$. If we denote the block element of matrix $\{\omega_{ij}\}$ 
where $\eb_i\in\cH^u$ and $\eb_j\in\cH^p$ as $\omega_{up}$ and similar for 
other eight block elements, matrix $\{\omega_{ij}\}$ is represented as 
the left hand side of eq.(\ref{omegadecomp}). 
This implies that by basis transformation corresponding 
to the ambiguity in $Q^u$ the inner product $\omega$ is decomposed 
as the right hand side 
\begin{equation}
 \{\omega_{ij}\}=\bp \omega_{uu}&\omega_{up}&\omega_{ut}\\
                     \omega_{pu}&\omega_{pp}&\omega_{pt}\\
                     \omega_{tu}&\omega_{tp}&\omega_{tt}\ep
                =\bp \omega_{uu}&\omega_{up}&\omega_{ut}\\
                     \omega_{pu}&\omega_{pp}& 0         \\
                     \omega_{tu}& 0         & 0         \ep\ 
 \lgraw          \bp  0         & 0         &\omega_{ut}\\
                      0         &\omega_{pp}& 0         \\
                     \omega_{tu}& 0         & 0         \ep\ . 
 \label{omegadecomp}
\end{equation}
These orthogonal decomposition of the inner product will be used 
in subsection \ref{ssec:53} when on-shell reduction of SFT is discussed. 
In closed string case the explicit orthogonal decomposition for string 
states is given in Appendix of \cite{Sbg2}.

$\bullet$\ \ {\it Graphical representation}

Here give some graphical representations for those defined above. 
It is helpful to realize intuitively, and used in subsection {\ref{ssec:53}} 
to simplify the arguments when using the Feynman graphs of SFT. 

Express $\la\omega|$ and $|\omega\ra$ as $\mms$ and $\mmp$, respectively. 
Moreover, vertices and the $A_\infty$-structure are expressed as follows\\
 \hspace*{2.0cm}
%WinTpicVersion3.08
\unitlength 0.1in
\begin{picture}( 38.2500,  8.9900)(  4.1500,-14.4900)
% LINE 2 0 3 0
% 2 3700 1000 3900 1000
% 
\special{pn 8}%
\special{pa 3700 1000}%
\special{pa 3900 1000}%
\special{fp}%
% LINE 2 0 3 0
% 2 3900 1000 4100 600
% 
\special{pn 8}%
\special{pa 3900 1000}%
\special{pa 4100 600}%
\special{fp}%
% LINE 2 0 3 0
% 2 3900 1000 4100 700
% 
\special{pn 8}%
\special{pa 3900 1000}%
\special{pa 4100 700}%
\special{fp}%
% LINE 2 0 3 0
% 2 3900 1000 4100 800
% 
\special{pn 8}%
\special{pa 3900 1000}%
\special{pa 4100 800}%
\special{fp}%
% LINE 2 0 3 0
% 2 3900 1000 4100 1400
% 
\special{pn 8}%
\special{pa 3900 1000}%
\special{pa 4100 1400}%
\special{fp}%
% LINE 2 2 3 0
% 2 4050 950 4050 1150
% 
\special{pn 8}%
\special{pa 4050 950}%
\special{pa 4050 1150}%
\special{dt 0.045}%
% STR 2 0 3 0
% 3 1000 900 1000 1000 5 0
% $\la V_k| = $
\put(10.0000,-10.0000){\makebox(0,0){$\la V_k| = $}}%
% STR 2 0 3 0
% 3 2400 900 2400 1000 5 0
% $,$
\put(24.0000,-10.0000){\makebox(0,0){$,$}}%
% STR 2 0 3 0
% 3 3400 900 3400 1000 5 0
% $m_k = $
\put(34.0000,-10.0000){\makebox(0,0){$m_k = $}}%
% DOT 0 0 3 0
% 2 3900 1000 3900 1000
% 
\special{pn 20}%
\special{sh 1}%
\special{ar 3900 1000 10 10 0  6.28318530717959E+0000}%
\special{sh 1}%
\special{ar 3900 1000 10 10 0  6.28318530717959E+0000}%
% LINE 2 0 3 0
% 2 1400 1000 1600 600
% 
\special{pn 8}%
\special{pa 1400 1000}%
\special{pa 1600 600}%
\special{fp}%
% LINE 2 0 3 0
% 2 1400 1000 1600 700
% 
\special{pn 8}%
\special{pa 1400 1000}%
\special{pa 1600 700}%
\special{fp}%
% LINE 2 0 3 0
% 2 1400 1000 1600 800
% 
\special{pn 8}%
\special{pa 1400 1000}%
\special{pa 1600 800}%
\special{fp}%
% LINE 2 0 3 0
% 2 1400 1000 1600 1400
% 
\special{pn 8}%
\special{pa 1400 1000}%
\special{pa 1600 1400}%
\special{fp}%
% LINE 2 2 3 0
% 2 1550 950 1550 1150
% 
\special{pn 8}%
\special{pa 1550 950}%
\special{pa 1550 1150}%
\special{dt 0.045}%
% DOT 0 0 3 0
% 2 1400 1000 1400 1000
% 
\special{pn 20}%
\special{sh 1}%
\special{ar 1400 1000 10 10 0  6.28318530717959E+0000}%
\special{sh 1}%
\special{ar 1400 1000 10 10 0  6.28318530717959E+0000}%
% CIRCLE 2 0 3 0
% 4 1650 600 1600 600 1700 600 1660 550
% 
\special{pn 8}%
\special{ar 1650 600 50 50  4.9097845 6.2831853}%
% CIRCLE 2 0 3 0
% 4 1650 1400 1600 1400 1660 1450 1700 1400
% 
\special{pn 8}%
\special{ar 1650 1400 50 50  6.2831853 6.2831853}%
\special{ar 1650 1400 50 50  0.0000000 1.3734008}%
% CIRCLE 2 0 3 0
% 4 1750 950 1750 1000 1700 950 1750 1000
% 
\special{pn 8}%
\special{ar 1750 950 50 50  1.5707963 3.1415927}%
% CIRCLE 2 0 3 0
% 4 1750 1050 1750 1000 1750 1000 1700 1050
% 
\special{pn 8}%
\special{ar 1750 1050 50 50  3.1415927 4.7123890}%
% LINE 2 0 3 0
% 2 1700 600 1700 950
% 
\special{pn 8}%
\special{pa 1700 600}%
\special{pa 1700 950}%
\special{fp}%
% LINE 2 0 3 0
% 2 1700 1400 1700 1050
% 
\special{pn 8}%
\special{pa 1700 1400}%
\special{pa 1700 1050}%
\special{fp}%
% STR 2 0 3 0
% 3 1850 900 1850 1000 5 0
% $k$
\put(18.5000,-10.0000){\makebox(0,0){$k$}}%
% CIRCLE 2 0 3 0
% 4 4140 599 4090 599 4190 599 4150 549
% 
\special{pn 8}%
\special{ar 4140 600 50 50  4.9097845 6.2831853}%
% CIRCLE 2 0 3 0
% 4 4140 1399 4090 1399 4150 1449 4190 1399
% 
\special{pn 8}%
\special{ar 4140 1400 50 50  6.2831853 6.2831853}%
\special{ar 4140 1400 50 50  0.0000000 1.3734008}%
% CIRCLE 2 0 3 0
% 4 4240 949 4240 999 4190 949 4240 999
% 
\special{pn 8}%
\special{ar 4240 950 50 50  1.5707963 3.1415927}%
% CIRCLE 2 0 3 0
% 4 4240 1049 4240 999 4240 999 4190 1049
% 
\special{pn 8}%
\special{ar 4240 1050 50 50  3.1415927 4.7123890}%
% LINE 2 0 3 0
% 2 4190 599 4190 949
% 
\special{pn 8}%
\special{pa 4190 600}%
\special{pa 4190 950}%
\special{fp}%
% LINE 2 0 3 0
% 2 4190 1399 4190 1049
% 
\special{pn 8}%
\special{pa 4190 1400}%
\special{pa 4190 1050}%
\special{fp}%
% STR 2 0 3 0
% 3 4340 899 4340 999 5 0
% $k$
\put(43.4000,-9.9900){\makebox(0,0){$k$}}%
% DOT 2 0 3 0
% 5 1600 600 1600 700 1600 800 1600 1400 1600 1400
% 
\special{pn 8}%
\special{sh 1}%
\special{ar 1600 600 10 10 0  6.28318530717959E+0000}%
\special{sh 1}%
\special{ar 1600 700 10 10 0  6.28318530717959E+0000}%
\special{sh 1}%
\special{ar 1600 800 10 10 0  6.28318530717959E+0000}%
\special{sh 1}%
\special{ar 1600 1400 10 10 0  6.28318530717959E+0000}%
\special{sh 1}%
\special{ar 1600 1400 10 10 0  6.28318530717959E+0000}%
% DOT 2 0 3 0
% 5 4100 600 4100 700 4100 800 4100 1400 4090 1400
% 
\special{pn 8}%
\special{sh 1}%
\special{ar 4100 600 10 10 0  6.28318530717959E+0000}%
\special{sh 1}%
\special{ar 4100 700 10 10 0  6.28318530717959E+0000}%
\special{sh 1}%
\special{ar 4100 800 10 10 0  6.28318530717959E+0000}%
\special{sh 1}%
\special{ar 4100 1400 10 10 0  6.28318530717959E+0000}%
\special{sh 1}%
\special{ar 4090 1400 10 10 0  6.28318530717959E+0000}%
\end{picture}%
\ .\\
In coalgebras or the operator representation, 
the operators act from left and accumulated on left. According to 
this order, we define the order of the operation for these graphs from 
right to left. 
The relations between the cyclic vertices and the $A_\infty$-structure 
are then represented as \\
\vspace*{0.1cm}\\
 \hspace*{-4.5cm}
%WinTpicVersion3.08
\unitlength 0.1in
\begin{picture}( 78.6909, 12.9429)(-16.0433,-16.3878)
% LINE 2 0 3 0
% 2 3200 799 3400 399
% 
\special{pn 8}%
\special{pa 3150 787}%
\special{pa 3347 393}%
\special{fp}%
% LINE 2 0 3 0
% 2 3200 799 3400 499
% 
\special{pn 8}%
\special{pa 3150 787}%
\special{pa 3347 492}%
\special{fp}%
% LINE 2 0 3 0
% 2 3200 799 3400 599
% 
\special{pn 8}%
\special{pa 3150 787}%
\special{pa 3347 590}%
\special{fp}%
% LINE 2 0 3 0
% 2 3200 799 3400 1199
% 
\special{pn 8}%
\special{pa 3150 787}%
\special{pa 3347 1181}%
\special{fp}%
% LINE 2 2 3 0
% 2 3350 749 3350 949
% 
\special{pn 8}%
\special{pa 3298 738}%
\special{pa 3298 935}%
\special{dt 0.045}%
% DOT 0 0 3 0
% 2 3200 799 3200 799
% 
\special{pn 20}%
\special{sh 1}%
\special{ar 3150 787 10 10 0  6.28318530717959E+0000}%
\special{sh 1}%
\special{ar 3150 787 10 10 0  6.28318530717959E+0000}%
% CIRCLE 2 0 3 0
% 4 3450 399 3400 399 3500 399 3460 349
% 
\special{pn 8}%
\special{ar 3396 393 50 50  4.9097845 6.2831853}%
% CIRCLE 2 0 3 0
% 4 3450 1199 3400 1199 3460 1249 3500 1199
% 
\special{pn 8}%
\special{ar 3396 1181 50 50  6.2831853 6.2831853}%
\special{ar 3396 1181 50 50  0.0000000 1.3734008}%
% CIRCLE 2 0 3 0
% 4 3550 749 3550 799 3500 749 3550 799
% 
\special{pn 8}%
\special{ar 3495 738 50 50  1.5707963 3.1415927}%
% CIRCLE 2 0 3 0
% 4 3550 849 3550 799 3550 799 3500 849
% 
\special{pn 8}%
\special{ar 3495 836 50 50  3.1415927 4.7123890}%
% LINE 2 0 3 0
% 2 3500 399 3500 749
% 
\special{pn 8}%
\special{pa 3445 393}%
\special{pa 3445 738}%
\special{fp}%
% LINE 2 0 3 0
% 2 3500 1199 3500 849
% 
\special{pn 8}%
\special{pa 3445 1181}%
\special{pa 3445 836}%
\special{fp}%
% STR 2 0 3 0
% 3 3750 699 3750 799 5 0
% $k+1$
\put(36.9094,-7.8642){\makebox(0,0){$k+1$}}%
% DOT 2 0 3 0
% 5 3400 399 3400 499 3400 599 3400 1199 3400 1199
% 
\special{pn 8}%
\special{sh 1}%
\special{ar 3347 393 10 10 0  6.28318530717959E+0000}%
\special{sh 1}%
\special{ar 3347 492 10 10 0  6.28318530717959E+0000}%
\special{sh 1}%
\special{ar 3347 590 10 10 0  6.28318530717959E+0000}%
\special{sh 1}%
\special{ar 3347 1181 10 10 0  6.28318530717959E+0000}%
\special{sh 1}%
\special{ar 3347 1181 10 10 0  6.28318530717959E+0000}%
% STR 2 0 3 0
% 3 4100 699 4100 799 5 0
% $=$
\put(40.3543,-7.8642){\makebox(0,0){$=$}}%
% LINE 2 0 3 0
% 2 6000 900 6200 600
% 
\special{pn 8}%
\special{pa 5906 886}%
\special{pa 6103 591}%
\special{fp}%
% LINE 2 0 3 0
% 2 6000 900 6200 1200
% 
\special{pn 8}%
\special{pa 5906 886}%
\special{pa 6103 1182}%
\special{fp}%
% LINE 2 0 3 0
% 2 6000 900 6200 680
% 
\special{pn 8}%
\special{pa 5906 886}%
\special{pa 6103 670}%
\special{fp}%
% LINE 2 0 3 0
% 2 6000 900 6200 760
% 
\special{pn 8}%
\special{pa 5906 886}%
\special{pa 6103 749}%
\special{fp}%
% LINE 2 2 3 0
% 2 6150 860 6150 1040
% 
\special{pn 8}%
\special{pa 6054 847}%
\special{pa 6054 1024}%
\special{dt 0.045}%
% DOT 2 0 3 0
% 4 6200 600 6200 680 6200 760 6200 760
% 
\special{pn 8}%
\special{sh 1}%
\special{ar 6103 591 10 10 0  6.28318530717959E+0000}%
\special{sh 1}%
\special{ar 6103 670 10 10 0  6.28318530717959E+0000}%
\special{sh 1}%
\special{ar 6103 749 10 10 0  6.28318530717959E+0000}%
\special{sh 1}%
\special{ar 6103 749 10 10 0  6.28318530717959E+0000}%
% DOT 2 0 3 0
% 2 6200 1200 6200 1200
% 
\special{pn 8}%
\special{sh 1}%
\special{ar 6103 1182 10 10 0  6.28318530717959E+0000}%
\special{sh 1}%
\special{ar 6103 1182 10 10 0  6.28318530717959E+0000}%
% LINE 2 0 3 0
% 2 6000 900 5900 900
% 
\special{pn 8}%
\special{pa 5906 886}%
\special{pa 5808 886}%
\special{fp}%
% LINE 2 0 3 0
% 2 5900 400 6200 400
% 
\special{pn 8}%
\special{pa 5808 394}%
\special{pa 6103 394}%
\special{fp}%
% CIRCLE 2 0 3 0
% 4 5900 650 5900 900 5900 400 5900 900
% 
\special{pn 8}%
\special{ar 5808 640 247 247  1.5707963 4.7123890}%
% LINE 2 0 3 0
% 2 5650 650 5550 650
% 
\special{pn 8}%
\special{pa 5562 640}%
\special{pa 5463 640}%
\special{fp}%
% CIRCLE 2 0 3 0
% 4 5530 650 5550 650 5550 650 5550 650
% 
\special{pn 8}%
\special{ar 5443 640 20 20  0.0000000 6.2831853}%
% DOT 2 0 3 0
% 2 6200 400 6200 400
% 
\special{pn 8}%
\special{sh 1}%
\special{ar 6103 394 10 10 0  6.28318530717959E+0000}%
\special{sh 1}%
\special{ar 6103 394 10 10 0  6.28318530717959E+0000}%
% DOT 0 0 3 0
% 2 6000 900 6000 900
% 
\special{pn 20}%
\special{sh 1}%
\special{ar 5906 886 10 10 0  6.28318530717959E+0000}%
\special{sh 1}%
\special{ar 5906 886 10 10 0  6.28318530717959E+0000}%
% CIRCLE 2 0 3 0
% 4 6250 1200 6200 1200 6260 1250 6300 1200
% 
\special{pn 8}%
\special{ar 6152 1182 50 50  6.2831853 6.2831853}%
\special{ar 6152 1182 50 50  0.0000000 1.3734008}%
% CIRCLE 2 0 3 0
% 4 6250 600 6200 600 6300 600 6260 550
% 
\special{pn 8}%
\special{ar 6152 591 50 50  4.9097845 6.2831853}%
% CIRCLE 2 0 3 0
% 4 6350 950 6400 950 6340 900 6300 950
% 
\special{pn 8}%
\special{ar 6251 936 50 50  3.1415927 4.5149934}%
% CIRCLE 2 0 3 0
% 4 6350 852 6400 852 6300 852 6340 902
% 
\special{pn 8}%
\special{ar 6251 839 50 50  1.7681919 3.1415927}%
% LINE 2 0 3 0
% 2 6300 1200 6300 950
% 
\special{pn 8}%
\special{pa 6201 1182}%
\special{pa 6201 936}%
\special{fp}%
% LINE 2 0 3 0
% 2 6300 600 6300 850
% 
\special{pn 8}%
\special{pa 6201 591}%
\special{pa 6201 837}%
\special{fp}%
% STR 2 0 3 0
% 3 6500 800 6500 900 5 0
% $k$
\put(63.9764,-8.8583){\makebox(0,0){$k$}}%
% LINE 2 0 3 0
% 2 1705 904 1905 673
% 
\special{pn 8}%
\special{pa 1679 890}%
\special{pa 1876 663}%
\special{fp}%
% LINE 2 0 3 0
% 2 1705 904 1905 750
% 
\special{pn 8}%
\special{pa 1679 890}%
\special{pa 1876 739}%
\special{fp}%
% LINE 2 0 3 0
% 2 1705 904 1905 1211
% 
\special{pn 8}%
\special{pa 1679 890}%
\special{pa 1876 1192}%
\special{fp}%
% LINE 2 2 3 0
% 2 1855 866 1855 1019
% 
\special{pn 8}%
\special{pa 1826 853}%
\special{pa 1826 1003}%
\special{dt 0.045}%
% DOT 0 0 3 0
% 2 1705 904 1705 904
% 
\special{pn 20}%
\special{sh 1}%
\special{ar 1679 890 10 10 0  6.28318530717959E+0000}%
\special{sh 1}%
\special{ar 1679 890 10 10 0  6.28318530717959E+0000}%
% ELLIPSE 2 0 3 0
% 4 1956 597 2006 635 2006 597 1966 558
% 
\special{pn 8}%
\special{ar 1926 588 50 38  4.9060110 6.2831853}%
% ELLIPSE 2 0 3 0
% 4 1956 1211 2006 1250 1966 1250 2006 1211
% 
\special{pn 8}%
\special{ar 1926 1192 50 39  6.2831853 6.2831853}%
\special{ar 1926 1192 50 39  0.0000000 1.3734008}%
% ELLIPSE 2 0 3 0
% 4 2056 866 2106 904 2006 866 2056 904
% 
\special{pn 8}%
\special{ar 2024 853 50 38  1.5707963 3.1415927}%
% ELLIPSE 2 0 3 0
% 4 2056 942 2106 981 2056 904 2006 942
% 
\special{pn 8}%
\special{ar 2024 928 50 39  3.1415927 4.7123890}%
% LINE 2 0 3 0
% 2 2006 597 2006 866
% 
\special{pn 8}%
\special{pa 1975 588}%
\special{pa 1975 853}%
\special{fp}%
% LINE 2 0 3 0
% 2 2006 1211 2006 942
% 
\special{pn 8}%
\special{pa 1975 1192}%
\special{pa 1975 928}%
\special{fp}%
% STR 2 0 3 0
% 3 2256 827 2256 904 5 0
% $k+1$
\put(22.2047,-8.8976){\makebox(0,0){$k+1$}}%
% DOT 2 0 3 0
% 5 1905 597 1905 673 1905 750 1905 1211 1905 1211
% 
\special{pn 8}%
\special{sh 1}%
\special{ar 1876 588 10 10 0  6.28318530717959E+0000}%
\special{sh 1}%
\special{ar 1876 663 10 10 0  6.28318530717959E+0000}%
\special{sh 1}%
\special{ar 1876 739 10 10 0  6.28318530717959E+0000}%
\special{sh 1}%
\special{ar 1876 1192 10 10 0  6.28318530717959E+0000}%
\special{sh 1}%
\special{ar 1876 1192 10 10 0  6.28318530717959E+0000}%
% LINE 2 0 3 0
% 2 200 800 400 800
% 
\special{pn 8}%
\special{pa 197 788}%
\special{pa 394 788}%
\special{fp}%
% LINE 2 0 3 0
% 2 400 800 600 400
% 
\special{pn 8}%
\special{pa 394 788}%
\special{pa 591 394}%
\special{fp}%
% LINE 2 0 3 0
% 2 400 800 600 500
% 
\special{pn 8}%
\special{pa 394 788}%
\special{pa 591 493}%
\special{fp}%
% LINE 2 0 3 0
% 2 400 800 600 600
% 
\special{pn 8}%
\special{pa 394 788}%
\special{pa 591 591}%
\special{fp}%
% LINE 2 0 3 0
% 2 400 800 600 1200
% 
\special{pn 8}%
\special{pa 394 788}%
\special{pa 591 1182}%
\special{fp}%
% LINE 2 2 3 0
% 2 550 750 550 950
% 
\special{pn 8}%
\special{pa 542 739}%
\special{pa 542 936}%
\special{dt 0.045}%
% DOT 0 0 3 0
% 2 400 800 400 800
% 
\special{pn 20}%
\special{sh 1}%
\special{ar 394 788 10 10 0  6.28318530717959E+0000}%
\special{sh 1}%
\special{ar 394 788 10 10 0  6.28318530717959E+0000}%
% CIRCLE 2 0 3 0
% 4 640 399 590 399 690 399 650 349
% 
\special{pn 8}%
\special{ar 630 393 50 50  4.9097845 6.2831853}%
% CIRCLE 2 0 3 0
% 4 640 1199 590 1199 650 1249 690 1199
% 
\special{pn 8}%
\special{ar 630 1181 50 50  6.2831853 6.2831853}%
\special{ar 630 1181 50 50  0.0000000 1.3734008}%
% CIRCLE 2 0 3 0
% 4 740 749 740 799 690 749 740 799
% 
\special{pn 8}%
\special{ar 729 738 50 50  1.5707963 3.1415927}%
% CIRCLE 2 0 3 0
% 4 740 849 740 799 740 799 690 849
% 
\special{pn 8}%
\special{ar 729 836 50 50  3.1415927 4.7123890}%
% LINE 2 0 3 0
% 2 690 399 690 749
% 
\special{pn 8}%
\special{pa 680 393}%
\special{pa 680 738}%
\special{fp}%
% LINE 2 0 3 0
% 2 690 1199 690 849
% 
\special{pn 8}%
\special{pa 680 1181}%
\special{pa 680 836}%
\special{fp}%
% STR 2 0 3 0
% 3 840 699 840 799 5 0
% $k$
\put(8.2677,-7.8642){\makebox(0,0){$k$}}%
% DOT 2 0 3 0
% 5 600 400 600 500 600 600 600 1200 590 1200
% 
\special{pn 8}%
\special{sh 1}%
\special{ar 591 394 10 10 0  6.28318530717959E+0000}%
\special{sh 1}%
\special{ar 591 493 10 10 0  6.28318530717959E+0000}%
\special{sh 1}%
\special{ar 591 591 10 10 0  6.28318530717959E+0000}%
\special{sh 1}%
\special{ar 591 1182 10 10 0  6.28318530717959E+0000}%
\special{sh 1}%
\special{ar 581 1182 10 10 0  6.28318530717959E+0000}%
% CIRCLE 2 0 3 0
% 4 1905 500 1905 600 1905 600 1905 400
% 
\special{pn 8}%
\special{ar 1876 493 99 99  4.7123890 6.2831853}%
\special{ar 1876 493 99 99  0.0000000 1.5707963}%
% LINE 2 0 3 0
% 2 2005 500 2105 500
% 
\special{pn 8}%
\special{pa 1974 493}%
\special{pa 2072 493}%
\special{fp}%
% CIRCLE 2 0 3 0
% 4 2125 500 2105 500 2105 500 2105 500
% 
\special{pn 8}%
\special{ar 2092 493 20 20  0.0000000 6.2831853}%
% SPLINE 2 0 3 0
% 4 1705 900 1875 610 1905 600 1905 600
% 
\special{pn 8}%
\special{pa 1679 886}%
\special{pa 1688 851}%
\special{pa 1698 816}%
\special{pa 1709 783}%
\special{pa 1721 751}%
\special{pa 1734 720}%
\special{pa 1749 691}%
\special{pa 1765 666}%
\special{pa 1785 642}%
\special{pa 1807 623}%
\special{pa 1832 607}%
\special{pa 1861 596}%
\special{pa 1876 591}%
\special{sp}%
% STR 2 0 3 0
% 3 1255 723 1255 800 5 0
% $=$
\put(12.3524,-7.8740){\makebox(0,0){$=$}}%
% STR 2 0 3 0
% 3 2750 900 2750 1000 5 0
% $,$
\put(27.0669,-9.8425){\makebox(0,0){$,$}}%
% STR 2 0 3 0
% 3 4700 1350 4700 1450 5 0
% $\la V_{k+1}|_{1\cdots}=\la V_{k+1}|_{b\cdots}\la\omega|_{1a}|\omega\ra_{ab}=\la\omega|_{12}|m_k(\ ,\cdots,\ )\ra_2$
\put(46.2598,-14.2717){\makebox(0,0){$\la V_{k+1}|_{1\cdots}=\la V_{k+1}|_{b\cdots}\la\omega|_{1a}|\omega\ra_{ab}=\la\omega|_{12}|m_k(\ ,\cdots,\ )\ra_2$}}%
% STR 2 0 3 0
% 3 1250 1350 1250 1450 5 0
% $|m_k\ra_1=\la V_{k+1}|_{a\cdots}|\omega\ra_{1a}$
\put(12.3031,-14.2717){\makebox(0,0){$|m_k\ra_1=\la V_{k+1}|_{a\cdots}|\omega\ra_{1a}$}}%
% LINE 2 0 3 0
% 2 1405 400 1905 400
% 
\special{pn 8}%
\special{pa 1383 394}%
\special{pa 1876 394}%
\special{fp}%
% LINE 2 0 3 0
% 2 4400 958 4601 775
% 
\special{pn 8}%
\special{pa 4331 943}%
\special{pa 4529 763}%
\special{fp}%
% LINE 2 0 3 0
% 2 4400 958 4601 835
% 
\special{pn 8}%
\special{pa 4331 943}%
\special{pa 4529 822}%
\special{fp}%
% LINE 2 0 3 0
% 2 4400 958 4601 1201
% 
\special{pn 8}%
\special{pa 4331 943}%
\special{pa 4529 1183}%
\special{fp}%
% LINE 2 2 3 0
% 2 4551 926 4551 1048
% 
\special{pn 8}%
\special{pa 4480 912}%
\special{pa 4480 1032}%
\special{dt 0.045}%
% DOT 0 0 3 0
% 2 4400 958 4400 958
% 
\special{pn 20}%
\special{sh 1}%
\special{ar 4331 943 10 10 0  6.28318530717959E+0000}%
\special{sh 1}%
\special{ar 4331 943 10 10 0  6.28318530717959E+0000}%
% ELLIPSE 2 0 3 0
% 4 4653 714 4704 744 4704 714 4663 683
% 
\special{pn 8}%
\special{ar 4580 703 51 30  4.8988759 6.2831853}%
% ELLIPSE 2 0 3 0
% 4 4653 1201 4704 1231 4663 1231 4704 1201
% 
\special{pn 8}%
\special{ar 4580 1183 51 30  6.2831853 6.2831853}%
\special{ar 4580 1183 51 30  0.0000000 1.3771743}%
% ELLIPSE 2 0 3 0
% 4 4754 926 4805 958 4704 926 4754 958
% 
\special{pn 8}%
\special{ar 4680 912 51 32  1.5707963 3.1415927}%
% ELLIPSE 2 0 3 0
% 4 4754 987 4805 1017 4754 958 4704 987
% 
\special{pn 8}%
\special{ar 4680 972 51 30  3.1415927 4.7123890}%
% LINE 2 0 3 0
% 2 4704 714 4704 926
% 
\special{pn 8}%
\special{pa 4630 703}%
\special{pa 4630 912}%
\special{fp}%
% LINE 2 0 3 0
% 2 4704 1201 4704 987
% 
\special{pn 8}%
\special{pa 4630 1183}%
\special{pa 4630 972}%
\special{fp}%
% STR 2 0 3 0
% 3 4956 896 4956 958 5 0
% $k+1$
\put(48.7795,-9.4291){\makebox(0,0){$k+1$}}%
% DOT 2 0 3 0
% 5 4601 714 4601 775 4601 835 4601 1201 4601 1201
% 
\special{pn 8}%
\special{sh 1}%
\special{ar 4529 703 10 10 0  6.28318530717959E+0000}%
\special{sh 1}%
\special{ar 4529 763 10 10 0  6.28318530717959E+0000}%
\special{sh 1}%
\special{ar 4529 822 10 10 0  6.28318530717959E+0000}%
\special{sh 1}%
\special{ar 4529 1183 10 10 0  6.28318530717959E+0000}%
\special{sh 1}%
\special{ar 4529 1183 10 10 0  6.28318530717959E+0000}%
% ELLIPSE 2 0 3 0
% 4 4601 637 4703 716 4601 716 4601 558
% 
\special{pn 8}%
\special{ar 4529 627 101 78  4.7123890 6.2831853}%
\special{ar 4529 627 101 78  0.0000000 1.5707963}%
% LINE 2 0 3 0
% 2 4703 637 4804 637
% 
\special{pn 8}%
\special{pa 4629 627}%
\special{pa 4729 627}%
\special{fp}%
% SPLINE 2 0 3 0
% 4 4400 955 4571 724 4601 716 4601 716
% 
\special{pn 8}%
\special{pa 4331 940}%
\special{pa 4345 907}%
\special{pa 4359 876}%
\special{pa 4374 844}%
\special{pa 4389 814}%
\special{pa 4407 787}%
\special{pa 4426 762}%
\special{pa 4448 742}%
\special{pa 4472 725}%
\special{pa 4501 713}%
\special{pa 4529 705}%
\special{sp}%
% ELLIPSE 2 0 3 0
% 4 4501 479 4601 558 4501 400 4501 558
% 
\special{pn 8}%
\special{ar 4431 472 99 78  1.5707963 4.7123890}%
% LINE 2 0 3 0
% 2 4501 558 4601 558
% 
\special{pn 8}%
\special{pa 4431 550}%
\special{pa 4529 550}%
\special{fp}%
% LINE 2 0 3 0
% 2 4501 400 4601 400
% 
\special{pn 8}%
\special{pa 4431 394}%
\special{pa 4529 394}%
\special{fp}%
% DOT 2 0 3 0
% 2 4601 400 4601 400
% 
\special{pn 8}%
\special{sh 1}%
\special{ar 4529 394 10 10 0  6.28318530717959E+0000}%
\special{sh 1}%
\special{ar 4529 394 10 10 0  6.28318530717959E+0000}%
% LINE 2 0 3 0
% 2 4400 479 4300 479
% 
\special{pn 8}%
\special{pa 4331 472}%
\special{pa 4233 472}%
\special{fp}%
% CIRCLE 2 0 3 0
% 4 4280 480 4300 480 4300 480 4300 480
% 
\special{pn 8}%
\special{ar 4213 473 20 20  0.0000000 6.2831853}%
% CIRCLE 2 0 3 0
% 4 4820 640 4800 640 4800 640 4810 640
% 
\special{pn 8}%
\special{ar 4745 630 20 20  0.0000000 6.2831853}%
% STR 2 0 3 0
% 3 5300 700 5300 800 5 0
% $=$
\put(52.1654,-7.8740){\makebox(0,0){$=$}}%
% STR 2 0 3 0
% 3 1250 1650 1250 1750 5 0
% $c^j_{i_1\cdots i_k}=(-1)^{\eb_l}\omega^{jl}\V_{li_1\cdots i_k}$
\put(12.3031,-17.2244){\makebox(0,0){$c^j_{i_1\cdots i_k}=(-1)^{\eb_l}\omega^{jl}\V_{li_1\cdots i_k}$}}%
% STR 2 0 3 0
% 3 4700 1650 4700 1750 5 0
% $\V_{i_1\cdots i_{k+1}}=\omega_{i_1j}\omega^{jl}\V_{li_2\cdots i_{k+1}}=(-1)^{\eb_{i_1}}\omega_{i_1l}c^l_{i_2\cdots i_{k+1}}$
\put(46.2598,-17.2244){\makebox(0,0){$\V_{i_1\cdots i_{k+1}}=\omega_{i_1j}\omega^{jl}\V_{li_2\cdots i_{k+1}}=(-1)^{\eb_{i_1}}\omega_{i_1l}c^l_{i_2\cdots i_{k+1}}$}}%
\end{picture}%
\\
\vspace*{0.1cm}\\
where in the second equality in the right hand side, 
the identity (\ref{poisson}) is written as 

\vspace*{0.5cm}

%\begin{figure}[h]
 \hspace*{-4.0cm}
%WinTpicVersion3.08
\unitlength 0.1in
\begin{picture}( 67.7657,  4.2815)(-20.5217, -7.8740)
% CIRCLE 2 0 3 0
% 4 3300 700 3300 600 3310 800 3300 600
% 
\special{pn 8}%
\special{ar 3249 689 99 99  4.7123890 6.2831853}%
\special{ar 3249 689 99 99  0.0000000 1.4711277}%
% CIRCLE 2 0 3 0
% 4 3200 500 3200 400 3200 400 3190 600
% 
\special{pn 8}%
\special{ar 3150 493 99 99  1.6704650 4.7123890}%
% LINE 2 0 3 0
% 2 3200 600 3300 600
% 
\special{pn 8}%
\special{pa 3150 591}%
\special{pa 3249 591}%
\special{fp}%
% LINE 2 0 3 0
% 2 3000 800 3300 800
% 
\special{pn 8}%
\special{pa 2953 788}%
\special{pa 3249 788}%
\special{fp}%
% LINE 2 0 3 0
% 2 3200 400 3500 400
% 
\special{pn 8}%
\special{pa 3150 394}%
\special{pa 3445 394}%
\special{fp}%
% DOT 2 0 3 0
% 3 3250 600 3500 400 3500 400
% 
\special{pn 8}%
\special{sh 1}%
\special{ar 3199 591 10 10 0  6.28318530717959E+0000}%
\special{sh 1}%
\special{ar 3445 394 10 10 0  6.28318530717959E+0000}%
\special{sh 1}%
\special{ar 3445 394 10 10 0  6.28318530717959E+0000}%
% STR 2 0 3 0
% 3 3800 500 3800 600 5 0
% $=$
\put(37.4016,-5.9055){\makebox(0,0){$=$}}%
% LINE 2 0 3 0
% 2 4200 600 4800 600
% 
\special{pn 8}%
\special{pa 4134 591}%
\special{pa 4725 591}%
\special{fp}%
% DOT 2 0 3 0
% 2 4800 600 4800 600
% 
\special{pn 8}%
\special{sh 1}%
\special{ar 4725 591 10 10 0  6.28318530717959E+0000}%
\special{sh 1}%
\special{ar 4725 591 10 10 0  6.28318530717959E+0000}%
% LINE 2 0 3 0
% 2 3400 700 3500 700
% 
\special{pn 8}%
\special{pa 3347 689}%
\special{pa 3445 689}%
\special{fp}%
% LINE 2 0 3 0
% 2 3100 500 3000 500
% 
\special{pn 8}%
\special{pa 3052 493}%
\special{pa 2953 493}%
\special{fp}%
% CIRCLE 2 0 3 0
% 4 2980 500 3000 500 3000 500 3000 500
% 
\special{pn 8}%
\special{ar 2934 493 20 20  0.0000000 6.2831853}%
% CIRCLE 2 0 3 0
% 4 3520 700 3500 700 3500 700 3500 700
% 
\special{pn 8}%
\special{ar 3465 689 20 20  0.0000000 6.2831853}%
% STR 2 0 3 0
% 3 2500 500 2500 600 5 0
% $\Lglraw$
\put(24.6063,-5.9055){\makebox(0,0){$\Lglraw$}}%
% STR 2 0 3 0
% 3 1200 350 1200 450 5 0
% $\la\omega|_{\cdot a}|\omega\ra_{a\cdot}= _\cdot\hspace*{-0.1cm}\1_\cdot$
\put(11.8110,-4.4291){\makebox(0,0){$\la\omega|_{\cdot a}|\omega\ra_{a\cdot}= _\cdot\hspace*{-0.1cm}\1_\cdot$}}%
% STR 2 0 3 0
% 3 1200 650 1200 750 5 0
% $\omega_{ij}\omega^{jk}=\delta_i^k$
\put(11.8110,-7.3819){\makebox(0,0){$\omega_{ij}\omega^{jk}=\delta_i^k$}}%
\end{picture}%
\ .
%\end{figure}

\vspace*{0.5cm}

Note that $\la\omega|$ and $|\omega\ra$ 
correspond to lowering and raising indices with $\omega_{ij}$ 
and $\omega^{ij}$ in their component language, 
and which correspond to reversing the outgoing lines in 
these graphs.

 \section{$A_\infty$-morphism and field transformation}
\label{sec:5}

For classical open SFT, an $A_\infty$-morphism corresponds 
to a field transformation and the Maurer-Cartan equation is the equation 
of motion of SFT. 
It has been explained in section \ref{ssec:23} that 
$A_\infty$-morphisms preserve the Maurer-Cartan equations 
for $A_\infty$-algebras. 
This means that if an $A_\infty$-morphism is given, 
we get a field transformation which preserves the equation of motion. 
Generally when two $A_\infty$-algebras are given, 
it is difficult to construct 
an $A_\infty$-morphism between them. 
However it is known that 
any $A_\infty$-algebras (with nonvanishing $m_1$) 
have a canonical $A_\infty$-quasi-isomorphism\cite{K,K2,KS} between the 
original $A_\infty$-algebra and another canonical $A_\infty$-algebra. 
The $A_\infty$-quasi-isomorphism can be constructed in a canonical way in 
terms of Feynman graphs\cite{KS}, and it fits for SFT very much. 
Here will explain the way of constructing the 
canonical $A_\infty$-quasi-isomorphism in terms of the Feynman graphs 
in subsection \ref{ssec:51}, 
clarify the identification between the Feynman graphs and that in SFTs 
in subsection \ref{ssec:5gf}. 
The arguments in subsection \ref{ssec:51} is then applied to SFT 
in subsection \ref{ssec:53} 
and we show that the canonical $A_\infty$-algebra is nothing but the 
on-shell S-matrix elements (Lem.\ref{lem:main}). 
From this result 
it is shown in (Thm.\ref{thm:main}) that 
every SFT constructed as explained in section \ref{sec:3} are 
quasi-isomorphic, and the quasi-isomorphism between those can be 
constructed in subsection \ref{ssec:main}. 
All the arguments are applicable also for classical closed SFT 
\ie $L_\infty$-algebras.

 \subsection{The minimal model theorem}
\label{ssec:51}

Let $(\cH,\m)$ be an $A_\infty$-algebra with $m_1\neq 0$. 
As mentioned above, 
for any $(\cH,\m)$, a canonical $A_\infty$-quasi-isomorphism $\ti\cF^p$ 
from another canonical $A_\infty$-algebra $(\cH^p,\ti\m^p)$ to 
the $A_\infty$-algebra $(\cH,\m)$ exists\cite{KS}. This is called 
{\it minimal model theorem}
\footnote{This naming has nothing to do with minimal model in the context of 
two-dimensional field theory directly. }. 

Here we construct the $A_\infty$-morphism $\{\ti{f}^p_k\}$ and 
$A_\infty$-structure $\{\ti{m}^p_k\}$ with $k\ge 2$ naturally 
as the problem of finding the solutions for 
the equation of motion (Maurer-Cartan equation) for SFT, and 
prove that the $(\cH^p,\ti\m^p)$ and $\ti\cF$ are indeed an 
$A_\infty$-algebra and a quasi-isomorphism, respectively
\footnote{This explanation of the minimal model theorem from the 
problem of finding the solutions for the Maurer-Cartan equation 
is motivated in the lecture by K.~Fukaya at Inst. of Tech. in Tokyo in 
December, 2000. }.
The procedure of finding the solution is quite natural and standard, 
and so similar procedures can be found in various problems. 
The procedure also relates to the way of finding some classical solutions 
in closed SFT \cite{MS,KZ} (see the next subsection) or 
constructing the tachyon potential\cite{MT} (see {\it tachyon condensation} 
in Discussions).

Consider solving the equation of motion for classical open SFT, 
\begin{equation}
 \sum_{k\ge 1}m_k(\Phi)=0 \label{eomsft}
\end{equation}
with $m_1=Q$. 
For $Q$ the coboundary operator of complex $(\cH, Q)$, 
we give the analogue of the Hodge-Kodaira decomposition
\begin{equation}
 QQ^++Q^+ Q+P=\1\ .
 \label{HKdecomp}
\end{equation}
In this section for simplicity we assume this identity gives just 
the the Hodge-Kodaira decomposition of $\cH$, that is, 
$Q^+$ is the adjoint of $Q$ and $P$ is the projection onto the 
harmonic form $(Ker Q)\cap(Ker Q^+)$. 
$Q^+$ can be regarded as the propagator for the SFT action and also 
plays the role of the gauge fixing. $P$ is related to the projection 
onto the physical states in SFT, so we denote $\cH^p:=P\cH$ in this 
subsection. These will be clarified in the next 
subsection, where the condition that $QQ^+$, $Q^+Q$, and $P$ are 
projections must be relaxed
\footnote{In addition we change the definition of $\cH^p$ after this section 
from $\cH^p:=P\cH$ to $\cH^p=P^p\cH$ where $P^p$ is defined as the precise 
projection onto physical states. }. 
Actually, the condition can be relaxed later in and after 
(Def.\ref{defn:minimal}) 
and only the identity (\ref{HKdecomp}) between $Q$, $Q^+$ and $P$ 
will be required. 

Since the solutions for eq.(\ref{eomsft}) are preserved 
under the gauge transformation 
$\delta_\alpha\Phi=Q(\alpha)+m_2(\alpha,\Phi)+m_2(\Phi,\alpha)+\cdots$, 
we will find the gauge fixed solutions $Q^+\Phi=0$. 
Here we assume that $\Phi$ is sufficiently `small', then 
the solution is almost the solution of $Q(\Phi)=0$. 
Express $\Phi$ as $\Phi=\Phi^p+\Phi^u$ 
where $\Phi^p\in \cH^p$ and $\Phi^u\in Q^+ Q\cH$. 
As will be explained below, $\Phi^u$ can be solved recursively 
with the power of $\Phi^p$. 
Because here we regard that $\Phi^p$ is `small', 
one can define a degree by the power of $\Phi^p$. 
Substituting $\Phi=\Phi^p+\Phi^u$ in e.o.m (\ref{eomsft}) leads 
\begin{equation}
 Q(\Phi^u)+\sum_{k\ge 2}m_k(\Phi^p+\Phi^u)=0\ ,\label{eom2}
\end{equation}
and acting $Q^+$ to both sides of this equation yields 
\begin{equation}
  \Phi^u=-\sum_{k\ge 2}Q^+ m_k(\Phi^p+\Phi^u)\ . \label{eom3}
\end{equation}
Here we get the $\Phi^u$ recursively by eq.(\ref{eom3}). 
However not all $\Phi=\Phi^p+\Phi^u$ expressed in terms of $\Phi^p$ give 
the solution of eq.(\ref{eomsft}) because eq.(\ref{eom3}) is derived from 
{\it $Q^+$ acting} eq.(\ref{eomsft}). 
In order to find $\Phi^u$ which is the solution of eq.(\ref{eomsft}), 
we substitute eq.(\ref{eom3}) to e.o.m(\ref{eomsft}) once again, 
\begin{equation}
 \begin{split}
 0&=Q(\Phi^p+\Phi^u)+ \sum_{k\ge 2}m_k(\Phi)\\
  &=(Q^+ Q+P-1)\sum_{k\ge 2}m_k(\Phi)+\sum_{k\ge 2}m_k(\Phi)\\
  &=Q^+ Q\sum_{k\ge 2}m_k(\Phi) +\sum_{k\ge 2}P m_k(\Phi)
 \end{split}
\label{eom4}
\end{equation}
and we can get a condition (obstruction) for $\Phi^p$. 
The first term in the third line of eq.(\ref{eom4}) vanishes due to 
e.o.m(\ref{eomsft}) 
because $Q\sum_{k\ge 2}m_k(\Phi)=-QQ(\Phi)=0$, 
and the condition for $\Phi^p$ is derived as 
\begin{equation}
 \sum_{k\ge 2}P m_k(\Phi^p+\Phi^u)=0\ .\label{mceq1}
\end{equation}
The above $\Phi^u$ can be represented recursively in terms of $\Phi^p$ 
by eq.(\ref{eom3}). 
This equation can be regarded as the Maurer-Cartan equation on $\cH^p$. 

The equation (\ref{eom3}) can be regarded as a nonlinear 
map from $\cH^p$ to $\cH$. 
Here we want to distinct the element of $\cH^p$ with that of $\cH$, 
so we rewrite $\Phi^p\in\cH^p$ as $\ti\Phi^p\in\cH^p$. 
If we write the map defined 
by eq.(\ref{eom3}) recursively as 
\begin{equation}
\Phi:=\ti{f}^p_1(\ti\Phi^p)+\ti{f}^p_2(\ti\Phi^p,\ti\Phi^p)
+\ti{f}^p_3(\ti\Phi^p,\ti\Phi^p,\ti\Phi^p)+\cdots
 \label{eom33}
\end{equation} 
with $\ti{f}^p_1$ the identity map (inclusion map), 
$\ti{f}^p_l(\ti\Phi^p)$ is given by 
connecting tree graphs corresponding to $\{-Q^+ m_k(\cdots)\}$ 
with all possible combination, 
summing up these, and picking up the term involving $l$ powers of 
$\ti\Phi^p$ (see (Def.\ref{defn:minimal}) below). 
Alternatively eq.(\ref{mceq1}) is also 
expressed as an equation for $\ti\Phi^p$. It is obtained by substituting 
eq.(\ref{eom3}) or eq.(\ref{eom33}) into eq.(\ref{mceq1}). 
Let us define $\ti{m}^p_l(\ti\Phi^p), l\ge 2$ 
as the term involving $l$ powers of $\ti\Phi^p$ in eq.(\ref{mceq1}). 
In other words we define them so that eq.(\ref{mceq1}) is rewritten as 
\begin{equation}
 \sum_{k\ge 2}\ti{m}^p_k(\ti\Phi^p)=0\ .\label{mceq2}
\end{equation}
Each $\ti{m}^p_l(\ti\Phi^p)$ for $l\ge 2$ is then given 
in the same way as $\ti{f}^p_l$ 
but replacing $-Q^+$ on the last outgoing line by $P$. 
Here denote the structure $\{\ti{m}^p_k\}_{k\ge 2}$ as $\ti\m^p$. The 
equation (\ref{mceq2}) is regarded as the Maurer-Cartan equation 
on $(\cH^p,\ti\m^p)$. 
Note that as will be proven, the $\ti\m^p$ defines an $A_\infty$-structure 
on $\cH^p$ and the nonlinear map (\ref{eom33}) defines the 
$A_\infty$-quasi-isomorphism from $(\cH^p,\ti\m^p)$ to $(\cH,\m)$. 
Thus the canonical $A_\infty$-algebra $(\cH^p,\ti\m^p)$ and 
the $A_\infty$-quasi-isomorphism $\ti\cF^p$ can be defined naturally 
in the problem of solving the Maurer-Cartan equations 
(the problem of constructing Kuranishi map in mathematical language, 
see (Rem.\ref{rem:inverse})). 
From SFT point of view the above result means that 
if the expectation value of physical states which satisfies 
the Maurer-Cartan equation (\ref{mceq2}) is given, 
the solution of e.o.m for SFT (\ref{eomsft}) 
is obtained by the $A_\infty$-quasi-isomorphism (\ref{eom33}). 
This statement will be explained more precisely in the next subsection.

Mention that as can be seen from eq.(\ref{eom4}) 
if we begin with $\ti\Phi\in\cH$ 
with $Q\ti\Phi\neq 0$ instead of $\ti\Phi^p$, 
the Maurer-Cartan equation (\ref{mceq2}) may be corrected 
by adding the term $\ti{m}_1(\ti\Phi)$ with $\ti{m}_1:=m_1=Q$. 
This case is also considered later (see (Rem.\ref{rem:twoAinfty})).

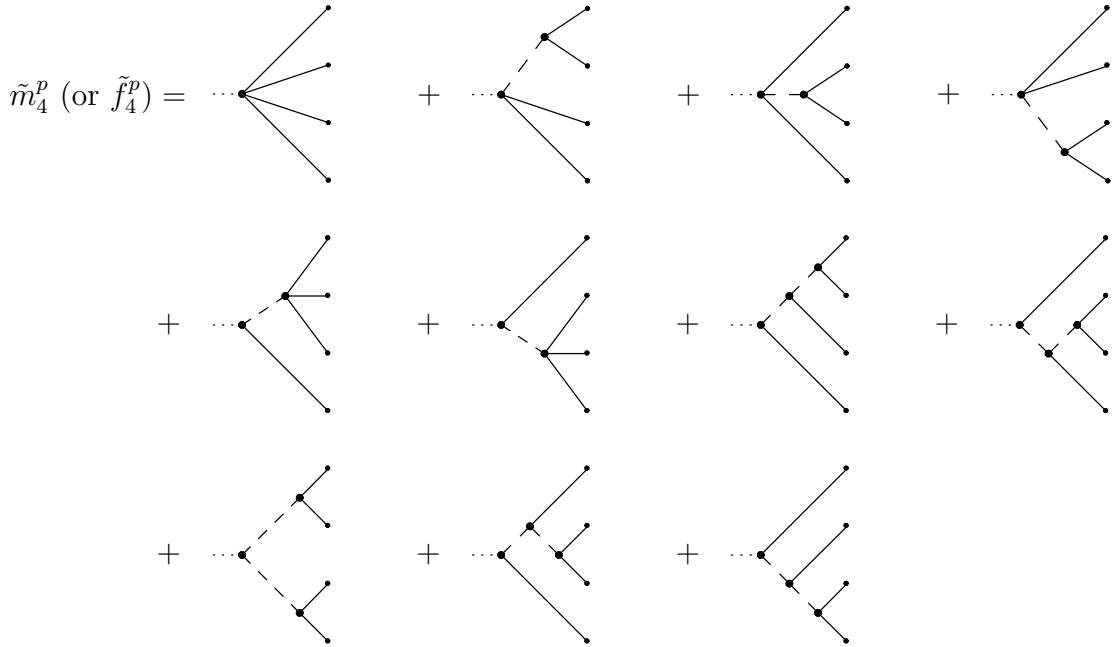
\begin{figure}[h]
 \hspace*{-2.5cm}
%WinTpicVersion3.08
\unitlength 0.1in
\begin{picture}( 69.6850, 33.2087)(-13.6220,-36.8996)
% STR 2 0 3 0
% 3 326 762 326 838 5 0
% $\ti{m}^p_4\ (\mbox{or}\ \ti{f}^p_4)=$
\put(3.2087,-8.2480){\makebox(0,0){$\ti{m}^p_4\ (\mbox{or}\ \ti{f}^p_4)=$}}%
% LINE 2 0 3 0
% 2 4309 2523 3850 2064
% 
\special{pn 8}%
\special{pa 4242 2484}%
\special{pa 3790 2032}%
\special{fp}%
% LINE 2 0 3 0
% 2 4309 2830 3850 3289
% 
\special{pn 8}%
\special{pa 4242 2786}%
\special{pa 3790 3238}%
\special{fp}%
% DOT 2 0 3 0
% 5 1552 375 1552 681 1552 988 1552 1294 1552 1294
% 
\special{pn 8}%
\special{sh 1}%
\special{ar 1528 370 10 10 0  6.28318530717959E+0000}%
\special{sh 1}%
\special{ar 1528 671 10 10 0  6.28318530717959E+0000}%
\special{sh 1}%
\special{ar 1528 973 10 10 0  6.28318530717959E+0000}%
\special{sh 1}%
\special{ar 1528 1274 10 10 0  6.28318530717959E+0000}%
\special{sh 1}%
\special{ar 1528 1274 10 10 0  6.28318530717959E+0000}%
% LINE 2 0 3 0
% 2 1552 375 1092 835
% 
\special{pn 8}%
\special{pa 1528 370}%
\special{pa 1075 822}%
\special{fp}%
% LINE 2 0 3 0
% 2 1552 1294 1092 835
% 
\special{pn 8}%
\special{pa 1528 1274}%
\special{pa 1075 822}%
\special{fp}%
% LINE 2 0 3 0
% 2 4309 1911 4156 1757
% 
\special{pn 8}%
\special{pa 4242 1881}%
\special{pa 4091 1730}%
\special{fp}%
% LINE 2 0 3 0
% 2 1552 3136 1399 2983
% 
\special{pn 8}%
\special{pa 1528 3087}%
\special{pa 1377 2937}%
\special{fp}%
% LINE 2 0 3 0
% 2 4309 2217 4003 1911
% 
\special{pn 8}%
\special{pa 4242 2183}%
\special{pa 3940 1881}%
\special{fp}%
% LINE 2 0 3 0
% 2 1552 3442 1399 3596
% 
\special{pn 8}%
\special{pa 1528 3388}%
\special{pa 1377 3540}%
\special{fp}%
% LINE 2 0 3 0
% 2 4309 3136 4003 3442
% 
\special{pn 8}%
\special{pa 4242 3087}%
\special{pa 3940 3388}%
\special{fp}%
% LINE 2 0 3 0
% 2 4309 3442 4156 3596
% 
\special{pn 8}%
\special{pa 4242 3388}%
\special{pa 4091 3540}%
\special{fp}%
% LINE 2 0 3 0
% 2 1552 681 1092 835
% 
\special{pn 8}%
\special{pa 1528 671}%
\special{pa 1075 822}%
\special{fp}%
% LINE 2 0 3 0
% 2 1552 988 1092 835
% 
\special{pn 8}%
\special{pa 1528 973}%
\special{pa 1075 822}%
\special{fp}%
% LINE 2 0 3 0
% 2 1552 2217 1322 1911
% 
\special{pn 8}%
\special{pa 1528 2183}%
\special{pa 1302 1881}%
\special{fp}%
% LINE 2 0 3 0
% 2 1552 1604 1322 1911
% 
\special{pn 8}%
\special{pa 1528 1579}%
\special{pa 1302 1881}%
\special{fp}%
% LINE 2 0 3 0
% 2 1552 1911 1322 1911
% 
\special{pn 8}%
\special{pa 1528 1881}%
\special{pa 1302 1881}%
\special{fp}%
% LINE 2 0 3 0
% 2 1092 2064 1552 2523
% 
\special{pn 8}%
\special{pa 1075 2032}%
\special{pa 1528 2484}%
\special{fp}%
% LINE 2 1 3 0
% 2 1092 2064 1322 1911
% 
\special{pn 8}%
\special{pa 1075 2032}%
\special{pa 1302 1881}%
\special{da 0.070}%
% LINE 2 0 3 0
% 2 2930 1911 2701 2217
% 
\special{pn 8}%
\special{pa 2884 1881}%
\special{pa 2659 2183}%
\special{fp}%
% LINE 2 0 3 0
% 2 2930 2523 2701 2217
% 
\special{pn 8}%
\special{pa 2884 2484}%
\special{pa 2659 2183}%
\special{fp}%
% LINE 2 0 3 0
% 2 2930 2217 2701 2217
% 
\special{pn 8}%
\special{pa 2884 2183}%
\special{pa 2659 2183}%
\special{fp}%
% LINE 2 0 3 0
% 2 2471 2064 2930 1604
% 
\special{pn 8}%
\special{pa 2433 2032}%
\special{pa 2884 1579}%
\special{fp}%
% LINE 2 1 3 0
% 2 2471 2064 2701 2217
% 
\special{pn 8}%
\special{pa 2433 2032}%
\special{pa 2659 2183}%
\special{da 0.070}%
% LINE 2 2 3 0
% 2 2471 2064 2318 2064
% 
\special{pn 8}%
\special{pa 2433 2032}%
\special{pa 2282 2032}%
\special{dt 0.045}%
% LINE 2 2 3 0
% 2 1092 2064 939 2064
% 
\special{pn 8}%
\special{pa 1075 2032}%
\special{pa 925 2032}%
\special{dt 0.045}%
% LINE 2 2 3 0
% 2 1092 3289 939 3289
% 
\special{pn 8}%
\special{pa 1075 3238}%
\special{pa 925 3238}%
\special{dt 0.045}%
% LINE 2 2 3 0
% 2 1092 835 939 835
% 
\special{pn 8}%
\special{pa 1075 822}%
\special{pa 925 822}%
\special{dt 0.045}%
% LINE 2 2 3 0
% 2 3850 3289 3696 3289
% 
\special{pn 8}%
\special{pa 3790 3238}%
\special{pa 3638 3238}%
\special{dt 0.045}%
% LINE 2 2 3 0
% 2 3850 2064 3696 2064
% 
\special{pn 8}%
\special{pa 3790 2032}%
\special{pa 3638 2032}%
\special{dt 0.045}%
% STR 2 0 3 0
% 3 2088 762 2088 838 5 0
% $+$
\put(20.5512,-8.2480){\makebox(0,0){$+$}}%
% STR 2 0 3 0
% 3 3467 762 3467 838 5 0
% $+$
\put(34.1240,-8.2480){\makebox(0,0){$+$}}%
% LINE 2 0 3 0
% 2 4309 1604 4156 1757
% 
\special{pn 8}%
\special{pa 4242 1579}%
\special{pa 4091 1730}%
\special{fp}%
% LINE 2 1 3 0
% 2 4156 1757 4003 1911
% 
\special{pn 8}%
\special{pa 4091 1730}%
\special{pa 3940 1881}%
\special{da 0.070}%
% LINE 2 1 3 0
% 2 4003 1911 3850 2064
% 
\special{pn 8}%
\special{pa 3940 1881}%
\special{pa 3790 2032}%
\special{da 0.070}%
% LINE 2 0 3 0
% 2 1552 2830 1399 2983
% 
\special{pn 8}%
\special{pa 1528 2786}%
\special{pa 1377 2937}%
\special{fp}%
% LINE 2 0 3 0
% 2 1552 3749 1399 3596
% 
\special{pn 8}%
\special{pa 1528 3690}%
\special{pa 1377 3540}%
\special{fp}%
% LINE 2 0 3 0
% 2 4309 3749 4156 3596
% 
\special{pn 8}%
\special{pa 4242 3690}%
\special{pa 4091 3540}%
\special{fp}%
% LINE 2 1 3 0
% 2 1399 3596 1092 3289
% 
\special{pn 8}%
\special{pa 1377 3540}%
\special{pa 1075 3238}%
\special{da 0.070}%
% LINE 2 1 3 0
% 2 1399 2983 1092 3289
% 
\special{pn 8}%
\special{pa 1377 2937}%
\special{pa 1075 3238}%
\special{da 0.070}%
% LINE 2 1 3 0
% 2 3850 3289 4003 3442
% 
\special{pn 8}%
\special{pa 3790 3238}%
\special{pa 3940 3388}%
\special{da 0.070}%
% LINE 2 1 3 0
% 2 4003 3442 4156 3596
% 
\special{pn 8}%
\special{pa 3940 3388}%
\special{pa 4091 3540}%
\special{da 0.070}%
% DOT 0 1 3 0
% 2 1092 835 1092 835
% 
\special{pn 20}%
\special{sh 1}%
\special{ar 1075 822 10 10 0  6.28318530717959E+0000}%
\special{sh 1}%
\special{ar 1075 822 10 10 0  6.28318530717959E+0000}%
% DOT 0 0 3 0
% 2 1322 1911 1322 1911
% 
\special{pn 20}%
\special{sh 1}%
\special{ar 1302 1881 10 10 0  6.28318530717959E+0000}%
\special{sh 1}%
\special{ar 1302 1881 10 10 0  6.28318530717959E+0000}%
% DOT 0 0 3 0
% 2 1092 2064 1092 2064
% 
\special{pn 20}%
\special{sh 1}%
\special{ar 1075 2032 10 10 0  6.28318530717959E+0000}%
\special{sh 1}%
\special{ar 1075 2032 10 10 0  6.28318530717959E+0000}%
% DOT 0 0 3 0
% 2 2471 2064 2471 2064
% 
\special{pn 20}%
\special{sh 1}%
\special{ar 2433 2032 10 10 0  6.28318530717959E+0000}%
\special{sh 1}%
\special{ar 2433 2032 10 10 0  6.28318530717959E+0000}%
% DOT 0 0 3 0
% 2 2701 2217 2701 2217
% 
\special{pn 20}%
\special{sh 1}%
\special{ar 2659 2183 10 10 0  6.28318530717959E+0000}%
\special{sh 1}%
\special{ar 2659 2183 10 10 0  6.28318530717959E+0000}%
% DOT 0 0 3 0
% 2 3850 3289 3850 3289
% 
\special{pn 20}%
\special{sh 1}%
\special{ar 3790 3238 10 10 0  6.28318530717959E+0000}%
\special{sh 1}%
\special{ar 3790 3238 10 10 0  6.28318530717959E+0000}%
% DOT 0 0 3 0
% 2 1092 3289 1092 3289
% 
\special{pn 20}%
\special{sh 1}%
\special{ar 1075 3238 10 10 0  6.28318530717959E+0000}%
\special{sh 1}%
\special{ar 1075 3238 10 10 0  6.28318530717959E+0000}%
% DOT 0 0 3 0
% 2 3850 2064 3850 2064
% 
\special{pn 20}%
\special{sh 1}%
\special{ar 3790 2032 10 10 0  6.28318530717959E+0000}%
\special{sh 1}%
\special{ar 3790 2032 10 10 0  6.28318530717959E+0000}%
% DOT 0 0 3 0
% 2 4156 1757 4156 1757
% 
\special{pn 20}%
\special{sh 1}%
\special{ar 4091 1730 10 10 0  6.28318530717959E+0000}%
\special{sh 1}%
\special{ar 4091 1730 10 10 0  6.28318530717959E+0000}%
% DOT 0 0 3 0
% 2 4003 1911 4003 1911
% 
\special{pn 20}%
\special{sh 1}%
\special{ar 3940 1881 10 10 0  6.28318530717959E+0000}%
\special{sh 1}%
\special{ar 3940 1881 10 10 0  6.28318530717959E+0000}%
% DOT 0 0 3 0
% 2 1399 2983 1399 2983
% 
\special{pn 20}%
\special{sh 1}%
\special{ar 1377 2937 10 10 0  6.28318530717959E+0000}%
\special{sh 1}%
\special{ar 1377 2937 10 10 0  6.28318530717959E+0000}%
% DOT 0 0 3 0
% 2 1399 3596 1399 3596
% 
\special{pn 20}%
\special{sh 1}%
\special{ar 1377 3540 10 10 0  6.28318530717959E+0000}%
\special{sh 1}%
\special{ar 1377 3540 10 10 0  6.28318530717959E+0000}%
% DOT 0 0 3 0
% 2 4003 3442 4003 3442
% 
\special{pn 20}%
\special{sh 1}%
\special{ar 3940 3388 10 10 0  6.28318530717959E+0000}%
\special{sh 1}%
\special{ar 3940 3388 10 10 0  6.28318530717959E+0000}%
% DOT 0 0 3 0
% 2 4156 3596 4156 3596
% 
\special{pn 20}%
\special{sh 1}%
\special{ar 4091 3540 10 10 0  6.28318530717959E+0000}%
\special{sh 1}%
\special{ar 4091 3540 10 10 0  6.28318530717959E+0000}%
% LINE 2 2 3 0
% 2 2318 838 2471 838
% 
\special{pn 8}%
\special{pa 2282 825}%
\special{pa 2433 825}%
\special{dt 0.045}%
% LINE 2 2 3 0
% 2 3696 838 3850 838
% 
\special{pn 8}%
\special{pa 3638 825}%
\special{pa 3790 825}%
\special{dt 0.045}%
% LINE 2 0 3 0
% 2 2930 379 2701 532
% 
\special{pn 8}%
\special{pa 2884 374}%
\special{pa 2659 524}%
\special{fp}%
% LINE 2 0 3 0
% 2 2930 685 2701 532
% 
\special{pn 8}%
\special{pa 2884 675}%
\special{pa 2659 524}%
\special{fp}%
% LINE 2 0 3 0
% 2 2930 992 2471 838
% 
\special{pn 8}%
\special{pa 2884 977}%
\special{pa 2433 825}%
\special{fp}%
% LINE 2 0 3 0
% 2 2930 1298 2471 838
% 
\special{pn 8}%
\special{pa 2884 1278}%
\special{pa 2433 825}%
\special{fp}%
% LINE 2 1 3 0
% 2 2701 532 2471 838
% 
\special{pn 8}%
\special{pa 2659 524}%
\special{pa 2433 825}%
\special{da 0.070}%
% DOT 0 0 3 0
% 2 2471 838 2471 838
% 
\special{pn 20}%
\special{sh 1}%
\special{ar 2433 825 10 10 0  6.28318530717959E+0000}%
\special{sh 1}%
\special{ar 2433 825 10 10 0  6.28318530717959E+0000}%
% DOT 0 0 3 0
% 2 2701 532 2701 532
% 
\special{pn 20}%
\special{sh 1}%
\special{ar 2659 524 10 10 0  6.28318530717959E+0000}%
\special{sh 1}%
\special{ar 2659 524 10 10 0  6.28318530717959E+0000}%
% DOT 2 0 3 0
% 2 2930 379 2930 379
% 
\special{pn 8}%
\special{sh 1}%
\special{ar 2884 374 10 10 0  6.28318530717959E+0000}%
\special{sh 1}%
\special{ar 2884 374 10 10 0  6.28318530717959E+0000}%
% DOT 2 0 3 0
% 2 2930 685 2930 685
% 
\special{pn 8}%
\special{sh 1}%
\special{ar 2884 675 10 10 0  6.28318530717959E+0000}%
\special{sh 1}%
\special{ar 2884 675 10 10 0  6.28318530717959E+0000}%
% DOT 2 0 3 0
% 2 2930 992 2930 992
% 
\special{pn 8}%
\special{sh 1}%
\special{ar 2884 977 10 10 0  6.28318530717959E+0000}%
\special{sh 1}%
\special{ar 2884 977 10 10 0  6.28318530717959E+0000}%
% DOT 2 0 3 0
% 2 2930 1298 2930 1298
% 
\special{pn 8}%
\special{sh 1}%
\special{ar 2884 1278 10 10 0  6.28318530717959E+0000}%
\special{sh 1}%
\special{ar 2884 1278 10 10 0  6.28318530717959E+0000}%
% LINE 2 0 3 0
% 2 3850 838 4309 1298
% 
\special{pn 8}%
\special{pa 3790 825}%
\special{pa 4242 1278}%
\special{fp}%
% LINE 2 0 3 0
% 2 3850 838 4309 379
% 
\special{pn 8}%
\special{pa 3790 825}%
\special{pa 4242 374}%
\special{fp}%
% LINE 2 1 3 0
% 2 3850 838 4079 838
% 
\special{pn 8}%
\special{pa 3790 825}%
\special{pa 4015 825}%
\special{da 0.070}%
% LINE 2 0 3 0
% 2 4079 838 4309 685
% 
\special{pn 8}%
\special{pa 4015 825}%
\special{pa 4242 675}%
\special{fp}%
% LINE 2 0 3 0
% 2 4079 838 4309 992
% 
\special{pn 8}%
\special{pa 4015 825}%
\special{pa 4242 977}%
\special{fp}%
% DOT 2 0 3 0
% 2 4309 379 4309 379
% 
\special{pn 8}%
\special{sh 1}%
\special{ar 4242 374 10 10 0  6.28318530717959E+0000}%
\special{sh 1}%
\special{ar 4242 374 10 10 0  6.28318530717959E+0000}%
% DOT 2 0 3 0
% 2 4309 685 4309 685
% 
\special{pn 8}%
\special{sh 1}%
\special{ar 4242 675 10 10 0  6.28318530717959E+0000}%
\special{sh 1}%
\special{ar 4242 675 10 10 0  6.28318530717959E+0000}%
% DOT 2 0 3 0
% 2 4309 992 4309 992
% 
\special{pn 8}%
\special{sh 1}%
\special{ar 4242 977 10 10 0  6.28318530717959E+0000}%
\special{sh 1}%
\special{ar 4242 977 10 10 0  6.28318530717959E+0000}%
% DOT 2 0 3 0
% 2 4309 1298 4309 1298
% 
\special{pn 8}%
\special{sh 1}%
\special{ar 4242 1278 10 10 0  6.28318530717959E+0000}%
\special{sh 1}%
\special{ar 4242 1278 10 10 0  6.28318530717959E+0000}%
% DOT 0 0 3 0
% 2 3850 838 3850 838
% 
\special{pn 20}%
\special{sh 1}%
\special{ar 3790 825 10 10 0  6.28318530717959E+0000}%
\special{sh 1}%
\special{ar 3790 825 10 10 0  6.28318530717959E+0000}%
% DOT 0 0 3 0
% 2 4079 838 4079 838
% 
\special{pn 20}%
\special{sh 1}%
\special{ar 4015 825 10 10 0  6.28318530717959E+0000}%
\special{sh 1}%
\special{ar 4015 825 10 10 0  6.28318530717959E+0000}%
% STR 2 0 3 0
% 3 4849 758 4849 835 5 0
% $+$
\put(47.7264,-8.2185){\makebox(0,0){$+$}}%
% DOT 2 0 3 0
% 2 5692 375 5692 375
% 
\special{pn 8}%
\special{sh 1}%
\special{ar 5603 370 10 10 0  6.28318530717959E+0000}%
\special{sh 1}%
\special{ar 5603 370 10 10 0  6.28318530717959E+0000}%
% DOT 2 0 3 0
% 2 5692 681 5692 681
% 
\special{pn 8}%
\special{sh 1}%
\special{ar 5603 671 10 10 0  6.28318530717959E+0000}%
\special{sh 1}%
\special{ar 5603 671 10 10 0  6.28318530717959E+0000}%
% DOT 2 0 3 0
% 2 5692 988 5692 988
% 
\special{pn 8}%
\special{sh 1}%
\special{ar 5603 973 10 10 0  6.28318530717959E+0000}%
\special{sh 1}%
\special{ar 5603 973 10 10 0  6.28318530717959E+0000}%
% DOT 2 0 3 0
% 2 5696 1298 5696 1298
% 
\special{pn 8}%
\special{sh 1}%
\special{ar 5607 1278 10 10 0  6.28318530717959E+0000}%
\special{sh 1}%
\special{ar 5607 1278 10 10 0  6.28318530717959E+0000}%
% STR 2 0 3 0
% 3 705 1983 705 2060 5 0
% $+$
\put(6.9390,-20.2756){\makebox(0,0){$+$}}%
% DOT 2 0 3 0
% 2 1548 1600 1548 1600
% 
\special{pn 8}%
\special{sh 1}%
\special{ar 1524 1575 10 10 0  6.28318530717959E+0000}%
\special{sh 1}%
\special{ar 1524 1575 10 10 0  6.28318530717959E+0000}%
% DOT 2 0 3 0
% 2 1548 1907 1548 1907
% 
\special{pn 8}%
\special{sh 1}%
\special{ar 1524 1877 10 10 0  6.28318530717959E+0000}%
\special{sh 1}%
\special{ar 1524 1877 10 10 0  6.28318530717959E+0000}%
% DOT 2 0 3 0
% 2 1548 2213 1548 2213
% 
\special{pn 8}%
\special{sh 1}%
\special{ar 1524 2179 10 10 0  6.28318530717959E+0000}%
\special{sh 1}%
\special{ar 1524 2179 10 10 0  6.28318530717959E+0000}%
% DOT 2 0 3 0
% 2 1548 2520 1548 2520
% 
\special{pn 8}%
\special{sh 1}%
\special{ar 1524 2481 10 10 0  6.28318530717959E+0000}%
\special{sh 1}%
\special{ar 1524 2481 10 10 0  6.28318530717959E+0000}%
% STR 2 0 3 0
% 3 2084 1983 2084 2060 5 0
% $+$
\put(20.5118,-20.2756){\makebox(0,0){$+$}}%
% DOT 2 0 3 0
% 2 2927 1600 2927 1600
% 
\special{pn 8}%
\special{sh 1}%
\special{ar 2881 1575 10 10 0  6.28318530717959E+0000}%
\special{sh 1}%
\special{ar 2881 1575 10 10 0  6.28318530717959E+0000}%
% DOT 2 0 3 0
% 2 2927 1907 2927 1907
% 
\special{pn 8}%
\special{sh 1}%
\special{ar 2881 1877 10 10 0  6.28318530717959E+0000}%
\special{sh 1}%
\special{ar 2881 1877 10 10 0  6.28318530717959E+0000}%
% DOT 2 0 3 0
% 2 2927 2213 2927 2213
% 
\special{pn 8}%
\special{sh 1}%
\special{ar 2881 2179 10 10 0  6.28318530717959E+0000}%
\special{sh 1}%
\special{ar 2881 2179 10 10 0  6.28318530717959E+0000}%
% DOT 2 0 3 0
% 2 2927 2520 2927 2520
% 
\special{pn 8}%
\special{sh 1}%
\special{ar 2881 2481 10 10 0  6.28318530717959E+0000}%
\special{sh 1}%
\special{ar 2881 2481 10 10 0  6.28318530717959E+0000}%
% LINE 2 2 3 0
% 2 5083 838 5236 838
% 
\special{pn 8}%
\special{pa 5003 825}%
\special{pa 5154 825}%
\special{dt 0.045}%
% LINE 2 0 3 0
% 2 5696 1298 5466 1145
% 
\special{pn 8}%
\special{pa 5607 1278}%
\special{pa 5380 1127}%
\special{fp}%
% LINE 2 0 3 0
% 2 5696 992 5466 1145
% 
\special{pn 8}%
\special{pa 5607 977}%
\special{pa 5380 1127}%
\special{fp}%
% LINE 2 0 3 0
% 2 5696 685 5236 838
% 
\special{pn 8}%
\special{pa 5607 675}%
\special{pa 5154 825}%
\special{fp}%
% LINE 2 0 3 0
% 2 5696 379 5236 838
% 
\special{pn 8}%
\special{pa 5607 374}%
\special{pa 5154 825}%
\special{fp}%
% LINE 2 1 3 0
% 2 5466 1145 5236 838
% 
\special{pn 8}%
\special{pa 5380 1127}%
\special{pa 5154 825}%
\special{da 0.070}%
% DOT 0 0 3 0
% 2 5236 838 5236 838
% 
\special{pn 20}%
\special{sh 1}%
\special{ar 5154 825 10 10 0  6.28318530717959E+0000}%
\special{sh 1}%
\special{ar 5154 825 10 10 0  6.28318530717959E+0000}%
% DOT 0 0 3 0
% 2 5466 1145 5466 1145
% 
\special{pn 20}%
\special{sh 1}%
\special{ar 5380 1127 10 10 0  6.28318530717959E+0000}%
\special{sh 1}%
\special{ar 5380 1127 10 10 0  6.28318530717959E+0000}%
% STR 2 0 3 0
% 3 3463 1983 3463 2060 5 0
% $+$
\put(34.0846,-20.2756){\makebox(0,0){$+$}}%
% DOT 2 0 3 0
% 2 4305 1600 4305 1600
% 
\special{pn 8}%
\special{sh 1}%
\special{ar 4238 1575 10 10 0  6.28318530717959E+0000}%
\special{sh 1}%
\special{ar 4238 1575 10 10 0  6.28318530717959E+0000}%
% DOT 2 0 3 0
% 2 4305 1907 4305 1907
% 
\special{pn 8}%
\special{sh 1}%
\special{ar 4238 1877 10 10 0  6.28318530717959E+0000}%
\special{sh 1}%
\special{ar 4238 1877 10 10 0  6.28318530717959E+0000}%
% DOT 2 0 3 0
% 2 4305 2213 4305 2213
% 
\special{pn 8}%
\special{sh 1}%
\special{ar 4238 2179 10 10 0  6.28318530717959E+0000}%
\special{sh 1}%
\special{ar 4238 2179 10 10 0  6.28318530717959E+0000}%
% DOT 2 0 3 0
% 2 4305 2520 4305 2520
% 
\special{pn 8}%
\special{sh 1}%
\special{ar 4238 2481 10 10 0  6.28318530717959E+0000}%
\special{sh 1}%
\special{ar 4238 2481 10 10 0  6.28318530717959E+0000}%
% STR 2 0 3 0
% 3 705 3209 705 3285 5 0
% $+$
\put(6.9390,-32.3327){\makebox(0,0){$+$}}%
% DOT 2 0 3 0
% 2 1548 2826 1548 2826
% 
\special{pn 8}%
\special{sh 1}%
\special{ar 1524 2782 10 10 0  6.28318530717959E+0000}%
\special{sh 1}%
\special{ar 1524 2782 10 10 0  6.28318530717959E+0000}%
% DOT 2 0 3 0
% 2 1548 3132 1548 3132
% 
\special{pn 8}%
\special{sh 1}%
\special{ar 1524 3083 10 10 0  6.28318530717959E+0000}%
\special{sh 1}%
\special{ar 1524 3083 10 10 0  6.28318530717959E+0000}%
% DOT 2 0 3 0
% 2 1548 3439 1548 3439
% 
\special{pn 8}%
\special{sh 1}%
\special{ar 1524 3385 10 10 0  6.28318530717959E+0000}%
\special{sh 1}%
\special{ar 1524 3385 10 10 0  6.28318530717959E+0000}%
% DOT 2 0 3 0
% 2 1548 3745 1548 3745
% 
\special{pn 8}%
\special{sh 1}%
\special{ar 1524 3687 10 10 0  6.28318530717959E+0000}%
\special{sh 1}%
\special{ar 1524 3687 10 10 0  6.28318530717959E+0000}%
% STR 2 0 3 0
% 3 3463 3209 3463 3285 5 0
% $+$
\put(34.0846,-32.3327){\makebox(0,0){$+$}}%
% DOT 2 0 3 0
% 2 4305 2826 4305 2826
% 
\special{pn 8}%
\special{sh 1}%
\special{ar 4238 2782 10 10 0  6.28318530717959E+0000}%
\special{sh 1}%
\special{ar 4238 2782 10 10 0  6.28318530717959E+0000}%
% DOT 2 0 3 0
% 2 4305 3132 4305 3132
% 
\special{pn 8}%
\special{sh 1}%
\special{ar 4238 3083 10 10 0  6.28318530717959E+0000}%
\special{sh 1}%
\special{ar 4238 3083 10 10 0  6.28318530717959E+0000}%
% DOT 2 0 3 0
% 2 4305 3439 4305 3439
% 
\special{pn 8}%
\special{sh 1}%
\special{ar 4238 3385 10 10 0  6.28318530717959E+0000}%
\special{sh 1}%
\special{ar 4238 3385 10 10 0  6.28318530717959E+0000}%
% DOT 2 0 3 0
% 2 4305 3745 4305 3745
% 
\special{pn 8}%
\special{sh 1}%
\special{ar 4238 3687 10 10 0  6.28318530717959E+0000}%
\special{sh 1}%
\special{ar 4238 3687 10 10 0  6.28318530717959E+0000}%
% LINE 2 2 3 0
% 2 5228 2064 5075 2064
% 
\special{pn 8}%
\special{pa 5146 2032}%
\special{pa 4996 2032}%
\special{dt 0.045}%
% DOT 0 0 3 0
% 2 5228 2064 5228 2064
% 
\special{pn 20}%
\special{sh 1}%
\special{ar 5146 2032 10 10 0  6.28318530717959E+0000}%
\special{sh 1}%
\special{ar 5146 2032 10 10 0  6.28318530717959E+0000}%
% STR 2 0 3 0
% 3 4841 1983 4841 2060 5 0
% $+$
\put(47.6476,-20.2756){\makebox(0,0){$+$}}%
% DOT 2 0 3 0
% 2 5684 1600 5684 1600
% 
\special{pn 8}%
\special{sh 1}%
\special{ar 5595 1575 10 10 0  6.28318530717959E+0000}%
\special{sh 1}%
\special{ar 5595 1575 10 10 0  6.28318530717959E+0000}%
% DOT 2 0 3 0
% 2 5684 1907 5684 1907
% 
\special{pn 8}%
\special{sh 1}%
\special{ar 5595 1877 10 10 0  6.28318530717959E+0000}%
\special{sh 1}%
\special{ar 5595 1877 10 10 0  6.28318530717959E+0000}%
% DOT 2 0 3 0
% 2 5684 2213 5684 2213
% 
\special{pn 8}%
\special{sh 1}%
\special{ar 5595 2179 10 10 0  6.28318530717959E+0000}%
\special{sh 1}%
\special{ar 5595 2179 10 10 0  6.28318530717959E+0000}%
% DOT 2 0 3 0
% 2 5684 2520 5684 2520
% 
\special{pn 8}%
\special{sh 1}%
\special{ar 5595 2481 10 10 0  6.28318530717959E+0000}%
\special{sh 1}%
\special{ar 5595 2481 10 10 0  6.28318530717959E+0000}%
% LINE 2 0 3 0
% 2 5688 1604 5228 2064
% 
\special{pn 8}%
\special{pa 5599 1579}%
\special{pa 5146 2032}%
\special{fp}%
% LINE 2 0 3 0
% 2 5381 2217 5688 2523
% 
\special{pn 8}%
\special{pa 5297 2183}%
\special{pa 5599 2484}%
\special{fp}%
% LINE 2 0 3 0
% 2 5535 2064 5688 1911
% 
\special{pn 8}%
\special{pa 5448 2032}%
\special{pa 5599 1881}%
\special{fp}%
% LINE 2 0 3 0
% 2 5535 2064 5688 2217
% 
\special{pn 8}%
\special{pa 5448 2032}%
\special{pa 5599 2183}%
\special{fp}%
% LINE 2 1 3 0
% 2 5228 2064 5381 2217
% 
\special{pn 8}%
\special{pa 5146 2032}%
\special{pa 5297 2183}%
\special{da 0.070}%
% LINE 2 1 3 0
% 2 5535 2064 5381 2217
% 
\special{pn 8}%
\special{pa 5448 2032}%
\special{pa 5297 2183}%
\special{da 0.070}%
% DOT 0 0 3 0
% 2 5381 2217 5381 2217
% 
\special{pn 20}%
\special{sh 1}%
\special{ar 5297 2183 10 10 0  6.28318530717959E+0000}%
\special{sh 1}%
\special{ar 5297 2183 10 10 0  6.28318530717959E+0000}%
% DOT 0 0 3 0
% 2 5535 2064 5535 2064
% 
\special{pn 20}%
\special{sh 1}%
\special{ar 5448 2032 10 10 0  6.28318530717959E+0000}%
\special{sh 1}%
\special{ar 5448 2032 10 10 0  6.28318530717959E+0000}%
% LINE 2 2 3 0
% 2 2471 3289 2318 3289
% 
\special{pn 8}%
\special{pa 2433 3238}%
\special{pa 2282 3238}%
\special{dt 0.045}%
% DOT 0 0 3 0
% 2 2471 3289 2471 3289
% 
\special{pn 20}%
\special{sh 1}%
\special{ar 2433 3238 10 10 0  6.28318530717959E+0000}%
\special{sh 1}%
\special{ar 2433 3238 10 10 0  6.28318530717959E+0000}%
% STR 2 0 3 0
% 3 2084 3209 2084 3285 5 0
% $+$
\put(20.5118,-32.3327){\makebox(0,0){$+$}}%
% DOT 2 0 3 0
% 2 2927 2826 2927 2826
% 
\special{pn 8}%
\special{sh 1}%
\special{ar 2881 2782 10 10 0  6.28318530717959E+0000}%
\special{sh 1}%
\special{ar 2881 2782 10 10 0  6.28318530717959E+0000}%
% DOT 2 0 3 0
% 2 2927 3132 2927 3132
% 
\special{pn 8}%
\special{sh 1}%
\special{ar 2881 3083 10 10 0  6.28318530717959E+0000}%
\special{sh 1}%
\special{ar 2881 3083 10 10 0  6.28318530717959E+0000}%
% DOT 2 0 3 0
% 2 2927 3439 2927 3439
% 
\special{pn 8}%
\special{sh 1}%
\special{ar 2881 3385 10 10 0  6.28318530717959E+0000}%
\special{sh 1}%
\special{ar 2881 3385 10 10 0  6.28318530717959E+0000}%
% DOT 2 0 3 0
% 2 2927 3745 2927 3745
% 
\special{pn 8}%
\special{sh 1}%
\special{ar 2881 3687 10 10 0  6.28318530717959E+0000}%
\special{sh 1}%
\special{ar 2881 3687 10 10 0  6.28318530717959E+0000}%
% LINE 2 0 3 0
% 2 2930 3749 2471 3289
% 
\special{pn 8}%
\special{pa 2884 3690}%
\special{pa 2433 3238}%
\special{fp}%
% LINE 2 0 3 0
% 2 2624 3136 2930 2830
% 
\special{pn 8}%
\special{pa 2583 3087}%
\special{pa 2884 2786}%
\special{fp}%
% LINE 2 0 3 0
% 2 2777 3289 2930 3442
% 
\special{pn 8}%
\special{pa 2734 3238}%
\special{pa 2884 3388}%
\special{fp}%
% LINE 2 0 3 0
% 2 2777 3289 2930 3136
% 
\special{pn 8}%
\special{pa 2734 3238}%
\special{pa 2884 3087}%
\special{fp}%
% LINE 2 1 3 0
% 2 2471 3289 2624 3136
% 
\special{pn 8}%
\special{pa 2433 3238}%
\special{pa 2583 3087}%
\special{da 0.070}%
% LINE 2 1 3 0
% 2 2777 3289 2624 3136
% 
\special{pn 8}%
\special{pa 2734 3238}%
\special{pa 2583 3087}%
\special{da 0.070}%
% DOT 0 0 3 0
% 2 2624 3136 2624 3136
% 
\special{pn 20}%
\special{sh 1}%
\special{ar 2583 3087 10 10 0  6.28318530717959E+0000}%
\special{sh 1}%
\special{ar 2583 3087 10 10 0  6.28318530717959E+0000}%
% DOT 0 0 3 0
% 2 2777 3289 2777 3289
% 
\special{pn 20}%
\special{sh 1}%
\special{ar 2734 3238 10 10 0  6.28318530717959E+0000}%
\special{sh 1}%
\special{ar 2734 3238 10 10 0  6.28318530717959E+0000}%
\end{picture}%
 \caption{For example $\ti{m}^p_4$ and $\ti{f}^p_4$ are given. 
The large dots represent the vertices $\{m_k\}$. 
The dashed lines denote the propagators and we attach 
$-Q^+$ on them. The dotted line on each graph indicates the outgoing 
external line. 
We attach $P$ for $\ti{m}^p_k$ and $-Q^+$ for $\ti{f}^p_k$. 
For $\ti{m}^p_4$ and $\ti{f}^p_4$, all such graphs with four 
incoming external states are summed up with weight $+1$. }
 \label{fig:m4f4}
\end{figure}

Here summarize the definition of the $A_\infty$-structure and 
the $A_\infty$-morphism derived above. 
\begin{defn}
 Let $(\cH,\m)$ be an $A_\infty$-algebra and assume that we have 
degree minus one operator $Q^+$ and degree zero operator $P$ 
which satisfies $QQ^++Q^+Q+P=\1$ on $\cH$. Then another 
$A_\infty$-algebra $(\cH^p,\ti\m^p)$ with $\cH^p:=P\cH$ and an 
$A_\infty$-morphism from $(\cH^p,\ti\m^p)$ to $(\cH,\m)$ are 
constructed. We define those with the following three equivalent 
expressions : 
\begin{itemize}
 \item \quad $\ti\m^p=\{\ti{m}^p_k\}_{k\ge 2}$ is given 
  by eq.(\ref{mceq1}) and eq.(\ref{mceq2}) : 
 \begin{equation*}
  (0=)\sum_{k\ge 2}P m_k(\ti\Phi^p+\ti\Phi^u)
  =\sum_{k\ge 2}\ti{m}^p_k(\ti\Phi^p)\ ,
 \end{equation*}
 together with eq.(\ref{eom3}): 
 $\ti\Phi^u=-\sum_{k\ge 2}Q^+ m_k(\ti\Phi^p+\ti\Phi^u)$ and the 
 $A_\infty$-morphism $\ti\cF^p=\{\ti{f}^p_k\}_{k\ge 2}$ is defined 
 by eq.(\ref{eom3}) : 
 \begin{equation*}
  \begin{split}
  \Phi&=\ti\Phi^p+\ti\Phi^u=\ti\Phi^p-\sum_{k\ge 2}Q^+ 
   m_k(\ti\Phi^p+\ti\Phi^u)\\
      &=\ti\Phi^p+\ti{f}^p_2(\ti\Phi^p,\ti\Phi^p)
 +\ti{f}^p_3(\ti\Phi^p,\ti\Phi^p,\ti\Phi^p)+\cdots\ .
  \end{split}
 \end{equation*}
 
 \item \quad $\{\ti{f}^p_k\}$ are defined recursively as 
 \begin{equation*}
  \ti{f}^p_k(\ti\Phi^p)=-Q^+\sum_{1\leq k_1<k_2\cdots <k_i=k}m_i(
\ti{f}^p_{k_1}(\ti\Phi^p),\ti{f}^p_{k_2-k_1}(\ti\Phi^p)
 ,\cdots,\ti{f}^p_{k-k_{i-1}}(\ti\Phi^p))
 \end{equation*}
 with $\ti{f}^p_1(\ti\Phi^p)=\ti\Phi^p$. 
 $\{\ti{m}^p_k\}_{k\ge 2}$ are then defined as 
 \begin{equation*}
  \ti{m}^p_k(\ti\Phi^p)=\sum_{1\leq k_1<k_2\cdots <k_i=k}P
 m_i(\ti{f}^p_{k_1}(\ti\Phi^p),\ti{f}^p_{k_2-k_1}(\ti\Phi^p)
 ,\cdots,\ti{f}^p_{k-k_{i-1}}(\ti\Phi^p))\ .
 \end{equation*}

 \item \quad $\ti{f}^p_l(\ti\Phi^p)$ is given by 
 connecting tree graphs $\{-Q^+ m_k(\cdots)\}$ 
 with all possible combination, 
 summing up these, and picking up the term involving 
 $l$ powers of $\ti\Phi^p$.
 $\ti{m}^p_l(\ti\Phi^p)$ is defined in the same way as $\ti{m}^p_l$ 
 but replacing $-Q^+$ on the last outgoing line by $P$. 

 Let $G_l$ be the set of the graphs with $l$ incoming states and an 
 element in it as $\Gamma_l\in G_l$. 
 For each $\Gamma_l$ associate the operator $\ti{m}^p_{\Gamma_l}$, 
 which is defined by attaching $m_k$ to each vertex with $k$ incoming legs 
 and one outgoing legs, attaching $-Q^+$ to each internal edges, 
 and connecting them (see (Fig.\ref{fig:m4f4})). 
 The derived $A_\infty$-structure and 
 $A_\infty$-quasi-isomorphism are then given as 
\begin{equation}
 \ti{m}^p_l=P\sum_{\Gamma_l\in G_l}\ti{m}^p_{\Gamma_l}\ ,\qquad 
 \ti{f}^p_l=-Q^+\sum_{\Gamma_l\in G_l}\ti{m}^p_{\Gamma_l}\ .
\end{equation}
\end{itemize}
Note that once getting $\ti{m}^p_k(\ti\Phi^p)$ and $\ti{f}^p_k(\ti\Phi^p)$, 
then $\ti{m}^p_k(\eb^p_1,\cdots,\eb^p_k)$ and 
$\ti{f}^p_k(\eb^p_1,\cdots,\eb^p_k)$ are immediately 
obtained by reading the coefficient of $\phi^k\cdots\phi^1$, 
where $\eb_i^p, i=1,\cdots k$ are the basis of $\cH^p$ and 
$\ti\Phi^p=\eb^p_i\phi^i$.
 \label{defn:minimal}
\end{defn}
The explicit example is given 
in (Fig.\ref{fig:m4f4}). 
In the order of the graphs in (Fig.\ref{fig:m4f4}), 
\begin{equation*}
 \begin{split}
\ti{m}_4^p(\eb^p_1,\eb^p_2,\eb^p_3,\eb^p_4)&
=Pm_4(\eb^p_1,\eb^p_2,\eb^p_3,\eb^p_4)+
Pm_3(-Q^+ m_2(\eb^p_1,\eb^p_2),\eb^p_3,\eb^p_4)\\&+
Pm_3(\eb^p_1,-Q^+ m_2(\eb^p_2,\eb^p_3),\eb^p_4)+
Pm_3(\eb^p_1,\eb^p_2,-Q^+ m_2(\eb^p_3,\eb^p_4))\\&+
Pm_2(-Q^+ m_3(\eb^p_1,\eb^p_2,\eb^p_3),\eb^p_4)+
Pm_2(\eb^p_1,-Q^+ m_3(\eb^p_2,\eb^p_3,\eb^p_4))\\&+
Pm_2(-Q^+ m_2(-Q^+ m_2(\eb^p_1,\eb^p_2),\eb^p_3),\eb^p_4)\\&+
Pm_2(\eb^p_1,-Q^+ m_2(-Q^+ m_2(\eb^p_2,\eb^p_3),\eb^p_4))\\&+
Pm_2(-Q^+ m_2(\eb^p_1,\eb^p_2),-Q^+ m_2(\eb^p_3,\eb^p_4))\\&+
Pm_2(-Q^+ m_2(\eb^p_1,-Q^+ m_2(\eb^p_2,\eb^p_3)),\eb^p_4)\\&+
Pm_2(\eb^p_1,-Q^+ m_2(\eb^p_2,-Q^+ m_2(\eb^p_3,\eb^p_4)))\ ,
 \end{split}
\end{equation*} 
and  $\ti{f}^p_4$ is obtained similarly but replaced each $P$ 
on the outgoing line to $-Q^+$.
\begin{rem}
Though an $A_\infty$-algebra $(\cH^p,\ti\m^p)$ and an 
$A_\infty$-quasi-isomorphism $\cF$ have been given here, 
we can get another $A_\infty$-algebra. 
It is obtained by replacing the $\cH^p$ of $(\cH^p,\ti\m^p)$ to $\cH$. 
We denote this $A_\infty$-algebra by $(\cH,\ti\m)$ or simply $\ti\cH$. 
The $A_\infty$-structure $\ti\m$ is that of $(\cH^p,\ti\m^p)$, but 
it has $\ti{m}_1=m_1=Q$. 
(This $\ti{m}_1$ vanishes trivially on $\cH^p$ so 
the $A_\infty$-structure of $(\cH^p,\ti\m^p)$ is 
$\ti\m^p=\{\ti{m}^p_k\}_{k\ge 2}$.) 
Thus the $A_\infty$-algebra $(\cH, \ti\m)$ is defined by 
(Def.\ref{defn:minimal}) but relaxing the restriction $\ti\Phi^p\in\cH^p$ 
as $\ti\Phi\in\cH$. 
$\ti\cF:=\ti\cF^p=\{\ti{f}^p_1=\Id,\ti{f}^p_k (k\ge 2)\}$ 
then defines the $A_\infty$-quasi-isomorphism 
from $(\cH,\ti\m)$ to $(\cH,\m)$
\footnote{Using them an explicit proof of (Rem.\ref{rem:inverse}) 
was presented by M.~Akaho at topology seminar in Univ. of Tokyo 
in July, 2001. 
}. 
Note that this is not only an $A_\infty$-quasi-isomorphism but 
also an $A_\infty$-isomorphism. 
The following diagram is obtained \\
%\begin{figure}[h]\
 \hspace*{3.0cm}
%WinTpicVersion3.08
\unitlength 0.1in
\begin{picture}( 27.3622,  7.0866)(  9.5472,-10.9744)
% STR 2 0 3 0
% 3 2200 500 2200 600 5 0
% $(\cH,\ti\m)$
\put(21.6535,-5.9055){\makebox(0,0){$(\cH,\ti\m)$}}%
% VECTOR 2 0 3 0
% 2 2800 600 3600 600
% 
\special{pn 8}%
\special{pa 2756 591}%
\special{pa 3544 591}%
\special{fp}%
\special{sh 1}%
\special{pa 3544 591}%
\special{pa 3478 571}%
\special{pa 3492 591}%
\special{pa 3478 611}%
\special{pa 3544 591}%
\special{fp}%
% STR 2 0 3 0
% 3 4200 500 4200 600 5 0
% $(\cH,\m)$
\put(41.3386,-5.9055){\makebox(0,0){$(\cH,\m)$}}%
% VECTOR 2 0 3 0
% 2 2150 800 2150 1000
% 
\special{pn 8}%
\special{pa 2117 788}%
\special{pa 2117 985}%
\special{fp}%
\special{sh 1}%
\special{pa 2117 985}%
\special{pa 2136 919}%
\special{pa 2117 933}%
\special{pa 2097 919}%
\special{pa 2117 985}%
\special{fp}%
% STR 2 0 3 0
% 3 2200 1100 2200 1200 5 0
% $(\cH^p,\ti\m^p)$
\put(21.6535,-11.8110){\makebox(0,0){$(\cH^p,\ti\m^p)$}}%
% STR 2 0 3 0
% 3 3200 380 3200 480 5 0
% \scalebox{0.7}[0.7]{$\ti\cF$}
\put(31.4961,-4.7244){\makebox(0,0){\scalebox{0.7}[0.7]{$\ti\cF$}}}%
% STR 2 0 3 0
% 3 3200 910 3200 1010 5 0
% \scalebox{0.7}[0.7]{$\ti\cF^p$}
\put(31.4961,-9.9409){\makebox(0,0){\scalebox{0.7}[0.7]{$\ti\cF^p$}}}%
% STR 2 0 3 0
% 3 2050 800 2050 900 5 0
% \scalebox{0.7}[0.7]{$P$}
\put(20.1772,-8.8583){\makebox(0,0){\scalebox{0.7}[0.7]{$P$}}}%
% VECTOR 2 0 3 0
% 2 2250 1000 2250 800
% 
\special{pn 8}%
\special{pa 2215 985}%
\special{pa 2215 788}%
\special{fp}%
\special{sh 1}%
\special{pa 2215 788}%
\special{pa 2195 854}%
\special{pa 2215 840}%
\special{pa 2235 854}%
\special{pa 2215 788}%
\special{fp}%
% STR 2 0 3 0
% 3 2350 800 2350 900 5 0
% $\iota$
\put(23.1299,-8.8583){\makebox(0,0){$\iota$}}%
% VECTOR 2 0 3 0
% 2 2800 1050 3600 750
% 
\special{pn 8}%
\special{pa 2756 1034}%
\special{pa 3544 739}%
\special{fp}%
\special{sh 1}%
\special{pa 3544 739}%
\special{pa 3476 744}%
\special{pa 3495 757}%
\special{pa 3490 780}%
\special{pa 3544 739}%
\special{fp}%
\end{picture}%
%\end{figure}

\vspace*{0.5cm}

\hspace*{-0.8cm}
where $\iota :\cH^p\raw\cH$ is the inclusion map. 
Explicitly $\ti\m^p$ and $\ti\cF^p$ are given by 
$\ti\m^p=P\circ\ti\m\circ\iota$ and $\ti\cF^p=\ti\cF\circ\iota$. 
We will consider these two $A_\infty$-algebras $(\cH^p,\ti\m^p)$ and 
$(\cH,\ti\m)$ later when an $A_\infty$-algebra $(\cH,\m)$ will be given. 
$(\cH^p,\ti\m^p)$ is used when a SFT is reduced to its on-shell physics 
in subsection \ref{ssec:53}, 
and $(\cH,\ti\m)$ is considered when the $\ti\cF$ is used as a 
field redefinition between two SFTs in subsection \ref{ssec:main} 
and section \ref{sec:6}. 
 \label{rem:twoAinfty}
\end{rem}
\begin{rem}
Both $A_\infty$-quasi-isomorphism $\ti\cF^p :(\cH^p,\ti\m^p)\raw (\cH,\m)$ 
and $\ti\cF : (\cH,\ti\m)\raw (\cH,\m)$ have their inverse 
$A_\infty$-quasi-isomorphisms. 
An quasi-inverse $(\ti\cF^p)^{-1} :(\cH,\m)\raw (\cH^p,\ti\m^p)$ 
is given simply by the projection $P$ 
\begin{equation*}
 \begin{array}{ccccc}
 (\ti\cF^p)^{-1}_* &:& \cH &\lgraw &\cH^p \\
              & & \Phi&\mapsto&\ti\Phi^p=P\Phi\ . 
 \end{array}
\end{equation*}
The inverse $(\ti\cF)^{-1} :(\cH,\m)\raw (\cH,\ti\m)$ is obtained by 
\begin{equation*}
 \begin{array}{ccccc}
 (\ti\cF)^{-1}_* &:& \cH &\lgraw &\cH \\
              & & \Phi&\mapsto&\ti\Phi=\Phi-\ti{f}(\Phi)
 \end{array}
\end{equation*}
where $\ti\Phi=\Phi+\ti{f}(\Phi)=\Phi-Q^+\sum_{k\ge 2} m_k(\Phi)$. 
This $(\ti\cF)^{-1}_*$ restricted to the solutions for the Maurer-Cartan 
equations is nothing but the Kuranishi map. 
 \label{rem:inverse}
\end{rem} 
\begin{rem}[Geometric realization]
The $A_\infty$-structures $\m$ and $\ti\m$ on $\cH$ are expressed as 
$\delta=\flpartial{\phi^j}c^j_i\phi^i
+\sum_{k=2}^\infty\flpartial{\phi^j}c^j_{i_1\cdots i_k}
 \phi^{i_k}\cdots\phi^{i_1}$ and 
$\ti\delta=\flpartial{\ti\phi^j}\ti{c}^j_i\ti\phi^i
+\sum_{k=2}^\infty\flpartial{\ti\phi^j}\ti{c}^j_{i_1\cdots i_k}
 \ti\phi^{i_k}\cdots\ti\phi^{i_1}$ in the dual picture. 
The leading coefficients of both formal 
vector fields are identical $c^j_i=\ti{c}^j_i$, because $m_1=\ti{m_1}=Q$. 
The $A_\infty$-quasi-isomorphism $\ti\cF$ gives the coordinate 
transformation $\ti\cF_*$ on $\cH$. Now by the definition of $\ti\m$, 
$\eb_j\ti{c}^j_{i_1\cdots i_k}\in\cH^p$ for $k\ge 2$ 
(and $\eb_j\ti{c}^j_i\in\cH^t$). 
Thus the $A_\infty$-quasi-isomorphism $\ti\cF$ is constructed so that 
the coefficients $\ti{c}^j_{i_1\cdots i_k}$ for $k\ge 2$ satisfies 
$\eb_j\ti{c}^j_{i_1\cdots i_k}\in\cH^p$. Then the 
$A_\infty$-structure $\ti\m$ on $\cH$ can be reduced to the one on $\cH^p$. 
Namely, 
the minimal model theorem says geometrically that for any graded vector 
space $\cH$ with a formal vector field $\delta$ which satisfies 
$\delta\cdot\delta=0$, there exists a formal coordinate transformation 
$\ti\cF_*$ so that 
the vector field 
$\sum_{k=2}^\infty\flpartial{\ti\phi^j}\ti{c}^j_{i_1\cdots i_k}
\ti\phi^{i_k}\cdots\ti\phi^{i_1}$ is along the $\cH^p$ direction. 
 \label{rem:geom}
\end{rem}

In the rest of this subsection, we shall give a proof that 
$(\cH^p,\ti\m^p)$ and $(\cH,\ti\m)$ are in fact $A_\infty$-algebras and 
$\ti\cF^{(p)}$ 
(by $\ti\cF^{(p)}$ we denote $\ti\cF$ or $\ti\cF^p$) 
is an $A_\infty$-morphism. 
The fact that $(\cH^p,\ti\m^p)$ is an $A_\infty$-algebra and $\ti\cF^p$ is 
an $A_\infty$-quasi-isomorphism between $(\cH^p,\ti\m^p)$ to $(\cH,\m)$ 
immediately follows from the fact that 
$(\cH,\ti\m)$ is an $A_\infty$-algebra and $\ti\cF$ is 
an $A_\infty$-quasi-isomorphism between $(\cH,\ti\m)$ to $(\cH,\m)$. 
The former is obtained by restricting the $\cH$ of $(\cH,\ti\m)$ to 
$\cH^p$ in the latter case. 
Therefore we will prove the latter fact. 
In order to see this, it is enough to confirm the following 
two fact : $\m\ti\cF=\ti\cF\ti\m$ and 
$(\ti\m)^2=0$ on $\cH$. 
We begin with a proof for the first statement. 
As has been seen in subsection \ref{ssec:23}, because 
$\ti\cF$ is the coalgebra homomorphism, $\ti\cF(e^{\ti\Phi}-\1)=e^\Phi-\1$ 
holds. Then $\m\ti\cF(e^{\ti\Phi})=\sum_{k\ge 1}m_k(\Phi)+\cdots$ for 
$\cdots\in \cH^{\otimes n\ge 2}$. 
$\sum_{k\ge 1}m_k(\Phi)$ can be rewritten similarly as eq.(\ref{eom4}), but 
now we consider generally when $\sum_{k\ge 1}m_k(\Phi)$ is not zero, 
and the rewriting leads the following equation, 
\begin{equation}
 \sum_{k\ge 1}m_k(\Phi)=\sum_{k\ge 1}\ti{m}_k(\ti\Phi)+
 Q^+ Q\sum_{k\ge 2}m_k(\Phi)\ .
 \label{cd1}
\end{equation}
We can show from now that this equation is in fact 
the $\cH^{\otimes 1}$ part of the equation 
$\m\ti\cF(e^{\ti\Phi})=\ti\cF\ti\m(e^{\ti\Phi})$. Note that it is sufficient 
for the condition $\m\ti\cF=\ti\cF\ti\m$ (by an inductive argument). 
By utilizing the $A_\infty$-condition for $\m$, the second term on the right 
hand side of eq.(\ref{cd1}) becomes 
\begin{equation*}
 Q^+ m_1\sum_{k\ge 2}m_k(\Phi)=
 -Q^+\sum_{l\ge 2}\l(\m_l(e^\Phi\sum_{k\ge 1}
 m_k(\Phi)e^\Phi) \Big|_{\cH^{\otimes 1}}\r)\ .
\end{equation*}
where $|_{\cH^{\otimes 1}}$ means picking up the $\cH^{\otimes 1}$ part 
from $C(\cH)$. Thus one gets the following recursive formula, 
\begin{equation}
 \m_*(e^\Phi)=\ti\m_*(e^{\ti\Phi})-Q^+
 \sum_{l\ge 2}\l(\m_l(e^\Phi \m_*(e^\Phi)e^\Phi) |_{\cH^{\otimes 1}}\r)\ .
 \label{cd2}
\end{equation}
Since $\Phi$ is represented in the expansion of the power of $\ti\Phi$, 
the above equation is satisfied separately in the homogeneous degree of 
$\ti\Phi$. Here we argue inductively and 
suppose that $(\m\ti\cF-\ti\cF\ti\m)e^{\ti\Phi}|_{\cH^{\otimes 1}}=0$ 
is satisfied at degree $(\ti\Phi)^k$ with 
$1\le k\le n-1$. Then consider the $(\ti\Phi)^n$ parts of this equation. 
The left hand side $\m_*(e^\Phi)$ involves $n$ powers of $\ti\Phi$, however 
the $\m_*(e^\Phi)$ has $(\ti\Phi)^k$ with $k\le n-1$ because the summention 
for $l$ begins at $l=2$. By the induction hypothesis, 
$e^\Phi \m_*(e^\Phi)e^\Phi$ in the second term on the right hand side of 
the equation (\ref{cd2}) can be rewritten when restricted to $(\ti\Phi)^n$ 
part as 
\begin{equation*}
 \begin{split}
 e^\Phi \m_*(e^\Phi)e^\Phi |_{\cH^{\otimes l\ge 2},(\ti\Phi)^n}
 &=\m\ti\cF(e^{\ti\Phi})|_{\cH^{\otimes l\ge 2},(\ti\Phi)^n}
 =\ti\cF\ti\m(e^{\ti\Phi})|_{\cH^{\otimes l\ge 2},(\ti\Phi)^n}\\
 &=\ti\cF\l(e^{\ti\Phi}\ti\m_*(e^{\ti\Phi})
 e^{\ti\Phi}\r)\Big|_{\cH^{\otimes l\ge 2},(\ti\Phi)^n}\ , 
 \end{split}
\end{equation*}
where the induction hypothesis is used in the second equality. 
Now the $(\ti\Phi)^n$ parts of the right hand side of eq.(\ref{cd2}) is 
\begin{equation*}
 \ti\m_*(e^{\ti\Phi})|_{(\ti\Phi)^n}-Q^+\sum_{l\ge 2}
 \l(\m_l\ti\cF \l(e^{\ti\Phi}\ti\m_*(e^{\ti\Phi})e^{\ti\Phi}\r)\r)
 \Big|_{\cH^{\otimes 1},(\ti\Phi)^n}\ ,
\end{equation*}
which is exactly equal to 
$\ti\cF\ti\m(e^{\ti\Phi})|_{\cH^{\otimes 1},(\ti\Phi)^n}$. 
This completes the proof by the induction that 
$(\m\ti\cF-\ti\cF\ti\m)(e^{\ti\Phi})|_{(\cH)^{\otimes 1}}=0$
\footnote{The coefficient of $\phi^n\cdots\phi^1$ of the equation reads 
the identity (\ref{amorphism}).}. 

Once getting $\m\ti\cF=\ti\cF\ti\m$, it is easy to show that 
$\ti\m$ defines an $A_\infty$-algebra. 
As was noted in (Rem.\ref{rem:inverse}), 
$\ti\cF$ has its inverse isomorphism $(\ti\cF)^{-1}$. 
Acting the $(\ti\cF)^{-1}$ on the both sides of $\m\ti\cF=\ti\cF\ti\m$ 
from left and $\ti\m$ can be expressed as $\ti\m=\ti\cF^{-1}\m\ti\cF$. 
We then get $(\ti\m)^2=0$ immediately from $(\m)^2=0$. 
Thus we has concreted the proof 
of the statements that $\ti\m$ gives an $A_\infty$-structure on 
$(\cH,\ti\m)$ and $\cF$ is 
an $A_\infty$-morphism between $(\cH,\ti\m)$ and $(\cH,\m)$. \qed

\begin{rem}
The fact proved above can be applied to show the following statement. 
Let $(\cH,\m)$ and $(\cH',\m')$ be two $A_\infty$-algebras, and 
let an $A_\infty$-quasi-isomorphism $\cF$ from $(\cH,\m)$ to $(\cH',\m')$ 
is given. Then there exists an inverse $A_\infty$-quasi-isomorphism 
$(\cF)^{-1} : (\cH',\m')\raw (\cH,\m)$. 
It can be now proved easily by applying the above results as follows. 
First, we can transform both $A_\infty$-algebras $(\cH,\m)$ and $(\cH',\m')$ 
to $(\cH^p,\ti\m^p)$ and $({\cH'}^p,\ti{\m'}^p)$ by the canonical 
$A_\infty$-quasi-isomorphisms $\ti\cF^p$ and $\ti{\cF'}^p$ 
in the above procedure. 
$\ti\cF^p$ and $\ti{\cF'}^p$ have their inverse quasi-isomorphisms, and 
the $A_\infty$-quasi-isomorphism from $(\cH^p,\ti\m^p)$ to 
$({\cH'}^p,\ti{\m'}^p)$ ($\star$)
is then given by the composition $(\ti{\cF'}^p)^{-1}\circ\cF\circ\ti\cF^p$ 
\begin{equation*}
\begin{CD}
 (\cH,\m) @>{\cF}>> (\cH',\m')\\
 @A{\ti\cF^p}AA     @A{\ti{\cF'}^p}AA   \\
 (\cH^p,\ti\m^p) @>{(\star)}>>   ({\cH'}^p,\ti{\m'}^p)
\end{CD}
\end{equation*}
so that the diagram commutes. 
Because (the leading of) the quasi-isomorphism $(\star)$ is isomorphism, 
$(\star)$ has its inverse, and one can obtain 
an $A_\infty$-quasi-isomorphism as 
$\ti\cF^p\circ(\star)^{-1}\circ(\ti{\cF'}^p)^{-1}$. 

In subsection \ref{ssec:main} and section \ref{sec:6} 
the canonical $A_\infty$-quasi-isomorphism 
and the canonical $A_\infty$-structure are applied to SFT in analogous 
way, though not precisely the same. 
 \label{rem:inverse}
\end{rem}

 \subsection{Minimal model theorem in gauge fixed SFT}
\label{ssec:5gf}

In the previous subsection, the analogue of 
the Hodge-Kodaira decomposition for the 
string Hilbert space $\cH$ was given in eq.(\ref{HKdecomp}), and 
an $A_\infty$-morphism $(\cH,\ti\m^{(p)})$ is constructed using $Q^+$. 
We claim that this $Q^+$ is the propagator in SFT. 
In this subsection we will explain the statement. 
Note that this means that the diagrams defined in order to construct 
the $A_\infty$-morphism $\ti\cF^{(p)}$ and 
$A_\infty$-structure $\ti\m^{(p)}$ 
are actually the Feynman diagram in SFT. 

In SFT, when one considers a propagator, first the action is gauge fixed. 
The propagator $Q^+$ is then defined as the inverse map of $Q$ 
onto the cokernel of $Q$ on the gauge. 
This propagator $Q^+$ actually gives the identity (\ref{HKdecomp}). 
Indeed define $P:=\1-\{Q, Q^+\}$ and the identity is obtained. 
Here for simplicity, we will argue the properties of the 
propagator mainly in the Siegel gauge.  

As was mentioned, the string field $\Phi$ includes both fields and 
antifields in the context of BV-formalism\cite{BV1,BV2,HT}. 
Let $c_0$ be a degree one operator. 
In order to express antifields explicitly, 
only in this subsection we represent the string field as 
$\Phi=\eb_i\phi^i+\eb_\ib\phi^\ib$ 
for $\eb_\ib:=c_0\eb_i$ and $\phi^\ib:=\phib^i$. 
$\{\phi\}$ is the fields and $\{\phib\}$ is the antifields. 
$\{\ebb\}:=\{\eb_\ib\}$ and $\{\eb\}:=\{\eb_i\}$ are the basis of 
the string Hilbert space $\cH$ which does and does 
not contain $c_0$, respectively. 
The degree of $\phi^i$ (resp. $\phi^\ib$) is defined to be minus the degree 
of $\eb_i$ (resp. $\eb_\ib$), so that the degree of $\Phi$ is zero. 
The degree of antifield $\phi^\ib$ is defined to be minus 
the degree of the corresponding field $\phi^i$ minus one in the context of 
BV-formalism. 
In particular in the usual oscillator representation, let $a_p$ be the matter 
oscillator of degree zero, $b_q$ and $c_r$ be the ghost oscillator of 
degree minus one and one, respectively. 
Then the `antistate' 
$\eb_\ib$ corresponding to a state 
$\eb_i\sim 
a_{-p_1}\cdots a_{-p_l}b_{-q_1}\cdots b_{-q_m}c_{-r_1}\cdots c_{-r_n}
|0\rangle$ is taken to be $\eb_\ib\sim a_{-p_1}\cdots a_{-p_l}
c_0c_{-q_1}\cdots c_{-q_m}b_{-r_1}\cdots b_{-r_n}|0\rangle$ with an 
appropriate normalization, 
where $p_k,\in\Z_{\ge 0}$, $q_k,r_k\in\Z_{>0}$ and $|0\rangle$ denotes the 
Fock space vacuum. The degree of the states are defined as the ghost number, 
that is, the number of $c_{-r}$ (including $c_0$) 
minus the number of $b_{-q}$ where the degree of the Fock vacuum $|0\ra$ is 
counted here as zero. The degree of $\eb_\ib$ is then minus the 
degree of $\eb_i$ plus one. Therefore the pair $\{\phi\}$ and 
$\{\phib\}$ has consistent degree in BV-formalism \cite{Th}
\footnote{ In \cite{Th} the explicit correspondence between the operator 
description for SFT and BV-formalism for field theories is found. }. 

The Siegel gauge fixing is then $b_0\Phi=0$, which 
restricts all the basis of $\cH$ to the basis $\{\eb\}$, 
in other words, restrict the space of fields to $\phib=0$. 
Express $Q$ manifestly such 
as the $c_0$ including part and $b_0$ including part and the rest part, 
\begin{equation*}
 Q=c_0L_0+b_0M+\Qt\ .
\end{equation*}
The kinetic term $\half\omega(\Phi,Q\Phi)|_{b_0\Phi=0}$ then reduces to 
$\half\omega(\eb_i\phi^i,c_0L_0\eb_j\phi^j)$ and the propagator is defined 
as $Q^+=b_0\ov{L_0}$, which acts as the inverse of $Q$ 
on the cokernel of $L_0$ in $\{\ebb\}$. 
Here define the projection onto the kernel of $L_0$ in $\cH$ as $P$. 
$Q^+$ is then extended to be the operator on $\cH$, which is written as 
$Q^+=b_0\ov{L_0}(\1-P)$. 
Since $Q$ commutes with Virasoro generators $L_m$, 
in particular with $L_0$, and $L_0$ does not 
include $c_0$ or $b_0$, $L_0$ commutes with $c_0L_0$, $b_0M$ and $\Qt$ 
independently. Therefore $Q^+$ anticommutes with $\Qt$, 
and also does trivially with $b_0M$. 
Then $\{Q,Q^+\}=\{c_0L_0,b_0\ov{L_0}(\1-P)\}=\1-P$, which is the desired 
form of the identity (\ref{HKdecomp}). 
Note that from this explicit form of $Q^+$ one can see that 
eq.(\ref{qqdag}) holds in the Siegel gauge (see for example \cite{N}).

More generally, denote the propagator in Schwinger representation 
with cut-off $\Lambda$ as 
\begin{equation}
 Q^+=b_0\int_0^\Lambda e^{-\tau L_0}d\tau \label{qdagsr}
\end{equation}
and we can define the identity (\ref{HKdecomp}). In this case $P$ which 
satisfies the identity (\ref{HKdecomp}) is the boundary term of $Q^+$
\begin{equation}
 P=e^{-\Lambda L_0}\ .\label{Preg}
\end{equation}
The cut-off $\Lambda$ is IR cut-off in target space theory, but 
short distance cut-off in string world sheet\cite{BR,N}. 

The $\ti\m$ in (Rem.\ref{rem:twoAinfty}) 
with $Q^+$ such a SFT propagator in fact defines an $A_\infty$-structure. 
No modification of the definition is needed. 
The proof that $\ti\m$ is an $A_\infty$-structure and 
$A_\infty$-quasi-isomorphic to $\m$ requires only the identity 
(\ref{HKdecomp}) and the condition that $QQ^+$, $Q^+Q$ or $P$ is a 
projection is not necessary. 
The construction of the $A_\infty$-algebra $(\cH,\ti\m)$ is thus 
considerably universal concept which is independent of some reguralization 
scheme. 
The fact holds true for $\ti\m^p$, but 
the story is a little different. 
Let us introduce $Q^u$ of degree minus one and 
denote the Hodge-Kodaira decomposition of $\cH$ as 
\begin{equation}
 QQ^u+Q^uQ+P^p=\1\ .
\end{equation}
In this expression $P^t:=QQ^u$, $P^u:=Q^uQ$ and $P^p$ are projections onto 
null states ($Q$-trivial states), unphysical states, 
physical states, respectively. 
Here define $\cH^p$ as physical state space  
\begin{equation}
 \cH^p:=P^p\cH\ .
\end{equation}
$P\cH$ is then different from $\cH^p$. Especially $P\cH$ which 
satisfies $QQ^++Q^+Q+P=\1$ includes unphysical states $\cH^u$. 
This makes some trouble when we define the on-shell 
$A_\infty$-algebra $(\cH^p,\ti\m^p)$ because the image of 
$\ti{m}^p_k$ on $(\cH^p)^{\otimes k}$ is not guaranteed to belong to 
$\cH^p$. However relating $\ti\m^p$ to string vertices provides us 
with the fact that actually $(\cH^p,\ti\m^p)$ is an $A_\infty$-algebra. 
This fact will be explained in the next subsection 
and Appendix \ref{ssec:B3}, 
where the relation between $\ti\m^p$ and on-shell string S-matrix elements 
is examined. 
Although the reduction to physical states $\cH^p$ only 
has been mentioned above, 
one can also include the $Q$-trivial states and define 
the reduced $A_\infty$-structure on $\cH^t\cup\cH^p$. It will also be 
explained there. 

As was stated in eq.(\ref{qdag}), we can also see 
from the above expression of $Q^+$ that $(Q^+)^2=0$ holds. 
$Q^+QQ^+=Q^+$ holds only when $P$ is a projection. 
$P$ of the form in eq.(\ref{Preg}) is not a projection and spoil the 
identity. 
However when some tree amplitudes are calculated with Feynman diagram 
in SFT, the $P$ in $\{Q,Q^+\}=\1-P$ should only contribute to the poles. 
When the external states are put so that some propagators get poles, 
the amplitudes are not well-defined. Therefore we ignore the case and then 
one may define as $\{Q,Q^+\}=\1$ between the vertices. 
This leads $Q^+QQ^+=Q^+$ even when $P$ is not defined as a projection. 
This identity will not be used 
in (Lem.\ref{lem:main}),(Thm.\ref{thm:main}) and (Prop.\ref{prop:Ap}). 
It is used to rewrite SFT actions by field redefinitions 
in subsection \ref{ssec:main} and section \ref{sec:6}, but 
they are justified by (Prop.\ref{prop:Ap}).
The origin of this problem can be found in Appendix \ref{ssec:B3}. 

Thus one can obtain the $A_\infty$-morphism $\ti\cF^{(p)}$ and 
$A_\infty$-structure $\ti\m^{(p)}$, 
constructed in the previous subsection, in 
the context of SFT in BV-formalism. The $A_\infty$-morphism constructed 
with $Q^+=b_0\ov{L_0}(1-P)$ is an $A_\infty$-morphism 
from $(\cH^{(p)},\ti\m^{(p)})$ to $(\cH,\m)$. 
Here on the Siegel gauge let us give more precise SFT interpretation of 
this $A_\infty$-morphism $\ti\cF^{(p)}$ and 
$A_\infty$-structure $\ti\m^{(p)}$ obtained here. 
It relates to the way of constructing some solutions of the 
equations of motions discussed for classical closed (non-polynomial) 
SFT in \cite{MS,KZ}. 
The Maurer-Cartan equation (\ref{eomsft}) corresponding to the equation 
of motion for the original SFT restricted on the Siegel gauge $\phib=0$ 
is now written as 
\begin{align}
 &c_0L_0\eb_j\phi^j+\eb_\ib\sum_{n\ge 2}
 m^\ib_n(\eb_{j_1}\phi^{j_1},\cdots,\eb_{j_n}\phi^{j_n})=0\ ,
 \label{eom21}\\
 &\Qt\eb_j\phi^j+\eb_i\sum_{n\ge 2}
 m^i_n(\eb_{j_1}\phi^{j_1},\cdots,\eb_{j_n}\phi^{j_n})=0\ ,
 \label{eom22}
\end{align}
where $m^\ib_n(\eb_{j_1}\phi^{j_1},\cdots,\eb_{j_n}\phi^{j_n})$ 
means the coefficient of $\eb_\ib$ 
for $m_n(\eb_{j_1}\phi^{j_1},\cdots,\eb_{j_n}\phi^{j_n})$ 
and similar for $m^i_n$. 
The first equation is the equation on $\{\ebb\}$ and 
it is the equation of motion for the Siegel gauge fixed 
action. The second one is that on $\{\eb\}$ and it 
means that the BRST transformation of the 
field $\{\phi\}$ which satisfies the equation of motion (the first equation) 
is zero on this Siegel gauge, where 
the gauge fixed BRST transformation acting on the fields $\{\phi\}$ 
is defined as
\footnote{By definition (\ref{brst}), 
the gauge fixed BRST-transformation is nilpotent and keeps the action 
invariant only up to the (gauge fixed) equation of motion, which are the 
standard facts in BV-formalism. 
} 
\begin{equation}
 \delta_{gf}=(\ ,S)|_{\phib=0}=
 \sum_i\flpartial{\phi^i}\frpart{S}{\phi^\ib}\Big|_{\phib=0}\ .
 \label{brst}
\end{equation}
One can easily see that eq.(\ref{eom22}) is nothing but 
the statement that $\frpart{S}{\phi^\ib}$ in the right hand side of the 
above equation is equal to zero. 
On the other hand, there exists the Maurer-Cartan equation on 
$(\cH^p,\ti\m^p)$. To represent the $\{\eb\}$ part and $\{\ebb\}$ part 
separately similarly as in eq.(\ref{eom21}) and (\ref{eom22}), these are 
of the form 
\begin{align}
 &\eb^p_\ib\sum_{n\ge 2}
 \ti{m}^{p,\ib}_n(\eb^p_{j_1}\ti\phi^{j_1},\cdots,
 \eb^p_{j_n}\ti\phi^{j_n})=0\ ,
 \label{eom21m}\\
 &\eb^p_i\sum_{n\ge 2}
 \ti{m}^{p,i}_n(\eb^p_{j_1}\ti\phi^{j_1},\cdots,
 \eb^p_{j_n}\ti\phi^{j_n})=0\ 
 \label{eom22m}
\end{align}
on the Siegel gauge. As was explained in subsection \ref{ssec:23}, 
$A_\infty$-morphisms preserve the solutions of 
Maurer-Cartan equations. Eq.(\ref{eom3}) reads that 
the $A_\infty$-morphism $\ti\cF^p$ is given by 
$\Phi|_{b_0\Phi=0}=\Phi^p|_{b_0\Phi=0}+\Phi^u|_{b_0\Phi=0}$ and 
\begin{equation}
 \begin{split}
 \Phi^u|_{b_0\Phi=0}
 &=-\sum_{k\ge 2}b_0\ov{L_0}(\1-P)m_k(\eb^p_j\ti\phi^j+\eb^u_l\phi^l)\\ 
  &=-\sum_{k\ge 2}\ov{L_0}(\1-P)\eb_i 
 m^\ib_k(\eb^p_j\ti\phi^j+\eb^u_l\phi^l)
 \end{split} 
 \label{AmorpS}
\end{equation}
on the Siegel gauge, where $\eb^p_j\in P^p\{\eb\}$ and 
$\eb^u_l\in P^u\{\eb\}$. 
Substituting $\Phi^u|_{b_0\Phi=0}=\eb^u_l\phi^l$ on the left hand side 
of the above equation into the right hand side recursively, we obtain 
the $A_\infty$-morphism $\ti\cF^p=\{\ti{f}^p_k\}_{k\ge 2}$ with 
$\ti{f}^p_k : (P^p\{\eb\})^{\otimes k}\raw \{\eb\}$. 
The fact that $\ti\cF^p$ preserves the Maurer-Cartan equations is then 
easily seen because substituting eq.(\ref{AmorpS}) into eq.(\ref{eom21}) and 
(\ref{eom22}) leads eq.(\ref{eom21m}) and (\ref{eom22m}), respectively. 
Note that the $A_\infty$-morphism 
$\Phi|_{b_0\Phi=0}=\Phi^p|_{b_0\Phi=0}+\ti{f}(\Phi^p|_{b_0\Phi=0})$ given by 
eq.(\ref{AmorpS}) is rewritten as 
\begin{equation}
 \begin{split}
 \Phi|_{b_0\Phi=0}&=\Phi^p|_{b_0\Phi=0}
 -\sum_{k\ge 2}\sum_i\ov{L_0}(\1-P)\eb_i\omega(\eb_i,
 m_k(\eb^p_j\ti\phi^j+\eb^u_l\phi^l))\\
 &=\Phi^p|_{b_0\Phi=0}
 -\sum_{k\ge 2}\sum_i\ov{L_0}(\1-P)\eb_i
 \ti\V_{k+1}(\eb_i,\eb^p_{j_1}\ti\phi^{j_1},\cdots, 
 \eb^p_{j_k}\ti\phi^{j_k})
  \end{split} \label{AmorpS2}
\end{equation}
where $\ti\V_{k+1}$ denotes the tree $k+1$ point (off-shell) 
correlation function. 
The second equality is justified in the next subsection.  
When $\Phi^p|_{b_0\Phi=0}$ satisfies the equation of motion 
(\ref{eom21m}), eq.(\ref{AmorpS2}) gives the solution of eq.(\ref{eom21}). 
In fact the $A_\infty$-morphism restricted on the Siegel gauge 
(\ref{AmorpS}) is derived by regarding only eq.(\ref{eom21}) as a 
Maurer-Cartan equation and applying the arguments in the previous 
subsection. The solutions derived in \cite{MS,KZ} for closed SFT are 
exactly this $\Phi|_{b_0\Phi=0}$ in eq.(\ref{AmorpS2}). 
$\Phi^p|_{b_0\Phi=0}=\eb^p_i\ti\phi^i$ express the condensation of marginal 
operators. 
In \cite{MS,KZ} the zero momentum dilaton condensation is discussed 
in closed SFT. 
The condensation of the zero momentum states corresponding 
to background $g$ and $B$ is also considered\cite{KZ}. 
In both case the obstruction 
eq.(\ref{eom21m}) is expected to vanish and all $\Phi^p|_{b_0\Phi=0}$ given 
by eq.(\ref{AmorpS2}) are the solutions, but generally there exists the 
obstruction (\ref{eom21m}). 
Furthermore, even if in the neighborhood of the origin 
the Siegel gauge is consistent in the sense 
in subsection \ref{ssec:42}, it is not necessarily true apart the origin. 
In order for the solutions to be the ones on which the space of 
the Siegel gauge condition $\phib=0$ is transversal to the gauge orbit, 
one must confirm that the solution actually satisfies eq.(\ref{eom22m}). 
This is equivalent to the condition that on 
the solution of eq.(\ref{eom21}) the gauge fixed BRST-transformation is 
zero. 
In {\it tachyon condensation} in Discussions, 
a few comments about the solution describing 
the tachyonic vacuum\cite{SZn} are presented from these viewpoints.

Though in the above argument the Siegel gauge is considered for simplicity, 
one can take another gauge for constructing $Q^+$. Generally 
in the context of BV-formalism the antifields are gauge fixed as 
$\phi^\ib=\fpart{\Psi(\phi)}{\phi^i}$, where $\Psi(\phi)$ has degree 
minus one in order for $\phi^\ib$ to have consistent degree 
and in this reason is called {\it gauge fixing fermion}. 
In order to keep the gauge fixed kinetic term quadratic, 
here we consider the quadratic gauge fixing fermion. It is generally 
of the form 
\begin{equation*}
 \Psi(\phi)=\phi^iM_{ij}\phi^j
\end{equation*}
where $M_{ij}$ is an appropriate $\C$ valued matrix. By this gauge fixing, 
the antifield is restricted to 
\begin{equation*} 
 \phi^\ib=M_{ij}\phi^j\ .
\end{equation*}
The string field $\Phi$ is then restricted to 
$\Phi|_{\gf}=\eb_i\phi^i+\eb_{\ib}M_{ij}\phi^j
=(\eb_i+\eb_{\jb}M_{ji})\phi^i$. 
Let $S_0|_{\gf}:=\half\omega(\Phi|_{\gf},Q\Phi|_{\gf})$ be 
the gauge fixed kinetic term and 
the propagator $Q^+$ is then defined as this inverse.
The most important thing here is that the propagator $Q^+$ with 
the identity (\ref{HKdecomp}) $QQ^++Q^+Q+P=\1$ always satisfies 
\begin{equation}
 P^p P=P^p\label{P}
\end{equation}
independent of the choice of the gauge. 
Note that for (Thm.\ref{thm:main}), only this identity is needed for $Q^+$. 
Indeed the $P$ in Siegel gauge (\ref{Preg}) satisfies this 
identity, and it is clear that in any other gauge 
$QQ^+$ or $Q^+Q$ can not detect 
the physical states similarly as $P^t=QQ^u$ and $P^u=Q^uQ$. 
In addition $(Q^+)^2=0$ holds, and 
$Q^+QQ^+=Q^+$ is assumed as argued above on the Siegel gauge. 

We will see in the next subsection that 
in any gauge as far as the identity (\ref{P}) holds 
$(\cH,\ti\m^p)$ defines an $A_\infty$-algebra. 
In addition we will propose there that for general configuration of 
fields of $\Phi|_{\gf}$ one may choose the appropriate gauge fixing 
and construct the $A_\infty$-morphism.

 \subsection{On-shell reduction of classical SFT}
\label{ssec:53}

In subsection \ref{ssec:51}, 
the $A_\infty$-structure $(\cH^{p}, \ti\m^{p})$ naturally 
induced from the $A_\infty$-algebra $(\cH,\m)$ was given. 
At the same time, an $A_\infty$-morphism $\ti\cF^p$ from $A_\infty$-algebra 
$(\cH^p, \ti\m^p)$ to $(\cH,\m)$ was constructed. 
Here let $(\cH,\m)$ be the $A_\infty$-algebra which defines a classical 
open SFT action $S(\Phi)$. 
Then by comparing the construction of SFT in section \ref{sec:3} and 
the construction of the $A_\infty$-structure $(\cH^p,\ti\m^p)$ 
in subsection \ref{ssec:51}, 
we can see that the $n$-point vertex defined by $A_\infty$-structure 
$\ti\m^p$ is nothing but the tree level $n$-point correlation function 
of open string(Lem.\ref{lem:main}). 
(Thm.\ref{thm:main}) in the next subsection 
immediately follows from this fact. 

Here will explain the fact. Because the construction of SFT as 
in section \ref{sec:3} guarantees that the scattering amplitudes of 
the SFT with $A_\infty$-structure $(\cH,\m)$ computed by the 
Feynman rule reproduce the correlation function of open string on-shell, 
what should be shown is that the scattering amplitudes of the SFT 
computed by the Feynman rule coincides with 
$-\ov{n}\omega(\eb^p_{i_1},
\ti{m}^p_{n-1}(\eb^p_{i_2},\cdots,\eb^p_{i_n}))$ where 
$\eb^p_{i_1},\cdots,\eb^p_{i_n}\in\cH^p$ 
are the external states of the amplitude. 

The $n$-point amplitude is computed with the Feynman rules as follows. 
Formally, when a SFT action $S=S_0+\V$ is given, the scattering amplitudes 
are computed with the partition function of SFT, 
\begin{equation}
 Z=\int\mes\Phi e^{-S}
 =\int\mes\Phi\l(\sum_{k=1}^\infty\ov{k!}(-\V_3-\V_4-\cdots)^k\r)e^{-S_0}\ .
 \label{sftpf}
\end{equation}
In the second equality, $e^{-\V}=e^{-(\V_3+\V_4+\cdots)}$ is expanded as 
a perturbation. 
Represent the vertex $\V_k$ as 
$\V_k=\ov{k}\la V_k| |\Phi\rangle_1\cdots |\Phi\rangle_k$. 
Fixing a gauge, constructing the propagator in the gauge, and its 
contraction between any two vertices 
$\langle V_{v_1}|$ and $\langle V_{v_2}|$ are described as 
\begin{equation}
 \langle V_{v_1}|\langle V_{v_2}|
 Q^+ |\omega\rangle_{ab}
 \label{vqv}
\end{equation}
where $|\omega\rangle_{ab}$ is the inverse reflection operator 
(\ref{poisson}), and 
the indices $a,b$ denote that the propagator connects the $a$-th legs of 
the vertices $\la V_{v_1}|$ with the $b$-th legs of the vertices for 
$1\le a\le v_1$ and $1\le b\le v_2$. 
Since the value of eq.(\ref{vqv}) does not rely on whether 
$Q^+$ operates on the ket $|\ \rangle_a$ or on the ket $|\ \rangle_b$, 
the index for $Q^+$ is omitted. 
The Feynman rule is then defined by the usual Wick contraction 
using eq.(\ref{vqv})
\footnote{Concerning the relation between this Feynman rule 
of the world sheet picture and the one of component field theory picture, 
the reference \cite{Th} also provides us with useful information. }. 
We are interested in tree amputated amplitudes. 
When the tree $n$-point amplitude with the external propagators is defined 
in the path integral form, 
the Wick contraction with the propagators can be divided 
into two processes : the contraction between the $n$ external states and 
the vertices, and the contraction between the vertices. 
Performing the latter process leads some function of $n$ powers of 
$\Phi$. Here define it as 
\begin{equation}
 -\ov{n}\la\ti{V_n}| |\Phi\ra\cdots |\Phi\ra
 =-\ov{n}\ti\V_n(\Phi,\cdots,\Phi)\ .
 \label{Vp}
\end{equation}
The former process, contracting the $\Phi$ in 
$\ti\V_n(\Phi,\cdots,\Phi)$ with $n$ external fields, 
finishes the calculation of the amplitude. 
Instead, the coefficient of $\phi^{i_n}\cdots\phi^{i_1}$ for 
eq.(\ref{Vp}) reads 
\begin{equation}
 -\ov{n}\la\ti{V_n}| |\eb_{i_1}\ra\cdots |\eb_{i_n}\ra
 =-\ov{n}\ti\V_{i_1\cdots i_n}\ ,
 \label{Vpoffshell}
\end{equation}
the amputated $n$-point amplitude with external states 
$\eb_{i_1}\cdots\eb_{i_n}$. 
Restricting the external states to physical states leads 
the on-shell $n$-point amplitude and express it as 
\begin{equation}
 -\ov{n}\la\ti{V_n}| |\eb^p_{i_1}\ra\cdots |\eb^p_{i_n}\ra
 =-\ov{n}\ti\V^p_{i_1\cdots i_n}\ .
 \label{Vponshell}
\end{equation}
We can also include the $Q$-trivial states 
for the external states but the on-shell amplitudes vanish 
even if one of the external states is $Q$-trivial. 
To avoid to increase the notation, here 
we argue in the case of physical state space $\cH^p$ only. 

More precisely, when computing the Feynman diagram 
we should comment about the gauge fixing . 
Generally let $\cO(\Phi^n)$ be any operators of $(\Phi)^n$ powers, 
then its expectation value is calculated as 
\begin{equation}
 \la\cO(\Phi^n)\ra\sim\int\mes\Phi\ \cO(\Phi^n)e^{-S}\Big|_{\gf}
 =\int\mes\Phi_{gf}\ \cO(\Phi^n_{\gf})
  e^{-\sum_{k\ge 3}\ov{k}\V_k(\Phi_{\gf},\cdots,\Phi_{\gf})}
  e^{-S_0|_{\gf}}
 \label{ev}
\end{equation}
where  $|_{\gf}$ denotes a gauge fixing discussed in the previous 
subsection, $\Phi_{\gf}$ denotes the gauge fixed $\Phi$, and 
$S_0|_{\gf}$ means $S_0$ but $\Phi$ in $S_0$ is replaced by $\Phi_{\gf}$. 
The propagator $Q^+$ is derived from this gauge fixed kinetic term 
$S_0|_{\gf}$. 
The expectation value 
$\la\cO(\Phi^n)\ra$ does not depend on the gauge fixing only when 
$\cO(\Phi^n)$ is a gauge invariant operator. 
The amputated $n$ point amplitudes 
$-\ov{n}\ti\V_{i_1\cdots i_n}$ generally depend on the gauge. 
It is calculated by using the propagator $Q^+$, which depends on the 
choice of the gauge. As will be seen, the vertex $\ti\V_{i_1\cdots i_n}$ 
relates to the $A_\infty$-structure $\ti\m$, 
which implies that the set of 
the vertices satisfies BV-master equation with an 
appropriate symplectic structure $\ti\omega$. 
Here in any gauge we can define an $A_\infty$-structure. 
Thus there are the ambiguities of the gauge choice 
when constructing the $A_\infty$-structure $\ti\m$. 
Mathematically the propagator 
is related to a homotopy operator. It is interesting that the ambiguities 
of homotopy operators are physically those of the propagators 
through the choice of the gauge, and are those of constructing 
higher vertices $\ti\V$ which satisfies BV-master equation. 
However from the expression in eq.(\ref{ev}), it is natural that 
the vertices $\ti\V_k(\Phi,\cdots,\Phi)$ is defined such that its value 
on $\Phi|_{\gf}$ is $\ti\V_k(\Phi_{\gf},\cdots,\Phi_{\gf})$ calculated 
by the propagator with the gauge $|_{\gf}$. 
We then propose this definition for the $A_\infty$-structure $\ti\m$ 
(and an $A_\infty$-quasi-isomorphism $\ti\cF$ which induces the 
transformation from the original $A_\infty$-structure $\m$ to $\ti\m$), 
though later arguments do not depend on this choice. 
Any way, the on-shell amplitudes $-\ov{n}\ti\V^p_{i_1\cdots i_n}$ is gauge 
invariant and is independent of the choice of the gauge. In this subsection 
the arguments are restricted to the on-shell physics and it is shown below 
that $\ti\V^p_{i_1\cdots i_n}$ defines the $A_\infty$-structure $\ti\m^p$. 
Hereafter in this paper $|_{gh}$ is omitted.

By construction, eq.(\ref{Vp}), (\ref{Vpoffshell}) or (\ref{Vponshell}) 
are defined as the sum of Feynman diagrams which is related to 
$\Gamma_{n-1}\in G_{n-1}$ in (Def.\ref{defn:minimal}). 
Here will show that the on-shell $n$-point amplitudes coincide with 
the amplitudes defined by $\ti{m}^p_{n-1}$. 
In order to see that, the following 
two statements must be confirmed : each Feynman graph 
in eq.(\ref{Vponshell}) coincides with some $\ti{m}_{\Gamma_{n-1}}$, 
and all the weight of these Feynman graphs are one because $\ti{m}_{n-1}$ 
is constructed by summing over each $\ti{m}_{\Gamma_{n-1}}$ with 
weight one. We shall confirm these two statements at the same time below.

\begin{figure}[h]
 \hspace*{1.7cm}\includegraphics{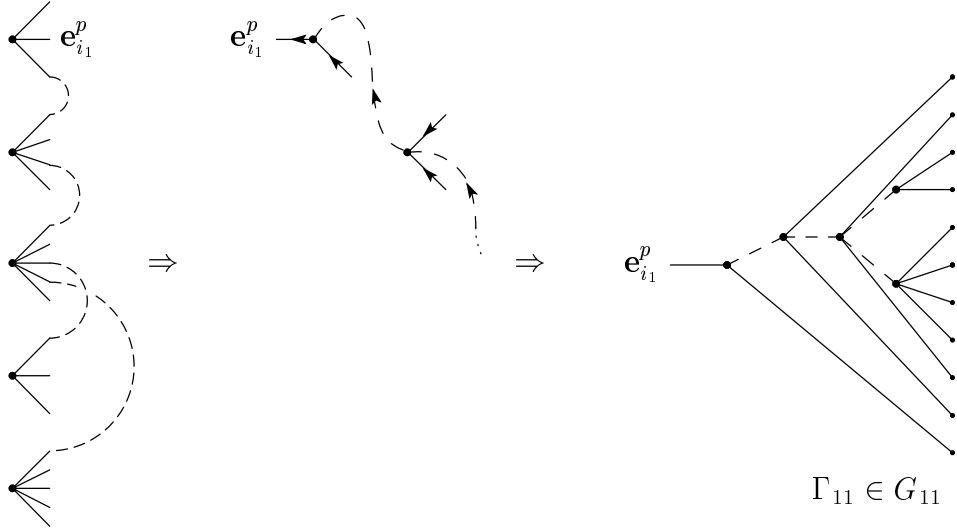}
 \caption{The graph on the left hand side is rewritten as the one 
on the right hand side through introducing the orientation on each edges 
and enjoying the graphical representations in subsection \ref{ssec:43}. 
Here on the left hand side, we adjust the outgoing legs of vertices 
to the top by employing 
the cyclic symmetry of the vertices. In this process for each vertex 
$v_m$ number of graphs are identified, 
and the factor $\ov{v_m}$ in front of $\la V_{v_m}|$ cancels. 
$\eb^p_{i_2},\cdots,\eb^p_{i_{12}}$ is assigned to the rest edges in 
cyclic order. }
 \label{fig:f1}
\end{figure}

As argued in section \ref{sec:3}, when the number of the propagators 
in one of these Feynman graphs is $I$, the number of the vertices is 
$I+1$, and we have an identity $n=\sum_{m=1}^{I+1}v_m-2I$, 
where $v_m$ is the number of the legs of the vertex. 
Because the tree diagrams are considered, there are not more than one 
propagator between any two vertices. 
Each term of eq.(\ref{Vp}) corresponding to each Feynman graph is then 
represented as 
\begin{equation}
 (sym.\ fac. )\ \frac{-1}{v_1}\la V_{v_1}|\cdots\frac{-1}{v_{I+1}}
 \la V_{v_{I+1}}|\ (Q^+|\omega\ra)^I (|\Phi\ra)^n
 \label{Vp2}
\end{equation}
where $(sym.\ fac. )$ means the symmetric factor appearing if 
$v_i=v_j$ for any $i\neq j$. Such factor comes from the coefficient of 
the Taylor expansion of $e^{-\V_{v_m}}$ in eq.(\ref{sftpf}). 
Both $Q^+|\omega\ra$ and $\Phi$ have degree zero, so eq.(\ref{Vp2}) does 
not depend on the order of them. 
Each $Q^+|\omega\ra$ connects any two external states of any two 
different vertices to each other. 
The simplest term is that of $I=0$, which is 
$-\ov{n}\la V_n|(|\Phi\ra)^n$. Therefore it is clear that 
$\la \ti{V}^p_n|=\la V_n|+\cdots$ where $\cdots$ are the terms of $I\ge 1$. 
Note that $\la\ti{V}_n|$ is cyclic-symmetrized by construction. 
Therefore for each term 
choose $\eb^p_{i_1}$ from $(\Phi)^n$, assign 
$\eb^p_{i_2}\cdots\eb^p_{i_n}$ to the rest $(\Phi)^{n-1}$ with cyclic order, 
and the on-shell amplitude eq.(\ref{Vponshell}) with external states 
$\eb^p_{i_1},\cdots,\eb^p_{i_n}$ is obtained by 
summing over all these graphs and dividing by $n$. 
This $n$ is represented 
explicitly in the expression (\ref{Vp}),(\ref{Vpoffshell}) 
and (\ref{Vponshell}). 
Let us concentrate on one of such $n$ point tree Feynman graph. 
We choose $\eb^p_{i_1}$ as the end points of the graph and introduce 
an orientation on the edge of the graphs as follows. 
Let the vertex one of whose external states is $\eb^p_{i_1}$ be 
$\langle V_{v_1}|$. 
On the edge between vertex $\langle V_{v_1}|$ and $\eb^p_{i_1}$, 
we introduce the orientation from $\langle V_{v_1}|$ to $\eb^p_{i_1}$
\footnote{An edge corresponds to a propagator $Q^+|\omega\ra_{ab}$. 
The $|\omega\ra_{ab}$ is symmetric with respect to the label $a, b$ and 
eq.(\ref{vqv}) does not rely on whether $Q^+$ operates on 
$|\ \rangle_a$ or on $|\ \rangle_b$. 
Therefore this orientation of the edges does 
not have any physical meaning.}. 
Other edges connected to $\langle V_{v_1}|$ 
are ordered so that the orientations on them flow into $\langle V_{v_1}|$. 
We write the edge connected to $\eb^p_{i_1}$ on the left hand side and 
the others on the right hand side of the vertex $\langle V_{v_1}|$ 
keeping its cyclic order (see the second step of 
an example in (Fig.\ref{fig:f1})). 
Some of the edges written on the right hand side connect to other 
vertices. We write the edges of those vertices, 
except the edge connecting to $\langle V_{v_1}|$, 
on the right hand side of those vertices keeping its cyclic order, 
and the orientations are ordered so that the flow of each edge is 
from right to left. 
Repeating this, we get a tree graph. An example of the Feynman graph 
in (Fig.\ref{fig:3}) in section \ref{sec:3} 
is figured in (Fig.\ref{fig:f1}). 
The above procedure gives the one-to-one correspondence between 
the $n$-point tree graphs of with external states fixed and 
the tree graphs 
$\Gamma_{n-1}\in G_{n-1}$ defined in (Def.\ref{defn:minimal}). 
The value corresponding to the Feynman graph is expressed 
in terms of $\langle V|$, $Q^+$ and $|\omega\rangle$ as in eq.(\ref{Vp2}), 
but we can rather express in terms of 
$m_k$, $Q^+$ and $|\omega\rangle$. It can be done using the relations in 
subsection \ref{ssec:43} such as 
\begin{equation}
 \langle V_{k+1}|\ |\omega\rangle=m_k\ .
 \label{vtom}
\end{equation}
The contribution of a graph $\Gamma_{n-1}$ to $\ti\V^p_{i_1\cdots i_n}$ 
in eq.(\ref{Vponshell}) is then evaluated as 
\begin{equation}
 -(sym. fac.)\cdot \ov{v_1\cdots v_{I+1}}(-1)^I(-1)^{I+1}
 \omega(\eb^p_{i_1},
\ti{m}^p_{\Gamma_{n-1}}(\eb^p_{i_2},\cdots,\eb^p_{i_n}))\ ,
\end{equation}
where $(-1)^{I+1}$ comes from the vertices and $(-1)^I$ appears because 
$Q^+$ in eq.(\ref{Vp2}) is replaced to $-Q^+$ and the above equation 
includes $I$ number of propagators $Q^+$. 
In practice, let us evaluate the value corresponding to the graph 
$\Gamma_{11}$ in (Fig.\ref{fig:f1}). It is 
\begin{equation*}
 \begin{split}
& -\l(\half\half\r)\frac{(-1)^4(-1)^5}{3\cdot 4\cdot 5\cdot 3\cdot 5}
 \la V_3|\la V_4|\la V_5|\la V_3|\la V_5|
 Q^+|\omega\ra_{(12)}Q^+|\omega\ra_{(23)}
 Q^+|\omega\ra_{(34)}Q^+|\omega\ra_{(35)}|\eb^p_{i_1}\ra
 \cdots |\eb^p_{i_{12}}\ra \\
 &=\l(\half\half\r)\ov{3\cdot 4\cdot 5\cdot 3\cdot 5}\cdot\\
 &\ \omega(\eb_{i_1},m_2(-Q^+m_3(\eb^p_{i_2},-Q^+m_4(\eb^p_{i_3}, 
 -Q^+m_2(\eb^p_{i_4},\eb^p_{i_5}),
 -Q^+m_4(\eb^p_{i_6},\eb^p_{i_7},\eb^p_{i_8},\eb^p_{i_9}),\eb^p_{i_{10}}), 
 \eb^p_{i_{11}}), \eb^p_{i_{12}}) \\
 &=:\l(\half\half\r)\ov{3\cdot 4\cdot 5\cdot 3\cdot 5}
 \omega(\eb^p_{i_1},\ti{m}^p_{\Gamma_{11}}
 (\eb^p_{i_2},\cdots,\eb^p_{i_{12}}))\ .
 \end{split}
\end{equation*}
On the first line of the above equation 
the vertices are labeled as $1\cdots 5$ in the order, 
and the indices $(ab)$ for $a, b=1\cdots 5$ denote that 
the propagator contracts vertices $a$ with $b$. 

On the other hand, the $n$ point vertex given by $\ti\m^p$ is 
$\omega(\eb^p_{i_1},\ti{m}_{n-1}^p(\eb^p_{i_2},\cdots,\eb^p_{i_n}))
=\sum_{\Gamma_{n-1}\in G_{n-1}}(\eb^p_{i_1},P\ti{m}^p_{\Gamma_{n-1}}
(\eb^p_{i_2},\cdots,\eb^p_{i_n}))$. 
The rest of the proof is then to confirm that 
the Feynman rule gives 
the contribution of a graph $\Gamma_{n-1}$ to $\ti\V^p_{i_1\cdots i_n}$ is 
$\omega(\eb^p_{i_1},\ti{m}^p_{\Gamma_{n-1}}(\eb^p_{i_2},\cdots,\eb^p_{i_n}))$ 
with weight $+1$ for each $\Gamma_{n-1}\in G_{n-1}$. 
First, the factor $\ov{v_m}$ in front of $\langle V_{v_m}|$ cancels 
because choosing one outgoing edges from $v_m$ legs 
creates the factor $v_m$ as stated in (Fig.\ref{fig:f1}).
Next, if the graph includes $k$ vertices with the same $v_m$, 
the exchanging of the vertices creates the factor $k!$, 
which cancels with 
the symmetric factor $(sym. fac.)$. 
Thus for each graph $\Gamma_{n-1}$ the same weight $+1$ is obtained and 
one can get 
\begin{equation}
 \ti\V^p_{i_1\cdots i_n}
 =\omega(\eb^p_{i_1},\sum_{\Gamma_{n-1}\in G_{n-1}}
 \ti{m}^p_{\Gamma_{n-1}}(\eb^p_{i_2},\cdots,\eb^p_{i_n}))\ .
 \label{vm}
\end{equation}
Here we claim that the above relation can also be written as 
\begin{equation}
 \ti\V^p_{i_1\cdots i_n}
 =\omega(\eb^p_{i_1},\ti{m}^p_{n-1}(\eb^p_{i_2},\cdots,\eb^p_{i_n}))\ .
 \label{vm2}
\end{equation}
It is guaranteed if $P$ would be the projection onto physical states, 
but as was seen in the previous subsection, it fails. 
However the claim still holds. 
It follows from the existence of the orthogonal 
decomposition (\ref{omegadecomp}) and 
$P^pP=P^p$ (eq.(\ref{P})). \qed
\begin{lem}
$(\cH^p,\ti\m^p)$ defines the $A_\infty$-algebra of string S-matrix elements 
on $\cH^p$, and 
the cohomomorphism $\ti\cF^p$ defined in (Def.\ref{defn:minimal}) is the 
$A_\infty$-quasi-isomorphism between the $A_\infty$-algebras 
$(\cH^p,\ti\m^p)$ and $(\cH,\m)$. 
 \label{lem:main}
\end{lem}
{\it proof.}\ \ The relation between $\ti\m^p$ and the string 
S-matrix elements was given in (\ref{vm2}). 
However it is necessary to show that $\ti\m^p$ actually 
defines an $A_\infty$-structure on physical states $\cH^p$. 
Because $P$ and $P^p$ are different, if we use $\ti\m^p$ naively as an 
$A_\infty$-structure, the image of $\ti{m}^p_k$ does not possibly belong 
to $\cH^p$, and if we define the $A_\infty$-structure on $\cH^p$ as 
the image of $\ti{m}^p_k$ is projected onto $\cH^p$, 
it is not guaranteed that it defines an $A_\infty$-condition (\ref{ainf}).
However fortunately it can be shown that 
\begin{equation*}
 \ti{m}^p_{n-1}(\eb^p_{i_2},\cdots,\eb^p_{i_{n}})\ \in\cH^t\cup\cH^p
\end{equation*}
for any $\eb^p_i\in\cH^p$. 
The proof is given in Appendix \ref{ssec:B3}. 
As was mentioned, in addition to 
$\eb^p_i\in\cH^p$, any $Q$-trivial states can be included and the same 
result is obtained. 
The structure $\ti\m^p$ defines the 
$A_\infty$-structure on on-shell states $\cH^t\cup\cH^p$. 
Furthermore as is shown in Appendix \ref{ssec:B3}, 
the $A_\infty$-structure can be reduced on $\cH^p$. 
We denote the $A_\infty$-algebra by $(\cH^p,\ti\m^p)$.
The following diagram is then obtained as the modified version 
in (Rem.\ref{rem:twoAinfty}) \\
%\begin{figure}[h]
 \hspace*{1.5cm}
%WinTpicVersion3.08
\unitlength 0.1in
\begin{picture}( 31.0039, 12.9921)(  5.9055,-16.8799)
% STR 2 0 3 0
% 3 2200 500 2200 600 5 0
% $(\cH,\ti\m)$
\put(21.6535,-5.9055){\makebox(0,0){$(\cH,\ti\m)$}}%
% VECTOR 2 0 3 0
% 2 2800 600 3600 600
% 
\special{pn 8}%
\special{pa 2756 591}%
\special{pa 3544 591}%
\special{fp}%
\special{sh 1}%
\special{pa 3544 591}%
\special{pa 3478 571}%
\special{pa 3492 591}%
\special{pa 3478 611}%
\special{pa 3544 591}%
\special{fp}%
% STR 2 0 3 0
% 3 4200 500 4200 600 5 0
% $(\cH,\m)$
\put(41.3386,-5.9055){\makebox(0,0){$(\cH,\m)$}}%
% VECTOR 2 0 3 0
% 2 2150 800 2150 1000
% 
\special{pn 8}%
\special{pa 2117 788}%
\special{pa 2117 985}%
\special{fp}%
\special{sh 1}%
\special{pa 2117 985}%
\special{pa 2136 919}%
\special{pa 2117 933}%
\special{pa 2097 919}%
\special{pa 2117 985}%
\special{fp}%
% STR 2 0 3 0
% 3 2200 1100 2200 1200 5 0
% $(\cH^t\cup\cH^p,\ti\m^p)$
\put(21.6535,-11.8110){\makebox(0,0){$(\cH^t\cup\cH^p,\ti\m^p)$}}%
% STR 2 0 3 0
% 3 3200 380 3200 480 5 0
% \scalebox{0.7}[0.7]{$\ti\cF$}
\put(31.4961,-4.7244){\makebox(0,0){\scalebox{0.7}[0.7]{$\ti\cF$}}}%
% STR 2 0 3 0
% 3 3200 710 3200 810 5 0
% \scalebox{0.7}[0.7]{$\ti\cF^p$}
\put(31.4961,-7.9724){\makebox(0,0){\scalebox{0.7}[0.7]{$\ti\cF^p$}}}%
% STR 2 0 3 0
% 3 1950 800 1950 900 5 0
% \scalebox{0.7}[0.7]{$P^t+P^p$}
\put(19.1929,-8.8583){\makebox(0,0){\scalebox{0.7}[0.7]{$P^t+P^p$}}}%
% STR 2 0 3 0
% 3 2050 1400 2050 1500 5 0
% \scalebox{0.7}[0.7]{$P^p$}
\put(20.1772,-14.7638){\makebox(0,0){\scalebox{0.7}[0.7]{$P^p$}}}%
% STR 2 0 3 0
% 3 2200 1700 2200 1800 5 0
% $(\cH^p,\ti\m^p)$
\put(21.6535,-17.7165){\makebox(0,0){$(\cH^p,\ti\m^p)$}}%
% VECTOR 2 0 3 0
% 2 2150 1400 2150 1600
% 
\special{pn 8}%
\special{pa 2117 1378}%
\special{pa 2117 1575}%
\special{fp}%
\special{sh 1}%
\special{pa 2117 1575}%
\special{pa 2136 1509}%
\special{pa 2117 1523}%
\special{pa 2097 1509}%
\special{pa 2117 1575}%
\special{fp}%
% VECTOR 2 0 3 0
% 2 2250 1000 2250 800
% 
\special{pn 8}%
\special{pa 2215 985}%
\special{pa 2215 788}%
\special{fp}%
\special{sh 1}%
\special{pa 2215 788}%
\special{pa 2195 854}%
\special{pa 2215 840}%
\special{pa 2235 854}%
\special{pa 2215 788}%
\special{fp}%
% VECTOR 2 0 3 0
% 2 2250 1600 2250 1400
% 
\special{pn 8}%
\special{pa 2215 1575}%
\special{pa 2215 1378}%
\special{fp}%
\special{sh 1}%
\special{pa 2215 1378}%
\special{pa 2195 1444}%
\special{pa 2215 1431}%
\special{pa 2235 1444}%
\special{pa 2215 1378}%
\special{fp}%
% STR 2 0 3 0
% 3 2350 800 2350 900 5 0
% $\iota$
\put(23.1299,-8.8583){\makebox(0,0){$\iota$}}%
% STR 2 0 3 0
% 3 2350 1400 2350 1500 5 0
% $\iota$
\put(23.1299,-14.7638){\makebox(0,0){$\iota$}}%
% STR 2 0 3 0
% 3 3200 1000 3200 1100 5 0
% \scalebox{0.7}[0.7]{$\ti\cF^p$}
\put(31.4961,-10.8268){\makebox(0,0){\scalebox{0.7}[0.7]{$\ti\cF^p$}}}%
% VECTOR 2 0 3 0
% 2 2800 1050 3600 750
% 
\special{pn 8}%
\special{pa 2756 1034}%
\special{pa 3544 739}%
\special{fp}%
\special{sh 1}%
\special{pa 3544 739}%
\special{pa 3476 744}%
\special{pa 3495 757}%
\special{pa 3490 780}%
\special{pa 3544 739}%
\special{fp}%
% VECTOR 2 0 3 0
% 2 2800 1500 3600 920
% 
\special{pn 8}%
\special{pa 2756 1477}%
\special{pa 3544 906}%
\special{fp}%
\special{sh 1}%
\special{pa 3544 906}%
\special{pa 3479 929}%
\special{pa 3501 937}%
\special{pa 3502 960}%
\special{pa 3544 906}%
\special{fp}%
\end{picture}%
%\end{figure}

\vspace*{0.55cm}

\hspace*{-0.77cm}
where $\ti\m$ and $\ti\cF$ are strictly the same ones 
in (Rem.\ref{rem:twoAinfty}), and 
$\iota :\cH^p\raw\cH^t\cup\cH^p$, $\iota :\cH^t\cup\cH^p\raw\cH$ 
are the inclusion maps. $\iota : \cH^p\raw\cH^t\cup\cH^p$ is extended 
to an $A_\infty$-quasi-isomorphism $\cF^\iota=\{f^\iota_k\}_{k\ge 1}$ as 
$f^\iota_1=\iota$ and $f^\iota_2=f^\iota_3=\cdots=0$. We represent 
$\cF^\iota$ simply as $\iota$. 
Similarly $\iota : \cH^t\cup\cH^p\raw\cH$, $P^p$ and $P^t+P^p$ are 
regarded as $A_\infty$-quasi-isomorphisms. 
The $A_\infty$-structure $\ti\m^p$ on $\cH^t\cup\cH^p$ and $\cH^p$ are 
then given by the composition $(P^t+P^p)\circ\ti\m\circ\iota$ and 
$P^p\circ\ti\m\circ\iota\circ\iota$, respectively. 
The $A_\infty$-quasi-isomorphisms 
$\ti\cF^p :(\cH^t\cup\cH^p,\ti\m^p)\raw (\cH,\m)$ 
and $\ti\cF^p :(\cH^p,\ti\m^p)\raw (\cH,\m)$ are also given as 
$\ti\cF^p=\ti\cF\circ\iota$ and $\ti\cF^p=\ti\cF\circ\iota\circ\iota$, 
respectively. $\ti\cF$ and these two $\ti\cF^p$ also have the inverse 
$A_\infty$-quasi-isomorphism similarly as (Rem.\ref{rem:inverse}). 

As will be seen, when the $A_\infty$-structure is applied together with the 
symplectic structure, the symplectic structure on the $A_\infty$-algebra 
$(\cH^t\cup\cH^p,\ti\m^p)$ is degenerate and 
the $A_\infty$-algebra $(\cH^p,\ti\m^p)$ is more convenient. 
\qed

We have been seen that there exists an $A_\infty$-structure 
$(\cH^p,\ti\m^p)$ on physical states $\cH^p$ (or on on-shell 
Hilbert space $\cH^t\cup\cH^p$ ) 
and the $n$-point amplitudes defined by $\ti\m^p$ 
coincides with the on-shell $n$-point tree amplitudes of open strings. 
The fact that the $n$ point correlation functions in two-dimensional theory 
possess the $A_\infty$-structure are essentially already known. 
It is described in \cite{WZ} that 
the $S^2$ tree amplitudes for closed strings has a 
$L_\infty$-structure, where the external states are restricted to 
physical states and therefore it has vanishing $Q$. This implies that 
the tree level closed string free energy satisfies the classical 
BV-master equation. The result is extended to 
quantum closed string, and it is shown that 
the free energy which consists of the closed string loop amplitudes 
satisfies the quantum BV-master equation\cite{V}
\footnote{In \cite{WZ,V} 
these structures are derived in the context of 2D-string theory, 
\ie the dimension of the target space is two. However they 
are in fact the general structures of the string world sheet. }. 
Moreover, in \cite{Z1}, for classical closed SFT 
the $\cM_k^0\raw\cM_k$ limits are considered, and it is shown that 
restricting the external states to physical states yields 
the $L_\infty$-structure found in \cite{WZ}. 
The open string version of this $L_\infty$-structure is nothing but 
the $A_\infty$-structure $\ti\m^p$ considered in this paper (and reviewed 
in Appendix \ref{ssec:B2}). 
What is obtained newly from the above result is that 
the $A_\infty$-algebras associated with SFTs and the 
$A_\infty$-algebra of two-dimensional theory are connected by an 
$A_\infty$-morphism, which preserves the equation of motions, 
and the explicit form of the $A_\infty$-morphism are realized with 
such a familiar language of the Feynman graph.

 \subsection{Field transformation between family of classical SFTs}
\label{ssec:main}

In section \ref{sec:3} we construct SFTs and which are defined more 
precisely in subsection \ref{ssec:43}. 
As their propagator we consider the one which satisfies eq.(\ref{P}). 
For the family of those well-defined SFTs, (Lem.\ref{lem:main}) leads 
the main claim of the present paper (Thm.\ref{thm:main}) described below. 
In this subsection after proving it, 
we show that they preserve the value of the actions in a certain subspace 
of $\cH$ using (Prop.\ref{prop:Ap}) presented later. 
The field transformations induce 
one-to-one correspondence of moduli spaces of classical solutions 
between such SFTs in the context of deformation theory. 
We will explain that the classical solutions are regarded as those 
corresponding to marginal deformations. 
Finally in this subsection a boundary SFT like action 
corresponding to the $A_\infty$-algebra 
$(\cH,\ti\m)$ defined in (Rem.\ref{rem:twoAinfty}) is proposed.

\begin{thm}
 All the well-defined classical SFTs which are constructed 
on a fixed conformal field theory are quasi-isomorphic to each other. 
 \label{thm:main}
\end{thm}
{\it proof.}\ \ 
As was explained in section \ref{sec:3}, 
when the decomposition of the moduli space of Riemann surfaces is 
given, then the vertices of SFT are determined, \ie SFT action 
is determined. 
Different decompositions of moduli space lead different SFTs. 
By construction the Feynman graphs for those SFTs should 
reproduce the string correlation functions on-shell. 
Here let $S(\Phi)$, $S'(\Phi')$ be such two SFT actions and 
$(\cH,\m)$, $(\cH,\m')$ be the corresponding $A_\infty$-algebras. 
(Lem.\ref{lem:main}) states that the set of the 
on-shell string correlation functions 
defines an $A_\infty$-algebra $(\cH^p,\ti\m^p)$ and 
the $A_\infty$-algebras of SFTs $(\cH,\m)$, $(\cH,\m')$ are 
$A_\infty$-quasi-isomorphic to $(\cH^p,\ti\m^p)$. 
Thus the $A_\infty$-quasi-isomorphism $\ti\cF^p$ from $(\cH^p,\ti\m^p)$ to 
$(\cH,\m)$ and $\ti{\cF'}^p$ from $(\cH^p,\ti\m^p)$ to $(\cH,\m')$ 
exist. 
Note that on a fixed conformal background any such $A_\infty$-algebras 
of SFTs are $A_\infty$-quasi-isomorphic to the same 
$A_\infty$-algebra $(\cH^p,\ti\m^p)$. 
The composition $\ti{\cF'}^p\circ (\ti\cF^p)^{-1}$ then defines the 
$A_\infty$-quasi-isomorphism from $(\cH,\m')$ to $(\cH,\m)$. 
This map is in fact a quasi-isomorphism because 
the inverse of quasi-isomorphism is a quasi-isomorphism and 
the composition of quasi-isomorphisms is a quasi-isomorphism. 
\qed

This result indicates that the equations of motions 
for any classical SFTs constructed from the
same on-shell S-matrix elements are transformed to each other by the 
above quasi-isomorphism $\ti{\cF'}^p\circ (\ti\cF^p)^{-1}$. 
However the quasi-isomorphism is not an isomorphism, 
it does not guarantee that there exists a field redefinition between them 
which preserves the value of the actions. 
In contrast, any diffeomorphisms $\cF$ on $\cH$ which preserve 
the value of the actions of the form 
\begin{equation}
 \Phi'=\cF_*(\Phi)=f_1(\Phi)+f_2(\Phi,\Phi)+\cdots\ 
 \label{fdiffeo}
\end{equation}
are $A_\infty$-isomorphisms, by defining 
the symplectic structures on $S(\Phi)$ and $S'(\Phi')$ so that the 
diffeomorphism $\cF^*$ preserves these symplectic structures\cite{GZ}. 
The statement that $\cF$ preserves the values of the actions 
is, in other words, that $\cF$ satisfies 
$S(\Phi)=\cF^*S'(\Phi'):=S'(\cF_*(\Phi))$. 
The SFTs, which are characterized by the pair $(S(\Phi),\omega)$, 
are called equivalent when the SFTs are connected by 
such a diffeomorphism preserving the action and the 
symplectic structures\cite{SZ1}. 
The fact that the diffeomorphism which connects equivalent SFTs is an 
$A_\infty$-morphism is clearly understood in the dual (component field) 
picture as follows. 
Express two string fields as $\Phi=\eb_i\phi^i$ and $\Phi'=\eb_i'{\phi'}^i$, 
and acting $\flpartial{\phi^i}\omega^{ij}\frpartial{\phi^j}$ on the 
identity $S(\Phi)=S'(\cF_*(\Phi))$ leads 
\begin{equation*}
\flpartial{\phi^i}\omega^{ij}\frpartial{\phi^j}S(\Phi)
 =\flpartial{\phi^i}\omega^{ij}\frpartial{\phi^j}S'(\cF_*(\Phi))\ .
\end{equation*}
The left hand sides is exactly $\delta$ : the dual description of the 
$A_\infty$-coderivative in eq.(\ref{dualD}). 
The right hand side is rewritten as 
\begin{equation*}
 \flpartial{{\phi'}^k}\flpart{{\phi'}^k}{\phi^i}\omega^{ij}
 \frpart{{\phi'}^l}{\phi^j}\frpartial{{\phi'}^l}S'(\Phi')\ .
\end{equation*}
Thus, if $\cF$ preserves the symplectic structure or equivalently 
the Poisson structure is preserved 
\begin{equation}
 \cF^*{\omega'}^{kl}=\flpart{{\phi'}^k}{\phi^i}\omega^{ij}
 \frpart{{\phi'}^l}{\phi^j}\ , \label{poipre}
\end{equation}
it is clear that the right hand side gives 
$\delta'=\flpartial{{\phi'}^k}{{\omega'}^{kl}}
\frpartial{{\phi'}^l}S'(\Phi')=(\ \ ,S'(\Phi'))$. 
To summarize and extend the above arguments, 
one can obtain the following fact : 
\begin{prop}
 When a cohomomorphism $\cF$ 
of the form in eq.(\ref{fdiffeo}) 
between two actions $S(\Phi)$ and $S'(\Phi')$ 
and two symplectic structures which are preserved 
by the cohomomorphism $\cF$ are given, 
then the following two statements are equivalent : 
\begin{itemize}
 \item $\cF$ is an $A_\infty$-morphism. 

 \item $\cF$ preserves the value of the action, that is, 
 $S(\Phi)=\cF^*S'(\Phi')$. 
\end{itemize}
 \label{prop:Ap}
\end{prop}
This equivalence follows from the fact that the symplectic structures 
on both sides are non-degenerate. 
Note that here $f_1$ may not be an isomorphism. 
(Prop.\ref{prop:Ap}) is well-defined and holds 
for general symplectic structures 
on $\cH$ which depend on $\{\phi\}$. 
Because in this paper mainly we deal with the constant symplectic structures 
associated with the BPZ-inner product, we avoid the explanation of 
the issue in this paper (see {\it cyclic algebra with BV-Poisson structure} 
in subsection \ref{ssec:43}). Of course when considering the 
graded commutative fields such as $U(1)$ gauge fields, the 
symplectic structure is that on supermanifold\cite{Sch} and is known to 
be well-defined.

Here we come back to the physical consequence of (Thm.\ref{thm:main}). 
We define the action $\ti{S}(\ti\Phi^p)$ on the $A_\infty$-algebra 
$(\cH^p,\ti\m^p)$ by summing up 
$\ov{n}\ti\V^p_n(\ti\Phi^p,\cdots,\ti\Phi^p)$ in eq.(\ref{Vp}) as 
\begin{equation}
 \ti{S}(\ti\Phi^p)=\sum_{k\ge 2}\ov{k+1}
 \omega(\ti\Phi^p,\ti{m}^p_k(\ti\Phi^p))\ .
 \label{Sp}
\end{equation}
Now we have an $A_\infty$-quasi-isomorphism 
$\ti{\cF'}^p\circ (\ti\cF^p)^{-1}$ 
between $(\cH,\m)$ and $(\cH,\m')$. 
Though it has not assumed in the proof of (Thm.\ref{thm:main}), 
the $\ti\cF^p$ and $\ti{\cF'}^p$ preserve the value of the actions, 
that is, the $A_\infty$-quasi-isomorphism $\ti\cF^p$ satisfies 
\begin{equation}
 \ti{S}(\ti\Phi^p)=(\ti\cF^p)^*S(\Phi)=S((\ti\cF^p)_*(\ti\Phi^p))
 \label{fppreserve}
\end{equation}
and similar for $\ti{\cF'}^p$. 
This fact yields that, on the subspace of $\cH$ 
which is the image of $\ti\cF^p$ from $(\cH^p,\ti\m^p)$, 
any actions $S(\Phi)$ are preserved. 
Eq.(\ref{fppreserve}) follows from (Prop.\ref{prop:Ap}) and the fact 
that the $\ti\cF^p$ preserves the symplectic structures. 
Let $\omega_{ij}$ be the symplectic structures on $(\cH,\m)$. 
Each term in $\ti{S}(\ti\Phi^p)$ was made of the $A_\infty$-structure 
$\ti{m}^p_k$ as $\ov{k+1}\omega(\ti\Phi^p,\ti{m}_k(\ti\Phi^p))$, 
so the symplectic structure on $(\cH^p,\ti\m^p)$ is $\omega_{ij}$ 
restricted on $\cH^p\subset\cH$. On the other hand, 
define $\ti\omega^p_{ij}$ as a symplectic structure on $(\cH^p,\ti\m^p)$ 
which is preserved under the transformation $\ti\cF^p$. 
$\ti\omega^p=(\ti\cF^p)^*\omega$ is written as 
\footnote{The equation below is actually equivalent to 
eq.(\ref{poipre}). 
The equivalence follows from $\omega_{ij}\omega^{jk}=\delta_i^k$ and 
$\ti\omega^p_{ij}\ti\omega^{p,jk}=\delta_i^k$. The 
argument is well-defined even for the non-constant symplectic structure.}
\begin{equation*}
 \ti\omega^p_{ij}
 =\frpart{\phi^k}{\ti\phi^{p,i}}\omega_{kl}\flpart{\phi^l}{\ti\phi^{p,j}}
 =(-1)^{\eb^p_i}
 \omega\l(\frpartial{\ti\phi^{p,i}}\Phi,\Phi\flpartial{\ti\phi^{p,j}}\r)\ .
\end{equation*}
Here we choose the basis so that the inner product $\omega$ is decomposed 
orthogonally as in eq.(\ref{omegadecomp}). 
Since $\Phi=\ti\Phi^p-Q^+\sum_{k\ge 2}m_k(\Phi)$ and 
the image of $Q^+$ vanishes in the symplectic inner product in the right 
hand side of the above equation by using eq.(\ref{qqdag}), 
the right hand side becomes 
$(-1)^{\eb^p_i}\omega(\frpartial{\ti\phi^{p,i}}\ti\Phi^p,
\ti\Phi^p\flpartial{\ti\phi^{p,j}})=\omega_{ij}$. 
Thus the $\ti\omega^p_{ij}$ coincides with the $\omega_{ij}$ restricted on 
$\cH^p$, the map $\ti\cF^p$ from $\ti{S}(\ti\Phi^p)$ to $S(\Phi)$ preserves 
the symplectic structures, and (Prop.\ref{prop:Ap}) leads that 
eq.(\ref{fppreserve}) holds
\footnote{In the proof the appropriate basis is chosen, but 
it can also be seen that the result is independent of the choice because 
$\omega_{ij}$ restricted on $\cH^p$ is independent of it. }. 

Although the proof of eq.(\ref{fppreserve}) has been concreted, 
it is interesting to observe $S((\ti\cF^p)_*(\Phi))$ directly by 
substituting $\Phi=\ti\cF^p_*(\ti\Phi^p)$ in $S(\Phi)$ 
\begin{equation}
 S((\ti\cF^p)_*(\Phi))
 =\half\omega\l(\ti{f}^p(\ti\Phi^p),Q\ti{f}^p(\ti\Phi^p)\r)
 +\sum_{k\ge 2}\ov{k+1}\omega\l(\ti\Phi^p+\ti{f}^p(\ti\Phi^p),
 m_k(\ti\Phi^p+\ti{f}^p(\ti\Phi^p))\r)
 \label{Sp2}
\end{equation}
and check that $S((\ti\cF^p)_*(\Phi))$ in fact coincides with 
$\ti{S}(\ti\Phi^p)$. 
The equation (\ref{Sp2}) is written as the power series of $\ti\Phi^p$ 
and let us observe the $(\ti\Phi^p)^n$ parts of eq.(\ref{Sp2}) for $n\ge 3$. 
For $n=3$ the first term in the right hand side of this equation 
drops out and it can be seen clearly that 
$S((\ti\cF^p)_*(\Phi))|_{(\ti\Phi^p)^{\otimes 3}}=\ov{3}
\omega(\ti\Phi^p,\ti{m}^p_2(\ti\Phi^p,\ti\Phi^p))$. 
Generally both first and second term contribute. 
Using $\ti{m}^p_{\Gamma_k}$ in (Def.\ref{defn:minimal}), 
the contributions of the first and second terms to the terms of 
$n$ powers of $\ti\Phi^p$ are 
\begin{equation}
 \sum_{\Gamma_{k_1\ge 2},\Gamma_{k_2\ge 2},k_1+k_2=n}
 -\half\omega(\ti{m}^p_{\Gamma_{k_1}}(\ti\Phi^p),
 -Q^+\ti{m}^p_{\Gamma_{k_2}}(\ti\Phi^p))
 \label{c1}
\end{equation}
and 
\begin{equation}
 \sum_{l\ge 3}^n\ov{l}\sum_{\substack{
 \Gamma_{k_1\ge 1},\cdots,\Gamma_{k_l\ge 1},\\
k_1+\cdots +k_l=n}}\omega\l(-Q^+\ti{m}^p_{\Gamma_{k_1}}(\ti\Phi^p),
 m_{l-1}(-Q^+\ti{m}^p_{\Gamma_{k_2}}(\ti\Phi^p),\cdots,
 -Q^+\ti{m}^p_{\Gamma_{k_l}}(\ti\Phi^p))\r)\ , 
 \label{c2}
\end{equation}
respectively. In eq.(\ref{c1}) $Q^+QQ^+=Q^+$ is used. 
Here in eq.(\ref{c2}) 
we denoted $\ti\Phi^p$ by $-Q^+\ti{m}^p_{\Gamma_1}(\ti\Phi^p)$. 
In the expression where the cyclicity of vertices  $\{m_{l-1}\}$ are 
emphasized, eq.(\ref{c1}) and (\ref{c2}) are graphically pictured as 

\vspace*{0.2cm}

 \hspace*{-2.0cm}
%WinTpicVersion3.08
\unitlength 0.1in
\begin{picture}( 63.0413,  8.4646)( -8.9075,-10.5315)
% LINE 2 0 3 0
% 2 1600 595 2000 595
% 
\special{pn 8}%
\special{pa 1575 586}%
\special{pa 1969 586}%
\special{fp}%
% STR 2 0 3 0
% 3 2200 495 2200 595 5 0
% $\Gamma_{k_2}$
\put(21.6535,-5.8563){\makebox(0,0){$\Gamma_{k_2}$}}%
% STR 2 0 3 0
% 3 1400 495 1400 595 5 0
% $\Gamma_{k_1}$
\put(13.7795,-5.8563){\makebox(0,0){$\Gamma_{k_1}$}}%
% STR 2 0 3 0
% 3 5700 505 5700 605 5 0
% $\Gamma_{k_1}$
\put(56.1024,-5.9547){\makebox(0,0){$\Gamma_{k_1}$}}%
% LINE 2 0 3 0
% 2 5100 605 5500 605
% 
\special{pn 8}%
\special{pa 5020 596}%
\special{pa 5414 596}%
\special{fp}%
% LINE 2 0 3 0
% 2 5100 605 5433 827
% 
\special{pn 8}%
\special{pa 5020 596}%
\special{pa 5348 814}%
\special{fp}%
% LINE 2 0 3 0
% 2 5100 605 5500 605
% 
\special{pn 8}%
\special{pa 5020 596}%
\special{pa 5414 596}%
\special{fp}%
% LINE 2 0 3 0
% 2 5100 605 5433 383
% 
\special{pn 8}%
\special{pa 5020 596}%
\special{pa 5348 377}%
\special{fp}%
% LINE 2 0 3 0
% 2 5100 607 5256 975
% 
\special{pn 8}%
\special{pa 5020 598}%
\special{pa 5174 960}%
\special{fp}%
% LINE 2 0 3 0
% 2 5100 607 5434 827
% 
\special{pn 8}%
\special{pa 5020 598}%
\special{pa 5349 814}%
\special{fp}%
% CIRCLE 2 2 3 0
% 4 5100 605 5100 405 5290 175 5010 1165
% 
\special{pn 8}%
\special{ar 5020 596 197 197  1.7301480 1.7901480}%
\special{ar 5020 596 197 197  1.9701480 2.0301480}%
\special{ar 5020 596 197 197  2.2101480 2.2701480}%
\special{ar 5020 596 197 197  2.4501480 2.5101480}%
\special{ar 5020 596 197 197  2.6901480 2.7501480}%
\special{ar 5020 596 197 197  2.9301480 2.9901480}%
\special{ar 5020 596 197 197  3.1701480 3.2301480}%
\special{ar 5020 596 197 197  3.4101480 3.4701480}%
\special{ar 5020 596 197 197  3.6501480 3.7101480}%
\special{ar 5020 596 197 197  3.8901480 3.9501480}%
\special{ar 5020 596 197 197  4.1301480 4.1901480}%
\special{ar 5020 596 197 197  4.3701480 4.4301480}%
\special{ar 5020 596 197 197  4.6101480 4.6701480}%
\special{ar 5020 596 197 197  4.8501480 4.9101480}%
\special{ar 5020 596 197 197  5.0901480 5.1284535}%
% STR 2 0 3 0
% 3 5600 815 5600 915 5 0
% $\Gamma_{k_2}$
\put(55.1181,-9.0059){\makebox(0,0){$\Gamma_{k_2}$}}%
% STR 2 0 3 0
% 3 5310 1055 5310 1155 5 0
% $\Gamma_{k_3}$
\put(52.2638,-11.3681){\makebox(0,0){$\Gamma_{k_3}$}}%
% STR 2 0 3 0
% 3 5560 195 5560 295 5 0
% $\Gamma_{k_l}$
\put(54.7244,-2.9035){\makebox(0,0){$\Gamma_{k_l}$}}%
% DOT 0 0 3 0
% 2 5100 605 5100 605
% 
\special{pn 20}%
\special{sh 1}%
\special{ar 5020 596 10 10 0  6.28318530717959E+0000}%
\special{sh 1}%
\special{ar 5020 596 10 10 0  6.28318530717959E+0000}%
% STR 2 0 3 0
% 3 400 500 400 600 5 0
% $-\ \sum_\Gamma\ (sym. fac.)$
\put(3.9370,-5.9055){\makebox(0,0){$-\ \sum_\Gamma\ (sym. fac.)$}}%
% STR 2 0 3 0
% 3 3950 500 3950 600 5 0
% $\sum_{l\ge 3}^n\sum_\Gamma\ (sym. fac.)$
\put(38.8780,-5.9055){\makebox(0,0){$\sum_{l\ge 3}^n\sum_\Gamma\ (sym. fac.)$}}%
% STR 2 0 3 0
% 3 2800 500 2800 600 5 0
% $+$
\put(27.5591,-5.9055){\makebox(0,0){$+$}}%
% DOT 0 0 3 0
% 2 1800 595 1800 595
% 
\special{pn 20}%
\special{sh 1}%
\special{ar 1772 586 10 10 0  6.28318530717959E+0000}%
\special{sh 1}%
\special{ar 1772 586 10 10 0  6.28318530717959E+0000}%
\end{picture}%
\qquad \ ,

\vspace*{0.4cm}

\hspace*{-0.8cm} where $\Gamma_k$ denotes that 
$\ti{m}^p_{\Gamma_k}(\ti\Phi^p)$ is in this place. 
The summations $\sum_{\Gamma}$ in the first and second term are 
the summations for $\Gamma_{k_i}$'s in eq.(\ref{c1}) and (\ref{c2}), 
respectively.  
The $(sym. fac. )$ in the first and second term are 
the symmetric factors with respect to the $\Gamma_{k_i}$'s. When 
$\Gamma_{k_i}\neq\Gamma_{k_j}$ for any $i\neq j$, then 
$(sym. fac.)=1$. In the first term when $\Gamma_{k_1}=\Gamma_{k_2}$ then 
$(sym. fac.)=\half$, and in the second term when $\Gamma_{k_i}=\Gamma_{k_j}$ 
for all $1\le i,j\le l$ then $(sym. fac.) =\ov{l}$. These factor comes from 
the coefficients in eq.(\ref{c1}) and (\ref{c2}). 
Each term contributes to the term 
$\omega(\ti\Phi^p,\ti{m}^p_{\Gamma_{n-1}}(\ti\Phi^p,\cdots,\ti\Phi^p))$ 
for some $\Gamma_{n-1}$. Here fix the tree graph $\Gamma_{n-1}$ and 
treat it as a cyclic graph, \ie 
a outgoing leg and $n-1$ incoming legs are not distinguished and 
the graphs which coincides with each other by moving cyclic are 
identified. Denote it by $\Gamma_{n-1}^{cyc}$ and 
let us observe 
the coefficient of 
$\omega(\ti\Phi^p,
\ti{m}^p_{\Gamma_{n-1}^{cyc}}(\ti\Phi^p,\cdots,\ti\Phi^p))$ 
from the above two contributions.
\begin{figure}[h]
 \hspace*{2.3cm}
%WinTpicVersion3.08
\unitlength 0.1in
\begin{picture}( 42.6181, 14.2224)( 14.9606,-25.2461)
% LINE 2 1 3 0
% 2 2200 1800 2500 1800
% 
\special{pn 8}%
\special{pa 2166 1772}%
\special{pa 2461 1772}%
\special{da 0.070}%
% LINE 2 0 3 0
% 2 2500 1800 2800 2100
% 
\special{pn 8}%
\special{pa 2461 1772}%
\special{pa 2756 2067}%
\special{fp}%
% LINE 2 0 3 0
% 2 2650 1650 2800 1800
% 
\special{pn 8}%
\special{pa 2609 1625}%
\special{pa 2756 1772}%
\special{fp}%
% LINE 2 1 3 0
% 2 2500 1800 2650 1650
% 
\special{pn 8}%
\special{pa 2461 1772}%
\special{pa 2609 1625}%
\special{da 0.070}%
% LINE 2 0 3 0
% 2 2650 1650 2800 1500
% 
\special{pn 8}%
\special{pa 2609 1625}%
\special{pa 2756 1477}%
\special{fp}%
% DOT 0 0 3 0
% 2 2200 1800 2200 1800
% 
\special{pn 20}%
\special{sh 1}%
\special{ar 2166 1772 10 10 0  6.28318530717959E+0000}%
\special{sh 1}%
\special{ar 2166 1772 10 10 0  6.28318530717959E+0000}%
% DOT 0 0 3 0
% 2 2500 1800 2500 1800
% 
\special{pn 20}%
\special{sh 1}%
\special{ar 2461 1772 10 10 0  6.28318530717959E+0000}%
\special{sh 1}%
\special{ar 2461 1772 10 10 0  6.28318530717959E+0000}%
% DOT 0 0 3 0
% 2 2650 1650 2650 1650
% 
\special{pn 20}%
\special{sh 1}%
\special{ar 2609 1625 10 10 0  6.28318530717959E+0000}%
\special{sh 1}%
\special{ar 2609 1625 10 10 0  6.28318530717959E+0000}%
% LINE 2 1 3 0
% 2 2200 1800 2050 1540
% 
\special{pn 8}%
\special{pa 2166 1772}%
\special{pa 2018 1516}%
\special{da 0.070}%
% LINE 2 0 3 0
% 2 2050 1540 2160 1130
% 
\special{pn 8}%
\special{pa 2018 1516}%
\special{pa 2126 1113}%
\special{fp}%
% LINE 2 0 3 0
% 2 1845 1485 1900 1280
% 
\special{pn 8}%
\special{pa 1816 1462}%
\special{pa 1871 1260}%
\special{fp}%
% LINE 2 1 3 0
% 2 2050 1540 1845 1485
% 
\special{pn 8}%
\special{pa 2018 1516}%
\special{pa 1816 1462}%
\special{da 0.070}%
% LINE 2 0 3 0
% 2 1845 1485 1640 1430
% 
\special{pn 8}%
\special{pa 1816 1462}%
\special{pa 1615 1408}%
\special{fp}%
% DOT 0 0 3 0
% 2 2050 1540 2050 1540
% 
\special{pn 20}%
\special{sh 1}%
\special{ar 2018 1516 10 10 0  6.28318530717959E+0000}%
\special{sh 1}%
\special{ar 2018 1516 10 10 0  6.28318530717959E+0000}%
% DOT 0 0 3 0
% 2 1845 1485 1845 1485
% 
\special{pn 20}%
\special{sh 1}%
\special{ar 1816 1462 10 10 0  6.28318530717959E+0000}%
\special{sh 1}%
\special{ar 1816 1462 10 10 0  6.28318530717959E+0000}%
% LINE 2 1 3 0
% 2 2200 1800 2050 2060
% 
\special{pn 8}%
\special{pa 2166 1772}%
\special{pa 2018 2028}%
\special{da 0.070}%
% LINE 2 0 3 0
% 2 2050 2060 1640 2170
% 
\special{pn 8}%
\special{pa 2018 2028}%
\special{pa 1615 2136}%
\special{fp}%
% LINE 2 0 3 0
% 2 2105 2265 1900 2320
% 
\special{pn 8}%
\special{pa 2072 2230}%
\special{pa 1871 2284}%
\special{fp}%
% LINE 2 1 3 0
% 2 2050 2060 2105 2265
% 
\special{pn 8}%
\special{pa 2018 2028}%
\special{pa 2072 2230}%
\special{da 0.070}%
% LINE 2 0 3 0
% 2 2105 2265 2160 2470
% 
\special{pn 8}%
\special{pa 2072 2230}%
\special{pa 2126 2432}%
\special{fp}%
% DOT 0 0 3 0
% 2 2050 2060 2050 2060
% 
\special{pn 20}%
\special{sh 1}%
\special{ar 2018 2028 10 10 0  6.28318530717959E+0000}%
\special{sh 1}%
\special{ar 2018 2028 10 10 0  6.28318530717959E+0000}%
% DOT 0 0 3 0
% 2 2105 2265 2105 2265
% 
\special{pn 20}%
\special{sh 1}%
\special{ar 2072 2230 10 10 0  6.28318530717959E+0000}%
\special{sh 1}%
\special{ar 2072 2230 10 10 0  6.28318530717959E+0000}%
% LINE 2 0 3 0
% 2 2200 1800 2540 2390
% 
\special{pn 8}%
\special{pa 2166 1772}%
\special{pa 2501 2353}%
\special{fp}%
% LINE 2 0 3 0
% 2 2540 1200 2200 1790
% 
\special{pn 8}%
\special{pa 2501 1182}%
\special{pa 2166 1762}%
\special{fp}%
% LINE 2 0 3 0
% 2 1520 1800 2201 1799
% 
\special{pn 8}%
\special{pa 1497 1772}%
\special{pa 2167 1771}%
\special{fp}%
% CIRCLE 2 2 3 0
% 4 2200 1800 2250 1800 2250 1800 2250 1800
% 
\special{pn 8}%
\special{ar 2166 1772 50 50  0.0000000 0.2400000}%
\special{ar 2166 1772 50 50  0.9600000 1.2000000}%
\special{ar 2166 1772 50 50  1.9200000 2.1600000}%
\special{ar 2166 1772 50 50  2.8800000 3.1200000}%
\special{ar 2166 1772 50 50  3.8400000 4.0800000}%
\special{ar 2166 1772 50 50  4.8000000 5.0400000}%
\special{ar 2166 1772 50 50  5.7600000 6.0000000}%
% LINE 2 1 3 0
% 2 5000 1800 5300 1800
% 
\special{pn 8}%
\special{pa 4922 1772}%
\special{pa 5217 1772}%
\special{da 0.070}%
% LINE 2 1 3 0
% 2 5000 1800 4800 1600
% 
\special{pn 8}%
\special{pa 4922 1772}%
\special{pa 4725 1575}%
\special{da 0.070}%
% LINE 2 1 3 0
% 2 4800 1600 4650 1450
% 
\special{pn 8}%
\special{pa 4725 1575}%
\special{pa 4577 1428}%
\special{da 0.070}%
% LINE 2 0 3 0
% 2 4650 1450 4650 1300
% 
\special{pn 8}%
\special{pa 4577 1428}%
\special{pa 4577 1280}%
\special{fp}%
% LINE 2 0 3 0
% 2 4650 1450 4500 1450
% 
\special{pn 8}%
\special{pa 4577 1428}%
\special{pa 4430 1428}%
\special{fp}%
% LINE 2 0 3 0
% 2 5000 1800 4450 1800
% 
\special{pn 8}%
\special{pa 4922 1772}%
\special{pa 4380 1772}%
\special{fp}%
% LINE 2 0 3 0
% 2 5300 1800 5850 1800
% 
\special{pn 8}%
\special{pa 5217 1772}%
\special{pa 5758 1772}%
\special{fp}%
% LINE 2 0 3 0
% 2 4800 1600 4480 1600
% 
\special{pn 8}%
\special{pa 4725 1575}%
\special{pa 4410 1575}%
\special{fp}%
% LINE 2 0 3 0
% 2 4800 1600 4800 1290
% 
\special{pn 8}%
\special{pa 4725 1575}%
\special{pa 4725 1270}%
\special{fp}%
% LINE 2 1 3 0
% 2 5000 1800 4800 2000
% 
\special{pn 8}%
\special{pa 4922 1772}%
\special{pa 4725 1969}%
\special{da 0.070}%
% LINE 2 0 3 0
% 2 4800 2000 4500 2000
% 
\special{pn 8}%
\special{pa 4725 1969}%
\special{pa 4430 1969}%
\special{fp}%
% LINE 2 0 3 0
% 2 4800 2000 4800 2250
% 
\special{pn 8}%
\special{pa 4725 1969}%
\special{pa 4725 2215}%
\special{fp}%
% LINE 2 1 3 0
% 2 5300 1800 5500 1600
% 
\special{pn 8}%
\special{pa 5217 1772}%
\special{pa 5414 1575}%
\special{da 0.070}%
% LINE 2 0 3 0
% 2 5500 1600 5800 1600
% 
\special{pn 8}%
\special{pa 5414 1575}%
\special{pa 5709 1575}%
\special{fp}%
% LINE 2 0 3 0
% 2 5500 1600 5500 1350
% 
\special{pn 8}%
\special{pa 5414 1575}%
\special{pa 5414 1329}%
\special{fp}%
% LINE 2 1 3 0
% 2 5300 1800 5500 2000
% 
\special{pn 8}%
\special{pa 5217 1772}%
\special{pa 5414 1969}%
\special{da 0.070}%
% LINE 2 1 3 0
% 2 5500 2000 5650 2150
% 
\special{pn 8}%
\special{pa 5414 1969}%
\special{pa 5562 2117}%
\special{da 0.070}%
% LINE 2 0 3 0
% 2 5650 2150 5650 2300
% 
\special{pn 8}%
\special{pa 5562 2117}%
\special{pa 5562 2264}%
\special{fp}%
% LINE 2 0 3 0
% 2 5650 2150 5800 2150
% 
\special{pn 8}%
\special{pa 5562 2117}%
\special{pa 5709 2117}%
\special{fp}%
% LINE 2 0 3 0
% 2 5500 2000 5820 2000
% 
\special{pn 8}%
\special{pa 5414 1969}%
\special{pa 5729 1969}%
\special{fp}%
% LINE 2 0 3 0
% 2 5500 2000 5500 2310
% 
\special{pn 8}%
\special{pa 5414 1969}%
\special{pa 5414 2274}%
\special{fp}%
% CIRCLE 2 2 3 0
% 4 5150 1800 5200 1800 5200 1800 5200 1800
% 
\special{pn 8}%
\special{ar 5069 1772 50 50  0.0000000 0.2400000}%
\special{ar 5069 1772 50 50  0.9600000 1.2000000}%
\special{ar 5069 1772 50 50  1.9200000 2.1600000}%
\special{ar 5069 1772 50 50  2.8800000 3.1200000}%
\special{ar 5069 1772 50 50  3.8400000 4.0800000}%
\special{ar 5069 1772 50 50  4.8000000 5.0400000}%
\special{ar 5069 1772 50 50  5.7600000 6.0000000}%
% DOT 0 2 3 0
% 2 5000 1800 5000 1800
% 
\special{pn 20}%
\special{sh 1}%
\special{ar 4922 1772 10 10 0  6.28318530717959E+0000}%
\special{sh 1}%
\special{ar 4922 1772 10 10 0  6.28318530717959E+0000}%
% DOT 0 2 3 0
% 2 4800 1600 4800 1600
% 
\special{pn 20}%
\special{sh 1}%
\special{ar 4725 1575 10 10 0  6.28318530717959E+0000}%
\special{sh 1}%
\special{ar 4725 1575 10 10 0  6.28318530717959E+0000}%
% DOT 0 2 3 0
% 2 4650 1450 4650 1450
% 
\special{pn 20}%
\special{sh 1}%
\special{ar 4577 1428 10 10 0  6.28318530717959E+0000}%
\special{sh 1}%
\special{ar 4577 1428 10 10 0  6.28318530717959E+0000}%
% DOT 0 2 3 0
% 2 4800 2000 4800 2000
% 
\special{pn 20}%
\special{sh 1}%
\special{ar 4725 1969 10 10 0  6.28318530717959E+0000}%
\special{sh 1}%
\special{ar 4725 1969 10 10 0  6.28318530717959E+0000}%
% DOT 0 2 3 0
% 2 5500 1600 5500 1600
% 
\special{pn 20}%
\special{sh 1}%
\special{ar 5414 1575 10 10 0  6.28318530717959E+0000}%
\special{sh 1}%
\special{ar 5414 1575 10 10 0  6.28318530717959E+0000}%
% DOT 0 2 3 0
% 2 5500 2000 5500 2000
% 
\special{pn 20}%
\special{sh 1}%
\special{ar 5414 1969 10 10 0  6.28318530717959E+0000}%
\special{sh 1}%
\special{ar 5414 1969 10 10 0  6.28318530717959E+0000}%
% DOT 0 2 3 0
% 2 5650 2150 5650 2150
% 
\special{pn 20}%
\special{sh 1}%
\special{ar 5562 2117 10 10 0  6.28318530717959E+0000}%
\special{sh 1}%
\special{ar 5562 2117 10 10 0  6.28318530717959E+0000}%
% STR 2 0 3 0
% 3 2200 2550 2200 2650 5 0
% (a)
\put(21.6535,-26.0827){\makebox(0,0){(a)}}%
% STR 2 0 3 0
% 3 5150 2550 5150 2650 5 0
% (b)
\put(50.6890,-26.0827){\makebox(0,0){(b)}}%
% DOT 0 0 3 0
% 2 5300 1800 5300 1800
% 
\special{pn 20}%
\special{sh 1}%
\special{ar 5217 1772 10 10 0  6.28318530717959E+0000}%
\special{sh 1}%
\special{ar 5217 1772 10 10 0  6.28318530717959E+0000}%
% DOT 2 0 3 0
% 2 2800 1500 2800 1500
% 
\special{pn 8}%
\special{sh 1}%
\special{ar 2756 1477 10 10 0  6.28318530717959E+0000}%
\special{sh 1}%
\special{ar 2756 1477 10 10 0  6.28318530717959E+0000}%
% DOT 2 0 3 0
% 2 2800 1800 2800 1800
% 
\special{pn 8}%
\special{sh 1}%
\special{ar 2756 1772 10 10 0  6.28318530717959E+0000}%
\special{sh 1}%
\special{ar 2756 1772 10 10 0  6.28318530717959E+0000}%
% DOT 2 0 3 0
% 2 2800 2100 2800 2100
% 
\special{pn 8}%
\special{sh 1}%
\special{ar 2756 2067 10 10 0  6.28318530717959E+0000}%
\special{sh 1}%
\special{ar 2756 2067 10 10 0  6.28318530717959E+0000}%
% DOT 2 0 3 0
% 2 2530 2380 2530 2380
% 
\special{pn 8}%
\special{sh 1}%
\special{ar 2491 2343 10 10 0  6.28318530717959E+0000}%
\special{sh 1}%
\special{ar 2491 2343 10 10 0  6.28318530717959E+0000}%
% DOT 2 0 3 0
% 2 2540 1190 2540 1190
% 
\special{pn 8}%
\special{sh 1}%
\special{ar 2501 1172 10 10 0  6.28318530717959E+0000}%
\special{sh 1}%
\special{ar 2501 1172 10 10 0  6.28318530717959E+0000}%
% DOT 2 0 3 0
% 2 2170 1120 2170 1120
% 
\special{pn 8}%
\special{sh 1}%
\special{ar 2136 1103 10 10 0  6.28318530717959E+0000}%
\special{sh 1}%
\special{ar 2136 1103 10 10 0  6.28318530717959E+0000}%
% DOT 2 0 3 0
% 2 1630 1430 1630 1430
% 
\special{pn 8}%
\special{sh 1}%
\special{ar 1605 1408 10 10 0  6.28318530717959E+0000}%
\special{sh 1}%
\special{ar 1605 1408 10 10 0  6.28318530717959E+0000}%
% DOT 2 0 3 0
% 2 1900 1270 1900 1270
% 
\special{pn 8}%
\special{sh 1}%
\special{ar 1871 1251 10 10 0  6.28318530717959E+0000}%
\special{sh 1}%
\special{ar 1871 1251 10 10 0  6.28318530717959E+0000}%
% DOT 2 0 3 0
% 2 1520 1800 1520 1800
% 
\special{pn 8}%
\special{sh 1}%
\special{ar 1497 1772 10 10 0  6.28318530717959E+0000}%
\special{sh 1}%
\special{ar 1497 1772 10 10 0  6.28318530717959E+0000}%
% DOT 2 0 3 0
% 2 1650 2170 1650 2170
% 
\special{pn 8}%
\special{sh 1}%
\special{ar 1625 2136 10 10 0  6.28318530717959E+0000}%
\special{sh 1}%
\special{ar 1625 2136 10 10 0  6.28318530717959E+0000}%
% DOT 2 0 3 0
% 2 1900 2320 1900 2320
% 
\special{pn 8}%
\special{sh 1}%
\special{ar 1871 2284 10 10 0  6.28318530717959E+0000}%
\special{sh 1}%
\special{ar 1871 2284 10 10 0  6.28318530717959E+0000}%
% DOT 2 0 3 0
% 2 2160 2470 2160 2470
% 
\special{pn 8}%
\special{sh 1}%
\special{ar 2126 2432 10 10 0  6.28318530717959E+0000}%
\special{sh 1}%
\special{ar 2126 2432 10 10 0  6.28318530717959E+0000}%
% DOT 2 0 3 0
% 2 4450 1800 4450 1800
% 
\special{pn 8}%
\special{sh 1}%
\special{ar 4380 1772 10 10 0  6.28318530717959E+0000}%
\special{sh 1}%
\special{ar 4380 1772 10 10 0  6.28318530717959E+0000}%
% DOT 2 0 3 0
% 2 5850 1800 5850 1800
% 
\special{pn 8}%
\special{sh 1}%
\special{ar 5758 1772 10 10 0  6.28318530717959E+0000}%
\special{sh 1}%
\special{ar 5758 1772 10 10 0  6.28318530717959E+0000}%
% DOT 2 0 3 0
% 2 5800 1600 5800 1600
% 
\special{pn 8}%
\special{sh 1}%
\special{ar 5709 1575 10 10 0  6.28318530717959E+0000}%
\special{sh 1}%
\special{ar 5709 1575 10 10 0  6.28318530717959E+0000}%
% DOT 2 0 3 0
% 2 5500 1350 5500 1350
% 
\special{pn 8}%
\special{sh 1}%
\special{ar 5414 1329 10 10 0  6.28318530717959E+0000}%
\special{sh 1}%
\special{ar 5414 1329 10 10 0  6.28318530717959E+0000}%
% DOT 2 0 3 0
% 2 4800 1300 4800 1300
% 
\special{pn 8}%
\special{sh 1}%
\special{ar 4725 1280 10 10 0  6.28318530717959E+0000}%
\special{sh 1}%
\special{ar 4725 1280 10 10 0  6.28318530717959E+0000}%
% DOT 2 0 3 0
% 2 4650 1300 4650 1300
% 
\special{pn 8}%
\special{sh 1}%
\special{ar 4577 1280 10 10 0  6.28318530717959E+0000}%
\special{sh 1}%
\special{ar 4577 1280 10 10 0  6.28318530717959E+0000}%
% DOT 2 0 3 0
% 2 4500 1450 4500 1450
% 
\special{pn 8}%
\special{sh 1}%
\special{ar 4430 1428 10 10 0  6.28318530717959E+0000}%
\special{sh 1}%
\special{ar 4430 1428 10 10 0  6.28318530717959E+0000}%
% DOT 2 0 3 0
% 2 4480 1600 4470 1600
% 
\special{pn 8}%
\special{sh 1}%
\special{ar 4410 1575 10 10 0  6.28318530717959E+0000}%
\special{sh 1}%
\special{ar 4400 1575 10 10 0  6.28318530717959E+0000}%
% DOT 2 0 3 0
% 2 4500 2000 4500 2000
% 
\special{pn 8}%
\special{sh 1}%
\special{ar 4430 1969 10 10 0  6.28318530717959E+0000}%
\special{sh 1}%
\special{ar 4430 1969 10 10 0  6.28318530717959E+0000}%
% DOT 2 0 3 0
% 2 4800 2250 4800 2250
% 
\special{pn 8}%
\special{sh 1}%
\special{ar 4725 2215 10 10 0  6.28318530717959E+0000}%
\special{sh 1}%
\special{ar 4725 2215 10 10 0  6.28318530717959E+0000}%
% DOT 2 0 3 0
% 2 5500 2300 5500 2300
% 
\special{pn 8}%
\special{sh 1}%
\special{ar 5414 2264 10 10 0  6.28318530717959E+0000}%
\special{sh 1}%
\special{ar 5414 2264 10 10 0  6.28318530717959E+0000}%
% DOT 2 0 3 0
% 2 5650 2300 5650 2300
% 
\special{pn 8}%
\special{sh 1}%
\special{ar 5562 2264 10 10 0  6.28318530717959E+0000}%
\special{sh 1}%
\special{ar 5562 2264 10 10 0  6.28318530717959E+0000}%
% DOT 2 0 3 0
% 2 5800 2150 5800 2150
% 
\special{pn 8}%
\special{sh 1}%
\special{ar 5709 2117 10 10 0  6.28318530717959E+0000}%
\special{sh 1}%
\special{ar 5709 2117 10 10 0  6.28318530717959E+0000}%
% DOT 2 0 3 0
% 2 5820 2000 5820 2000
% 
\special{pn 8}%
\special{sh 1}%
\special{ar 5729 1969 10 10 0  6.28318530717959E+0000}%
\special{sh 1}%
\special{ar 5729 1969 10 10 0  6.28318530717959E+0000}%
\end{picture}%
 \caption{Suppose that $\Gamma_{n-1}^{cyc}$ is the graph (a). 
For each vertex $\bullet$ 
in (a) eq.(\ref{c2}) contributes and for each propagator (dashed line) in 
(a) eq.(\ref{c2}) contributes. The term corresponding to the vertex 
surrounded by the circle of dotted line has $(sym. fac. )=\ov{3}$. 
On the other hand, consider the case when 
$\Gamma_{n-1}^{cyc}$ is the graph in (b). This is the exceptional case. 
The graph has $(sym. fac. )=\half$ with respect to the propagator marked 
by the circle. }
 \label{fig:Sp3}
\end{figure}
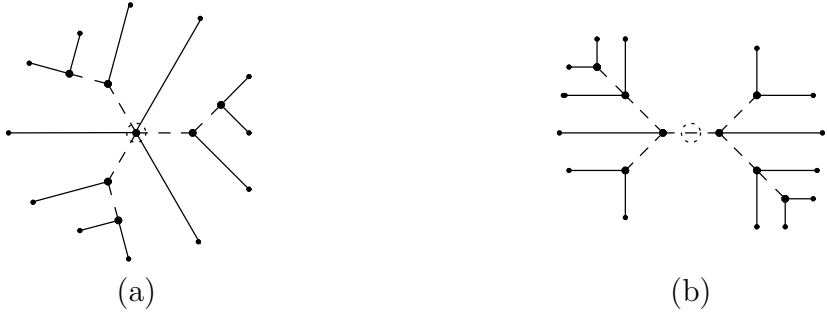 
As was seen in section \ref{sec:3} and the previous subsection, 
if $\Gamma_{n-1}^{cyc}$ has $I$ propagators, it contains $I+1$ vertices. 
We label the vertices in $\Gamma_{n-1}^{cyc}$ as $v_i,\ i=1,\cdots,I+1$ and 
the propagators as $j=1,\cdots,I$. Without an exceptional case explained 
later, the first terms contribute for each propagators $j$ in 
$\Gamma_{n-1}^{cyc}$ 
and the second terms contribute for each vertex $v_i$ in 
$\Gamma_{n-1}^{cyc}$. The coefficient for 
$\omega(\ti\Phi^p,
\ti{m}^p_{\Gamma_{n-1}^{cyc}}(\ti\Phi^p,\cdots,\ti\Phi^p))$ 
is then 
\begin{equation}
 -\sum_{j=1}^I(sym. fac. )_j +\sum_{i=1}^{I+1}(sym. fac.)_{v_i}\ .
 \label{Spcoef}
\end{equation}
Note that if $(sym. fac.)_{v_i}\neq 1$ for certain $i$, then 
$(sym. fac.)_{v_{i'}}=1$ for all $i'\neq i$
(this fact can be read easily from the graph). 
In the same way if $(sym. fac. )_j=\half$ for certain $j$, then 
the other $(sym. fac. )_{j'}$ is equal to one. 
Thus in the case when $(sym. fac. )_j=1$ for all propagators $j$, 
eq.(\ref{Spcoef}) becomes 
$-I+\sum_{i=1}^{I+1}(sym. fac.)_{v_i}=(sym. fac.)_{\Gamma_{n-1}^{cyc}}$ 
where $(sym. fac.)_{\Gamma_{n-1}^{cyc}}$ denotes the cyclic symmetric factor 
for $\Gamma_{n-1}^{cyc}$ which, if not equal to one, 
comes from $(sym. fac.)_{v_i}$ for certain $i$. 
The case when $(sym. fac. )_j=\half$ for certain $j$ is the exceptional case 
mentioned above (see for example the graph (b) in (Fig.\ref{fig:Sp3})), 
and the coefficient for 
$\omega(\ti\Phi^p,
\ti{m}^p_{\Gamma_{n-1}^{cyc}}(\ti\Phi^p,\cdots,\ti\Phi^p))$ 
is not given by eq.(\ref{Spcoef}). In this case $I+1$ is even and 
the graph $\Gamma_{n-1}^{cyc}$ is 
symmetric with respect to the propagator $j$, so the overcounting must be 
divided in eq.(\ref{Spcoef}). The coefficient is then 
$-\half(I-1)-\half+\half(I+1)$ because in this case $(sym. fac.)_{v_i}=1$ 
for all $i$. This is equal to $\half$, which is exactly the cyclic 
symmetric factor for $\Gamma_{n-1}^{cyc}$ also in this case. 
{}From the above result, eq.(\ref{Sp2}) is rewritten as 
\begin{equation}
 S((\ti\cF^p)_*(\Phi))=\sum_{n\ge 3,\Gamma_{n-1}^{cyc}}
 (sym. fac. )_{\Gamma_{n-1}^{cyc}}
 \omega(\ti\Phi^p,\ti{m}^p_{\Gamma_{n-1}^{cyc}}(\ti\Phi^p,\cdots,\ti\Phi^p))
\end{equation}
where the summation for $\Gamma_{n-1}^{cyc}$ runs over 
the $\Gamma_{n-1}$'s which are identified by the cyclic symmetry. 
Therefore re-symmetrizing the sum reproduces 
the desired form of the action (\ref{Sp}). 
It is interesting and worth emphasizing that for each edge of 
$\Gamma_{n-1}^{cyc}$ the first term (\ref{c1}) contributes, 
for each vertex of $\Gamma_{n-1}^{cyc}$ 
the second term (\ref{c2}) contributes, and 
the overcounting of the graphs are just canceled. 
\qed

The action $\ti{S}(\ti\Phi^p)$ is an effective action 
in the following sense. 
$\ti{S}(\ti\Phi^p)$ is obtained by substituting $\Phi=\ti\cF^p(\ti\Phi^p)$ 
into $S(\Phi)$ as explained above. When we express 
$\Phi=\Phi^p+\Phi^u$ where 
$\Phi^p$ and $\Phi^u$ denotes the physical and unphysical modes of $\Phi$, 
respectively, the substituting means $\Phi^p=\ti\Phi^p$ and 
$\Phi^u=f(\ti\Phi^p)$. 
As is seen from eq.(\ref{eom3}), 
the latter is nothing but the equation of motion for $\Phi^u$. 
Moreover $\ti{S}(\ti\Phi^p)$ is related to $S(\Phi)$ by integrating out 
$\Phi^u$ at tree level through a gauge fixing as 
\begin{equation}
 \int\mes\Phi^u e^{-S(\Phi)}=e^{-\ti{S}(\ti\Phi^p)} \ .
\end{equation}
In this sense the action $\ti{S}(\ti\Phi^p)$ is an effective action.

Here we clarify the properties of the solutions of the equations of 
motions. As was explained in subsection \ref{ssec:23}, 
because the $A_\infty$-morphism preserves the solutions 
of Maurer-Cartan equations, 
there is one-to-one correspondence between 
the set of the equations of motions for different SFTs on the same 
conformal background.  
However in the context of deformation theory, the solution for the 
Maurer-Cartan equation are assumed to be of the form 
$\Phi=\epsilon\ti\Phi^p+\cO(\epsilon^2)$ for $\epsilon$ a `small' 
formal deformation parameter. The argument in subsection \ref{ssec:51} is 
just the case where the small parameter is thought to be included in 
$\ti\Phi^p$. Such solutions are expressed as $\Phi=\ti\cF^p_*(\ti\Phi^p)$ 
where $\ti\Phi^p$ is a solution for the Maurer-Cartan equation 
$\ti\m^p_*(e^{\ti\Phi^p})=0$. 
The solutions can be regarded as those corresponding to the marginal 
deformation as will be explained below. 
Infinitesimally around the origin of 
$(\cH^p,\ti\m^p)$ the Maurer-Cartan equation is the quadratic form 
$\ti{m}^p_2(\ti\Phi,\ti\Phi)\sim 0$ and generally the path of the solutions 
which flows from the origin exists
\footnote{Such path exists if the Hessian has the eigenvalues of 
opposite sign with respect to the states $\{\eb^p_i\}$. }. 
Consider the continuous deformation of the solutions on this path. 
Near the origin $\ti\Phi^p=0$ \ie $\Phi=0$, the equation of motion is 
$Q\Phi=0$ so the solution is the one corresponding to 
the marginal deformation. 
Note that at the origin the value of the action $S(\Phi)$ is zero. 
Since the continuous family of the solutions which connects to 
the origin is now considered, $S(\Phi)$ is kept to be zero. 
Next, apart from the origin, we consider the physics around a solution 
$\Phi_{bg}$ on the path. Expanding the action $S(\Phi)$ around $\Phi_{bg}$, 
another action $S'(\Phi'):=S(\Phi_{bg}+\Phi')$ is obtained where 
$\Phi=\Phi_{bg}+\Phi'$ and $\Phi'\in\cH'$ : the string Hilbert space 
on another conformal background. 
As will be explained later in Discussions about the background independence, 
$S'(\Phi')$ also has an $A_\infty$-structure. 
Here represent the kinetic term of $S'(\Phi')$ as 
\begin{equation*}
 S'(\Phi')=\half\omega'(\Phi',Q'\Phi')+\cdots\ .
\end{equation*}
The infinitesimal deformation along the path from $\Phi_{bg}\in\cH$ then 
corresponds to the solution of $Q'\Phi'=0$ on $\cH'$ because 
$S'(\Phi')\sim\half\omega'(\Phi',Q'\Phi')=0$. 
Assuming the background independence of the action $S(\Phi)$, 
$Q'$ is regarded as the BRST operator on another conformal background 
(up to the isomorphism of the vector space $\cH'$), and therefore 
the infinitesimal deformation on the path 
can be regarded as the marginal deformation 
even if finitely apart from the origin.

In the above arguments we obtained a quasi-isomorphism $\ti\cF^p$ and 
discussed various meaning it has. 
However some additional input (from world sheet picture) 
may derive more strong results. 
For instance in \cite{HZ} it is shown that for closed SFT 
the infinitesimal variation of 
the way of the decomposition of the moduli space leads the infinitesimal 
field redefinition preserving the value of the actions and 
the BV-symplectic structures. 
Similarly for an one parameter family of classical open SFTs, 
the infinitesimal field redefinition 
preserving the actions is found\cite{N}, which is discussed 
in the next section. 
If one wishes to find a field redefinition preserving 
the value of the action, we must consider an isomorphism 
instead of the quasi-isomorphism $\ti\cF^p$. 
{}From the general arguments in this paper, 
there exists only one candidate for the 
isomorphism, which is the 
$A_\infty$-isomorphism $\ti\cF : (\cH,\ti\m)\raw (\cH,\m)$. 
However when any two actions 
$S(\Phi)$ and $S'(\Phi')$ on the same conformal background 
are given, generally $\ti\cF^*S(\Phi)$ and $\ti{\cF'}^*S'(\Phi')$, 
both of which are the functional of $\ti\Phi$, do not coincide 
off-shell. 
Therefore we cannot apply the isomorphism $\ti\cF$ in order to 
construct a field redefinition preserving the value of the action. 
Only on-shell $\ti\cF^*S(\Phi)$ and $\ti{\cF'}^*S'(\Phi')$ coincide 
and the above argument has held.

Finally we comment about the SFT action $\ti\cF^*S(\Phi)$, which is the one 
obtained by substituting the field redefinition $\Phi=\ti\cF(\ti\Phi)$ into 
the original SFT action $S(\Phi)$. The form of the action is derived 
directly in the same way as $\ti{S}(\ti\Phi^p)=(\ti\cF^p)^*S(\Phi)$
\begin{equation}
 \begin{split}
 \ti{S}(\ti\Phi):=\ti\cF^*S(\Phi)&=\half\omega(\ti\Phi,Q\ti\Phi)+
 \sum_{k\ge 2}\ov{k+1}\omega(\ti\Phi,\ti{m}^{cyc}_k(\ti\Phi))\\ 
 &-\omega(Q^+Q\ti\Phi,\sum_{k\ge 2}\ti{m}^{cyc}_k(\ti\Phi))\ .
 \end{split}
 \label{tiS}
\end{equation}
Note that $Q^+Q\ti\Phi$ is almost equal to $P^u\ti\Phi$. 
$\ti{m}^{cyc}_k$ denotes the one which is 
obtained by removing $P$ on the outgoing line of $\ti{m}_k$ defined 
in (Def.\ref{defn:minimal}) and (Rem.\ref{rem:twoAinfty}). 
$\ti{m}^{cyc}_k=\sum_{\Gamma_k\in G_k}\ti{m}_{\Gamma_k}$ and 
$P\ti{m}^{cyc}_k=\ti{m}_k$ holds. 
Note that these $\{\ti{m}^{cyc}_k\}_{k\ge 2}$ with $\ti{m}^{cyc}_1:=Q$ 
do not define an $A_\infty$-structure. Instead, 
$\ti\V(\ ,\cdots,\ )=\omega(\ ,\ti{m}^{cyc}_k(\ ,\cdots,\ ))$ 
has the cyclic symmetry similarly as $m_k$ or $\ti{m}^p_k$ does. 
The second term is derived in the same way as the on-shell action 
$\ti{S}(\ti\Phi^p)$ in eq.(\ref{Sp}). 
Here in addition the kinetic term and the third term appear. 
These vanish when $\ti\Phi$ is restricted to $\ti\Phi^p\in\cH^p$, so 
it can be seen that $\ti\cF^*S(\Phi)$ reduces to $\ti{S}(\ti\Phi^p)$ 
on-shell. Both the kinetic term and the third term in the right hand side 
of eq.(\ref{tiS}) come from the kinetic term of the action $S(\Phi)$, 
which vanish in eq.(\ref{Sp2}) because the fields $\ti\Phi^p$ is restricted 
on-shell. 

Thus we get an off-shell action which coincides with the string S-matrix 
elements on-shell. On this action, we define the symplectic structure 
$\ti\omega$, which is different from $\omega$, so that 
the $A_\infty$-isomorphism $\ti\cF$ from $(\cH,\ti\m,\ti\omega)$ to 
$(\cH,\m,\omega)$ preserves the symplectic structures. 
This $\ti\omega$ is written as 
\begin{equation}
  \ti\omega_{ij}
 =\frpart{\phi^k}{\ti\phi^i}\omega_{kl}\flpart{\phi^l}{\ti\phi^j}
 =(-1)^{\eb_i}
 \omega\l(\frpartial{\ti\phi^i}\Phi,\Phi\flpartial{\ti\phi^j}\r)\ ,
\end{equation}
which coincides with $\omega_{ij}$ when $\eb_i, \eb_j$ are restricted 
on-shell, but generally different from $\omega_{ij}$ off-shell. 
Thus $\ti\omega$ is a field dependent symplectic form. 
By (Prop.\ref{prop:Ap}), 
\begin{equation*}
 \ti\delta=\flpartial{\ti\phi^i}\ti\omega^{ij}\frpartial{\ti\phi^j}
 \ti{S}(\ti\Phi))
\end{equation*}
coincides with the dual of the $A_\infty$-structure $\ti\m$. 
Moreover the fact that this $\ti\delta$ defines an $A_\infty$-structure 
on $(\cH,\ti\m)$ implies that the action $\ti{S}(\ti\Phi)$ satisfies 
the BV-master equation with respect to the symplectic structure $\ti\omega$. 
Consequently, an action $\ti{S}(\ti\Phi)$, which coincides with the string 
correlation functions on-shell and satisfies the BV-master equation, 
is obtained. 
In this sense, this action $\ti{S}(\ti\Phi)$ can be regarded as 
one definition of boundary SFT\cite{W3} 
on the neighborhood of the origin of two-dimensional theory space $\cH$. 
It is interesting that, 
although rather formally, the action $\ti{S}(\ti\Phi)$ is related to the 
original open classical SFT action by the field redefinition $\ti\cF$. 
This $\ti\cF$ is nothing but the coordinate transformation on a 
formal noncommutative supermanifold in (Rem.\ref{rem:geom}). 
See also {\it boundary string field theory} in Discussions.

 \section{RG-flow and Field redefinition}
\label{sec:6}

In this section we discuss the arguments in subsection \ref{ssec:main} 
on a more explicit description of SFT : the classical open SFT discussed 
in \cite{N}. 

The most explicit way of creating a SFT action is based on 
the variation of the cut-off length of the propagator as in 
\cite{SZ,BdA,BR} for closed SFT, and in \cite{Z2,N} for open SFT. 
The cut-off of the propagator can be regarded as an UV-reguralization 
on target space, 
and the variation of the cut-off length has been discussed in these 
literatures in the context of Polchinski's renormalization group\cite{P}. 
One can consider the one parameter family of SFTs 
in this procedure. 
Let $\zeta$ be the cut-off length of the propagator 
and $S(\Phi^\zeta;\zeta)$ be the SFT action in this scale. 
This $\zeta$ parameterizes an one parameter family of 
SFT $S(\Phi^\zeta;\zeta)$. 
The action $S(\Phi^\zeta;\zeta)$ is $\zeta$-dependent in two means : 
explicit $\zeta$-dependence of $\V_{i_1\cdots i_n}$ or equivalently $\m$
(and $\omega^{ij}$), and $\zeta$-dependence through the 
$\zeta$-dependence of $\Phi^\zeta$. 
In classical SFT case, the renormalization group flow is then defined so 
that the total $\zeta$-dependence of $S(\Phi^\zeta;\zeta)$ cancels 
\begin{equation}
0= \fd{\zeta}S(\Phi^\zeta;\zeta)=\fpart{S(\Phi^\zeta;\zeta)}{\zeta}
 +\fpart{\Phi^\zeta}{\zeta}\fpart{S(\Phi^\zeta;\zeta)}{\Phi^\zeta}\ .
 \label{RGflow}
\end{equation}

Here we concentrate on the classical open SFT which possesses an 
$A_\infty$-structure\cite{N}. 
In \cite{N} the infinitesimal field redefinition $\fpart{\Phi^\zeta}{\zeta}$ 
which satisfies the above equation is derived (eq.(\ref{RGflow2})).  
We restrict the arguments on the Siegel gauge in this section. 
First 
in (Def.\ref{defn:mzeta}) we define the $A_\infty$-structure explicitly. 
In (Prop.\ref{prop:6Ap}) it is shown that 
this field redefinition is an $A_\infty$-isomorphism on the Siegel gauge. 
Next the action $S(\Phi^\zeta;\zeta)$ are transformed to 
another action $(\ti\cF^\zeta)^*S(\Phi^\zeta;\zeta)$ by the 
$A_\infty$-isomorphism $\ti\cF^\zeta$ and 
its properties are observed. 
The action $(\ti\cF^\zeta)^*S(\Phi^\zeta;\zeta)$ is the one 
which coincides with the string S-matrix on-shell (\ref{tiS}). 
Using this, it is shown that 
on the subspace of $\cH$ which is the image of $(\ti\cF^\zeta)^p$ from 
$(\cH^p,\ti\m^{\zeta,p})$, the finite field redefinition 
$\ti\cF^{\zeta',p}\circ(\ti\cF^{\zeta,p})^{-1}$ from 
$S(\Phi^\zeta; \zeta)$ to $S(\Phi^{\zeta'};\zeta')$ 
reduces to the above $A_\infty$-isomorphism 
when its infinitesimal limit is taken. 
Finally various pictures are summarized on this explicit model.

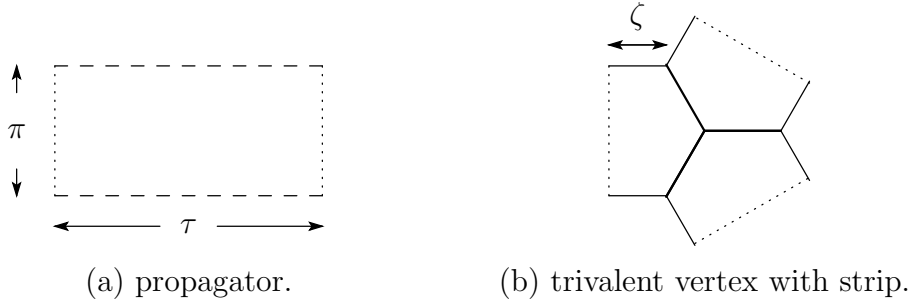
\begin{figure}[h]
 \hspace*{1.8cm} 
%WinTpicVersion3.08
\unitlength 0.1in
\begin{picture}( 43.7500, 14.0000)(  9.7500,-29.1500)
% LINE 1 0 3 0
% 2 4800 2200 5200 2200
% 
\special{pn 13}%
\special{pa 4800 2200}%
\special{pa 5200 2200}%
\special{fp}%
% LINE 1 0 3 0
% 2 4800 2200 4600 1854
% 
\special{pn 13}%
\special{pa 4800 2200}%
\special{pa 4600 1854}%
\special{fp}%
% LINE 1 0 3 0
% 2 4800 2200 5200 2200
% 
\special{pn 13}%
\special{pa 4800 2200}%
\special{pa 5200 2200}%
\special{fp}%
% LINE 1 0 3 0
% 2 4800 2200 4600 2546
% 
\special{pn 13}%
\special{pa 4800 2200}%
\special{pa 4600 2546}%
\special{fp}%
% LINE 1 0 3 0
% 2 4800 2200 4600 2546
% 
\special{pn 13}%
\special{pa 4800 2200}%
\special{pa 4600 2546}%
\special{fp}%
% LINE 2 0 3 0
% 2 4600 2540 4300 2540
% 
\special{pn 8}%
\special{pa 4600 2540}%
\special{pa 4300 2540}%
\special{fp}%
% LINE 2 0 3 0
% 2 4600 1860 4300 1860
% 
\special{pn 8}%
\special{pa 4600 1860}%
\special{pa 4300 1860}%
\special{fp}%
% LINE 2 0 3 0
% 2 4600 1860 4750 1600
% 
\special{pn 8}%
\special{pa 4600 1860}%
\special{pa 4750 1600}%
\special{fp}%
% LINE 2 0 3 0
% 2 5200 2200 5350 1940
% 
\special{pn 8}%
\special{pa 5200 2200}%
\special{pa 5350 1940}%
\special{fp}%
% LINE 2 0 3 0
% 2 4600 2540 4750 2800
% 
\special{pn 8}%
\special{pa 4600 2540}%
\special{pa 4750 2800}%
\special{fp}%
% LINE 2 0 3 0
% 2 5200 2200 5350 2460
% 
\special{pn 8}%
\special{pa 5200 2200}%
\special{pa 5350 2460}%
\special{fp}%
% LINE 2 2 3 0
% 2 4300 1860 4300 2540
% 
\special{pn 8}%
\special{pa 4300 1860}%
\special{pa 4300 2540}%
\special{dt 0.045}%
% LINE 2 2 3 0
% 2 4760 1610 5349 1950
% 
\special{pn 8}%
\special{pa 4760 1610}%
\special{pa 5350 1950}%
\special{dt 0.045}%
% LINE 2 2 3 0
% 2 4750 2790 5339 2450
% 
\special{pn 8}%
\special{pa 4750 2790}%
\special{pa 5340 2450}%
\special{dt 0.045}%
% LINE 2 1 3 0
% 2 2800 2540 1400 2540
% 
\special{pn 8}%
\special{pa 2800 2540}%
\special{pa 1400 2540}%
\special{da 0.070}%
% LINE 2 1 3 0
% 2 2800 1860 1400 1860
% 
\special{pn 8}%
\special{pa 2800 1860}%
\special{pa 1400 1860}%
\special{da 0.070}%
% LINE 2 2 3 0
% 2 2800 1860 2800 2540
% 
\special{pn 8}%
\special{pa 2800 1860}%
\special{pa 2800 2540}%
\special{dt 0.045}%
% LINE 2 2 3 0
% 2 1400 1860 1400 2540
% 
\special{pn 8}%
\special{pa 1400 1860}%
\special{pa 1400 2540}%
\special{dt 0.045}%
% VECTOR 2 0 3 0
% 2 1200 2000 1200 1860
% 
\special{pn 8}%
\special{pa 1200 2000}%
\special{pa 1200 1860}%
\special{fp}%
\special{sh 1}%
\special{pa 1200 1860}%
\special{pa 1180 1928}%
\special{pa 1200 1914}%
\special{pa 1220 1928}%
\special{pa 1200 1860}%
\special{fp}%
% VECTOR 2 0 3 0
% 2 1200 2400 1200 2540
% 
\special{pn 8}%
\special{pa 1200 2400}%
\special{pa 1200 2540}%
\special{fp}%
\special{sh 1}%
\special{pa 1200 2540}%
\special{pa 1220 2474}%
\special{pa 1200 2488}%
\special{pa 1180 2474}%
\special{pa 1200 2540}%
\special{fp}%
% STR 2 0 3 0
% 3 1200 2100 1200 2200 5 0
% $\pi$
\put(12.0000,-22.0000){\makebox(0,0){$\pi$}}%
% VECTOR 2 0 3 0
% 2 1950 2700 1400 2700
% 
\special{pn 8}%
\special{pa 1950 2700}%
\special{pa 1400 2700}%
\special{fp}%
\special{sh 1}%
\special{pa 1400 2700}%
\special{pa 1468 2720}%
\special{pa 1454 2700}%
\special{pa 1468 2680}%
\special{pa 1400 2700}%
\special{fp}%
% VECTOR 2 0 3 0
% 2 2250 2700 2800 2700
% 
\special{pn 8}%
\special{pa 2250 2700}%
\special{pa 2800 2700}%
\special{fp}%
\special{sh 1}%
\special{pa 2800 2700}%
\special{pa 2734 2680}%
\special{pa 2748 2700}%
\special{pa 2734 2720}%
\special{pa 2800 2700}%
\special{fp}%
% STR 2 0 3 0
% 3 2100 2600 2100 2700 5 0
% $\tau$
\put(21.0000,-27.0000){\makebox(0,0){$\tau$}}%
% VECTOR 2 0 3 0
% 2 4600 1750 4300 1750
% 
\special{pn 8}%
\special{pa 4600 1750}%
\special{pa 4300 1750}%
\special{fp}%
\special{sh 1}%
\special{pa 4300 1750}%
\special{pa 4368 1770}%
\special{pa 4354 1750}%
\special{pa 4368 1730}%
\special{pa 4300 1750}%
\special{fp}%
% VECTOR 2 0 3 0
% 2 4300 1750 4600 1750
% 
\special{pn 8}%
\special{pa 4300 1750}%
\special{pa 4600 1750}%
\special{fp}%
\special{sh 1}%
\special{pa 4600 1750}%
\special{pa 4534 1730}%
\special{pa 4548 1750}%
\special{pa 4534 1770}%
\special{pa 4600 1750}%
\special{fp}%
% STR 2 0 3 0
% 3 4450 1500 4450 1600 5 0
% $\zeta$
\put(44.5000,-16.0000){\makebox(0,0){$\zeta$}}%
% STR 2 0 3 0
% 3 2100 2900 2100 3000 5 0
% (a)\ propagator.
\put(21.0000,-30.0000){\makebox(0,0){(a)\ propagator.}}%
% STR 2 0 3 0
% 3 4800 2900 4800 3000 5 0
% (b)\ trivalent vertex with strip.
\put(48.0000,-30.0000){\makebox(0,0){(b)\ trivalent vertex with strip.}}%
\end{picture}%
 \caption{(a).\ $\tau$ runs between $0\le\tau\le\infty$ 
in Witten's cubic open SFT. 
When Cutting-off the propagator with length $2\zeta$, 
$\tau$ runs from $2\zeta$ to $\infty$. The moduli which is not reproduced 
by connecting Witten's trivalent vertices with such propagators equal to 
the moduli which is not reproduced by connecting 
the modified trivalent vertices (b) with the usual 
propagator ($0\le\tau\le\infty$).}
 \label{fig:pv}
\end{figure}

We begin with a brief review of the construction of one 
parameter family of classical open SFT\cite{N,Z2}. It is argued in \cite{N} 
very clear. 
The main idea was explained in section \ref{sec:3}, but 
this procedure relies on the fact that all moduli space of disks 
with $n$ punctures can be reproduced by connecting 
Witten's type trivalent vertex\cite{W1} with propagators. 
We simply represent the propagator $Q^+$ in the Siegel gauge : 
$b_0\Phi=0$ as $b_0\ov{L_0}$ since the arguments 
in this section do not depend on the detail. 
In the Schwinger representation it is 
\begin{equation}
 b_0\ov{L_0}=b_0\int_0^\infty e^{-\tau L_0}\ .
\end{equation} 
One can interpret $e^{-\tau L_0}$ as the evolution operator for 
the open string.
The width of the propagator is set to be $\pi$ (Fig.\ref{fig:pv}.(a).). 
When cutting-off the propagator with length $2\zeta$, 
the length of the propagator $\tau$ runs from $\tau=2\zeta$ to 
$\tau=\infty$. Therefore the subspace of moduli space, which has 
been reproduced by connecting trivalent vertices with the propagator with
length $0\le\tau\le2\zeta$ in $2\zeta=0$ (no-cut-off) theory, can not 
be reproduced by the trivalent vertices 
in the theory of cut-off length $2\zeta$. 
Such diagram must be add to the SFT action as higher vertices. 
Vertices in the $2\zeta$ cut-off theory are constructed recursively 
in this way. By construction, 
the width of the external legs of $n$ point vertex with $n\ge 3$ 
are of course $\pi$. 

Cutting-off the propagator with length $2\zeta$ can be replaced by sewing 
the strip with width $\pi$ and length $\zeta$ to all external legs of 
the vertices. Such trivalent vertex is pictured in (Fig.\ref{fig:pv}.(b).) 
for example. It is carried out by attaching the evolution operator 
$e^{-\zeta L_0}$ to each external legs.  
Then the length of the propagator $\tau$ runs from 
$\tau=0$ to $\tau=\infty$ and the modification for the propagator 
does not need.

Let $\m^\zeta$ be the $A_\infty$-structure corresponding to the 
vertices in the $2\zeta$-cut-off theory. The $A_\infty$-structure for 
Witten's cubic SFT is described as 
$\m^0=\{m^0_1=Q,m^0_2,m^0_3=m^0_4=\cdots=0\}$. 
Recalling the arguments in subsection \ref{ssec:5gf} and \ref{ssec:53} 
yields that $\m^\zeta$ is given essentially as $\ti\m$ 
by replacing $Q^+$ and $P$  
in the definition of $\ti\m^{(p)}$ in (Def.\ref{defn:minimal}) 
to $Q^{\zeta,+}$ and $P^\zeta$ defined below 
in the context of the present paper. Here explicitly present it as follows. 
\begin{defn}[$A_\infty$-structure $\m^\zeta$]
Define $Q^{\zeta,+}:=b_0\int_0^{2\zeta}e^{-\tau L_0}d\tau$ and 
$P^\zeta:=e^{-2\zeta L_0}$. These satisfy the following identity
\begin{equation*}
 \{Q,Q^{\zeta,+}\}+P^\zeta=\1\ .
\end{equation*}
Let us define as an intermediate step 
$\{f^{\ti\zeta}_k\}$ and $\{m^{\ti\zeta}_k\}_{k\ge 2}$ recursively by 
 \begin{equation*}
  f^{\ti\zeta}_k(\Phi^{\ti\zeta}):=-Q^{\zeta,+}\sum_{1\leq k_1<k_2=k}m^0_2(
 f^{\ti\zeta}_{k_1}(\Phi^{\ti\zeta}),f^{\ti\zeta}_{k_2-k_1}(\Phi^{\ti\zeta}))
 \end{equation*}
 with $f^{\ti\zeta}_1(\Phi^{\ti\zeta})=\Phi^{\ti\zeta}$ and  
 \begin{equation*}
  m^{\ti\zeta}_k(\Phi^{\ti\zeta})
 :=\sum_{1\leq k_1<k_2=k}P^\zeta
 m^0_2(f^{\ti\zeta}_{k_1}(\Phi^{\ti\zeta}),
 f^{\ti\zeta}_{k_2-k_1}(\Phi^{\ti\zeta}))\ .
 \end{equation*}
for $k\ge 2$. By shifting the field as 
$\Phi^{\ti\zeta}=e^{-\zeta L_0}\Phi^\zeta$, 
the $A_\infty$-structure $\m^\zeta$ is defined as 
\begin{equation}
 \begin{split}
 m^\zeta_k(\Phi^\zeta)
 &:=e^{\zeta L_0}m^{\ti\zeta}_k(e^{-\zeta L_0}\Phi^\zeta)\\
 &=e^{-\zeta L_0}\sum_{1\leq k_1<k_2=k}
 m^0_2(f^{\ti\zeta}_{k_1}(e^{-\zeta L_0}\Phi^\zeta),
 f^{\ti\zeta}_{k_2-k_1}(e^{-\zeta L_0}\Phi^\zeta))\ .
 \end{split}
\end{equation}
 \label{defn:mzeta}
\end{defn}
The fact that actually $\m^\zeta$ defines an $A_\infty$-structure 
follows from this construction and the fact that $\m^0$ defines an 
$A_\infty$-structure. The fact that the vertices have 
cyclic symmetry is also clear by construction. 

For example when $k=2$, the three point vertex $m^\zeta_2$ is 
$e^{-\zeta L_0}m^0_2(e^{-\zeta L_0}\ \ , e^{-\zeta L_0}\ \ )$, which is 
just the one in (Fig.\ref{fig:pv}.(b)). For any $n\ge 3$, one can 
see that $m^\zeta_n$ includes $n-3$ propagators all of which have length 
$0\le\tau\le 2\zeta$. They cannot be reproduced by connecting 
lower vertices $m^\zeta_k\ (k\le n-1)$ with the propagators. 

In this situation, the flow of $\Phi^\zeta$ which satisfies 
eq.(\ref{RGflow}) is defined in \cite{N} as 
\begin{equation}
 \fpart{\Phi^\zeta}{\zeta}=b_0\sum_{k\ge 2}m^\zeta_k(\Phi^\zeta)
 \label{RGflow2}
\end{equation}
in the Siegel gauge. 
The fact that this infinitesimal field redefinition preserves the value of 
the actions can be checked directly by substituting this 
into eq.(\ref{RGflow}) 
because now the variation of $m^\zeta_k$ with respect to $\zeta$ is derived 
directly by the explicit construction of $\m^\zeta$ in 
(Def.\ref{defn:mzeta}). 
\begin{prop}
 This field redefinition (\ref{RGflow2}) 
is an $A_\infty$-isomorphism on the Siegel gauge. 
 \label{prop:6Ap}
\end{prop}
The field redefinition (\ref{RGflow2}) is defined only on 
the Siegel gauge $b_0\Phi^\zeta=0$. Therefore this proposition claims, 
in other words, that this field redefinition can be extended to the 
infinitesimal neighborhood of the submanifold $b_0\Phi^\zeta=0$ 
so that the derivative with respect to ${\bar \phi}$ can be defined. \\
{\it proof.}\ \ Now the symplectic structure for $S(\Phi^\zeta;\zeta)$ is 
$\omega$ and is independent of $\zeta$. 
Because this field redefinition $\fpart{\Phi^\zeta}{\zeta}$ preserves 
the value of the action, by (Prop.\ref{prop:Ap}) 
it is sufficient for the proof to show that 
the field redefinition preserves the symplectic form $\omega$. 
Let us consider the infinitesimal field transformation 
$\cF^{\delta_\epsilon}$ defined by $\epsilon$ as follows 
\begin{equation}
 \begin{split}
 (\cF^{\delta_\epsilon})^*a(\phi^{\zeta'})
 &:=a(\phi^\zeta)+\delta_\epsilon a(\phi^\zeta)\\
                 &:=a(\phi^\zeta)+(a(\phi^\zeta),\epsilon(\phi^\zeta))\ ,
 \end{split}
 \label{canfredef}
\end{equation}
where $a(\phi^\zeta)$ is a functions of $\phi^\zeta$ and 
$\epsilon(\phi^\zeta)$ is a infinitesimal function of $\phi^\zeta$ which 
determines the infinitesimal transformation. 
$(\ ,\ )$ is the BV-Poisson structure with respect to $\omega$. 
Such infinitesimal transformation is called 
{\it canonical transformation} of (BV-)symplectic structure and 
is applied to the infinitesimal field redefinition in closed SFT 
in \cite{HZ}. 
{}
This transformation preserves the Poisson structure, that is, 
\begin{equation}
 (a(\phi^\zeta),b(\phi^\zeta))+\delta_\epsilon (a(\phi^\zeta),b(\phi^\zeta))
 =(a(\phi^\zeta)+\delta_\epsilon a(\phi^\zeta),
 b(\phi^\zeta)+\delta_\epsilon b(\phi^\zeta))
 \label{cantransf}
\end{equation}
up to $(\epsilon)^2$. It immediately follows from the Jacobi identity of 
$(\ ,\ )$. Now $(\ ,\ )$ is defined on the theory with parameter $\zeta$. 
On the other hand, if the symplectic structure $(\ ,\ )_{\zeta'}$ 
on the theory with parameter $\zeta'$ is defined so that 
$\cF^{\delta_\epsilon}$ preserves the symplectic structures, 
we have the following identity 
\begin{equation}
 (\cF^{\delta_\epsilon})^*(a(\phi^{\zeta'}),b(\phi^{\zeta'}))_{\zeta'}
 =((\cF^{\delta_\epsilon})^*a(\phi^{\zeta'}),
 (\cF^{\delta_\epsilon})^*b(\phi^{\zeta'}))\ .
 \label{cantransf2}
\end{equation}
The right hand side exactly coincides with the right hand side 
in eq.(\ref{cantransf}), therefore the symplectic structure 
$(\ ,\ )_{\zeta'}$ induced by the transformation $\cF^{\delta\epsilon}$ 
is determined by the equality between the left hand sides 
in eq.(\ref{cantransf}) and eq.(\ref{cantransf2}). 
When we set $a(\phi^{\zeta'})=\phi^{\zeta',i}$ and 
$b(\phi^{\zeta'})=\phi^{\zeta',j}$, the equality becomes 
\begin{equation*}
 (\cF^{\delta_\epsilon})^*\omega^{\zeta',ij}=
 \omega^{ij}+\delta_\epsilon\omega^{ij}
\end{equation*}
where $\omega^{\zeta',ij}$ denotes the Poisson tensor of $(\ ,\ )_{\zeta'}$. 
This equality implies that if $\omega^{ij}$ is constant, 
$\omega^{\zeta',ij}$ is equal to $\omega^{ij}$. 
Thus it is shown that the constant symplectic structure $\omega^{ij}$ 
is preserved under the infinitesimal field redefinition of the form 
in eq.(\ref{canfredef}). 
Here the field redefinition $\fpart{\Phi^\zeta}{\zeta}$ can be rewritten 
in the form (\ref{canfredef}) as $\epsilon(\phi^\zeta)
=\omega(\Phi^\zeta,b_0\sum_{k\ge 2}m^\zeta_k(\Phi^\zeta))$. 
In fact on the Siegel gauge, 
\begin{equation*}
 (\Phi^\zeta,\epsilon)|_{b_0\Phi^\zeta=0}
 =b_0\sum_{k\ge 2}m^\zeta_k(\Phi^\zeta)|_{b_0\Phi^\zeta=0}
\end{equation*}
holds. This completes the proof of (Prop.\ref{prop:6Ap}).  \qed

Next let us observe the action $(\ti\cF^\zeta)^*S(\Phi)$. As was seen in 
eq.(\ref{tiS}), it is of the form 
\begin{equation}
 \begin{split}
 \ti{S}(\ti\Phi^\zeta):=(\ti\cF^\zeta)^*S(\Phi^\zeta;\zeta)
 &=\half\omega(\ti\Phi^\zeta,Q\ti\Phi^\zeta)+
 \sum_{k\ge 2}\ov{k+1}
 \omega(\ti\Phi^\zeta,\ti{m}^{\zeta,cyc}_k(\ti\Phi^\zeta))\\ 
 &-\omega(Q^+Q\ti\Phi^{\zeta},
 \sum_{k\ge 2}\ti{m}^{\zeta,cyc}_k(\ti\Phi^\zeta))\ ,
 \end{split}\label{tiS6}
\end{equation}
where $\ti{m}^{\zeta,cyc}_k$ is the one related to $m_k^\zeta$ 
as in eq.(\ref{tiS}). 
At the same time we have the field redefinition 
$\Phi^\zeta=\ti\cF^\zeta_*(\ti\Phi^\zeta)
=\ti\Phi^\zeta+\ti{f}^\zeta(\ti\Phi)$. 
By construction, comparing with $\ti{m}^{0,cyc}$, 
$\ti{m}^{\zeta,cyc}_k$ has its external legs of length $\zeta$, 
which indicates 
\begin{equation*}
 \ti{m}^{\zeta,cyc}_k(\ ,\cdots,\ )
=e^{-\zeta L_0}\ti{m}^{0,cyc}_k(e^{-\zeta L_0}\ ,\cdots,e^{-\zeta L_0}\ )\ .
\end{equation*}
In the same way the following relation between $\ti{f}^\zeta_k$ 
with different $k$ holds, 
\begin{equation*}
 \ti{f}^\zeta_k(\ ,\cdots,\ )
=e^{-\zeta L_0}\ti{f}^0_k(e^{-\zeta L_0}\ ,\cdots,e^{-\zeta L_0}\ )\ , 
\end{equation*}
where we used the $Q^+=b_0\ov{L_0}$ on the outgoing 
states of $\ti{f}^\zeta_k$ and $e^{-\zeta L_0}$ commute. 
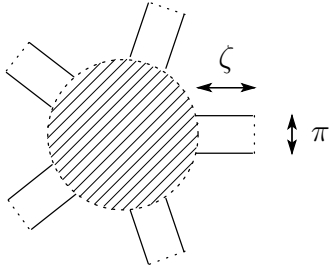
\begin{figure}[h]
 \begin{center}
%WinTpicVersion3.08
\unitlength 0.1in
\begin{picture}( 15.0098, 13.6811)( 27.3622,-16.6339)
% CIRCLE 2 2 3 0
% 4 3400 1000 3800 1000 3800 1000 3800 1000
% 
\special{pn 8}%
\special{ar 3347 985 394 394  0.0000000 0.0300000}%
\special{ar 3347 985 394 394  0.1200000 0.1500000}%
\special{ar 3347 985 394 394  0.2400000 0.2700000}%
\special{ar 3347 985 394 394  0.3600000 0.3900000}%
\special{ar 3347 985 394 394  0.4800000 0.5100000}%
\special{ar 3347 985 394 394  0.6000000 0.6300000}%
\special{ar 3347 985 394 394  0.7200000 0.7500000}%
\special{ar 3347 985 394 394  0.8400000 0.8700000}%
\special{ar 3347 985 394 394  0.9600000 0.9900000}%
\special{ar 3347 985 394 394  1.0800000 1.1100000}%
\special{ar 3347 985 394 394  1.2000000 1.2300000}%
\special{ar 3347 985 394 394  1.3200000 1.3500000}%
\special{ar 3347 985 394 394  1.4400000 1.4700000}%
\special{ar 3347 985 394 394  1.5600000 1.5900000}%
\special{ar 3347 985 394 394  1.6800000 1.7100000}%
\special{ar 3347 985 394 394  1.8000000 1.8300000}%
\special{ar 3347 985 394 394  1.9200000 1.9500000}%
\special{ar 3347 985 394 394  2.0400000 2.0700000}%
\special{ar 3347 985 394 394  2.1600000 2.1900000}%
\special{ar 3347 985 394 394  2.2800000 2.3100000}%
\special{ar 3347 985 394 394  2.4000000 2.4300000}%
\special{ar 3347 985 394 394  2.5200000 2.5500000}%
\special{ar 3347 985 394 394  2.6400000 2.6700000}%
\special{ar 3347 985 394 394  2.7600000 2.7900000}%
\special{ar 3347 985 394 394  2.8800000 2.9100000}%
\special{ar 3347 985 394 394  3.0000000 3.0300000}%
\special{ar 3347 985 394 394  3.1200000 3.1500000}%
\special{ar 3347 985 394 394  3.2400000 3.2700000}%
\special{ar 3347 985 394 394  3.3600000 3.3900000}%
\special{ar 3347 985 394 394  3.4800000 3.5100000}%
\special{ar 3347 985 394 394  3.6000000 3.6300000}%
\special{ar 3347 985 394 394  3.7200000 3.7500000}%
\special{ar 3347 985 394 394  3.8400000 3.8700000}%
\special{ar 3347 985 394 394  3.9600000 3.9900000}%
\special{ar 3347 985 394 394  4.0800000 4.1100000}%
\special{ar 3347 985 394 394  4.2000000 4.2300000}%
\special{ar 3347 985 394 394  4.3200000 4.3500000}%
\special{ar 3347 985 394 394  4.4400000 4.4700000}%
\special{ar 3347 985 394 394  4.5600000 4.5900000}%
\special{ar 3347 985 394 394  4.6800000 4.7100000}%
\special{ar 3347 985 394 394  4.8000000 4.8300000}%
\special{ar 3347 985 394 394  4.9200000 4.9500000}%
\special{ar 3347 985 394 394  5.0400000 5.0700000}%
\special{ar 3347 985 394 394  5.1600000 5.1900000}%
\special{ar 3347 985 394 394  5.2800000 5.3100000}%
\special{ar 3347 985 394 394  5.4000000 5.4300000}%
\special{ar 3347 985 394 394  5.5200000 5.5500000}%
\special{ar 3347 985 394 394  5.6400000 5.6700000}%
\special{ar 3347 985 394 394  5.7600000 5.7900000}%
\special{ar 3347 985 394 394  5.8800000 5.9100000}%
\special{ar 3347 985 394 394  6.0000000 6.0300000}%
\special{ar 3347 985 394 394  6.1200000 6.1500000}%
\special{ar 3347 985 394 394  6.2400000 6.2700000}%
% LINE 2 0 3 0
% 2 3790 900 4090 900
% 
\special{pn 8}%
\special{pa 3731 886}%
\special{pa 4026 886}%
\special{fp}%
% LINE 2 0 3 0
% 2 3790 1100 4090 1100
% 
\special{pn 8}%
\special{pa 3731 1083}%
\special{pa 4026 1083}%
\special{fp}%
% LINE 2 0 3 0
% 2 3630 648 3725 363
% 
\special{pn 8}%
\special{pa 3573 638}%
\special{pa 3667 358}%
\special{fp}%
% LINE 2 0 3 0
% 2 3440 585 3535 300
% 
\special{pn 8}%
\special{pa 3386 576}%
\special{pa 3480 296}%
\special{fp}%
% LINE 2 0 3 0
% 2 2780 669 3018 852
% 
\special{pn 8}%
\special{pa 2737 659}%
\special{pa 2971 839}%
\special{fp}%
% LINE 2 0 3 0
% 2 2902 510 3140 693
% 
\special{pn 8}%
\special{pa 2857 502}%
\special{pa 3091 683}%
\special{fp}%
% LINE 2 0 3 0
% 2 2914 1503 3149 1317
% 
\special{pn 8}%
\special{pa 2869 1480}%
\special{pa 3100 1297}%
\special{fp}%
% LINE 2 0 3 0
% 2 2790 1346 3025 1160
% 
\special{pn 8}%
\special{pa 2747 1325}%
\special{pa 2978 1142}%
\special{fp}%
% LINE 2 0 3 0
% 2 3440 1403 3535 1688
% 
\special{pn 8}%
\special{pa 3386 1381}%
\special{pa 3480 1662}%
\special{fp}%
% LINE 2 0 3 0
% 2 3630 1340 3725 1624
% 
\special{pn 8}%
\special{pa 3573 1319}%
\special{pa 3667 1599}%
\special{fp}%
% VECTOR 2 0 3 0
% 2 3800 750 4100 750
% 
\special{pn 8}%
\special{pa 3741 739}%
\special{pa 4036 739}%
\special{fp}%
\special{sh 1}%
\special{pa 4036 739}%
\special{pa 3970 719}%
\special{pa 3984 739}%
\special{pa 3970 758}%
\special{pa 4036 739}%
\special{fp}%
% VECTOR 2 0 3 0
% 2 4100 750 3800 750
% 
\special{pn 8}%
\special{pa 4036 739}%
\special{pa 3741 739}%
\special{fp}%
\special{sh 1}%
\special{pa 3741 739}%
\special{pa 3807 758}%
\special{pa 3793 739}%
\special{pa 3807 719}%
\special{pa 3741 739}%
\special{fp}%
% STR 2 0 3 0
% 3 3950 500 3950 600 5 0
% $\zeta$
\put(38.8780,-5.9055){\makebox(0,0){$\zeta$}}%
% VECTOR 2 0 3 0
% 2 4300 900 4300 1100
% 
\special{pn 8}%
\special{pa 4233 886}%
\special{pa 4233 1083}%
\special{fp}%
\special{sh 1}%
\special{pa 4233 1083}%
\special{pa 4252 1017}%
\special{pa 4233 1031}%
\special{pa 4213 1017}%
\special{pa 4233 1083}%
\special{fp}%
% VECTOR 2 0 3 0
% 2 4300 1100 4300 900
% 
\special{pn 8}%
\special{pa 4233 1083}%
\special{pa 4233 886}%
\special{fp}%
\special{sh 1}%
\special{pa 4233 886}%
\special{pa 4213 952}%
\special{pa 4233 938}%
\special{pa 4252 952}%
\special{pa 4233 886}%
\special{fp}%
% STR 2 0 3 0
% 3 4450 900 4450 1000 5 0
% $\pi$
\put(43.7992,-9.8425){\makebox(0,0){$\pi$}}%
% LINE 3 0 3 0
% 36 3700 740 3140 1300 3730 770 3170 1330 3750 810 3210 1350 3770 850 3250 1370 3780 900 3300 1380 3800 940 3340 1400 3800 1000 3400 1400 3790 1070 3470 1390 3770 1150 3550 1370 3670 710 3110 1270 3640 680 3080 1240 3600 660 3060 1200 3560 640 3040 1160 3520 620 3020 1120 3470 610 3010 1070 3420 600 3000 1020 3360 600 3000 960 3280 620 3020 880
% 
\special{pn 4}%
\special{pa 3642 729}%
\special{pa 3091 1280}%
\special{fp}%
\special{pa 3672 758}%
\special{pa 3121 1310}%
\special{fp}%
\special{pa 3691 798}%
\special{pa 3160 1329}%
\special{fp}%
\special{pa 3711 837}%
\special{pa 3199 1349}%
\special{fp}%
\special{pa 3721 886}%
\special{pa 3249 1359}%
\special{fp}%
\special{pa 3741 926}%
\special{pa 3288 1378}%
\special{fp}%
\special{pa 3741 985}%
\special{pa 3347 1378}%
\special{fp}%
\special{pa 3731 1054}%
\special{pa 3416 1369}%
\special{fp}%
\special{pa 3711 1132}%
\special{pa 3495 1349}%
\special{fp}%
\special{pa 3613 699}%
\special{pa 3062 1251}%
\special{fp}%
\special{pa 3583 670}%
\special{pa 3032 1221}%
\special{fp}%
\special{pa 3544 650}%
\special{pa 3012 1182}%
\special{fp}%
\special{pa 3504 630}%
\special{pa 2993 1142}%
\special{fp}%
\special{pa 3465 611}%
\special{pa 2973 1103}%
\special{fp}%
\special{pa 3416 601}%
\special{pa 2963 1054}%
\special{fp}%
\special{pa 3367 591}%
\special{pa 2953 1004}%
\special{fp}%
\special{pa 3308 591}%
\special{pa 2953 945}%
\special{fp}%
\special{pa 3229 611}%
\special{pa 2973 867}%
\special{fp}%
% LINE 2 2 3 0
% 2 4100 900 4100 1100
% 
\special{pn 8}%
\special{pa 4036 886}%
\special{pa 4036 1083}%
\special{dt 0.045}%
% LINE 2 2 3 0
% 2 3540 1690 3730 1620
% 
\special{pn 8}%
\special{pa 3485 1664}%
\special{pa 3672 1595}%
\special{dt 0.045}%
% LINE 2 2 3 0
% 2 3730 360 3550 300
% 
\special{pn 8}%
\special{pa 3672 355}%
\special{pa 3495 296}%
\special{dt 0.045}%
% LINE 2 2 3 0
% 2 2780 660 2900 510
% 
\special{pn 8}%
\special{pa 2737 650}%
\special{pa 2855 502}%
\special{dt 0.045}%
% LINE 2 2 3 0
% 2 2790 1350 2900 1500
% 
\special{pn 8}%
\special{pa 2747 1329}%
\special{pa 2855 1477}%
\special{dt 0.045}%
\end{picture}%
 \end{center}
  \caption{$\ti{m}^{\zeta,cyc}_4$ is figured. 
The interior of the circle denotes that all tree five point 
Feynman graphs are summed up and which means that the integral runs the 
whole moduli space $\cM_5$. Comparing to $\ti{m}^{0,cyc}_4$ 
the additional strips with length $\zeta$ are attached. }
  \label{fig.61}
\end{figure}

Let us restrict the external states $\ti\Phi$ to $\ti\Phi^p\in\cH^p$. 
In this case $\ti{m}^{\zeta,cyc}_k$ and $\ti{f}^\zeta_k$ are replaced by 
$\ti{m}^{\zeta,p}_k$ and $\ti{f}^{\zeta,p}_k$. Then 
$\ti{m}^{\zeta,p}_k$ coincides with $\ti{m}^{0,p}_k$ because 
$e^{-\zeta L_0}=1$ on $\ti\Phi^p$. Thus the on-shell effective action 
$\ti{S}(\ti\Phi^{\zeta,p}):=(\ti\cF^{\zeta,p})^*S(\Phi^\zeta)$ 
is independent of $\zeta$. 
On the other hand, the situation is not the same for $\ti{f}^\zeta_k$, 
because the outgoing states of $f_k^\zeta$ do not belong to $\cH^p$ but 
the image of $Q^+$. The outgoing legs has its length 
$\zeta$ and then the propagator acts on it.  
The facts leads 
\begin{equation}
 \ti{f}^{\zeta,p}_k=e^{-\zeta L_0}\ti{f}^{0,p}_k\ .
\end{equation}
Note that $\ti{f}^{0,p}_k$ has no $\zeta$-dependence. 
We can then consider the infinitesimal variation of the field redefinition 
$\Phi^\zeta=\ti\Phi^{\zeta,p}+\ti{f}^{\zeta,p}(\ti\Phi^p)$, 
\begin{equation*}
 \begin{split}
 \fpart{\Phi^\zeta}{\zeta}&=-L_0e^{-\zeta L_0}\ti{f}^{0,p}(\ti\Phi^{0,p})\\
                          &=-L_0\ti{f}^{\zeta,p}(\ti\Phi^{\zeta,p})\ 
 \end{split}
\end{equation*}
where $\ti{f}^{\zeta,p}(\ti\Phi^{\zeta,p})
=\sum_{k\ge 2}\ti{f}^{\zeta,p}_k(\ti\Phi^{\zeta,p})$. 
By definition, one can rewrite $\ti{f}^{\zeta,p}_k(\ti\Phi^{\zeta,p})$ as 
\begin{equation*}
 \ti{f}^{\zeta,p}(\ti\Phi^{\zeta,p})
 =-Q^+\sum_{k\ge 2}m^\zeta_k(\Phi^\zeta) 
\end{equation*}
where $\Phi^\zeta$ in the right hand side is the image of 
$\ti\cF^{\zeta,p}_*$ from $\ti\Phi^p\in\cH^p$. 
Recalling $Q^+=b_0\ov{L_0}$, 
the infinitesimal field redefinition is derived as 
\begin{equation*}
 \fpart{\Phi^\zeta}{\zeta}=b_0\sum_{k\ge 2}m^\zeta_k(\Phi^\zeta)\ ,
\end{equation*}
which exactly coincides with the renormalization group flow (\ref{RGflow2}).
The finite field redefinition from $S(\Phi^\zeta;\zeta)$ to 
$S(\Phi^{\zeta'};\zeta')$ on this subspace is given by 
\begin{equation*}
 \begin{split}
 \Phi^{\zeta'}&=\Phi^\zeta+e^{-\zeta' L_0}f^{\zeta'}(\Phi^{0,\zeta'})
                          -e^{-\zeta L_0}f^{\zeta'}(\Phi^{0,\zeta})\\
              &=\Phi^\zeta
                +b_0\int_0^{\zeta'-\zeta}e^{-\tau L_0}d\tau
                 \sum_{k\ge 2}m^\zeta_k(\Phi^\zeta)\ .
 \end{split}
\end{equation*}

Finally the various SFT action obtained here and their relation between 
each other are summarized. 
\begin{figure}[h]
 \hspace*{2.0cm}
%WinTpicVersion3.08
\unitlength 0.1in
\begin{picture}( 40.7972, 14.6161)( 10.4331,-17.7165)
% STR 2 0 3 0
% 3 1600 300 1600 400 5 0
% $(\cH,\m^0)$
\put(15.7480,-3.9370){\makebox(0,0){$(\cH,\m^0)$}}%
% VECTOR 2 0 3 0
% 2 1600 600 5200 600
% 
\special{pn 8}%
\special{pa 1575 591}%
\special{pa 5119 591}%
\special{fp}%
\special{sh 1}%
\special{pa 5119 591}%
\special{pa 5053 571}%
\special{pa 5066 591}%
\special{pa 5053 611}%
\special{pa 5119 591}%
\special{fp}%
% STR 2 0 3 0
% 3 2800 300 2800 400 5 0
% $(\cH,\m^\zeta)$
\put(27.5591,-3.9370){\makebox(0,0){$(\cH,\m^\zeta)$}}%
% STR 2 0 3 0
% 3 3400 300 3400 400 5 0
% $(\cH,\m^{\zeta'})$
\put(33.4646,-3.9370){\makebox(0,0){$(\cH,\m^{\zeta'})$}}%
% STR 2 0 3 0
% 3 5200 300 5200 400 5 0
% $(\cH^\infty,\m^\infty)$
\put(51.1811,-3.9370){\makebox(0,0){$(\cH^\infty,\m^\infty)$}}%
% VECTOR 2 0 3 0
% 2 5200 1800 5200 610
% 
\special{pn 8}%
\special{pa 5119 1772}%
\special{pa 5119 601}%
\special{fp}%
\special{sh 1}%
\special{pa 5119 601}%
\special{pa 5099 667}%
\special{pa 5119 653}%
\special{pa 5138 667}%
\special{pa 5119 601}%
\special{fp}%
% STR 2 0 3 0
% 3 5600 1100 5600 1200 5 0
% $(\cH,\ti\m^{\zeta'})$
\put(55.1181,-11.8110){\makebox(0,0){$(\cH,\ti\m^{\zeta'})$}}%
% STR 2 0 3 0
% 3 5600 1300 5600 1400 5 0
% $(\cH,\ti\m^\zeta)$
\put(55.1181,-13.7795){\makebox(0,0){$(\cH,\ti\m^\zeta)$}}%
% DOT 1 0 3 0
% 2 5200 1400 5200 1400
% 
\special{pn 13}%
\special{sh 1}%
\special{ar 5119 1378 10 10 0  6.28318530717959E+0000}%
\special{sh 1}%
\special{ar 5119 1378 10 10 0  6.28318530717959E+0000}%
% DOT 1 0 3 0
% 2 5200 1200 5200 1200
% 
\special{pn 13}%
\special{sh 1}%
\special{ar 5119 1182 10 10 0  6.28318530717959E+0000}%
\special{sh 1}%
\special{ar 5119 1182 10 10 0  6.28318530717959E+0000}%
% DOT 1 0 3 0
% 2 2800 600 2800 600
% 
\special{pn 13}%
\special{sh 1}%
\special{ar 2756 591 10 10 0  6.28318530717959E+0000}%
\special{sh 1}%
\special{ar 2756 591 10 10 0  6.28318530717959E+0000}%
% DOT 1 0 3 0
% 2 3400 600 3400 600
% 
\special{pn 13}%
\special{sh 1}%
\special{ar 3347 591 10 10 0  6.28318530717959E+0000}%
\special{sh 1}%
\special{ar 3347 591 10 10 0  6.28318530717959E+0000}%
% VECTOR 2 0 3 0
% 2 5080 1360 2920 640
% 
\special{pn 8}%
\special{pa 5001 1339}%
\special{pa 2875 630}%
\special{fp}%
\special{sh 1}%
\special{pa 2875 630}%
\special{pa 2931 670}%
\special{pa 2925 647}%
\special{pa 2943 632}%
\special{pa 2875 630}%
\special{fp}%
% VECTOR 2 0 3 0
% 2 5080 1160 3520 640
% 
\special{pn 8}%
\special{pa 5001 1142}%
\special{pa 3465 630}%
\special{fp}%
\special{sh 1}%
\special{pa 3465 630}%
\special{pa 3521 670}%
\special{pa 3515 647}%
\special{pa 3534 632}%
\special{pa 3465 630}%
\special{fp}%
% STR 2 0 3 0
% 3 3950 1100 3950 1200 5 0
% $\ti\cF^{\zeta}$
\put(38.8780,-11.8110){\makebox(0,0){$\ti\cF^{\zeta}$}}%
% STR 2 0 3 0
% 3 4500 700 4500 800 5 0
% $\ti\cF^{\zeta'}$
\put(44.2913,-7.8740){\makebox(0,0){$\ti\cF^{\zeta'}$}}%
% DOT 1 0 3 0
% 2 1600 600 1600 600
% 
\special{pn 13}%
\special{sh 1}%
\special{ar 1575 591 10 10 0  6.28318530717959E+0000}%
\special{sh 1}%
\special{ar 1575 591 10 10 0  6.28318530717959E+0000}%
\end{picture}%
\end{figure}
$(\cH,\m^0)$ is the cubic open SFT. On the horizontal line the 
one parameter family of SFT $(\cH,\m^\zeta)$ is defined and 
there exists the infinitesimal field redefinition on it. 
This field redefinition preserves the $A_\infty$-structure \ie 
the BRST-symmetry on the Siegel gauge, and 
formally by integrating it there 
exists a field redefinition between any two SFTs on this one parameter 
family. 
Alternatively, for each $(\cH,\m^\zeta)$ there exists an equivalent SFT 
$(\cH,\ti\m^\zeta)$. These are related to each other by $\ti\cF^\zeta$. 
The field transformations on the horizontal line and $\ti\cF^\zeta$ are 
compatible. 
When $(\cH,\ti\m^\zeta)$ is restricted to physical states, 
the reduced theory does not depend on $\zeta$ and 
coincides with $(\cH^p,\ti\m^p)$. 
In this subspace the composition 
$\ti\cF^{\zeta',p}\circ(\ti\cF^\zeta)^{-1}$ defines the finite field 
transformation between $(\cH,\m^\zeta)$ and $(\cH,\m^{\zeta'})$. 
Its infinitesimal version actually coincides with the field redefinition 
given in \cite{N}. 

Note that in the limit $\zeta\raw\infty$, 
each vertex of the action on $(\cH,\m^{\infty})$ has the whole moduli 
and it coincides with the 
correlation function itself. 
$(\cH,\m^\infty)$ then coincides with $(\cH,\ti\m^\infty)$ and 
its restriction onto the physical state space $\cH^p$ is just 
$(\cH,\ti\m^p)$. 
This is the open string version of the argument given in \cite{Z1}, 
where the $L_\infty$-structure of closed SFT is reduced to the 
$L_\infty$-structure in string world sheet theory\cite{WZ,V}. 
This argument is reviewed in Appendix \ref{ssec:B2}. 

Thus we have two one parameter families of SFT $(\cH,\m^\zeta)$ and 
$(\cH,\ti\m^\zeta)$. The flow on $(\cH,\m^\zeta)$ is discussed from 
the viewpoint of the renormalization group in \cite{N}. 
The flow on $(\cH,\ti\m^\zeta)$ might also be interpreted as 
the renormalization flow of boundary SFT.

 \section{Conclusions and Discussions}
\label{CD}

We discussed classical open SFTs with cyclic vertices and 
argued that what extent their structure is governed by their general 
properties. 
These SFTs have the structure of $A_\infty$-algebras, and 
it is shown that the $A_\infty$-algebras of them 
are $A_\infty$-quasi-isomorphic to the $A_\infty$-algebra of on-shell 
S-matrix elements. 
Moreover applying this, it is shown that 
any such SFTs which differ in the decomposition of moduli space are 
$A_\infty$-quasi-isomorphic to each other. This implies that 
between any such SFTs 
there is one-to-one correspondence of the solutions of 
the equations of motions which describe marginal deformations. 
In subspaces which relate to physical state space $\cH^p$ 
the finite field transformation between two SFTs are described 
in terms of the Feynman graph. 

We discuss the above arguments explicitly on 
the one parameter family of classical open SFTs on the Siegel gauge. 
On this one parameter family, there exists a field redefinition 
preserving the value of the actions. 
We showed that the infinitesimal field redefinition 
is an $A_\infty$-isomorphism. 
It was observed that the infinitesimal version of 
the above finite field transformation in the subspace coincides 
with the $A_\infty$-isomorphism preserving the actions. 

Through the explanation of the above statements, 
various expressions for $A_\infty$-algebras of SFTs are used and 
their relations were summarized. 
The sign attending on the degree (ghost number) is uniformed 
self-consistently. The relation to the conventional definition of the sign 
is not denoted explicitly in this paper, but one can easily obtain 
the relation by comparing 
the two definitions of 
$A_\infty$-algebras the elements of which have their degree differ by one 
(see (Rem.\ref{rem:degree}) and, for example, \cite{GJ}. 
Essentially the relation can be read from the 
relation between the convention 
in first half and the latter half of \cite{N}.  ). 

The problem of taking dual of coalgebras has some subtlety 
when the graded vector space is infinite dimensional. 
SFT is just the case. However SFTs are field theories. Therefore 
as far as assuming that the SFT is well-defined as field theory 
the dual of the coalgebras should be able to be taken. 
Moreover the dual language is introduced in the present paper only for 
intuitive and geometric understanding. 
All the arguments on the dual are rearranged in coalgebra language 
and then hold even in the model where 
the decomposition of fields and basis is difficult. 

The convergence was not discussed. 
The finite field redefinitions or the solutions 
for the Maurer-Cartan equations for SFTs, which are formally preserved 
under the field redefinitions, are defined by polynomials 
of infinite powers. 
Of course many of the arguments in this paper make sense 
as formal power series. For instance each coefficient of the 
Maurer-Cartan equations for the canonical $A_\infty$-algebra defines 
each on-shell S-matrix element. 
However, it should be checked that when the solutions converge, etc. . 
Their seems no ways to confirm the convergences 
instead of doing some numerical analysis explicitly. 
However, the finite field redefinitions and the equations of motions are 
defined by the Feynman graph of SFT. Therefore the problem of the 
convergences 
relates to the problem of the original SFT itself. 
Looking for some `good' model on a conformal background might be a 
good issue. 
Alternatively, one can also argue these on an appropriate subspace, 
due to, for instance, the momentum conservation of the vertices. 
Therefore we think that some well-defined field redefinitions or 
the solutions of the equations of motions are obtained in the subspace. 
One such example will be commented below 
in {\it tachyon condensation} in Discussions.

We ends with presenting the following related topics or future directions.

\vspace*{0.2cm}

$\bullet$\ \ {\it the background independence.}\quad

SFTs have mainly two directions of deformations : 
changing the decomposition of moduli space of Riemann surfaces as 
discussed in this paper, and transferring to other backgrounds. 

The issue of the background independence can be treated in the category 
of weak $A_\infty$-algebras as follows. 
Consider two points $x$ and $y$ on CFT theory space. 
$x$ and $y$ denote two conformal backgrounds. 
Let $\cH_x$ and $\cH_y$ be two string Hilbert spaces on the conformal 
backgrounds and 
let $\Phi_x\in\cH_x$ and $\Phi_y\in\cH_y$ be the string fields. 
Generally the field transformation $\cF_y$ is of the form 
\begin{equation*}
 \Phi_y:=\cF_{y,*}(\Phi_x)=\Phi_{bg} +\cF_*(\Phi_x)
 =\Phi_{bg}+f_1(\Phi_x)+f_2(\Phi_x,\Phi_x)+\cdots\ ,
\end{equation*}
where $\Phi_{bg}=\eb_i\Phi^i_{bg}$ denotes a background in $\cH_y$. 
Only the degree zero part of $\Phi^i_{bg}$ can be nonzero, 
because the vacuum expectation value of the action should belong to 
$\R$. 
The case $\Phi_{bg}=0$ reduces to the problem on the same conformal 
background. 

If a SFT gives a background independent formulation of string theory, 
each solution $\Phi_{bg}$ of the equation of motion for the SFT 
$S_y(\Phi_y)$ on $\cH_y$ describes a conformal background. 
Moreover $S(\Phi)$ re-expanded around the equation of motion $\Phi_{bg}$ 
should define a SFT action on the conformal background of $\cH_x$. 

Let $S_x(\Phi_x)$ be a SFT action on $\cH_x$. 
Suppose that these two actions $S_y(\Phi_y)$ and $S_x(\Phi_x)$ 
satisfy the BV-master equations on their conformal backgrounds. 
Then if there exists $\cF_y$ which preserves the symplectic structures 
and satisfies $\cF_y^*S_y(\Phi_y)=S_y(\cF_{y,*}(\Phi_x))=S_x(\Phi_x)$, 
the action $S_y(\Phi_y)$ is certainly background independent. 

Here let us express this $\cF_y$ as the composition 
$\cF_y^*=\cF^*\circ\cF^*_{bg}$ where 
\begin{equation*}
\Phi_y=\cF_{bg,*}(\Phi')=\Phi_{bg}+\Phi'\ ,\qquad 
\Phi'=\cF_*(\Phi_x)\ . 
\end{equation*}
By $\cF_{bg}$ the SFT $S_y(\Phi_y)$ on $\cH_y$ is transformed to a SFT 
on $\cH_x$ which is regarded as a SFT on the conformal background 
corresponds to $\Phi_{bg}$. 
It is known that when the $\Phi_{bg}$ denotes an equation of motion 
for $S_y(\Phi_y)$ the action expanded around $\Phi_{bg}$, 
$\cF_{bg}^*S_y(\Phi_y)=S_y(\Phi_{bg}+\Phi')$, also has an 
$A_\infty$-structure\cite{Sinfty} with symplectic structure unchanged. 
It is explained in {\it weak $A_\infty$} in Appendix \ref{ssec:A2} 
in the dual picture. Denote by $\m_y$ an $A_\infty$-structure on 
$\Phi_y\in\cH_y$, define $\cF_{bg}$ as a cohomomorphism, 
and the induced $A_\infty$-structure $\m'$ on $\cH_x$ is given by 
(\ref{wainf})
\footnote{These arguments are treated on the category of 
weak $A_\infty$-algebras. Actually when $\Phi_{bg}$ does not satisfy 
the equation of motion, $\m'$ in eq.(\ref{Abg}) defines a weak 
$A_\infty$-structure on $\cH_x$. }
\begin{equation}
 \m'=\m_y\circ\cF_{bg}\ . \label{Abg}
\end{equation}
Generally this induced 
$A_\infty$-structure $\m'$ is not quasi-isomorphic to the original one, 
since $m'_1=:Q'$, and especially its cohomology class, is changed. 

Next we consider the field redefinition 
\begin{equation}
 \cF_*(\Phi_x)=f_1(\Phi_x)+f_2(\Phi_x,\Phi_x)+\cdots\ .
\end{equation} 
In order for $\Phi_{bg}\in\cH_y$ describes the conformal background $x$, 
$f_1$ should be an isomorphism. 
We use this $f$ in order for the kinetic term of 
$S_y(\Phi_{bg}+\Phi')$ to coincides with that of $S_x(\Phi_x)$. 
Note that by (Prop.\ref{prop:Ap}) as far as $\cF$ preserves the 
symplectic structures, $\cF$ is an $A_\infty$-isomorphism between 
two $A_\infty$-algebras. 

Locally $\cH_y$ can be regarded as a fiber on $y$, and 
the total space can be viewed as a vector bundle. 
In \cite{SZ1} for classical closed SFT the infinitesimal 
background independence is proved by utilizing the CFT theory space 
connection \cite{RSZ} (and the argument is extended for quantum closed SFT 
in \cite{SZ2}). 
Here the `infinitesimal' means that the two conformal backgrounds which 
relate to each other by infinitesimal marginal deformation are considered. 
In this case $\Phi_y-\Phi_x$ is infinitesimal and the infinitesimal 
deformation of $f_1$ corresponds to the connection on the vector bundle. 
$f_2,f_3,\cdots$ preserving the symplectic structures 
are constructed in \cite{SZ1}. 

Generally it is difficult to construct $\{f_k\}_{k\ge 2}$. 
Then giving up constructing $\{f_k\}_{k\ge 2}$ and one 
can consider the reduction of the shifted action 
$S_y(\Phi_{bg}+f_1(\Phi''))-S_y(\Phi_{bg})$ to ``minimal'' as in 
section \ref{sec:5}. 
By construction the shifted action satisfies the BV-master 
equation. Therefore if the corresponding ``minimal'' action coincides with 
the on-shell S-matrix elements, the shifted action might be regarded as 
the SFT action on $x$, and the action $S_y(\Phi_y)$ is background 
independent. 
This argument is an reformulation of the arguments in \cite{Sbg2}, 
where infinitesimal marginal deformation $\Phi_{bg}$ is discussed.

\vspace*{0.2cm}

$\bullet$\ \ {\it tachyon condensation.}\quad
The solution describing the tachyonic nonperturbative vacuum in cubic 
open SFT\cite{SZn} is 
one of the solution of the e.o.m (\ref{eom21}) 
with the condition (\ref{eom22}). 
The issue can be viewed as a toy model for applying 
the argument in the present paper, but a modification is needed. 

Consider the tachyonic solution in the Siegel gauge. 
Since we are interested in the Lorentz invariant solution 
with twist symmetry, 
the solution is non-zero only for even level scalar fields 
which is constant in space-time. 
Each corresponding state is then not physical state because for 
the basis corresponding to constant (zero-momentum) fields, 
the eigenvalues of $L_0$ for the base of level $0$, $2$, $4$ $\cdots$ are 
$-1$, $1$, $3$, $\cdots$, respectively. 
The field corresponding to the level zero state is the constant 
tachyon field, and we denote it by $t$. 
In order to obtain the tachyonic solution for $t$ 
and the value of the action at the solution, 
the tachyon effective potential $V(t)$ has been needed. 
Its equation of motion is $\fpartial{t}V(t)=0$. 
We want to relate $V(t)$ with $\ti{S}(\ti\Phi^p)$ in eq.(\ref{Sp}) and 
$\fpartial{t}V(t)=0$ with $\sum_{k\ge 2}\ti{m}^p(\ti\Phi^p)=0$ (\ref{mceq2}) 
: the Maurer-Cartan equation on $\cH^p$. 
Other fields corresponding to the states of level $2, 4,\cdots$ 
have been expressed as the power series of the tachyon field $t$. 
It is regarded as the field redefinition $\ti\cF^p$ from $(\cH^p,\ti\m^p)$ 
to $(\cH,\m)$, 
that is, we want to regard the tachyon field as $\ti\Phi^p\in\cH^p$ and 
other fields as $\ti\Phi^u=f(\ti\Phi^p)\in\cH^u$. 
However a modification is needed because the tachyon is not physical and 
the tachyonic solution can not be obtained by marginal deformation. 
We then modify the definition of $P$, which was essentially the projection 
onto the physical state. 
Let the new $P$ be the projection onto the tachyon $t$. 
The $P$ in the propagator 
$Q^+=b_0\ov{L_0}(1-P)$ is also replaced 
to this $P$. 
Then beginning with the equations (\ref{eom21}) and (\ref{eom22}) 
with $m_3=m_4=\cdots=0$, the field redefinition $\ti\cF^p$ and 
the Maurer-Cartan equations (\ref{eom21m}) and (\ref{eom22m}) 
with $P$ replaced are obtained. Then the solution of eq.(\ref{eom21m}) is 
the one we are looking for. The corresponding effective potential 
is exactly what is mentioned in \cite{KoSa} and 
presented explicitly using Feynman graphs in \cite{MT}. 
However in order that the solution is well-defined on the Siegel gauge, 
it should be BRST-invariant. In \cite{HS} it is checked in level $(2, 6)$ 
approximation with good agreement. The condition is nothing but 
eq.(\ref{eom22m}) with replaced $P$. 
By discussing the issue from this viewpoints, 
some symmetry around this tachyon effective potential can be seen and 
it might give some insight also for the problem of the 
exact solution and the physics around it.

\vspace*{0.2cm}

$\bullet$\ \ {\it boundary string field theory.}\quad

In subsection \ref{ssec:main} and section \ref{sec:6} 
a boundary SFT like action is obtained in eq.(\ref{tiS}) and (\ref{tiS6}). 
Each vertex in the action coincides with the string correlation function 
on-shell, and is extended off-shell in the similar notion as \cite{Samuel}. 
The action relates to 
the original SFT action by a finite field redefinition constructed by 
the Feynman graphs of SFT and satisfies the classical BV-master equation. 
The field redefinition can be realized as the coordinate transformation 
on a formal noncommutative supermanifold in (Rem.\ref{rem:geom}). 
However no relation to other literature has been clarified. 
The property of the BV-BRST transformation $\ti\delta$ and the symplectic 
form $\ti\omega$ should be investigated further. 
This argument is rather formal but the definition of 
boundary SFT\cite{W3} is also formal, so relating it to the ordinary SFT 
might help us to realize the structure of boundary SFT. 

Note that action $\ti{S}(\ti\Phi)$ reproduces 
the string S-matrix on-shell even if it is treated as the fundamental 
action of field theory. 
Since in eq.(\ref{tiS}) $Q^+Q\ti\Phi$ is almost $P^u\ti\Phi$, 
let us identify it with unphysical fields. 
The action $\ti{S}(\ti\Phi)$ then does not contain the terms 
which is linear for unphysical fields.
Therefore  when computing the on-shell amplitudes the exchanging 
diagrams do not appear and they indeed coincide 
with the on-shell string S-matrices.

\vspace*{0.2cm}

$\bullet$\ \ {\it other SFT.}\quad
In this paper we deal with the classical open SFT and its 
$A_\infty$-structure. The argument holds true also 
for the classical closed SFT, 
because by commutative-symmetrizing the arguments on 
$A_\infty$-structures reduce to $L_\infty$-algebras and it is known that 
the classical closed SFT is described by $L_\infty$-algebras\cite{Z1}. 
(The definition of the symplectic structure etc. in subsection \ref{ssec:43} 
is necessarily modified. )
In other case, like quantum closed, classical and quantum open closed case, 
the algebraic structures are governed by the BV-algebra in any case. 
However in order to extend the use of the quasi-isomorphism in the 
minimal model theorem in these case, it is necessary to refine 
the algebraic structure in more detail. 
Such study might makes clear the general structure of these SFTs. 
Some study for the algebraic structure of quantum closed SFT are 
found in \cite{SZ2,SZ3,Markl}.

%\newpage

\begin{center}
\noindent{\large \textbf{Acknowledgments}}
\end{center}

I am very grateful to A.~Kato for helpful discussions and advice.
I would like to thank H.~Hata, Y.~Kazama and T.~Uesugi 
for useful comments and discussions. 
For many discussions and informations about mathematical aspects 
I would like to thank M.~Akaho, K.~Fukaya, T.~Gocho and Y.~Terashima. 
I also acknowledge the Summer Institute 2001 at Fuji-Yoshida 
and its organizer who gave me a chance to talk. 
The first half of this paper is based on the talk. 
The author is supported by JSPS Research Fellowships for Young
Scientists.

\appendix

 \section{Dual description of homotopy algebras}
\label{sec:A}

We shall give the dual description of $A_\infty$-algebras. 
In this picture, $A_\infty$-algebras are understood more geometrically. 
The definition of the dual of a coalgebra in the present paper is 
given in subsection \ref{ssec:A1}, and its geometrical point of view 
is explained in subsection \ref{ssec:A2}, where we deal with a 
formal noncommutative supermanifold. 
These arguments hold similarly for $L_\infty$-algebras.

 \subsection{The definition of the dual of a coalgebra}
\label{ssec:A1}

Let $\cH$ be a graded vector space, and 
$C(\cH):=\oplus_{n=1}^\infty\left(\cH^{\otimes n}\right)$ be its tensor 
algebra. The basis of $\cH$ is denoted by $\{\eb_i\}$, and 
here we define the dual basis of 
$\eb_{i_1}\cdots\eb_{i_k}\in\cH^{\otimes k}$ with an inner 
product as follows. 
At first, denote the dual basis of $\{\eb_i\}$ by $\{\eb^i\}$, and 
define an inner product between $\cH$ and $\cH^*$ as 
\begin{equation}
 \langle \eb^i |\eb_j\rangle =\delta^i_j\ .
 \label{ip1}
\end{equation}
We represent an elements of $C(\cH)$ as 
$g=\sum_{k=1}^\infty g^{i_k\cdots i_1}\eb_{i_1}\cdots\eb_{i_k}$, 
and an element of $C(\cH)^*$, the dual of $C(\cH)$ as  
$a=\sum_{k=1}^\infty a_{i_1\cdots i_k}\eb^{i_k}\cdots\eb^{i_1}$. 
Generalizing the above inner product between $\cH$ and $\cH^*$ (\ref{ip1}), 
here the inner product between $C(\cH)$ and $C(\cH)^*$ is defined as 
\begin{equation}
 \langle \eb^{i_k}\cdots\eb^{i_1} | \eb_{j_1}\cdots\eb_{j_l}\rangle 
 =\epsilon^{i_1\cdots i_k}_{j_1\cdots j_l}\label{ip2}, 
\end{equation}
where $\epsilon^{i_1\cdots i_k}_{j_1\cdots j_l}$ equal zero for $k\neq l$, 
and if $k=l$, $\epsilon^{i_1\cdots i_k}_{j_1\cdots j_k}
=\delta^{i_1}_{j_1}\cdots\delta^{i_k}_{j_k}$. 
Moreover, for $a_1,\cdots,a_n\in C(\cH)^*$ and $g_1,\cdots,g_n\in C(\cH)$, 
the inner product of $n$-tensor is given by 
\begin{equation*}
 \la a_1\otimes\cdots\otimes a_n| g_1\otimes\cdots\otimes g_n\ra
 =\la a_1| g_1\ra\cdots\la a_n| g_n\ra\ .
 \label{ipn}
\end{equation*}

Now we have obtained the inner product 
between $C(\cH)$ and its dual $C(\cH)^*$, 
we will translate operations on $C(\cH)$ into those on $C(\cH)^*$. 
For the coproducts $\triangle$ on $C(\cH)$, 
the product $m$ on $C(\cH)^*$ is defined as 
\begin{equation}
 \langle m(a\otimes b) | g\rangle=\langle a\otimes b | \triangle g\rangle\ ,
 \label{defm}
\end{equation}
the derivation $\delta$ corresponding to the coderivation $\m$ is defined as 
\begin{equation}
 \langle \delta(a) | g\rangle=\langle a | \m(g)\rangle\ ,
 \label{defD}
\end{equation}
and homomorphism $\fb$ corresponds to the cohomomorphism $\cF$ from $C(\cH)$ 
to another tensor algebra $C(\cH')$ is 
determined as 
\begin{equation}
 \langle \fb(a) | g\rangle=\langle a | \cF(g)\rangle\ .
 \label{deff}
\end{equation}
Because $g\in C(\cH)$ and $a\in C(\cH')^*$, the homomorphism $\fb$ is a 
map from $C(\cH')^*$ to $C(\cH)^*$. Therefore $\fb$ can be regarded as 
$\cF^*$ : the pullback of $\cF$. 
Here we write the elements of $C(\cH)$ on the left hand side and 
the elements of $C(\cH)^*$ on the right hand side. 
The operations on $C(\cH')$ or $C(\cH')^*$ are distinguished by 
attaching $'$ to them. 
The above definitions of the operations on $C(\cH)^*$ translate 
various conditions for the operations on $C(\cH)$ into those on $C(\cH)^*$ 
as follows. 
The coassociativity of $\tri$ is equivalent to the associativity of $m$ : 
\begin{equation}
 \begin{array}{ccc}
 \langle m(m(a\otimes b)\otimes c) | g\rangle&=&
   \langle a\otimes b\otimes c | (\tri\otimes\1)\tri(g)\rangle\\
                       &&\rotatebox[origin=c]{90}{=}\\
   \langle m(a\otimes m(b\otimes c))
    | g\rangle
 &=&\langle a\otimes b\otimes c|(\1\otimes\tri)\triangle g\rangle
 \end{array}\ .\label{asso}
\end{equation}
The condition that $\m$ is the coderivation is translated 
into the Leibniz rule for $\delta$ :  
\begin{equation}
 \begin{array}{ccc}
 \langle \delta\cdot m(a\otimes b) | g\rangle&=&
   \langle a\otimes b | \tri\cdot \m (g)\rangle\\
                       &&\rotatebox[origin=c]{90}{=}\\
   \langle m(\delta\otimes{\bf 1}+{\bf 1}\otimes \delta)(a\otimes b)
    | g\rangle
 &=&\langle a\otimes b|(\m\otimes{\bf 1}+{\bf 1}\otimes \m)\triangle g\rangle
 \end{array}\ .\label{QD}
\end{equation}
The condition that $\cF : C(\cH)\raw C(\cH')$ 
is a cohomomorphism is rewritten as the one that 
$\fb : C(\cH')^*\raw C(\cH)^*$ is a homomorphism : 
\begin{equation}
 \begin{array}{ccc}
 \langle \fb\cdot m'(a\otimes b) | g\rangle&=&
   \langle a\otimes b | \tri'\cdot \cF(g)\rangle\\
                       &&\rotatebox[origin=c]{90}{=}\\
   \langle m(\fb(a)\otimes \fb(b)) | g\rangle
 &=&\langle a\otimes b|(\cF\otimes\cF)\triangle g\rangle
 \end{array}\ .\label{Ff}
\end{equation}
$(\cH, \m)$ is an $A_\infty$-algebra means 
that $(C(\cH)^*, \delta)$ is a complex on the dual : 
\begin{equation}
 0=\langle \delta\cdot\delta(a) | g\rangle
 =\langle a | \m\cdot\m (g)\rangle=0\ .\label{QQDD}
\end{equation} 
Finally the condition that $\cF$ is an $A_\infty$-morphism is tranlated 
into the equivariance of $\fb$ : 
\begin{equation}
 \begin{array}{ccc}
 \langle \delta\cdot \fb(a) | g\rangle&=&
   \langle a | \cF\cdot \m(g)\rangle\\
                       &&\rotatebox[origin=c]{90}{=}\\
   \langle \fb\cdot \delta'(a) | g\rangle
 &=&\langle a | \m'\cdot\cF(g)\rangle
 \end{array}\ .
 \label{QFDf}
\end{equation}

The above statement will be realized with some graphs
\footnote{The graphs used below is different from that in the body of this 
paper. In fact, a line denotes the flow of an element of $\cH$ in the 
body of this paper, but the line used below denotes an element of $C(\cH)$. 
}. 
In the above explanation, the elements of $C(\cH)$ are written in the left 
hand side of the inner products (ket), and the elements 
of the dual algebra $C(\cH)^*$ are in the right hand side (bra). 
Here, for the algebra on the left hand side, we represent the product $m$, 
the derivation $\delta$, and the homomorphism $\fb$ as  
$m=\mm$, $\delta=\D$, $\fb=\f$. According to the operations of the algebra 
from left, the lines of the graphs are connected to the right direction. 
In other words, the operations on the algebra $C(\cH^*)$ 
in the left hand side from left yields the flow from the left to the right 
on the lines of the graphs. 
Next, for the coalgebra $C(\cH)$ in the right hand side, we represent 
the coproduct $\triangle$, the coderivation $\m$, 
and the cohomomorphism $\cF$ as $\triangle=\mm$, $\m=\Q$, $\cF=\F$, 
and define the orientation of the operation from the right to the left 
on the lines of the graphs. 
Lastly, in order to distinguish the left and right in the inner products, 
we introduce $\langle\ |\ \rangle$ between the algebra $C(\cH)$ and 
the coalgebra $C(\cH)^*$. 

The definition of the algebra $C(\cH)^*$ dual to the coalgebra $C(\cH)$ 
(\ref{defm})(\ref{defD})(\ref{deff}) are written graphically as follows :
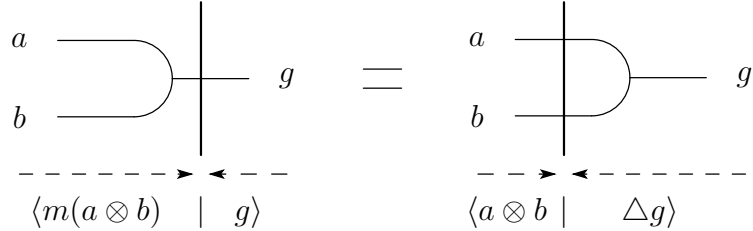
\begin{figure}[h]
 \begin{center}
%WinTpicVersion3.08
\unitlength 0.1in
\begin{picture}( 44.3500, 10.1500)( 11.7000,-18.1500)
% LINE 2 0 3 0
% 2 2005 1000 2405 1000
% 
\special{pn 8}%
\special{pa 2006 1000}%
\special{pa 2406 1000}%
\special{fp}%
% LINE 2 0 3 0
% 2 2005 1400 2405 1400
% 
\special{pn 8}%
\special{pa 2006 1400}%
\special{pa 2406 1400}%
\special{fp}%
% CIRCLE 2 0 3 0
% 4 2405 1200 2405 1400 2405 1400 2405 1000
% 
\special{pn 8}%
\special{ar 2406 1200 200 200  4.7123890 6.2831853}%
\special{ar 2406 1200 200 200  0.0000000 1.5707963}%
% LINE 2 0 3 0
% 2 2605 1200 3005 1200
% 
\special{pn 8}%
\special{pa 2606 1200}%
\special{pa 3006 1200}%
\special{fp}%
% STR 2 0 3 0
% 3 1805 900 1805 1000 5 0
% $a$
\put(18.0500,-10.0000){\makebox(0,0){$a$}}%
% STR 2 0 3 0
% 3 1805 1300 1805 1400 5 0
% $b$
\put(18.0500,-14.0000){\makebox(0,0){$b$}}%
% STR 2 0 3 0
% 3 3205 1100 3205 1200 5 0
% $g$
\put(32.0500,-12.0000){\makebox(0,0){$g$}}%
% LINE 2 0 3 0
% 2 3605 1150 3805 1150
% 
\special{pn 8}%
\special{pa 3606 1150}%
\special{pa 3806 1150}%
\special{fp}%
% LINE 2 0 3 0
% 2 3605 1250 3805 1250
% 
\special{pn 8}%
\special{pa 3606 1250}%
\special{pa 3806 1250}%
\special{fp}%
% LINE 2 0 3 0
% 2 4400 995 4800 995
% 
\special{pn 8}%
\special{pa 4400 996}%
\special{pa 4800 996}%
\special{fp}%
% LINE 2 0 3 0
% 2 4400 1395 4800 1395
% 
\special{pn 8}%
\special{pa 4400 1396}%
\special{pa 4800 1396}%
\special{fp}%
% CIRCLE 2 0 3 0
% 4 4800 1195 4800 1395 4800 1395 4800 995
% 
\special{pn 8}%
\special{ar 4800 1196 200 200  4.7123890 6.2831853}%
\special{ar 4800 1196 200 200  0.0000000 1.5707963}%
% LINE 2 0 3 0
% 2 5000 1195 5400 1195
% 
\special{pn 8}%
\special{pa 5000 1196}%
\special{pa 5400 1196}%
\special{fp}%
% STR 2 0 3 0
% 3 4200 895 4200 995 5 0
% $a$
\put(42.0000,-9.9500){\makebox(0,0){$a$}}%
% STR 2 0 3 0
% 3 4200 1295 4200 1395 5 0
% $b$
\put(42.0000,-13.9500){\makebox(0,0){$b$}}%
% STR 2 0 3 0
% 3 5600 1095 5600 1195 5 0
% $g$
\put(56.0000,-11.9500){\makebox(0,0){$g$}}%
% LINE 1 0 3 0
% 2 2755 800 2755 1600
% 
\special{pn 13}%
\special{pa 2756 800}%
\special{pa 2756 1600}%
\special{fp}%
% LINE 1 0 3 0
% 2 4655 800 4655 1600
% 
\special{pn 13}%
\special{pa 4656 800}%
\special{pa 4656 1600}%
\special{fp}%
% VECTOR 2 1 3 0
% 2 1805 1700 2705 1700
% 
\special{pn 8}%
\special{pa 1806 1700}%
\special{pa 2706 1700}%
\special{da 0.070}%
\special{sh 1}%
\special{pa 2706 1700}%
\special{pa 2638 1680}%
\special{pa 2652 1700}%
\special{pa 2638 1720}%
\special{pa 2706 1700}%
\special{fp}%
% VECTOR 2 1 3 0
% 2 3205 1700 2805 1700
% 
\special{pn 8}%
\special{pa 3206 1700}%
\special{pa 2806 1700}%
\special{da 0.070}%
\special{sh 1}%
\special{pa 2806 1700}%
\special{pa 2872 1720}%
\special{pa 2858 1700}%
\special{pa 2872 1680}%
\special{pa 2806 1700}%
\special{fp}%
% VECTOR 2 1 3 0
% 2 4205 1700 4605 1700
% 
\special{pn 8}%
\special{pa 4206 1700}%
\special{pa 4606 1700}%
\special{da 0.070}%
\special{sh 1}%
\special{pa 4606 1700}%
\special{pa 4538 1680}%
\special{pa 4552 1700}%
\special{pa 4538 1720}%
\special{pa 4606 1700}%
\special{fp}%
% VECTOR 2 1 3 0
% 2 5605 1700 4705 1700
% 
\special{pn 8}%
\special{pa 5606 1700}%
\special{pa 4706 1700}%
\special{da 0.070}%
\special{sh 1}%
\special{pa 4706 1700}%
\special{pa 4772 1720}%
\special{pa 4758 1700}%
\special{pa 4772 1680}%
\special{pa 4706 1700}%
\special{fp}%
% STR 2 0 3 0
% 3 2755 1800 2755 1900 5 0
% $|$
\put(27.5500,-19.0000){\makebox(0,0){$|$}}%
% STR 2 0 3 0
% 3 4655 1800 4655 1900 5 0
% $|$
\put(46.5500,-19.0000){\makebox(0,0){$|$}}%
% STR 2 0 3 0
% 3 2205 1800 2205 1900 5 0
% $\langle m(a\otimes b)$
\put(22.0500,-19.0000){\makebox(0,0){$\langle m(a\otimes b)$}}%
% STR 2 0 3 0
% 3 3005 1800 3005 1900 5 0
% $g\rangle$
\put(30.0500,-19.0000){\makebox(0,0){$g\rangle$}}%
% STR 2 0 3 0
% 3 4355 1800 4355 1900 5 0
% $\langle a\otimes b$
\put(43.5500,-19.0000){\makebox(0,0){$\langle a\otimes b$}}%
% STR 2 0 3 0
% 3 5105 1800 5105 1900 5 0
% $\triangle g\rangle$
\put(51.0500,-19.0000){\makebox(0,0){$\triangle g\rangle$}}%
\end{picture}%
 \end{center}
 \caption[subsection]{$\langle m(a\otimes b) | g\rangle
  =\langle a\otimes b | \triangle g\rangle$\ \ \ eq.(\ref{defm})}
 \label{fig:coalg1}
\end{figure}
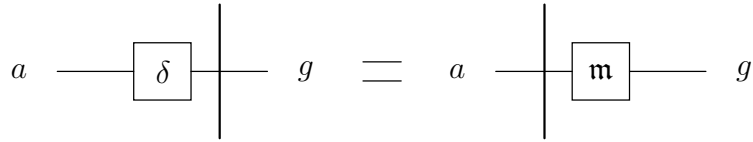
\begin{figure}[h]
 \begin{center}
%WinTpicVersion3.08
\unitlength 0.1in
\begin{picture}( 37.9500,  7.0000)( 14.7000,-11.5000)
% LINE 2 0 3 0
% 2 1805 800 2205 800
% 
\special{pn 8}%
\special{pa 1806 800}%
\special{pa 2206 800}%
\special{fp}%
% BOX 2 0 3 0
% 2 2205 650 2505 950
% 
\special{pn 8}%
\special{pa 2206 650}%
\special{pa 2506 650}%
\special{pa 2506 950}%
\special{pa 2206 950}%
\special{pa 2206 650}%
\special{fp}%
% LINE 2 0 3 0
% 2 2505 800 2905 800
% 
\special{pn 8}%
\special{pa 2506 800}%
\special{pa 2906 800}%
\special{fp}%
% STR 2 0 3 0
% 3 1605 700 1605 800 5 0
% $a$
\put(16.0500,-8.0000){\makebox(0,0){$a$}}%
% STR 2 0 3 0
% 3 3105 700 3105 800 5 0
% $g$
\put(31.0500,-8.0000){\makebox(0,0){$g$}}%
% LINE 2 0 3 0
% 2 3405 750 3605 750
% 
\special{pn 8}%
\special{pa 3406 750}%
\special{pa 3606 750}%
\special{fp}%
% LINE 2 0 3 0
% 2 3405 850 3605 850
% 
\special{pn 8}%
\special{pa 3406 850}%
\special{pa 3606 850}%
\special{fp}%
% LINE 2 0 3 0
% 2 4100 800 4500 800
% 
\special{pn 8}%
\special{pa 4100 800}%
\special{pa 4500 800}%
\special{fp}%
% BOX 2 0 3 0
% 2 4500 650 4800 950
% 
\special{pn 8}%
\special{pa 4500 650}%
\special{pa 4800 650}%
\special{pa 4800 950}%
\special{pa 4500 950}%
\special{pa 4500 650}%
\special{fp}%
% LINE 2 0 3 0
% 2 4800 800 5200 800
% 
\special{pn 8}%
\special{pa 4800 800}%
\special{pa 5200 800}%
\special{fp}%
% STR 2 0 3 0
% 3 3900 700 3900 800 5 0
% $a$
\put(39.0000,-8.0000){\makebox(0,0){$a$}}%
% STR 2 0 3 0
% 3 5400 700 5400 800 5 0
% $g$
\put(54.0000,-8.0000){\makebox(0,0){$g$}}%
% LINE 1 0 3 0
% 2 2655 450 2655 1150
% 
\special{pn 13}%
\special{pa 2656 450}%
\special{pa 2656 1150}%
\special{fp}%
% LINE 1 0 3 0
% 2 4355 450 4355 1150
% 
\special{pn 13}%
\special{pa 4356 450}%
\special{pa 4356 1150}%
\special{fp}%
% STR 2 0 3 0
% 3 2360 700 2360 800 5 0
% $\delta$
\put(23.6000,-8.0000){\makebox(0,0){$\delta$}}%
% STR 2 0 3 0
% 3 4650 700 4650 800 5 0
% $\m$
\put(46.5000,-8.0000){\makebox(0,0){$\m$}}%
\end{picture}%
 \end{center}
 \caption[subsection]
{$\langle \delta(a) | g\rangle=\langle a | \m (g)\rangle$
  \ \ \ eq.(\ref{defD})}
 \label{fig:coalg2}
\end{figure}
\begin{figure}[h]
 \begin{center}
%WinTpicVersion3.08
\unitlength 0.1in
\begin{picture}( 37.9500,  7.0000)( 14.7000,-11.5000)
% LINE 2 0 3 0
% 2 1805 800 2205 800
% 
\special{pn 8}%
\special{pa 1806 800}%
\special{pa 2206 800}%
\special{fp}%
% BOX 2 0 3 0
% 2 2205 650 2505 950
% 
\special{pn 8}%
\special{pa 2206 650}%
\special{pa 2506 650}%
\special{pa 2506 950}%
\special{pa 2206 950}%
\special{pa 2206 650}%
\special{fp}%
% LINE 2 0 3 0
% 2 2505 800 2905 800
% 
\special{pn 8}%
\special{pa 2506 800}%
\special{pa 2906 800}%
\special{fp}%
% STR 2 0 3 0
% 3 1605 700 1605 800 5 0
% $a$
\put(16.0500,-8.0000){\makebox(0,0){$a$}}%
% STR 2 0 3 0
% 3 3105 700 3105 800 5 0
% $g$
\put(31.0500,-8.0000){\makebox(0,0){$g$}}%
% LINE 2 0 3 0
% 2 3405 750 3605 750
% 
\special{pn 8}%
\special{pa 3406 750}%
\special{pa 3606 750}%
\special{fp}%
% LINE 2 0 3 0
% 2 3405 850 3605 850
% 
\special{pn 8}%
\special{pa 3406 850}%
\special{pa 3606 850}%
\special{fp}%
% LINE 2 0 3 0
% 2 4100 800 4500 800
% 
\special{pn 8}%
\special{pa 4100 800}%
\special{pa 4500 800}%
\special{fp}%
% BOX 2 0 3 0
% 2 4500 650 4800 950
% 
\special{pn 8}%
\special{pa 4500 650}%
\special{pa 4800 650}%
\special{pa 4800 950}%
\special{pa 4500 950}%
\special{pa 4500 650}%
\special{fp}%
% LINE 2 0 3 0
% 2 4800 800 5200 800
% 
\special{pn 8}%
\special{pa 4800 800}%
\special{pa 5200 800}%
\special{fp}%
% STR 2 0 3 0
% 3 3900 700 3900 800 5 0
% $a$
\put(39.0000,-8.0000){\makebox(0,0){$a$}}%
% STR 2 0 3 0
% 3 5400 700 5400 800 5 0
% $g$
\put(54.0000,-8.0000){\makebox(0,0){$g$}}%
% LINE 1 0 3 0
% 2 2655 450 2655 1150
% 
\special{pn 13}%
\special{pa 2656 450}%
\special{pa 2656 1150}%
\special{fp}%
% LINE 1 0 3 0
% 2 4355 450 4355 1150
% 
\special{pn 13}%
\special{pa 4356 450}%
\special{pa 4356 1150}%
\special{fp}%
% STR 2 0 3 0
% 3 2360 700 2360 800 5 0
% $\fb$
\put(23.6000,-8.0000){\makebox(0,0){$\fb$}}%
% STR 2 0 3 0
% 3 4650 700 4650 800 5 0
% $\cF$
\put(46.5000,-8.0000){\makebox(0,0){$\cF$}}%
\end{picture}%
 \end{center}
 \caption[subsection]{$\langle \fb(a) | g\rangle=\langle a | \cF(g)\rangle$
  \ \ \ eq.(\ref{deff})}
 \label{fig:coalg3}
\end{figure}
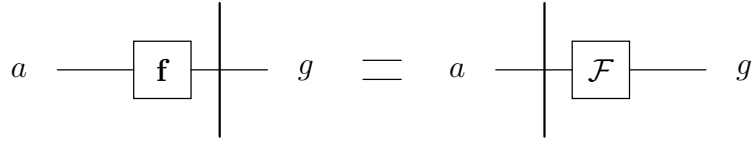
The graphs in both sides of the equations represent the 
$\C$ valued inner products. 
The arrow on the dashed line in (Fig.\ref{fig:coalg1}) 
denotes the orientation of the operations in both sides. 
The $m$ is defined so that the inner product is invariant 
when the $|$ on the right hand side of (Fig.\ref{fig:coalg1}) 
is moved to the left. Then the $\mm$ is $m$ on the left of $|$, 
and it becomes $\tri$ on the right of $|$. 
Similarly, in (Fig.\ref{fig:coalg2}), 
the $\D$ on the left of $|$ becomes $\Q$ on the right and 
the $\Q$ is $\D$ when is transferred to the left. 
The situation is similar for $\f$ and $\F$ (Fig.\ref{fig:coalg3}). 

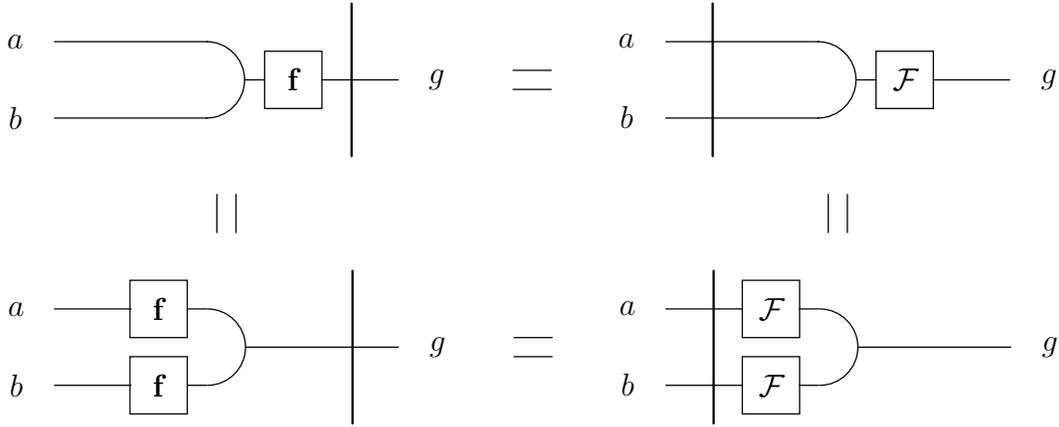
\begin{figure}[h]
 \begin{center}
%WinTpicVersion3.08
\unitlength 0.1in
\begin{picture}( 54.1500, 22.0000)(  8.6000,-34.0000)
% CIRCLE 2 0 3 0
% 4 2000 3000 2000 3200 2000 3200 2000 2800
% 
\special{pn 8}%
\special{ar 2000 3000 200 200  4.7123890 6.2831853}%
\special{ar 2000 3000 200 200  0.0000000 1.5707963}%
% STR 2 0 3 0
% 3 1000 2700 1000 2800 5 0
% $a$
\put(10.0000,-28.0000){\makebox(0,0){$a$}}%
% STR 2 0 3 0
% 3 1000 3100 1000 3200 5 0
% $b$
\put(10.0000,-32.0000){\makebox(0,0){$b$}}%
% STR 2 0 3 0
% 3 3205 2900 3205 3000 5 0
% $g$
\put(32.0500,-30.0000){\makebox(0,0){$g$}}%
% LINE 2 0 3 0
% 2 3605 2950 3805 2950
% 
\special{pn 8}%
\special{pa 3606 2950}%
\special{pa 3806 2950}%
\special{fp}%
% LINE 2 0 3 0
% 2 3605 3050 3805 3050
% 
\special{pn 8}%
\special{pa 3606 3050}%
\special{pa 3806 3050}%
\special{fp}%
% STR 2 0 3 0
% 3 4200 2695 4200 2795 5 0
% $a$
\put(42.0000,-27.9500){\makebox(0,0){$a$}}%
% STR 2 0 3 0
% 3 4200 3095 4200 3195 5 0
% $b$
\put(42.0000,-31.9500){\makebox(0,0){$b$}}%
% LINE 1 0 3 0
% 2 2760 2600 2760 3400
% 
\special{pn 13}%
\special{pa 2760 2600}%
\special{pa 2760 3400}%
\special{fp}%
% LINE 2 0 3 0
% 2 1895 2800 1995 2800
% 
\special{pn 8}%
\special{pa 1896 2800}%
\special{pa 1996 2800}%
\special{fp}%
% LINE 2 0 3 0
% 2 1895 3200 1995 3200
% 
\special{pn 8}%
\special{pa 1896 3200}%
\special{pa 1996 3200}%
\special{fp}%
% BOX 2 0 3 0
% 2 1595 2650 1895 2950
% 
\special{pn 8}%
\special{pa 1596 2650}%
\special{pa 1896 2650}%
\special{pa 1896 2950}%
\special{pa 1596 2950}%
\special{pa 1596 2650}%
\special{fp}%
% STR 2 0 3 0
% 3 1750 2700 1750 2800 5 0
% $\fb$
\put(17.5000,-28.0000){\makebox(0,0){$\fb$}}%
% BOX 2 0 3 0
% 2 1595 3050 1895 3350
% 
\special{pn 8}%
\special{pa 1596 3050}%
\special{pa 1896 3050}%
\special{pa 1896 3350}%
\special{pa 1596 3350}%
\special{pa 1596 3050}%
\special{fp}%
% STR 2 0 3 0
% 3 1750 3100 1750 3200 5 0
% $\fb$
\put(17.5000,-32.0000){\makebox(0,0){$\fb$}}%
% LINE 2 0 3 0
% 2 2200 3000 3000 3000
% 
\special{pn 8}%
\special{pa 2200 3000}%
\special{pa 3000 3000}%
\special{fp}%
% CIRCLE 2 0 3 0
% 4 5205 3000 5205 3200 5205 3200 5205 2800
% 
\special{pn 8}%
\special{ar 5206 3000 200 200  4.7123890 6.2831853}%
\special{ar 5206 3000 200 200  0.0000000 1.5707963}%
% STR 2 0 3 0
% 3 6410 2900 6410 3000 5 0
% $g$
\put(64.1000,-30.0000){\makebox(0,0){$g$}}%
% LINE 1 0 3 0
% 2 4650 2600 4650 3400
% 
\special{pn 13}%
\special{pa 4650 2600}%
\special{pa 4650 3400}%
\special{fp}%
% LINE 2 0 3 0
% 2 5100 2800 5200 2800
% 
\special{pn 8}%
\special{pa 5100 2800}%
\special{pa 5200 2800}%
\special{fp}%
% LINE 2 0 3 0
% 2 5100 3200 5200 3200
% 
\special{pn 8}%
\special{pa 5100 3200}%
\special{pa 5200 3200}%
\special{fp}%
% BOX 2 0 3 0
% 2 4800 2650 5100 2950
% 
\special{pn 8}%
\special{pa 4800 2650}%
\special{pa 5100 2650}%
\special{pa 5100 2950}%
\special{pa 4800 2950}%
\special{pa 4800 2650}%
\special{fp}%
% STR 2 0 3 0
% 3 4955 2700 4955 2800 5 0
% $\cF$
\put(49.5500,-28.0000){\makebox(0,0){$\cF$}}%
% BOX 2 0 3 0
% 2 4800 3050 5100 3350
% 
\special{pn 8}%
\special{pa 4800 3050}%
\special{pa 5100 3050}%
\special{pa 5100 3350}%
\special{pa 4800 3350}%
\special{pa 4800 3050}%
\special{fp}%
% STR 2 0 3 0
% 3 4955 3100 4955 3200 5 0
% $\cF$
\put(49.5500,-32.0000){\makebox(0,0){$\cF$}}%
% LINE 2 0 3 0
% 2 5405 3000 6205 3000
% 
\special{pn 8}%
\special{pa 5406 3000}%
\special{pa 6206 3000}%
\special{fp}%
% LINE 2 0 3 0
% 2 1200 2800 1600 2800
% 
\special{pn 8}%
\special{pa 1200 2800}%
\special{pa 1600 2800}%
\special{fp}%
% LINE 2 0 3 0
% 2 1200 3200 1600 3200
% 
\special{pn 8}%
\special{pa 1200 3200}%
\special{pa 1600 3200}%
\special{fp}%
% LINE 2 0 3 0
% 2 4400 2800 4800 2800
% 
\special{pn 8}%
\special{pa 4400 2800}%
\special{pa 4800 2800}%
\special{fp}%
% LINE 2 0 3 0
% 2 4400 3200 4800 3200
% 
\special{pn 8}%
\special{pa 4400 3200}%
\special{pa 4800 3200}%
\special{fp}%
% CIRCLE 2 0 3 0
% 4 1995 1600 1995 1800 1995 1800 1995 1400
% 
\special{pn 8}%
\special{ar 1996 1600 200 200  4.7123890 6.2831853}%
\special{ar 1996 1600 200 200  0.0000000 1.5707963}%
% STR 2 0 3 0
% 3 995 1300 995 1400 5 0
% $a$
\put(9.9500,-14.0000){\makebox(0,0){$a$}}%
% STR 2 0 3 0
% 3 995 1700 995 1800 5 0
% $b$
\put(9.9500,-18.0000){\makebox(0,0){$b$}}%
% STR 2 0 3 0
% 3 3200 1500 3200 1600 5 0
% $g$
\put(32.0000,-16.0000){\makebox(0,0){$g$}}%
% LINE 2 0 3 0
% 2 3600 1550 3800 1550
% 
\special{pn 8}%
\special{pa 3600 1550}%
\special{pa 3800 1550}%
\special{fp}%
% LINE 2 0 3 0
% 2 3600 1650 3800 1650
% 
\special{pn 8}%
\special{pa 3600 1650}%
\special{pa 3800 1650}%
\special{fp}%
% STR 2 0 3 0
% 3 4195 1295 4195 1395 5 0
% $a$
\put(41.9500,-13.9500){\makebox(0,0){$a$}}%
% STR 2 0 3 0
% 3 4195 1695 4195 1795 5 0
% $b$
\put(41.9500,-17.9500){\makebox(0,0){$b$}}%
% LINE 1 0 3 0
% 2 2755 1200 2755 2000
% 
\special{pn 13}%
\special{pa 2756 1200}%
\special{pa 2756 2000}%
\special{fp}%
% STR 2 0 3 0
% 3 6405 1500 6405 1600 5 0
% $g$
\put(64.0500,-16.0000){\makebox(0,0){$g$}}%
% LINE 1 0 3 0
% 2 4645 1200 4645 2000
% 
\special{pn 13}%
\special{pa 4646 1200}%
\special{pa 4646 2000}%
\special{fp}%
% LINE 2 0 3 0
% 2 1200 1400 2000 1400
% 
\special{pn 8}%
\special{pa 1200 1400}%
\special{pa 2000 1400}%
\special{fp}%
% LINE 2 0 3 0
% 2 1200 1800 2000 1800
% 
\special{pn 8}%
\special{pa 1200 1800}%
\special{pa 2000 1800}%
\special{fp}%
% LINE 2 0 3 0
% 2 2200 1600 2300 1600
% 
\special{pn 8}%
\special{pa 2200 1600}%
\special{pa 2300 1600}%
\special{fp}%
% BOX 2 0 3 0
% 2 2300 1450 2600 1750
% 
\special{pn 8}%
\special{pa 2300 1450}%
\special{pa 2600 1450}%
\special{pa 2600 1750}%
\special{pa 2300 1750}%
\special{pa 2300 1450}%
\special{fp}%
% STR 2 0 3 0
% 3 2455 1500 2455 1600 5 0
% $\fb$
\put(24.5500,-16.0000){\makebox(0,0){$\fb$}}%
% LINE 2 0 3 0
% 2 2600 1600 3000 1600
% 
\special{pn 8}%
\special{pa 2600 1600}%
\special{pa 3000 1600}%
\special{fp}%
% CIRCLE 2 0 3 0
% 4 5195 1600 5195 1800 5195 1800 5195 1400
% 
\special{pn 8}%
\special{ar 5196 1600 200 200  4.7123890 6.2831853}%
\special{ar 5196 1600 200 200  0.0000000 1.5707963}%
% LINE 2 0 3 0
% 2 4400 1400 5200 1400
% 
\special{pn 8}%
\special{pa 4400 1400}%
\special{pa 5200 1400}%
\special{fp}%
% LINE 2 0 3 0
% 2 4400 1800 5200 1800
% 
\special{pn 8}%
\special{pa 4400 1800}%
\special{pa 5200 1800}%
\special{fp}%
% LINE 2 0 3 0
% 2 5400 1600 5500 1600
% 
\special{pn 8}%
\special{pa 5400 1600}%
\special{pa 5500 1600}%
\special{fp}%
% BOX 2 0 3 0
% 2 5500 1450 5800 1750
% 
\special{pn 8}%
\special{pa 5500 1450}%
\special{pa 5800 1450}%
\special{pa 5800 1750}%
\special{pa 5500 1750}%
\special{pa 5500 1450}%
\special{fp}%
% STR 2 0 3 0
% 3 5655 1500 5655 1600 5 0
% $\cF$
\put(56.5500,-16.0000){\makebox(0,0){$\cF$}}%
% LINE 2 0 3 0
% 2 5800 1600 6200 1600
% 
\special{pn 8}%
\special{pa 5800 1600}%
\special{pa 6200 1600}%
\special{fp}%
% LINE 2 0 3 0
% 2 5250 2200 5250 2400
% 
\special{pn 8}%
\special{pa 5250 2200}%
\special{pa 5250 2400}%
\special{fp}%
% LINE 2 0 3 0
% 2 5350 2200 5350 2400
% 
\special{pn 8}%
\special{pa 5350 2200}%
\special{pa 5350 2400}%
\special{fp}%
% LINE 2 0 3 0
% 2 2050 2200 2050 2400
% 
\special{pn 8}%
\special{pa 2050 2200}%
\special{pa 2050 2400}%
\special{fp}%
% LINE 2 0 3 0
% 2 2150 2200 2150 2400
% 
\special{pn 8}%
\special{pa 2150 2200}%
\special{pa 2150 2400}%
\special{fp}%
\end{picture}%
 \end{center}
 \caption[subsection]{$\langle \fb\cdot m(a\otimes b) | g\rangle
  = \langle m(\fb(a)\otimes \fb(b)) | g\rangle$\ \ \ eq.(\ref{Ff})}
 \label{fig:coalgex}
\end{figure}

By the benefit of the above rewriting, the following facts can be 
understood naturally 
with these graphs : the equivalence between that $\m$ is a coderivative and 
that the $\delta$ is a derivative (\ref{QD}), 
the one between that the $\cF$ is a cohomomorphism and 
that the $\fb$ is a homomorphism (\ref{Ff}), the one between 
that the $(\cH, \m)$ is an $A_\infty$-algebra and that 
the $(C(\cH)^*, \delta)$ is a complex(\ref{QQDD}), and the one between 
the $\cF$ is an $A_\infty$-morphism and that the $\fb$ is 
$\delta$-equivariant (\ref{QFDf}). 
For instance, eq.(\ref{Ff}) is shown as (Fig.\ref{fig:coalgex}).  

 \subsection{The geometry on $C(\cH)^*$ : formal noncommutative 
supermanifold}
\label{ssec:A2}

In this subsection, 
we represent explicitly $m$, $\delta$ and $\fb$, which correspond to 
$\triangle$, $\m$ and $\cF$, respectively, and realize them geometrically 
on the algebra $C(\cH)^*$ dual to the $C(\cH)$. 
For the coassociative coproduct
\begin{equation*}
 \triangle(\eb_1\cdots \eb_n)=\sum_{k=1}^{n-1}
  (\eb_1\cdots \eb_k)
  \otimes (\eb_{k+1}\cdots\eb_n)\ , 
\end{equation*} 
the corresponding associative product $m$ 
defined in eq.(\ref{defm}) are written as 
\begin{equation}
 m((\eb^{i_k}\cdots\eb^{i_1})\otimes(\eb^{j_l}\cdots\eb^{j_1}) )
 =\eb^{j_l}\cdots\eb^{j_1}\eb^{i_k}\cdots\eb^{i_1}\ .
 \label{Apro}
\end{equation}
For $a=\sum_{k=1}^\infty a_{i_1\cdots i_k}\eb^{i_k}\cdots\eb^{i_1}$ and 
$b=\sum_{l=1}^\infty b_{j_1\cdots j_l}\eb^{j_l}\cdots\eb^{j_1}$, 
$m(a\otimes b)$ becomes 
\begin{equation*} 
 \begin{split}
 &m((\sum_{k=1}^\infty a_{i_1\cdots i_k}\eb^{i_k}\cdots\eb^{i_1})\otimes
  (\sum_{l=1}^\infty b_{j_1\cdots j_l}\eb^{j_l}\cdots\eb^{j_1}) )
 =\sum_{n}(a\cdot b)_{m_1\cdots m_n}\eb^{m_n}\cdots\eb^{m_1}\\
 &(a\cdot b)_{m_1\cdots m_n}=\sum_{p=1}^{n-1}
 \epsilon_{m_1\cdots m_n}^{i_1\cdots i_pj_1\cdots j_{n-p}}
 a_{i_1\cdots i_p}b_{j_1\cdots j_{n-p}}\ .
 \end{split}
\end{equation*}
It is easily seen that by the above definition of $m$, 
$(a\cdot b)_{m_1\cdots m_n}=
\langle m(a\otimes b)| \eb_{m_1}\cdots\eb_{m_n}\rangle
=\langle a\otimes b| \triangle(\eb_{m_1}\cdots\eb_{m_n})\ra$ holds. 

$a, b\in C(\cH)^*$ can be regarded as the polynomial functions 
on the graded vector space $\cH$. 
Consider $\Phi=\eb_i\phi^i\in\cH$. $\phi^i$ is a coordinate of $\cH$, 
and its degree is set to be minus the degree of $\eb_i$ in order for $\Phi$ 
to have its degree zero. $\{\phi^i\}$ is isomorphic to $\cH^*$, and so 
it is identified with $\cH^*$. Actually, introduce a natural pairing $(\ )$ 
between $\cH$ and $\cH^*$
\footnote{The inner product is the same type that is defined 
in eq.(\ref{ip1}) and different from that in eq.(\ref{ip2}).}, 
and one can define $a(\Phi)$ as 
\begin{equation*}
 a(\Phi):=\sum_{k=1}^\infty a_{i_1\cdots i_k}
(\eb^{i_k}(\Phi))\cdots(\eb^{i_1}(\Phi))
=\sum_{k=1}^\infty a_{i_1\cdots i_k}\phi^{i_k}\cdots\phi^{i_1}\ .
\end{equation*}
This $a(\Phi)\in C(\cH)^*$ is nothing but a polynomial function on the
graded vector space $\cH$ because $\{\phi^i\}$ is the coordinate on $\cH$. 
The pair of the graded vector space and the algebra of formal power series 
of the coordinates on the graded vector space is called as 
{\it formal supermanifold}\cite{AKSZ,K}. 
In this case $a(\Phi=0)=0$ for any $a\in C(\cH)^*$, 
and the coordinates are associative but noncommutative, so 
this is a formal noncommutative pointed supermanifold. 
We can translate $L_\infty$-algebras in the language of the 
formal supermanifold, too. In this situation the coordinates are graded 
commutative which reflects the cocommutativity of the $L_\infty$-algebra, 
and we get a formal (commutative) pointed supermanifold. 

\begin{rem}
 Note that the product (\ref{Apro}) is defined so as to satisfy the 
following compatibility 
\begin{equation}
 \begin{array}{ccc}
 \tri(\Phi^{\otimes n})&=&\sum_{k=1}^n(\Phi^{\otimes k})\otimes 
 (\Phi^{\otimes n-k})\\
  \rotatebox[origin=c]{90}{=} & & \rotatebox[origin=c]{90}{=}\\
  \tri(\eb_{i_1}\cdots\eb_{i_n})\phi^{i_n}\cdots\phi^{i_1}
   & & \sum_{k=1}^{n-1}(\eb_{i_1}\cdots\eb_{i_k})\phi^{i_k}\cdots\phi^{i_1}
 \otimes (\eb_{i_{k+1}}\cdots\eb_{i_n})\phi^{i_n}\cdots\phi^{i_{k+1}}\ .
 \end{array}
 \label{comp}
\end{equation}
Recall that $\tri(\eb_{i_1}\cdots\eb_{i_n})=\sum_{k=1}^n
(\eb_{i_1}\cdots\eb_{i_k})\otimes(\eb_{i_{k+1}}\cdots\eb_{i_n})$. 
Here we identify $\phi^{i}$ and $\eb^i$. 
Then the equality between the two terms 
in the second line in eq.(\ref{comp}) 
means that the operation of $\phi^i$' in the tensor coalgebra $C(\cH)$ 
are defined by eq.(\ref{Apro}).
Thus the coalgebra $C(\cH)$ can be defined as 
a coalgebra of $C(\cH)^*$-module. 
\end{rem}

\vspace*{0.2cm}

 $\bullet$\ \ {\it coderivation}

\vspace*{0.2cm}

Next, for a coderivation $\m=\m_1+\m_2+\cdots$,  
\begin{equation}
 \m_k(\eb_1\cdots\eb_n)=\sum_{p=1}^{n-k}
 (-1)^{\eb_1+\cdots+\eb_p}\eb_1\cdots\eb_{p-1}
 m_k(\eb_p\cdots\eb_{p+k-1})
 \eb_{p+k}\cdots\eb_n \ ,\quad \eb_i\in\cH\ ,
 \label{Am}
\end{equation}
we construct $\delta$ which corresponds to $\m$. 
By the definition of $\delta$ (\ref{defD}), 
one sees that a derivation corresponding to the coderivative 
may be constructed separately for $k$. 
Express $m_k : \cH^k\raw \cH$ as 
\begin{equation}
 m_k(\eb_{i_1}\cdots\eb_{i_k})=\eb_jc_{i_1\cdots i_k}^j
 \label{lc}
\end{equation}
and $\delta_k : C(\cH)^* \raw C(\cH)^*$, 
\[
 \delta_k=\flpartial{\phi^j}c^j_{i_1\cdots i_k}\phi^{i_k}\cdots\phi^{i_1}
\]
is a derivation. Here we identify the coordinate $\{\phi^i\}$ with 
$\cH^*$ and replace $\eb^i$ to $\phi^i$. The derivation $\delta$ is 
constructed as $\delta=\delta_1+\delta_2+\cdots$. It is regarded as an 
(odd) {\it formal vector field} on the formal noncommutative pointed 
supermanifold. 
Note that the condition that $\m_k$ is a coderivation is replaced to that 
$\delta_k$ satisfies the Leibniz rule on the polynomials of $\phi^i$'s. 
Moreover, as will be seen explicitly, $\delta^2=0$ holds iff $\m$ define 
an $A_\infty$-algebra. The formal manifold with such $\delta$ is called 
{\it $Q$-manifold} in \cite{AKSZ}
\footnote{This $Q$ does not correspond to the BRST operator $Q$ in the 
body of this paper but $\delta$ : the BRST-generator in gauge theory. 
$\delta$ in this paper is written as $Q$ in \cite{AKSZ}. }. 
\begin{rem}
The operation of $\m_k$ is compatible with the decomposition 
of the supercoordinates in the following sense. 
Here compute $\m_k(\Phi^{\otimes n})$ in two ways. One way is 
acting $\m_k$ after rewriting 
$\Phi^{\otimes n}=(\eb_{i_1}\cdots\eb_{i_n})\phi^{i_n}\cdots\phi^{i_1}$ 
and we get the result in eq.(\ref{Am}) as the coefficient of 
$\phi^{i_n}\cdots\phi^{i_1}$. 
Another way is computing $\m_k(\Phi^{\otimes n})$ as 
\begin{equation}
 \begin{split}
 \m_k(\Phi^{\otimes n})&=\sum_{p=1}^{n-k}\Phi^{\otimes p}m_k(\Phi) 
 \Phi^{\otimes n-k-p}\\
   &=\sum_{p=1}^{n-k}(-1)^{\eb_{i_1}+\cdots+\eb_{i_{p_1}}}
(\eb_{i_1}\cdots\eb_{i_{p-1}})
 m_k(\eb_{i_p}\cdots\eb_{i_{p+k-1}})
 (\eb_{i_{p+k}}\cdots\eb_{i_n})\phi^{i_n}\cdots\phi^{i_1}\ ,
 \end{split}
 \label{comp2}
\end{equation}
and picking up the coefficient of $\phi^{i_n}\cdots\phi^{i_1}$. 
One can see that this leads 
the same results as in (\ref{Am}) and these arguments are compatible. 
In the second equality of eq.(\ref{comp2}), one gets 
the sign $(-1)^{\eb_{i_1}+\cdots+\eb_{i_{p_1}}}$ because 
$\phi^{i_1},\cdots,\phi^{i_{p-1}}$ pass through $m_k$ which has 
degree one. 
\end{rem}
\begin{rem}
When $\m$ satisfies $\m\cdot \m=0$, we have relations between $\m_k$, 
Rewriting $m_k$ using eq.(\ref{lc}) yields 
relations between $c^j_{i_1\cdots i_k}$. 
On the other hand, in the dual language, 
the condition $\m\cdot \m=0$ is $\delta\cdot\delta=0$. 
Calculating $\delta\cdot\delta$ and concentrating on the term of 
$n$ powers of $\phi^i$ leads 
\begin{equation*}
 \begin{split}
 &\sum_{k+l=n+1}\delta_k\cdot \delta_l=
 \l(\flpartial{\phi^i}c^i_{i_1\cdots i_k}\phi^{i_k}\cdots\phi^{i_1}\r)
 \flpartial{\phi^{j}}c^{j}_{j_1\cdots j_l}\phi^{j_l}\cdots\phi^{j_1}\\
 &\qquad=\flpartial{\phi^i}
 \sum_{k+l=n+1}\sum_{m=1}^k(-1)^{\eb_{i_1}+\cdots+\eb_{i_{m-1}}}
 c^i_{i_1\cdots i_k}c^{i_m}_{j_1\cdots j_l}
 \phi^{i_k}\cdots\phi^{i_{m+1}}\l(
 \phi^{j_l}\cdots\phi^{j_1}\r)\phi^{i_{m-1}}\cdots\phi^{i_1}\ .
 \end{split}
\end{equation*}
The coefficient of $\phi^n\cdots\phi^1$ then reads 
\begin{equation}
0= \sum_{\substack{k+l=n+1\\m=0,\cdots,k-1}}
 {(-1)^{\eb_1+\cdots+\eb_m}}
 c^i_{1\cdots m,i_m,m+l+1\cdots n}c^{i_m}_{m+1\cdots m+l}\ .
\end{equation}
This is exactly the relation $\m\cdot\m=0$ (or eq.(\ref{ainf})) rewritten 
with $\{c^i_{i_1\cdots i_k}\}$. 
 \label{rem:Qmanifold}
\end{rem} 

\vspace*{0.2cm}

$\bullet$\ \ {\it cohomomorphism}

\vspace*{0.2cm}

In the terminology of the formal supermanifold, a homomorphism corresponding 
to a cohomomorphism $\cF$ are constructed as follows. 
Let $\cH, \cH'$ be two graded vector space and 
$\cF_n\ :\cH^{\otimes n}\lgraw \cH'$.  
A cohomomorphism $\cF$ from $C(\cH)$ to $C(\cH')$ is now given by 
\begin{equation}
 \begin{split}
 \cF&=\cF^1+\cF^2+\cF^3+\cdots\ ,\qquad
  \cF^l\ :\ C(\cH)\lgraw {\cH'}^{\otimes l}\\
 \cF^l&(\eb_1\cdots \eb_n)=
  \sum_{\substack{n_1,\cdots,n_l\ge 1\\ n_1+\cdots+n_l=n}}
 f_{n_1}(\eb_1\cdots\eb_{n_1})
 \otimes\cdots\otimes f_{n_l}(\eb_{n-n_l+1}\cdots\eb_n)\ .
 \end{split} \label{A:cohom} 
\end{equation}
Now we express $f_n$ as 
\[
 f_n(\eb_{i_1}\cdots\eb_{i_n})=\eb_{j'}f^{j'}_{i_1\cdots i_n}\ .
\]
The homomorphism $\fb$ gives the pullback from $C(\cH')^*$, 
the formal power series ring on $\cH'$, $C(\cH)^*$. 
In this reason we can write as $\fb=\cF^*$. 
Let $\{\phi^i\}$ and $\{\phi^{i'}\}$ be the coordinates 
on $\cH$ and $\cH'$, respectively, and take an element of $C(\cH')^*$ 
:$a(\phi'):=\sum_{k=1}^\infty 
a_{i_1\cdots i_k}\phi^{i'_k}\cdots\phi^{i'_1}$. Then 
$\fb : C(\cH')^*\raw C(\cH)^*$ is induced from $\cF_*$ : 
\begin{equation}
 \begin{array}{cccc}
  \cF_* :&\cH&\raw & \cH' \\
     & \phi&\mapsto& \phi'=\cF_*(\phi)
 \end{array}\ ,\qquad 
 \phi^{j'}=\cF_*^{j'}(\phi)
 =f^{j'}_i\phi^i+f^{j'}_{i_1i_2}\phi^{i_2}\phi^{i_1}
 +\cdots+f^{j'}_{i_1\cdots i_n}\phi^{i_n}\cdots\phi^{i_1}+\cdots
\end{equation}
as $\fb(a(\phi'))=a(\cF_*(\phi))$. 
One can see that the cohomomorphism $\cF$ is, in the dual geometric 
picture, a nonlinear map $\cF_*$ from a formal supermanifold $\cH$ to 
$\cH'$ preserving the origin.

\vspace*{0.2cm}

$\bullet$\ \ {\it $A_\infty$-morphism}

\vspace*{0.2cm}

The condition that this $\cF$ is an $A_\infty$-morphism 
is equivalent to the statement 
that this map $\cF_*$ between two formal supermanifolds 
is compatible with the action of $\delta$ and $\delta'$ on both sides, \ie 
$\cF_*$ is a morphism between $Q$-manifolds. 
For any $a(\phi')\in C(\cH')^*$, the condition is 
\begin{equation}
 \fb\delta'(a(\phi'))=\delta\fb a(\phi')\ ,
\end{equation}
and is written explicitly as 
\begin{equation*}
 \fb\l( a(\phi')\flpartial{\phi^{j'}}c^{j'}(\phi') \r)=
 a(\cF_*(\phi))
 \flpartial{\phi^j}c^j(\phi)
\end{equation*}
where we expressed $\delta=\flpartial{\phi^j}c^j(\phi)$.  
Because $a(\cF_*(\phi))\flpartial{\phi^j}c^j(\phi)=
\fb\l(a(\phi')\flpartial{\phi^{j'}}\r)\flpart{\phi^{j'}}{\phi^j}c^j(\phi)$ 
in the right hand side, we get 
\begin{equation}
 \fb\l(c^{j'}(\phi')\r)=\flpart{\phi^{j'}}{\phi^j}c^j(\phi)\ .
 \label{cc'}
\end{equation}
We can see that when $\delta$ and $\fb$ are given and $\fb$ has its inverse, 
then $\delta'$ is induced as 
$c^{j'}(\phi')=\fb^{-1}\l(\flpart{\phi^{j'}}{\phi^j}c^j(\phi)\r)$. 

\vspace*{0.2cm}

$\bullet$\ \ {\it weak $A_\infty$}

\vspace*{0.2cm}

Let us add the term $\flpartial{\phi^j}c^j$ to $\delta$ where $c^j$ is a 
constant and write it as 
$\delta_w$. 
Explicitly $\delta_w$ is of the form 
\begin{equation*}
 \delta_w=\flpartial{\phi^j}c^j+\flpartial{\phi^j}c^j(\phi)=
\sum_{k=0}^\infty
\flpartial{\phi^j}c^j_{i_1\cdots i_k}\phi^{i_k}\cdots\phi^{i_1}\ .
\end{equation*}
Acting it on $C(\cH)^*$ yields constant term generally. 
Thus let us enlarge $C(\cH)^*$ as $C(\cH)^*_w$, which denotes the space 
of $C(\cH)^*$ plus constant, \ie $C(\cH)^*_w=\C\oplus C(\cH)^*$. 
$C(\cH)^*_w$ is regarded as the space of 
functions on $\cH$ which do not vanish at the origin generally. 
Similarly consider the map $\fb_w :C(\cH')^*_w\raw C(\cH)^*_w$ induced from 
$\cF_{w,*}$ defined as 
\begin{equation*}
  \phi^{j'}=\cF_{w,*}^{j'}(\phi)
 =f^{j'}+f^{j'}_i\phi^i+f^{j'}_{i_1i_2}\phi^{i_2}\phi^{i_1}
 +\cdots+f^{j'}_{i_1\cdots i_n}\phi^{i_n}\cdots\phi^{i_1}+\cdots\ 
\end{equation*}
where $f^{j'}\in\C$. 
It can be seen that this map does not preserve the origin. 
In this extended situation, we can again consider the following conditions 
\begin{equation}
 \delta_w\cdot\delta_w=0
 \label{wAinfty}
\end{equation}
\begin{equation}
 \fb_w\delta'_w=\delta_w\fb_w\ .
 \label{wAinfmorp}
\end{equation}
The condition (\ref{wAinfty}) and (\ref{wAinfmorp}) are 
dual version of the definition of 
weak $A_\infty$-algebras and weak $A_\infty$-morphisms, respectively. 

Let us consider a weak $A_\infty$-isomorphism of the form 
\begin{equation*}
  \phi^{j'}=\cF_{w,*}^{j'}(\phi)
 =f^{j'}+f^{j'}_i\phi^i\ ,
\end{equation*}
where $f^{j'}\in\C$ and $f^{j'}_i$ has its inverse. 
For simplicity let $f^{j'}_i=\delta^{j'}_i$. 
The weak $A_\infty$ version of eq.(\ref{cc'}) then becomes 
\begin{equation*}
 c_w^j(\phi^i)=c_w'(f^i+\phi^i)\ .
\end{equation*}
The weak $A_\infty$-structure $\delta_w$ on $C(\cH)^*$ 
is naturally induced from $\delta'_w$ on $C(\cH')^*$ by $\fb_w$. 
Note that when $\delta'_w$ defines strictly an $A_\infty$-structure, that 
is, the constant part $c_w'$ vanishes, 
and $f^i$ is the solution of the Maurer-Cartan equation on 
$C(\cH')^*$, then $c_w^j\in\C$ 
vanishes and the induced $\delta_w$ is also an 
$A_\infty$-algebra. 
Express the coalgebra representation corresponding to $\delta_w$ and 
$\delta'_w$ as $\m_w$ and $\m'_w$, respectively, and $\m_w$ is given by 
\begin{equation}
 \m_w(e^\Phi)=\m_w'(e^{\Phi_{bg}+\Phi})
 \label{wainf}
\end{equation}
where $\Phi_{bg}=\eb_if^i$. 
Note that since $f^{j'}\in\C$, the corresponding $\eb_i$ has degree zero.

%%%%%%%%%%%%%%  Appendix B   %%%%%%%%%%%%%%

 \section{Some relations on vertices in SFT}
\label{sec:B}

In this section some properties of vertices in SFT are derived 
using identities and notations in the body of this paper.

 \subsection{The recursion relation}
\label{ssec:B1}

In section \ref{sec:3} it was explained that in order for the Feynman rule 
to reproduce the single covered moduli spaces 
of Riemann surfaces, the vertices in SFT must satisfy the string 
factorization equations (\ref{sfeq}),  
\begin{equation}
 0=\partial(\V_n)+\sum_{\substack{k_1+k_2=n+2\\k_1,k_2\ge 3}}
 \half(\V_{k_1})\partial(\-)(\V_{k_2})\ . 
 \label{sfeq2}
\end{equation}
Here let us define the (off-shell) $n$ point tree amplitude using 
that defined in eq.(\ref{Vp}) as 
\begin{equation}
 \ov{n}\int_{\cM_n}\la\Omega|(|\ti\Phi\ra)^n:=
 \ov{n}\ti\V_n(\ti\Phi)=\ov{n}\la\ti{V}_n|(|\Phi\ra)^n \ .
 \label{bamp}
\end{equation}
In this subsection we shall derive 
the recursion relation (\ref{sfeq2}) or 
equivalently the classical master equation by 
employing only the following two relations : 
\begin{equation}
 0=\int_{\partial\cM_n}\la\Omega|(|\ti\Phi\ra)^n
 =\partial(\la\ti{V}_n|)(\ti\Phi)^n\ , \label{bfirst}
\end{equation}
where $\partial(\la\ti{V}_n|_{1\cdots n})
=\la\ti{V}_n|_{1\cdots k\cdots n}\sum_{k=1}^n Q^{(k)}$ and 
\begin{equation}
 \{Q,Q^+\}=\1-P\sim \1\ . \label{bid}
\end{equation}
The first equality in eq.(\ref{bfirst}) follows from $\partial\cM_n=0$. 
As argued in subsection \ref{ssec:5gf}, the $P$ in eq.(\ref{bid}) 
just contribute to the poles in eq.(\ref{bamp}). Thus when the external 
states $\ti\Phi$ are set so that the propagators in $\la\ti{V}_n|$ have 
the pole, eq.(\ref{bfirst}) itself is not 
well defined. Therefore including this case we define $\{Q,Q^+\}=\1$ 
between vertices $\{\la V_k|\}$. 
Originally the recursion relations (\ref{sfeq2}) are defined as the 
relations between (the subspaces of) the moduli spaces. 
The problem here arises from attaching the value 
$\la\ti{V}_n|\in (\cH^*)^{\otimes n}$ 
to each the moduli space $\cM_n$. 
 
Expanding eq.(\ref{bfirst}) with respect to the number of the propagators 
$Q^+$ leads 
\begin{equation}
 \partial(\la V_n|)+\sum_{\substack{k_1+k_2=n+2\\k_1,k_2\ge 3}}
 \half\partial(\la V_{k_1}|\la V_{k_2}|\ -Q^+|\omega\ra)
 |Q\Phi\ra(|\Phi\ra)^{n-1}+\cdots\ . \label{bfirst2}
\end{equation}
The origin of the minus in front of $Q^+$ can be found 
by recalling the calculation using Feynman diagram in subsection 
\ref{ssec:53}. 
Each of the second term is rewritten as 
\begin{equation*}
 \begin{split}
 \partial(\la V_{k_1}|\la V_{k_2}|\ -Q^+|\omega\ra) 
 &=(\partial\la V_{k_1}|)\la V_{k_2}|\ -Q^+|\omega\ra\\
 &+\la V_{k_1}|(\partial\la V_{k_2}|)\ -Q^+|\omega\ra\\
 &-\la V_{k_1}|\la V_{k_2}|\ \{Q,-Q^+\}|\omega\ra
 \end{split}
\end{equation*}
and the identity (\ref{bid}) leads the third term is 
$\la V_{k_1}|\la V_{k_2}|\ |\omega\ra$.
As was explained in section \ref{sec:3}, eq.(\ref{bfirst2}) is equivalent 
to 
\begin{equation*}
 \partial(\la V_n|)+\sum_{\substack{k_1+k_2=n+2\\k_1,k_2\ge 3}}
 \half\la V_{k_1}|\la V_{k_2}|\ |\omega\ra=0 \ .
\end{equation*}
Thus the recursion relation (\ref{sfeq}) is derived. 
This identity can be rewritten as 
\begin{equation*}
 (S_0,\V_n)+\sum_{\substack{k_1+k_2=n+2\\k_1,k_2\ge 3}}\half(\V_k,\V_l)=0
\end{equation*}
and summing up this equation for $n\ge 3$ leads 
the classical BV-master equation (\ref{meq}).

 \subsection{On-shell reduction of the vertices I}
\label{ssec:B2}

Let us consider the $\cM_n^0\raw \cM_n$ limit for each vertex as was 
mentioned in subsection \ref{ssec:53} and restrict their external states 
on-shell. This should give the on-shell string correlation functions. 
In \cite{Z1} such arguments are given in order to derive 
the $L_\infty$-structure for string world sheet theory found in 
\cite{WZ,V}. 
Here it is reviewed in open string case and derive the on-shell 
$A_\infty$-structure from the $A_\infty$-structure of string field vertices. 

In order for $\la V_n|$ to cover the whole moduli $\cM_n$ 
of $\dim\cM_n=n-3$, 
$$\sum_{\substack{k_1+k_2=n+2\\k_1,k_2\ge 3}}
\half\la V_{k_1}|\la V_{k_2}|\ -Q^+|\omega\ra$$ 
must cover the subspace of $\cM_n$ whose dimension is less than $n-3$. 
Because the sum of the dimensions of the moduli spaces corresponding to 
$\la V_{k_1}|$ and $\la V_{k_2}|$ is 
$(k_1-3)+(k_2-3)=n-4$, the propagator $Q^+$ must not 
create one more dimension. 
Consequently, each vertex has infinite length strips for their 
external states. 
$A_\infty$-structure $\m^{\zeta\raw\infty}$ in section \ref{sec:6} is 
just the case and the vertex 
$\V_n=\omega(\Phi, m_{n-1}^\infty(\Phi,\cdots,\Phi))$ 
is the string correlation function when the external states are 
strictly restricted on-shell (or on physical states). 

Now $\V_n$ is of the form $\V_n=\int_{\cM_n}\la\Omega|(|\Phi\ra)^n$ 
and $\partial\cM_n=0$, the recursion relation (\ref{sfeq2}) is 
saturated separately as 
\begin{equation}
 \partial(\la V_n|)=0\ ,\qquad \sum_{\substack{k_1+k_2=n+2\\k_1,k_2\ge 3}}
\half\la V_{k_1}|\la V_{k_2}||\omega\ra=0\ .
 \label{sfeqd}
\end{equation}
The second identity is just the condition of $A_\infty$-structures with 
$Q=0$. By employing the first identity, it is reduced to on-shell 
$A_\infty$-structure as follows. 

{}From eq.(\ref{reflexp}) and (\ref{opmv}) the relation between 
$A_\infty$-structure $m_{n-1}$ and $\V_n$ can be read as 
\begin{equation}
 m_{n-1}(\eb_{i_2},\cdots,\eb_{i_n})=
(-1)^{\eb_k}\eb_j\omega^{jk}
\V_n(\eb_k,\eb_{i_2},\cdots,\eb_{i_n})\ .
 \label{mvonshell}
\end{equation}
Here we restrict the external states as 
$\eb_{i_2},\cdots,\eb_{i_n}\in\cH^p\cup\cH^t$. Choosing the orthogonal 
basis as in eq.(\ref{omegadecomp}), $m_{n-1}$ is then decomposed as 
\begin{equation}
 \begin{split}
 m_{n-1}(\eb_{i_2},\cdots,\eb_{i_{n}})
 &=(-1)^{\eb_k}P^p\eb_j\omega^{jk}\V_n(P^p\eb_k,
\eb_{i_2},\cdots,\eb_{i_{n}})\\
 &+(-1)^{\eb_k}P^t\eb_j\omega^{jk}\V_n(P^u\eb_k,
\eb_{i_2},\cdots,\eb_{i_{n}})\\
 &+(-1)^{\eb_k}P^u\eb_j\omega^{jk}\V_n(P^t\eb_k,
\eb_{i_2},\cdots,\eb_{i_{n}})\ .
 \end{split} \label{mdecomp}
\end{equation}
The term on the third line vanishes due to the first identity 
in eq.(\ref{sfeqd}). Because $P^t\eb_k$ is $Q$-exact, write this as 
$P^t\eb_k=Q(Q^u\eb_k)$, and 
\begin{equation}
\V_n(Q(Q^u\eb_k),\eb_{i_2},\cdots,\eb_{i_{n}})=0
 \label{vanish}
\end{equation} 
follows from $(\partial\V_n)(Q^u\eb_k,\eb_{i_2},\cdots,\eb_{i_{n}})=0$. 
The fact that eq.(\ref{vanish}) is hold is an expected result since 
the string correlation function vanishes even if one $Q$-trivial 
external state is included. 

Thus it is shown that 
\begin{equation}
 m_{n-1}(\eb_{i_2},\cdots,\eb_{i_{n}}) \in\cH^p\cup\cH^t
 \label{onshell}
\end{equation}
for any $\eb_{i_2},\cdots,\eb_{i_n}\in\cH^p\cup\cH^t$ 
and the $A_\infty$-structure $m_{n-1}$ can be reduced on-shell 
$\cH^p\cup\cH^t$. 

Furthermore one can see that even if one of the external states $\eb_{i_k}$ 
for $2\le k\le n$ belongs to $\cH^t$, 
$\V_n(P^p\eb_k,\eb_{i_2},\cdots,\eb_{i_{n}})$ vanishes in the same 
reason as above and only the term on the first line in 
eq.(\ref{mdecomp}) survives. 

On the other hand, the $A_\infty$-condition corresponding to the second 
identity in eq.(\ref{sfeqd}) is eq.(\ref{ainf}) with $Q=0$ : 
\begin{equation}
\sum_{\substack{k+l=n+1,\ k,l\ge 2\\j=0,\cdots,k-1}}
{(-1)^{\eb_1+\cdots+\eb_j}
 m_k(\eb_1,\cdots,\eb_j,m_l(\eb_{j+1},\cdots,\eb_{j+l}),
 \eb_{j+l+1},\cdots,\eb_n)}=0
 \label{ainfred}
\end{equation}
with $k=k_1-1$ and $l=k_2-1$. 
Acting $P^p$ on left and restricting the external states 
$\eb_1,\cdots,\eb_n$ on physical states $\cH^p$ then leads 
\begin{equation}
 \sum_{\substack{k+l=n+1,\ k,l\ge 2\\j=0,\cdots,k-1}}
{(-1)^{\eb^p_1+\cdots+\eb^p_j}
 P^p m_k(\eb^p_1,\cdots,\eb^p_j,
P^p m_l(\eb^p_{j+1},\cdots,\eb^p_{j+l}),
 \eb^p_{j+l+1},\cdots,\eb^p_n)}=0
 \label{ainfred2}
\end{equation}
The reason why 
$m_l$ can be replaced by $P^pm_l$ is that 
the contribution of $P^tm_l(\eb^p_{j+1},\cdots,\eb^p_{j+l})$ to 
$m_k(\ ,\cdots,\ )$ necessarily belongs to $\cH^t$ as was stated above 
and is projected out by $P^p$ acting in front of $m_k$. 

This concretes the proof that the $A_\infty$-structure of string vertices 
can be reduced to the $A_\infty$-structure $\{P^pm_k\}_{k\ge 2}$ 
in string world sheet theory.

 \subsection{On-shell reduction of the vertices II}
\label{ssec:B3}

In this subsection it will be shown that  
$\ti\m^p$ defined in (Def.\ref{defn:minimal}) indeed define an 
$A_\infty$-structure on-shell and it can be reduced to an 
$A_\infty$-structure on physical state space $\cH^p$, 
which was postponed in (Lem.\ref{lem:main}). 

The $\ti\m^p$ defines the on-shell S-matrix elements and the 
corresponding vertices $\ti\V^p_n$ satisfies $\partial(\la\ti V^p_n|)=0$ 
because of $\partial\cM_n=0$ similarly 
as the first identity in eq.(\ref{sfeqd}). 
Thus the proof is almost the same as that in the above subsection. 
Replacing $m_{n-1}$ in the previous subsection by $\ti{m}^p_{n-1}$ in 
subsection \ref{ssec:53} and repeating the argument 
from eq.(\ref{mvonshell}) to eq.(\ref{ainfred2}) give us the proof. 
Only one different point is that $\ti\m^p_n$ in (Def.\ref{defn:minimal}) 
contains $P$ and the identity corresponding to eq.(\ref{mvonshell}) is 
\begin{equation*}
 \ti{m}^p_{n-1}(\eb^p_{i_2},\cdots,\eb^p_{i_{n}})
 =(-1)^{\eb_k}P\eb_j\omega^{jk}
 \ti\V_n(\eb_k,\eb^p_{i_2},\cdots,\eb^p_{i_{n}})
\end{equation*} 
where the external states $\eb^p_{i_2},\cdots,\eb^p_{i_{n}}$ are 
now restricted on physical state space $\cH^p$. 
In the orthogonal basis, this decompose as 
\begin{equation*}
 \begin{split}
 \ti{m}^p_{n-1}(\eb^p_{i_2},\cdots,\eb^p_{i_{n}})
 &=(-1)^{\eb_k}PP^p\eb_j\omega^{jk}\ti\V_n(P^p\eb_k,
\eb^p_{i_2},\cdots,\eb^p_{i_{n}})\\
 &+(-1)^{\eb_k}PP^t\eb_j\omega^{jk}\ti\V_n(P^u\eb_k,
\eb^p_{i_2},\cdots,\eb^p_{i_{n}})\\
 &+(-1)^{\eb_k}PP^u\eb_j\omega^{jk}\ti\V_n(P^t\eb_k,
\eb^p_{i_2},\cdots,\eb^p_{i_{n}})
 \end{split}
\end{equation*}
and the term in the third line of the above equation vanishes 
for the same reason as in eq.(\ref{mdecomp}). 
Here recall that $PP^p=P^p$ and note that the identity 
$QQ^++Q^+Q+P=\1$ leads $Q$ and $P$ commute to each another. 
The above equation is then rewritten as 
\begin{equation}
 \begin{split}
 \ti{m}^p_{n-1}(\eb^p_{i_2},\cdots,\eb^p_{i_{n}})
 &=(-1)^{\eb_k}P^p\eb_j\omega^{jk}\ti\V_n(P^p\eb_k,
\eb^p_{i_2},\cdots,\eb^p_{i_{n}})\\
 &+(-1)^{\eb_k}Q(PQ^u\eb_j)\omega^{jk}\ti\V_n(P^u\eb_k,
\eb^p_{i_2},\cdots,\eb^p_{i_{n}})
 \end{split} \label{mdecomp2}
\end{equation}
where in the second line $PP^t=PQQ^u=QPQ^u$ is used. 
Thus it has been shown that the image of $\ti\m^p$ indeed belongs to 
on-shell $\cH^p\cup\cH^t$ similarly as the previous subsection. 
It is easily seen that this result does not change when 
the elements in $\cH^t$ are included as the external states. 
Furthermore, because the term in the second line in eq.(\ref{mdecomp2}) 
belong to $\cH^t$, the above $\ti{m}^p_{n-1}$ can be reduced to the 
$A_\infty$-structure on $\cH^p$ similarly as in eq.(\ref{ainfred2}). 
Let $\iota : \cH^p\raw\cH^p\cup\cH^t$ be the inclusion map and 
the reduced $A_\infty$-structure is given as 
\begin{equation*}
 P^p\circ\ti\m^p\circ\iota\ ,
\end{equation*}
where $P^p$ and $\iota$ is extended naturally as $A_\infty$-morphisms. 
This is equal to $P^p\m$ derived in eq.(\ref{ainfred2}). 
The reduced $A_\infty$-algebra is denoted as $(\cH^p,\ti\m^p)$ again 
to avoid increasing notations. 

Thus we complete the proof that $\ti\m^p$ defines an 
$A_\infty$-structure on physical states $\cH^p$.

%%%%%%%%%%%%%%%%%%%%%%%%%%%%%%%%%%%%%%%%%%%%%%%

\end{document}